\documentclass[traditabstract]{aa}  
\usepackage{graphicx}
\usepackage{txfonts}
\usepackage{verbatim}
\usepackage{natbib}
\usepackage{enumerate}
\usepackage{lscape}
\usepackage{subfigure}
\usepackage{hyperref}
\newcommand{\ergs}{\mbox{$\mathrm{erg\,s^{-1}}$}}

\newcommand{\Ha}{\mbox{${\mathrm H\alpha}$}}
\newcommand{\Hb}{\mbox{${\mathrm H\beta}$}}
\newcommand{\HeI}{\mbox{${\mathrm {He\,I}}$}}
\newcommand{\HeII}{\mbox{${\mathrm {He\,II}}$}}
\newcommand{\ergcms}{\mbox{$\mathrm{erg\,cm^{-2}s^{-1}}$}}
\newcommand{\Nh}{\mbox{$\mathrm{N_H}$}}
\newcommand{\Msun}{\mbox{$\mathrm{M}_{\odot}$}}
\newcommand{\Lx}{L$_{\mathrm X}$}
\newcommand{\lnls}{log\,N($>$S)\,--\,log\,S}
\newcommand{\Pid}{\mbox{$\mathrm{P_{id}}$}}

\begin{document}
\title{The XMM-Newton SSC survey of the Galactic Plane\thanks{Based on observations obtained with XMM-Newton, an ESA science mission with instruments and contributions directly funded by ESA Member States and NASA. Based on observations carried out at the European Southern Observatory, La Silla and Paranal, Chile under program Nos 69.D-0143, 70.D-0227, 71.D-0296, 71.D-0552. Based on observations obtained at the Canada-France-Hawaii Telescope (CFHT) which is operated by the National Research Council of Canada, the Institut National des Sciences de l'Univers of the Centre National de la Recherche Scientifique of France, and the University of Hawaii. Based on observations obtained at the Observatoire de Haute Provence which is operated by the Centre National de la Recherche Scientifique of France. }}
\authorrunning{Nebot G\'omez-Mor\'an et al.}
\titlerunning{The XMM-Newton SSC survey of the Galactic Plane}
\author{
A.~Nebot G\'omez-Mor\'an\inst{1},
C.~Motch\inst{1},
X.~Barcons\inst{2},
F.~J.~Carrera\inst{2},
M.~T.~Ceballos\inst{2},
M.~Cropper\inst{3},
N.~Grosso\inst{1},
P.~Guillout\inst{1},
O.~H\'erent\inst{1},
S.~Mateos\inst{2},
L.~Michel\inst{1},
J.~P.~Osborne\inst{4},
M.~Pakull\inst{1},
F.-X.~Pineau\inst{1},
J.~P.~Pye \inst{4},
T.~P.~Roberts\inst{5},
S.~R.~Rosen\inst{4},
A.~D.~Schwope\inst{6},
M.~G.~Watson\inst{4},
N.~Webb\inst{7}
}
\institute{
Observatoire Astronomique de Strasbourg, Universit\'e de Strasbourg, CNRS, UMR 7550, 11 rue de l'Universit\'e, 67000 Strasbourg, France.\\
\email{ada.nebot@astro.unistra.fr}
\and
Instituto de Fisica de Cantabria (CSIC-UC), Avenida de los Castros, 39005 Santander, Spain.
\and
Mullard Space Science Laboratory, University College London, Holmbury St Mary, DorkingSurrey RH5 6NT.
\and
Space Research Centre, Department of Physics \& Astronomy, University of Leicester, Leicester LE1 7RH, UK.
\and
Department of Physics, Durham University, South Road, Durham, DH1 3LE, UK.
\and
Leibniz-Institut f\"ur Astrophysik Potsdam (AIP), An der Sternwarte 16, 14482, Potsdam, Germany.
\and
Institut de Recherche en Astrophysique and Plan\'etologie (IRAP), Universit\'e de Toulouse, UPS, 9 Avenue du colonel Roche, 31028 Toulouse Cedex 4, France.
}

\date{}

\abstract
{
Many different classes of X-ray sources contribute to the Galactic landscape at high energies. Although the nature of the most luminous X-ray emitters is now fairly well understood, the population of low-to-medium X-ray luminosity (\Lx\,=\,$10^{27-34}$\,\ergs) sources remains much less studied, our knowledge being mostly based on the observation of local members. The advent of wide field and high sensitivity X-ray telescopes such as XMM-Newton now offers the opportunity to observe this low-to-medium \Lx\ population at large distances. 
We report on the results of a Galactic plane survey conducted by the XMM-Newton Survey Science Centre (SSC). Beyond its astrophysical goals, this survey aims at gathering a representative sample of identified X-ray sources at low latitude that can be used later on to statistically identify the rest of the serendipitous sources discovered in the Milky Way. The survey is based on 26 XMM-Newton observations, obtained at $|b|\,<\,20$\,deg, distributed over a large range in Galactic longitudes and covering a summed area of 4\,deg$^2$. The flux limit of our survey is $2\times10^{-15}$\,\ergcms\ in the soft (0.5\,--\,2\,keV) band and $1\times10^{-14}$\,\ergcms\ in the hard (2\,--\,12\,keV) band. We detect a total of 1319 individual X-ray sources. Using optical follow-up observations supplemented by cross-correlation with a large range of multi-wavelength archival catalogues we identify 316 X-ray sources. This constitutes the largest group of spectroscopically identified low latitude X-ray sources at this flux level.
The majority of the identified X-ray sources are active coronae with spectral types in the range A\,--\,M at maximum distances of $\sim1$\,kpc. The number of identified active stars increases towards late spectral types, reaching a maximum at K. Using infrared colours we classify 18\% of the stars as giants. The observed distributions of F$_{\mathrm{X}}$/F$_\mathrm{V}$, X-ray and infrared colours indicates that our sample is dominated by a young (100 Myr) to intermediate (600 Myr) age population with a small contribution of close main sequence or evolved binaries. We find other interesting objects such as cataclysmic variables ($d\,\sim\,0.6\,-\,2$\,kpc), low luminosity high mass stars (likely belonging to the class of $\gamma$-Cas-like systems, $d\sim1.5\,-\,7$\,kpc), T~Tauri and Herbig-Ae stars. A handful of extragalactic sources located in the highest Galactic latitude fields could be optically identified. For the 20 fields observed with the EPIC pn camera, we have constructed \lnls\ curves in the soft and hard  bands. In the soft band, the majority of the sources are positively identified with active coronae and the fraction of stars increases by about one order of magnitude from $b\,=\,60^\circ$ to $b\,=\,0^\circ$ at an X-ray flux of $2\times10^{-14}$\,\ergcms. The hard band is dominated by extragalactic sources, but there is a small contribution from a hard Galactic population formed by CVs, HMXB candidates or $\gamma$-Cas-like systems and by some active coronae that are also detected in the soft band. At $b\,=\,0^\circ$ the surface density of hard sources brighter than $1\times10^{-13}$\,\ergcms\ steeply increases by one order of magnitude from $l\,=\,20^\circ$ to the Galactic centre region ($l\,=\,0.9^\circ$). 
}
\maketitle
\section{Introduction}
\label{sec:intro}
Non-solar X-ray emission was discovered in the early 1960s using collimating instruments. The low spatial resolution and high background inherent in these detectors only allowed the observations of bright, mostly Galactic X-ray sources. The launch of focusing X-ray telescopes in the '80s has paved the way for the study of fainter high energy sources and opened the high energy window to virtually all types of astrophysical objects from comets to the most remote AGNs. 

The first imaging Galactic X-ray survey was conducted by the Einstein satellite \citep{giacconietal79-1,hertz+grindlay84-1}. A decade later, the ROSAT \citep{truemper82-1} all-sky survey and pointed observations mapped the entire soft ($<$2\,keV) X-ray content of the Galaxy \citep[see e.\,g.][]{motchetal97-1,morleyetal01-1} and allowed the discovery of many new species of soft X-ray sources. The ASCA \citep{tanakaetal94-1} Galactic Plane Survey \citep{sugizakietal01-1} was the first to explore the hard ($>2$\,keV) X-ray content of the Galaxy at X-ray fluxes much lower than achievable with collimating instruments.   
  
The launch of the XMM-Newton \citep{jansenetal01-1} and Chandra \citep{weisskopfetal02-1} X-ray observatories has opened the possibility to carry out large surveys by analysing the properties of the X-ray sources serendipitously detected around the observation's main target. Although based on a field by field approach, these surveys benefit from the high quality of the parameters, position, spectral indices, etc.. derived for each source and a several-fold improved sensitivity, achievable thanks to the large collecting area of the telescopes. In particular, the much improved position accuracy of the detected X-ray sources facilitates their identification at other wavelengths, thus opening the way for detailed source characterisation. The ChaMPlane project, based on Chandra observations, has been described in \cite{grindlayetal05-1}. It will eventually cover about 8 deg$^2$ down to limiting fluxes a few times fainter than can be achieved with XMM-Newton. Recent results on the Galactic centre area have been reported in \citet{vandenberg2012,hongetal12-1,hong12-1}. XMM-Newton has conducted a shallow low latitude X-ray survey of relatively large area \citep{handsetal04-1}. Optical identifications and properties of its brightest sources carried out in the framework of the XMM-Newton Survey Science Centre \citep[SSC,][]{watsonetal01-1}, have been discussed in \cite{motchetal10-1}. Finally, the nature of the overall low latitude XMM-Newton source population has recently been investigated by \cite{warwick2011} and \cite{motch+pakull12-1} using cross-correlations with large optical and X-ray catalogues.  

Many different classes of unresolved sources contribute to the X-ray content of the Galaxy: early and late-type stars, interacting binaries such as cataclysmic variables (CVs), RS CVns, white dwarfs, neutron stars or black holes with low (LMXB) or high mass companion stars (HMXB), isolated neutron stars and possibly isolated black holes \citep[for a review, see][]{motch06-1}.

At low X-ray luminosities, between $10^{27}$ and $10^{31}$\,\ergs, and soft energies (kT\,$<\,2$\,keV) the X-ray sky is dominated by relatively nearby active stellar coronae \citep[see e.\,g.][and references therein]{motchetal97-1}. In contrast, at high X-ray luminosities, $>10^{35}$\,\ergs, the make-up of the Galaxy is dominated by HMXBs and LMXBs \citep{grimmetal02-1,gilfanov04-1}. However, at intermediate X-ray luminosities, the nature of the hard ($>2$\,keV) X-ray sources is still poorly understood. Although CVs and stellar coronae are expected to contribute significantly to this population \citep{sazonovetal06-1,hong12-1}, other interesting objects such as magnetic OB stars \citep{gagneetal11-1}, single and binary Wolf Rayet stars \citep{skinneretal10-1,kogure09-1}, $\gamma$-Cas like objects \citep{motchetal07-1,oliveiraetal10-1} and X-ray transient binaries in quiescent state have also been identified in this luminosity range. 

The scientific interest of optically identified Galactic X-ray surveys cannot be overemphasised. The low to medium X-ray luminosity point source population may only be resolved in our Galaxy and to some extent in the Magellanic Clouds due to the currently available combination of spatial resolution and sensitivity delivered by the most efficient X-ray observatories in operation such as Chandra and XMM-Newton. In addition, the optical identification process, which is a mandatory step when studying the detailed nature of these X-ray populations can only be performed on a large number of sources in our Galaxy due to the still relatively large X-ray error circles. Similar to other wavelength ranges, flux limited X-ray surveys allow us to gather large and homogeneous samples of different species of high energy sources such as stars or AGN \citep[see e.g.][]{guilloutetal99-1,barconsetal02-1}, while they have proved to be instrumental for discovering rare or elusive X-ray emitters such as isolated neutron stars \citep[see e.g.][for results based on XMM-Newton data]{pires2009}. 

So far, observations provide very few constraints on evolutionary theories of low and high mass X-ray binaries. For instance, we do not detect the long-lived wind accreting low X-ray luminosity stages preceding or following the bright phase during which they become conspicuous. The common envelope  spiral-in  creation channel for low-mass X-ray binaries predicts the existence of pre-LMXBs which could radiate as much as $\sim$ 10$^{32}$\,\ergs\ in hard X-rays through accretion of stellar wind onto the neutron star or its magnetosphere \citep[see e.g.][]{tauris2006}. Up to 10$^{4-5}$ of these objects could be currently present in the Galaxy \citep{willems2003}. Likewise, evolution theories of high mass X-ray binary foresee that about 10$^{6}$ wind accreting binaries made of a main sequence star and of a neutron star or a black hole populate the Galaxy \citep{pfahl2002}.

The relative census of the different species of Galactic X-ray sources strongly depends on the Galactic structure considered. Not unexpectedly, massive X-ray emitting stars, either in accreting binaries or alone, concentrate in the Galactic disc \citep[see e.\,g. the concentration of INTEGRAL HMXBs in the Norma arm][]{walteretal04-1}, while a large population of CVs and low-mass X-ray binaries seem to gather in the predominantly old Galactic Bulge. Likewise, the very central regions of the Galaxy harbour a heavily concentrated population of low to medium \Lx\ sources \citep{munoetal09-1,hongetal09-1} whose exact nature, CVs, stars, remains highly uncertain. X-ray sources are thus useful tracers of their parent stellar populations and their study can shed light on the evolutionary mechanisms of single and binary stars in remote parts of our Galaxy. They may as well help to constrain the past stellar formation rate, in particular the early formation stages during which many massive compact remnants were created.

Solar type stars emit X-rays from magnetically heated coronae. Therefore, their X-ray luminosity strongly depends on differential rotation \citep{pallavicini1981}. Since magnetic braking efficiently decreases stellar rotation \citep[see e.g.][]{kawaler1988,matt2012} both X-ray luminosities and thin thermal plasma temperatures \citep{guedeletal97-1} strongly decay during pre main sequence and early main sequence stages. Old close binaries in which rotation is maintained by orbital motion are also known to contribute significantly to the X-ray emitting stellar population. One can take advantage of the marked \Lx\ dependency upon age to discriminate with high efficiency young stars from the background of older populations. In this respect, X-ray selection is more practical than any selection based on weak optical proxy spectral signatures such as, for instance, re-emission in the Ca II H\&K lines \cite[see e.g][]{schrijver1987} which arises from the stellar chromosphere. Comparing the observed properties of stellar X-ray surveys with population models can provide important constraints on the evolution of Galactic scale height with age \citep{guilloutetal96-1} and reveal large scale local structures, e.g. the late type component of the Gould Belt.
\citep{guillout1998}. 
      
In this paper we present results from an optical identification campaign conducted in the Galactic Plane, at low and intermediate Galactic latitudes ($|b|\,<\,20^\circ$) and covering a wide range of Galactic longitudes, by the XMM-Newton Survey Science Centre \citep{watsonetal01-1}. We report the optical identification of over 300 Galactic X-ray sources, most of them being classified on the basis of optical spectroscopy. This constitutes the largest sample of spectroscopically identified X-ray sources at low Galactic latitudes. 

Beyond its astrophysical motivations, this project also aims at gathering a large and representative sample of identified low-latitude sources which can be used as a learning sample for identifying and classifying in a statistical manner serendipitous XMM-Newton sources detected in the Milky Way. First attempts to automatically classify XMM-Newton X-ray sources at high Galactic latitudes have been reported in \cite{pineau2010}. SSC surveys covering other Galactic directions have been already presented by \cite{barconsetal02-1} and \cite{dellacecaetal04-1}, for high Galactic latitude sources at medium and bright fluxes, and by \cite{motchetal10-1} for low Galactic latitude sources and bright fluxes.

The structure of the paper is as follows. In Section~\ref{sec:xray-data} we present the XMM-Newton data, followed by the optical observations in Section~\ref{sec:optical-data}, and optical and infrared cataloge identifications in Section~\ref{sec:optical-counterparts}. We present the source classification in Section~\ref{sec:identifications} and the stellar population content of the survey in Section~\ref{sec:stel_pop}. We discuss the overall properties and characteristics of the sample in Section~\ref{sec:properties} and conclude in Section~\ref{sec:concl}.

\section{XMM-Newton data}
\subsection{Observations and data reduction}
\label{sec:xray-data}
\begin{figure}[t!]
\begin{center}
\includegraphics[width=\linewidth,angle=0]{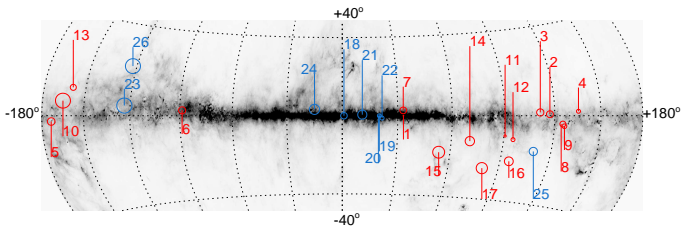}
\caption{Galactic positions of the XMM-Newton fields over-plotted on top of the Galactic extinction map from \cite{schlegeletal98-1} shown in Aitoff projection running in Galactic longitude from $-180^\circ$ to $180^\circ$. The optically bright and faint samples (see Section~\ref{sec:optical-data}) are shown in red and blue respectively. The symbol size is proportional to the number of XMM-Newton sources detected in each field. Numbers indicate the field as listed in Table~\ref{t:iden}. \label{g:fields}}
\end{center}
\end{figure}

\begin{table*}
\begin{center}
\caption{Target fields and source classification.}\label{t:iden}
\begin{tabular}{rlrrrrrrrrrrr}
\hline\hline
\noalign{\smallskip}
\# & Field          & l [$^\circ$]     & b  [$^\circ$]   & \Nh \dag              & Observation  & Epic pn  & N$_{\mathrm{tot}}$  & Active  & HMXB & Pre-MS   & EG    & unID\\
   &                &                  &                 & [$\mathrm{cm^{-2}}$]  & ID           & exp [ks] &                 & Coronae &     &     &          &     \\ 
\noalign{\smallskip}
\hline
\noalign{\smallskip}
\multicolumn{13}{c}{Optically bright sample, Rlim$\sim$17}\\
\noalign{\smallskip}
 1 & Ridge\,1           & 33.1  & +0.0 & $1.4\times10^{23}$ & 0017740401 & 26    & 4       & 1  (25\%)     & 0   & 0       & 0        &  3  \\ 
 2 & RX~J0002+6246     & 117.0 & +0.5 & $4.2\times10^{21}$ & 0016140101 & [17]  & 40      & 7  (18\%)     & 1   & 0       & 0        & 32  \\
 3 & LHB-3              & 111.1 & +1.1 & $1.3\times10^{22}$ & 0203130201 & 19 & 42      & 13 (31\%)     & 0   & 0       & 0        & 29  \\
 4 & GRB010220          & 135.1 & +1.4 & $4.8\times10^{21}$ & 0128530401 &  9 & 23      & 12 (52\%)     & 0   & 0       & 0        & 11  \\
 5 & Saturn             & 187.0 & -1.6 & $6.2\times10^{21}$ & 0089370501 & 21 & 44      & 6  (14\%)     & 0   & 0       & 0        & 38  \\
 6 & RXJ0925.7-4758  & 271.4 & +1.9 & $1.4\times10^{22}$ & 0111150201 & 57 & 42      & 9  (21\%)     & 0   & 0       & 0        & 33  \\
 7 & Ridge~2           & 33.0  & +2.0 & $1.5\times10^{22}$ & 0017740601 & 20 & 39      & 12 (31\%)     & 0   & 1       & 0        & 26  \\
 8 & HTCas            & 125.2 & -2.7 & $3.9\times10^{21}$ & 0111310101 & 18 & 34      & 6 (18\%)      & 0   & 0       & 0        & 28  \\
 9 & PSRJ0117+5914    & 126.3 & -3.4 & $3.4\times10^{21}$ & 0112200201 &  6 & 27      & 4 (14\%)      & 1   & 0       & 0        & 22  \\
10 & Geminga            & 195.1 & +4.3 & $3.0\times10^{21}$ & 0111170101 & [86]  & 88      & 9 (10\%)      & 0   & 0       & 0        & 79  \\
11 & SS~Cyg              & 90.7  & -7.1 & $2.5\times10^{21}$ & 0111310201 & [12]  & 13      & 5 (38\%)      & 0   & 0       & 2        &  6  \\
12 & ARLac              & 95.6  & -8.3 & $1.9\times10^{21}$ & 0111370101 & [20]  & 24      & 5 (22\%)      & 0   & 0       & 0        & 20  \\
13 & PSR0656+14         & 201.1 & +8.3 & $5.0\times10^{20}$ & 0112200101 & [21]  & 36      & 6 (17\%)      & 0   & 0       & 0        & 30  \\
14 & PSRJ2043+2740      & 70.6  & -9.2 & $1.4\times10^{21}$ & 0037990101 & [17]  & 54      & 10(19\%)      & 0   & 0       & 1        & 43  \\
15 & AXJ2019+112        & 53.6  &-13.5 & $1.2\times10^{21}$ & 0112960301 & 12 & 69      & 6 (9\%)       & 0   & 0       & 1        & 62  \\
16 & 3C449              & 95.4  &-15.9 & $8.9\times10^{20}$ & 0002970101 & 18 & 52      & 9 (17\%)      & 0   & 1       & 3        & 39  \\
17 & 3C436              & 80.2  &-18.8 & $5.0\times10^{20}$ & 0201230101 & 29 & 65      & 5 (8\%)       & 0   & 0       & 1        & 59  \\
\noalign{\smallskip}
\hline
\noalign{\smallskip}
& All                &       &      &                    &            &    &  696    & 125       & 2   & 2       & 8        & 560 \\
&                    &       &      &                    &            &    &         & 17.9\%    & 0.3\% & 0.3\% & 1.1\%    & 80.3\% \\
\noalign{\smallskip}
\hline
\noalign{\smallskip}
\multicolumn{13}{c}{Optically faint sample, Rlim$\sim$21}\\
\noalign{\smallskip}
18 & GC2                &  0.9  & +0.0  & $4.0\times10^{23}$ & 0112970201 & 13 & 38      & 15 (39\%)     & 0     & 0       & 0        & 23 \\
19 & Ridge~3            &  20.0 & +0.0  & $1.1\times10^{23}$ & 0104460301 &  8 & 22      & 11 (54\%)     & 0     & 1       & 0        & 10 \\
20 & Ridge~4            &  20.4 & +0.0  & $1.1\times10^{23}$ & 0104460401 &  5 & 12      & 2  (17\%)     & 0     & 0       & 0        & 10 \\
21 & WR110              &  10.8 & +0.4  & $5.5\times10^{22}$ & 0024940201 & 23 & 52      & 21 (40\%)     & 1     & 1       & 0        & 28 \\
22 & G21.5-09 offset 2  &  21.5 & -0.9  & $5.0\times10^{22}$ & 0122700301 & 25 & 24      & 5 (22\%)      & 0     & 0       & 0        & 19 \\
23 & PKS~0745-19-offset & 236.4 & +3.3  & $2.3\times10^{21}$ & 0105870201 & 36 & 88      & 24 (27\%)     & 0     & 0       & 18       & 46 \\
24 & GROJ1655-40       & 345.0 & +2.4  & $6.1\times10^{21}$ & 0112460201 & 21 & 59      & 15 (25\%)     & 0     & 0       & 1        & 43 \\
25 & Z~And              & 109.0 &-12.1  & $8.9\times10^{20}$ & 0093552701 & 17 & 48      & 13 (27\%)     & 0     & 0       & 0        & 35 \\ 
26 & GRB001025          & 237.4 &+16.3  & $3.3\times10^{20}$ & 0128530301 & 26 & 87      & 4  (5\%)      & 0     & 0       & 6        & 77 \\
\noalign{\smallskip}
\hline
\noalign{\smallskip}
&  All                  &       &       &                 &            &    & 430     & 110             & 1     & 2       & 25       & 291\\
&                       &       &       &                 &            &    &         & 25.8\%          & 0.2\% & 0.4\%   & 5.8\%    & 67.7\% \\
\noalign{\smallskip}
\hline
\hline
& Total                 &       &       &                 &            &    & 1126    & 235             & 3     & 4       & 33       & 851  \\       
\noalign{\smallskip}
\noalign{\smallskip}
\hline
\end{tabular}
\end{center}
\textbf{Notes.} Field, Galactic coordinates, observation ID, EPIC-pn exposure time (EPIC-MOS exposure time in brackets), total number of 2XMMi-DR3 sources detected, number of identified active coronae (AC), High-mass X-ray binaries (HMXB), pre-main sequence stars (pre-MS), extragalactic sources (EG), and unidentified sources (unID). We restricted to sources with $\mathrm{sum\_flag = 0}$. \dag Calculated hydrogen column density. We used the \cite{schlegeletal98-1} maps to estimate the color excess E(B--V) at each Galactic position. We transformed into optical extinction A$_V$ using the relation from \cite{savage+mathis79-1}, and we used the empirical relation from \cite{predehl+schmitt95-1} to calculate \Nh. We note that for very low Galactic latitudes ($|b|<5^\circ$) values could be wrong, but we use the values for orientative purposes. 
\end{table*}

The X-ray Multi-Mirror mission \citep[XMM-Newton,][]{jansenetal01-1} was launched in December 1999 by the European Space Agency (ESA). The XMM-Newton satellite has three X-ray telescopes and is equipped with a set of CCD detectors, which constitutes the European Photon Imaging Cameras (EPIC). There are two MOS-CCD arrays \citep[MOS camera,][]{turner2001}, and one pn-CCD \citep[pn camera,][]{struederetal01-1}. The two MOS cameras are located behind the two telescopes equipped with gratings which divert about half of the light towards the Reflecting Grating Spectrometers \citep[RGS,][]{denherder2001}, so only about 40\% of the light reaches the MOS cameras. The pn camera is located behind the third telescope, receiving all the incident light \citep{struederetal01-1}. The field of view is about 30\arcmin, it covers the energy range from 0.15 to 15\,keV, with spectral resolution ($\frac{E}{dE}$)\,=\,20-50 and angular resolution of 6\arcsec.

The 26 XMM-Newton fields presented in this work were selected shortly after the launch of the XMM-Newton satellite. They are all at low and intermediate Galactic latitudes ($|b|\,<\,20^\circ$) and cover a wide range in Galactic longitudes (see Fig.~\ref{g:fields}). Observations were selected so as to be void of extended diffuse emission and represent as much as possible typical Galactic fields. We therefore excluded stellar clusters and star forming regions in general. Very bright target sources were also avoided if possible. Fields are divided in two samples: the \emph{optically bright} and the \emph{optically faint} samples, depending on the limiting magnitude reached by the combination of telescope and instrument used for the optical identification of the X-ray sources (see Section~\ref{sec:optical-data}). In Table~\ref{t:iden} we list the fields, their Galactic coordinates, observation IDs and EPIC pn exposure times. Among the 26 fields, 20 were observed with the three EPIC cameras, while for six we have only MOS detections.  Exposure times range from 5 to 57 ks, reaching flux limits of around $2\times10^{-15}$\,\ergcms\ and $1\times10^{-14}$\,\ergcms\ in the 0.5\,--\,2\,keV and in the 2\,--\,12\,keV energy bands respectively. The area covered by this survey is $\sim 4$ deg$^2$ ($\sim 3$ deg$^2$ for the pn camera). This survey is ten times deeper than the ROSAT medium sensitivity survey of the Galactic Plane from \cite{morleyetal01-1} and the ASCA faint X-ray survey from \cite{sugizakietal01-1}. Although 10 times shallower than the Chandra deep Galactic Plane survey of \cite{ebisawaetal05-1} and than the ChaMPlane survey of \cite{grindlayetal05-1} our survey covers a much larger area. The XMM-Newton SSC survey of the Galactic Plane can be considered a medium sensitivity survey slightly deeper than that of \cite{handsetal04-1}. 

\begin{figure*}
\begin{center}
\begin{minipage}{\linewidth}
\begin{minipage}{0.48\linewidth}
\includegraphics[width=\linewidth,angle=0]{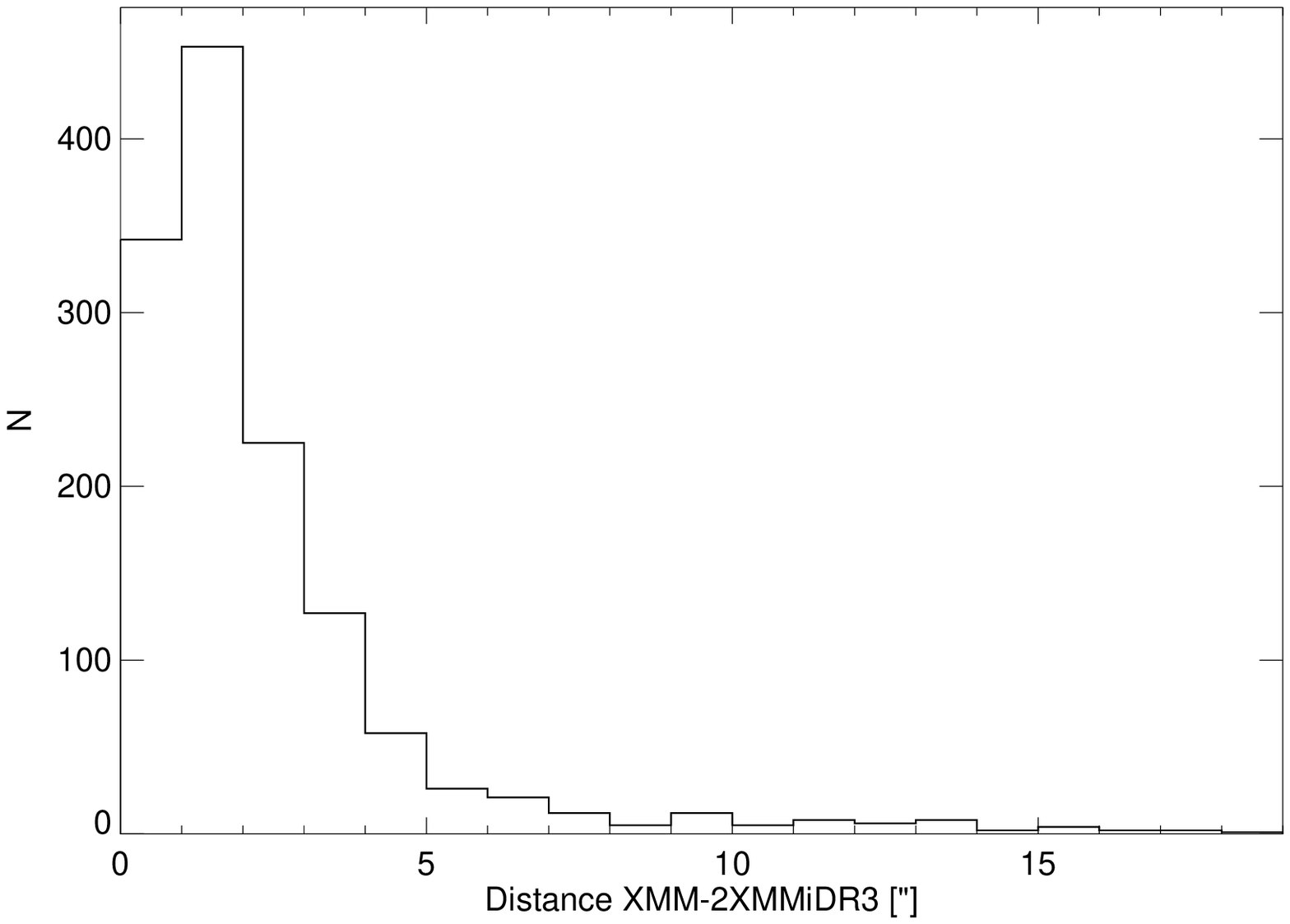}
\caption{Distance between the original XMM-Newton source positions and the 2XMMi-DR3 counterparts within 20 arcsec. Most of the counterparts are found within 3-5 arcsec.
\label{g:distance}}
\end{minipage}\hfill
\begin{minipage}{0.48\linewidth}
\includegraphics[width=\linewidth,angle=0]{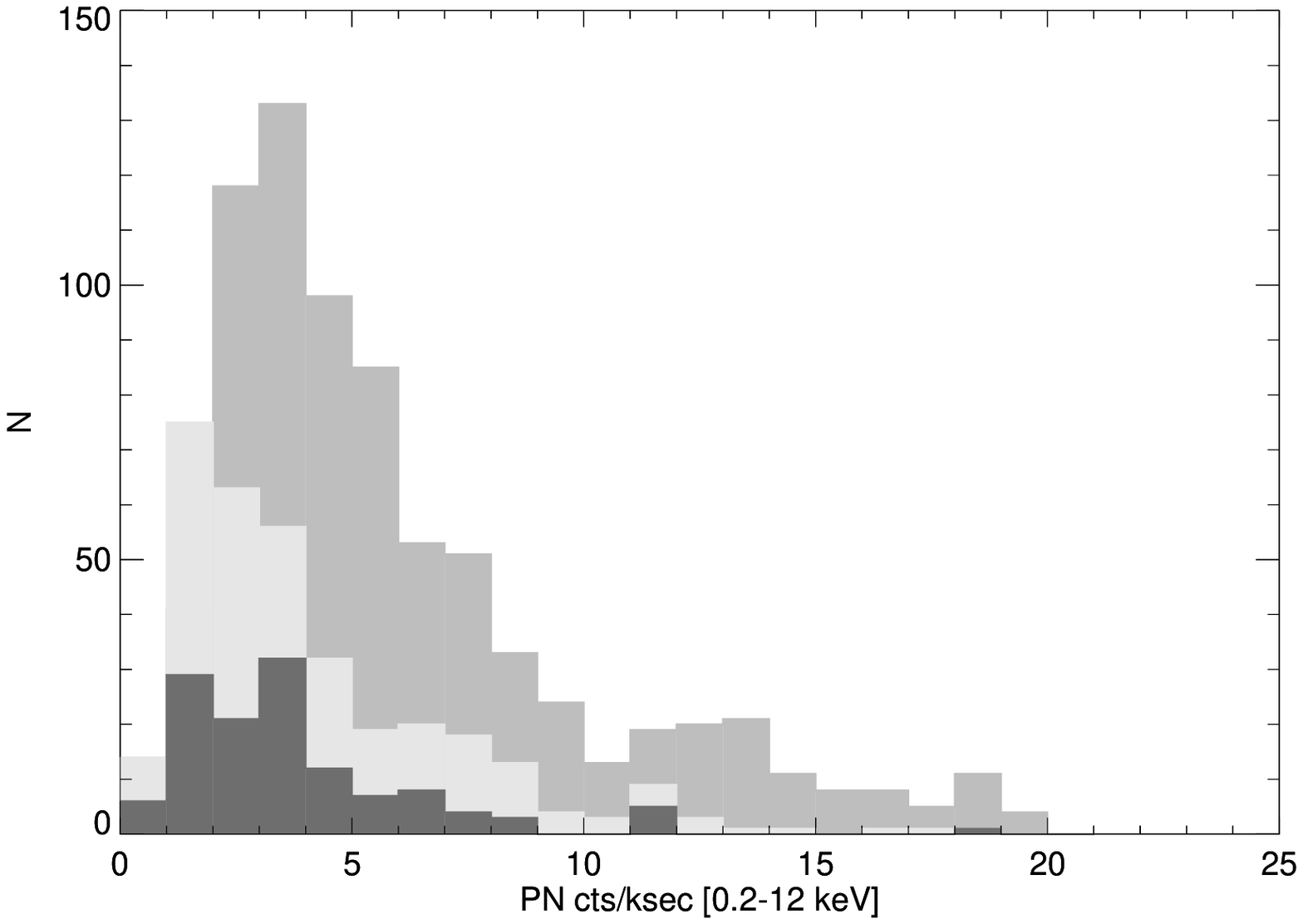}
\caption{EPIC pn count rate (0.2\,--\,12\,keV) distribution for the original unscreened XMM-Newton sources with (medium grey) and without (light grey) 2XMMi-DR3 counterpart, highlighting with a dark grey histogram those detections originally flagged as spurious or dubious. \label{g:pntot}}
\end{minipage}
\end{minipage}
\end{center}
\end{figure*}

The original source list contained a total of 2353 X-ray sources, detected in either the pn or MOS cameras, using the Science Analysis System (SAS)\footnote{\fontsize{7}{7}{\url{http://xmm.esa.int/sas/}}} version 5.0 or earlier.
In all cases we excluded the target of the observation. All observations were visually screened in order to discard spurious or dubious detections before scheduling sources for optical identification at ground-based telescopes, leaving us with a total of about 1800 "good" sources. Optical spectra were taken using the X-ray source positions derived in this first analysis of the X-ray data. In most cases, optical targets were prioritised using the broadband X-ray flux, starting with the brightest sources. Since then, other improved SAS processing versions became available, and in order to have up-to-date and homogeneous X-ray parameters we cross-matched our source lists with the 2XMMi-DR3 catalogue\footnote{\fontsize{7}{7}{{\url{http://xmmssc-www.star.le.ac.uk/Catalogue/xcat_public_2XMMi-DR3.html}}}}. 
To evaluate the success rate of our visual screening we performed the cross-match for all the 2353 X-ray sources and found 1319 sources in the 2XMMi-DR3 within a 20 arcsec radius (1126 with the best XMM-Newton quality, i.e. $\mathrm{sum\_flag\,=\,0}$\footnote{\fontsize{7}{7}{The summary flag $\mathrm{sum\_flag}$ contains information about the flags set automatically and manually for each source, a value equal to zero assures that there are no negative flags for the detection of the source.}}). Extending the radius to larger values did not yield many more matches. In Fig.~\ref{g:distance} we show the distribution of the distance between the original XMM-Newton sources and the 2XMMi-DR3 sources. The majority of the matches are found to be within 3-5\arcsec. About 50\% of the unrecovered sources in the 2XMMi-DR3 catalogue had been flagged by us as being spurious or dubious based on our visual screening.  Unrecovered sources have lower EPIC-pn count rates (0.2\,--\,12\,keV) than those with counterparts (see Fig.~\ref{g:pntot}) and are thus likely to be spurious sources with low detection likelihood \citep[][Section 4.4]{watsonetal09-1}. Hereinafter we limit our analysis to the 1319 sources with 2XMMi-DR3 counterpart and X-ray parameters are based on 2XMMi-DR3 catalogue values.

\subsection{Modelled X-ray spectra}
\label{sec:xray-models}
Making use of XSPEC\footnote{\fontsize{7}{7}{\url{http://heasarc.nasa.gov/xanadu/xspec/}}} we created model X-ray spectra for the following different types of objects: 
\begin{enumerate}[i)]
\item three different populations of active coronae: a young population of 70 Myr, an intermediate population of 300 Myr, and an old population of 1.9 Gyr. We assumed 2\,--\,temperature thermal emission \citep{guedeletal97-1}, one component representing the hot plasma (T$_1$) and a second hotter component (T$_2$). Temperatures used for the coronae are (kT$_1$, kT$_2$)\,=\,(0.2, 0.8), (0.12, 0.55), and (0.09, 0.32)\,keV, and X-ray emission measure ratios are EM$_2$/EM$_1\,=\,1$, EM$_2$/EM$_1\,=\,1$, and EM$_2$/EM$_1\,=\,0.6$ for the young, intermediate, and old populations respectively. 
\item for AY Cet, a typical BY\,Dra binary \citep{dempseyetal97-1}, with (kT$_1$, kT$_2$)\,=\,(0.2,1.38)\,keV and EM$_2$/EM$_1\,=\,3.75$, and for the RS\,CVn star WW\,Dra \citep{dempseyetal93-1}, with (kT$_1$, kT$_2$)\,=\,(0.2, 2.1)\,keV and EM$_2$/EM$_1\,=\,6.77$. 
\item accreting binaries, assuming two different power laws with photon indices $\Gamma$ of 0 and 2. 
\item active galactic nuclei (AGN) assuming a power law with photon index $\Gamma\,=\,1.9$. 
\end{enumerate}
We used the models \emph{mekal} and \emph{powerlaw}, alongside \emph{phabs} for the photoelectric absorption, for our simulations. They were performed for a wide range of Galactic absorption (\Nh\,$=\,10^{20-23}$ cm$^{-2}$) and with a step of 0.1 in $\log$(\Nh). Simulated hardness ratios\footnote{\fontsize{7}{7}{$\mathrm{HR_{i}} = \frac{\mathrm{C_{i+1}-C_{i}}}{\mathrm{C_{i}+C_{i+1}}}$, being $\mathrm{C_{i}}$ the count rate in band $\mathrm{i}$, where $\mathrm{i}=0.2-0.5,0.5-1.0,1.0-2.0,2.0-4.5,4.5-12.0$\,keV.}} (HR) were computed for each model with the aim of comparing with the observed values.

\section{Optical observations}
\label{sec:optical-data}
Photometry was carried out at the Isaac Newton Telescope (INT), at the Canada France Hawaii Telescope (CFHT) and at the 2.2m ESO telescope. INT images were bias subtracted and flat-field corrected using the WFC pipeline, following the instructions described by the Cambridge Astronomy Survey Unit\footnote{\fontsize{7}{7}{\url{http://www.ast.cam.ac.uk/~wfcsur}}}. CFHT images were reduced using the ELIXIR\footnote{\fontsize{7}{7}{\url{http://www.cfht.hawaii.edu/science/CFHTLS/cfhtlsdataflow.html}}} pipeline. ESO images were reduced using the GaBoDS pipeline\footnote{\fontsize{7}{7}{\url{http://www.astro.uni-bonn.de/~heckmill/Software/GABODS.html}}}.
Images were used to find the optical counterpart of the X-ray sources, to measure the magnitude of the faintest candidate counterparts, and to prioritise the target selection for spectroscopic identification. Whenever a bright extended optical object was detected in the X-ray error box we considered the source as being extragalactic and did not obtain optical spectra. 

Optical spectroscopic observations were carried out between the years 2000 and 2003 with the Very Large Telescope (VLT-UT4) and with the ESO-3.6\,m telescope at ESO La Silla/Paranal Observatory, with the William Herschel Telescope (WHT) at the Observatory Roque de los Muchachos (La Palma), and with the 1.9\,m telescope at the Observatoire de Haute-Provence (OHP). A large part of the optical data were acquired in the framework of the AXIS project\footnote{\fontsize{7}{7}{\url{http://venus.ifca.unican.es/~xray/AXIS/}}}. Telescope, instrument, instrumental setup, wavelength coverage and spectral resolution are listed in Table~\ref{t:spec_log}. 
Depending on the telescope/instrument used for each observed field the limiting magnitude is \mbox{R$\sim17$} or \mbox{R$\sim21$}. Seventeen fields were observed reaching a magnitude limit of R$\sim17$, belonging to the optically bright sample. The remaining nine fields were observed with a limiting magnitude R$\sim21$ defining the optically faint sample (see Table~\ref{t:iden}). 

Spectra were bias corrected, flat-fielded, and extracted using standard MIDAS\footnote{\fontsize{7}{7}{\url{http://www.eso.org/sci/software/esomidas}}} procedures. We used arc-lamps to calibrate in wavelength and spectrophotometric standard stars to flux calibrate \citep[see][for details]{motchetal10-1}.

Observations were carried out using the original (see Section~\ref{sec:xray-data}) positions of the X-ray sources, available at the time of the observing runs. In the Galactic Plane, optical crowding often prevents a clear identification of X-ray sources. It is therefore very important to have a good knowledge of the X-ray positional errors. The XMM-Newton error circle is the combination of two errors. The first one is the statistical uncertainty on the centroid of the PSF determined by the detection algorithm, $\sigma_{radec}$ typically $\sim1-2\arcsec$. The second one is the systematic error introduced by the uncertainty of the satellite's attitude, with values $\sigma_{systematic}\sim1\arcsec$. In the less crowded fields, systematic errors were reduced to $\sim\,0.35\arcsec$ by matching the XMM-Newton positions with the USNO-B1.0 \citep{monetetal03-1} catalogue making use of the SAS task \emph{eposcorr}. Assuming a two-dimensional Gaussian distribution, the 90\% confidence-level radius is given by:
\begin{equation}
r_{90} = 2.15\times\sqrt{\sigma_{radec}^2+\sigma_{systematic}^2}
\end{equation}
In many cases the X-ray source position error circle contained more than one optical candidate brighter than our limiting magnitude. We obtained optical spectra of all candidates present in the 90\% confidence error circle down to our limiting magnitude, sorting objects by decreasing R brightness. A source was positively identified when specific spectral signatures such as emission lines were detected. 

Spectroscopic classification of stellar spectra was carried out as in \cite{motchetal10-1}. In summary, template spectra from \cite{jacobyetal84-1,pickles98-1}, STELIB \citep{leborgneetal03-1} and from the NASA/JPL NStars project\footnote{\fontsize{7}{7}{\url{http://stellar.phys.appstate.edu/}}}, degraded to the resolution of our observations (Table~\ref{t:spec_log}), were fitted to the observed spectra by adjusting the mean flux and the interstellar absorption. Spectral classification of the sources is given in the online version of the paper and statistics on the identifications are discussed in Section~\ref{sec:identifications}.

\addtolength{\tabcolsep}{-4pt}
\begin{table}
\begin{center}
\caption{Spectroscopic settings.}\label{t:spec_log}
\begin{tabular}{llcccc}
\hline 
\hline 
\noalign{\smallskip}
Telescope & Instr.  & Slit            & Spectral    & Spectral    & Grisms \\
          &         & width (\arcsec) & range (\AA) & res. (\AA)  &        \\
\noalign{\smallskip}
\hline 
\noalign{\smallskip}
VLT-UT4   & FORS2   & 1.0             & 3850--7500  & 15       & GRIS\_300V   \\
WHT       & WYFOS   & 2.7             & 3900--7100  & 6--7     &              \\
WHT       & ISIS    & 1.2--2.0        & 3500--8500  & 3.0--3.3 &              \\
ESO-3.6   & EFOSC2  & 1.5             & 3185--10940 & 59       & Grism\,\#1   \\
ESO-3.6   & EFOSC2  & 1.5             & 3860--8070  & 17.3     & Grism\,\#6   \\
ESO-3.6   & EFOSC2  & 1.5             & 3095--5085  & 7.1      & Grism\,\#14  \\
OHP       & CARELEC & 1.0--2.0        & 3760--6840  & 5--7     & 133\AA/mm    \\
\noalign{\smallskip}
\hline
\end{tabular}
\end{center}
\end{table}
\section{Optical and infrared catalogue identifications}
\label{sec:optical-counterparts}
\begin{center}
\begin{figure*}[t!]
\includegraphics[width=0.5\linewidth,angle=0]{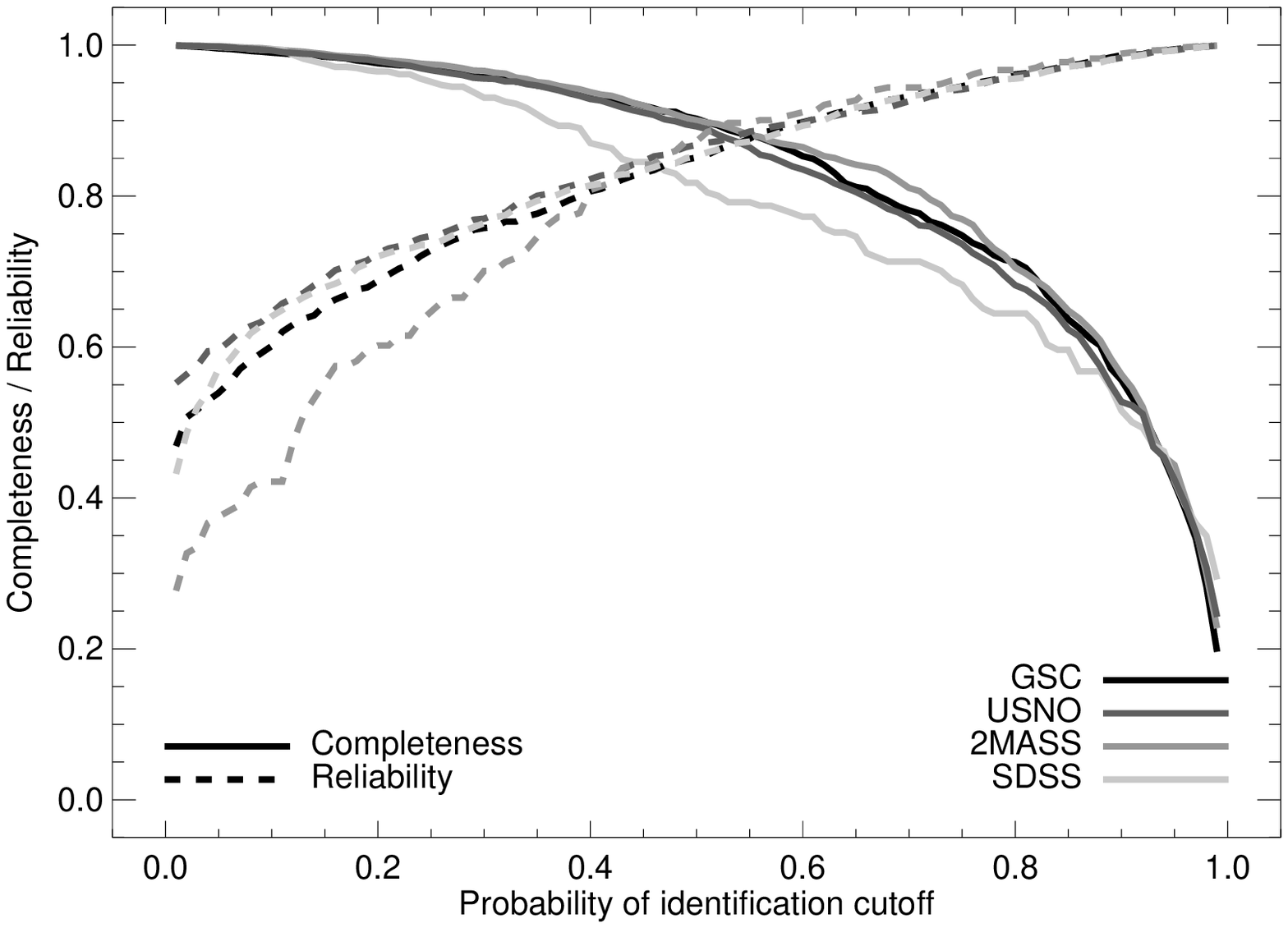}
\includegraphics[width=0.5\linewidth,angle=0]{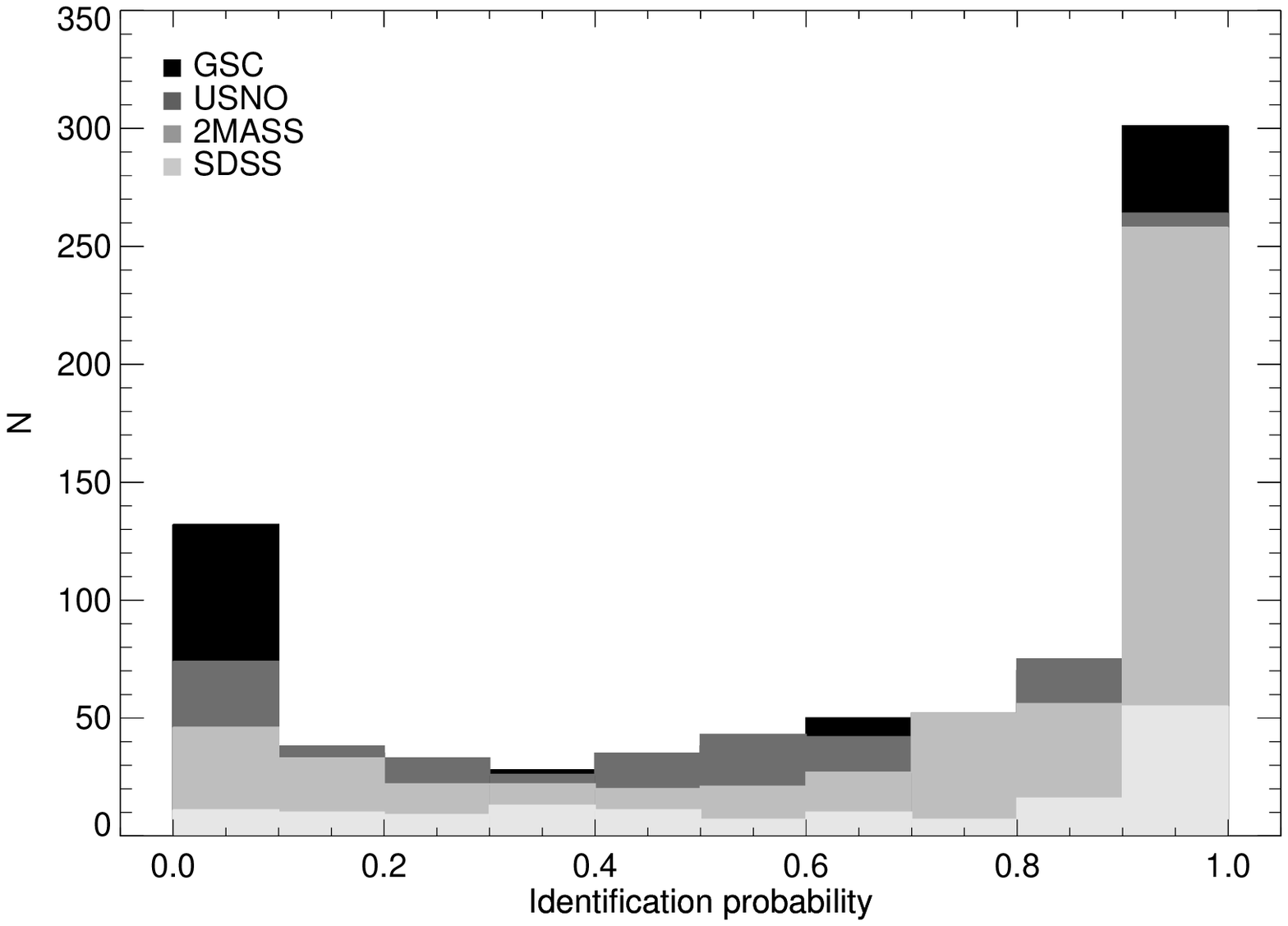}
\caption{Completeness and reliability of the sample as a function of the threshold in the identification probability (left). Distribution of the probability of being the true counterpart associated with each X-ray source with entry in the 2MASS, USNO-B1.0, GSC~2.3 and SDSS-DR7 catalogues (right).}\label{g:probabilities}
\end{figure*}
\end{center}
As stated above, source identification can be difficult in the Galactic plane since several optical candidates are often found within the X-ray error circle and spectral signatures obtained at the telescope can be sometimes ambiguous. We therefore helped our optical identification process by cross-correlating X-ray source positions with the following archival catalogues: 
2MASS \citep{cutrietal03-1}, USNO-B1.0 \citep{monetetal03-1}, GSC~2.3 \citep{laskeretal08-1} and SDSS-DR7 \citep{abazajianetal09-1} catalogues as in \cite{motchetal10-1}. All these cross-matches are provided through the SSC public XCat-DB interface\footnote{\fontsize{7}{7}{\url{http://xcatdb.u-strasbg.fr/2xmmidr3/}}} \citep{micheletal04-1}.
\subsection{Cross-correlation method}
The cross-correlation process is described in \cite{pineauetal08-1} and \cite{pineauetal11-1}. 
In brief, it is based on the classical ratio between the likelihood of the X-ray and catalogue sources to be at the same position, and the likelihood of being a spurious association. This likelihood ratio (LR) depends on the probability of an X-ray source to have a counterpart in the considered catalogue, probability which depends on the distribution and characteristics of the different populations that enter the sample. To estimate the probability of being an spurious identification the method uses a geometrical approach. 
The process searches for all possible counterparts within a radius corresponding to a 99.7\% ($3\sigma$) completeness, which for a two dimensional Gaussian distribution corresponds to 3.44 times the $1\sigma$ combined positional error. The combined radius is computed by adding in quadrature the X-ray and the catalogue 1$\sigma$ errors. For each possible counterpart, the probability of it being the true counterpart to the X-ray source (\Pid) is given by the ratio between the total number of observed counterparts and the number of estimated spurious associations. Above a given threshold of this probability we calculate the sample reliability, i.e. the expected fraction of correct identifications among all the matches, from which we derive the number of spurious associations, and sample completeness. The sample completeness is defined as the ratio between the number of true associations recovered above that probability threshold and the number of true associations expected in the survey (see left panel Fig.~\ref{g:probabilities}). As we increase the cutoff in the individual identification probability we reduce the completeness of the survey, but increase the reliability, reducing the fraction of possible spurious identifications. 

\subsection{Cross-correlation results}
\begin{table}
\begin{center}
\caption{2XMMi-DR3 matches in infrared/optical catalogues.}\label{t:stats_cats}
\begin{tabular}{rrr}
\hline
\hline
\noalign{\smallskip}
Catalogue & N$_{total}$$^\dag$ & N$_{P>90\%}$$^\ddag$ \\
\noalign{\smallskip}
\hline
\noalign{\smallskip}
2MASS     & 557        & 258 \\
GSC~2.3   & 741        & 301 \\
USNO-B1.0 & 682        & 264 \\
SDSS-DR7  & 149        & 55  \\
\noalign{\smallskip}
\hline
\noalign{\smallskip}
Optical       & 801 & 329 \\
IR or Optical & 829 & 350 \\
\noalign{\smallskip}
\hline
\end{tabular}
\end{center}
Notes: $^\dag$ Total number of matches. $^\ddag$ Number of matches with individual identification probability higher than 90\%.
\end{table}
\begin{figure*}
\begin{center}
\includegraphics[width=0.7\linewidth,angle=0]{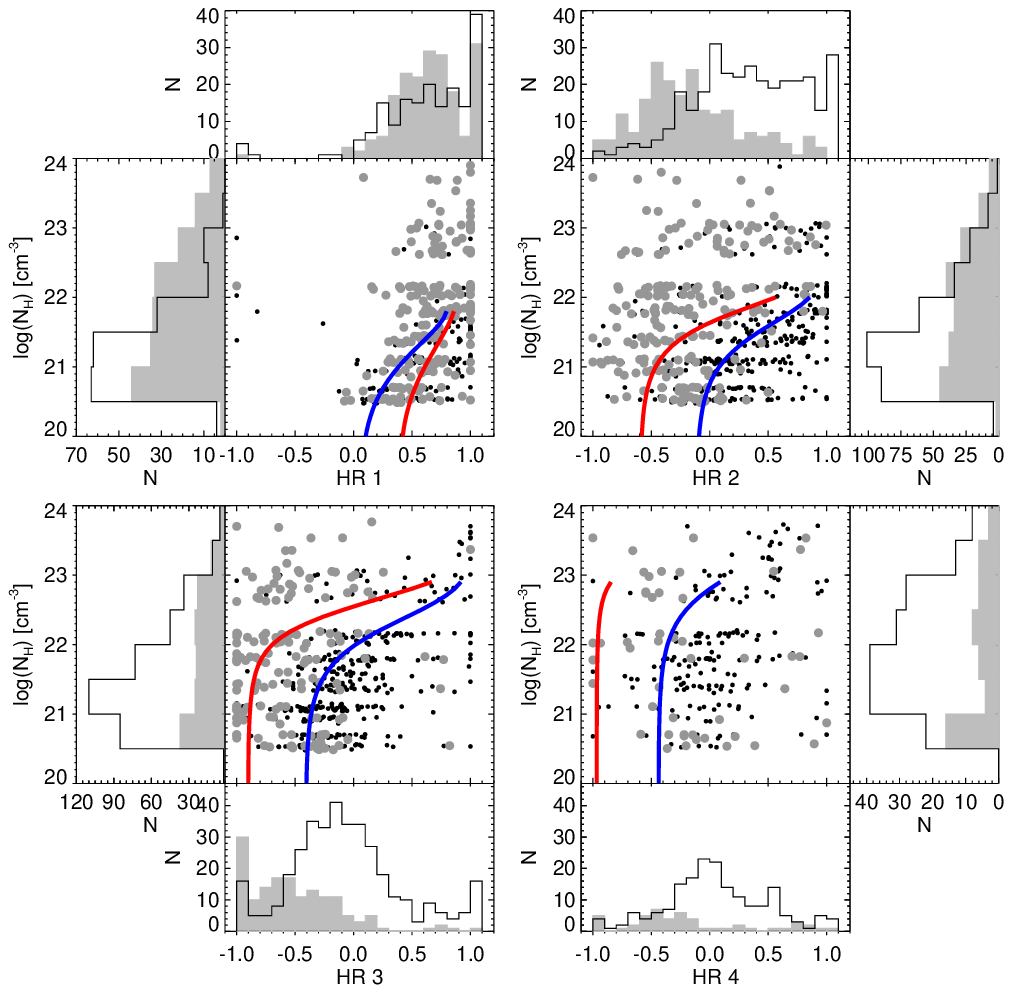} 
\caption{Hardness ratio and Galactic N$_\mathrm{H}$ distribution of X-ray sources with optical or infrared catalogue counterparts with identification probability $>0.9$ (large grey dots and grey-filled histograms), and of X-ray sources without catalogue counterpart or with identification probability below 0.9 (small black dots and empty black histograms). We only considered sources with good X-ray quality (\mbox{sum\_flag\,=\,0}) and with errors smaller than 0.3 in X-ray colours. Blue and red lines show the expected position of AGNs ($\Gamma\,=\,1.9$) and stars (70 Myr old population, see Section~\ref{sec:xray-models} for details) respectively. Most of the objects with no infrared or optical counterpart follow a distribution similar to that of AGNs.}\label{g:hrs_counterparts}
\end{center}
\end{figure*}
Cross-correlation statistics are given in Table~\ref{t:stats_cats}. The much lower number of X-ray sources with SDSS-DR7 counterpart is due to the fact that only five fields (LHB\,3, PSR\,J2043+2740, 3C449, Saturn and Z~And) are covered by the SDSS-DR7 footprint. Probabilities of being the true counterpart are shown in the right panel of Fig.~\ref{g:probabilities}. There are a total of 258 ($\sim20\%$) and 329 ($\sim25\%$) X-ray sources with infrared and optical (in either SDSS-DR7, GSC~2.3 or USNO-B1.0) counterparts respectively above the 90\% identification probability. The number of expected spurious matches with individual identification probability above 90\% is below 2\% and we are recovering about 55\% among all true associations between the XMM-Newton and the different catalogues (see left panel in Fig.~\ref{g:probabilities}). Hereinafter we will only consider matches with individual identification probability higher than 90\%. 

\subsection{X-ray properties of catalogue counterparts}
Stars are soft X-ray emitters, therefore it should be easy to distinguish them from the hard X-ray extragalactic and accretion-powered sources on the basis of their X-ray colours. We analysed the hardness ratio distributions of X-ray sources with bright optical or infrared counterparts and compared it with those of sources with faint or no catalogue counterpart. 

A total of 350 X-ray sources have either an infrared or an optical catalogue counterpart above the 90\% identification probability, among which seven sources exclusively come from the 2XMMi-DR3/SDSS-DR7 crossmatch. Since the limiting magnitude for the SDSS is much deeper (\emph{g}\,$\sim\,23$) than that for the GSC~2.3 (B$\sim$21) and the USNO-B1.0 (V$\sim$20) we discarded these sources for the following analysis. We limit to sources with the best X-ray quality (sum\_flag\,=\,0) and with hardness ratio errors lower than 0.3. The HR distributions of the remaining sources (230) is shown in Fig.~\ref{g:hrs_counterparts} and compared to those of X-ray sources without a counterpart or with a probability of identification lower than 90\% (474). 

As expected, X-ray sources with optical or infrared counterparts have lower (softer) HRs than those without counterpart, consistent with the fact that most soft sources are stars, while hard sources are likely to be extragalactic. For each sources we estimated the total \Nh\, in the light of sight using the colour excess E(B-V) calculated from the dust maps from \cite{schlegeletal98-1} and their IDL code\footnote{\tiny{\url{http://www.astro.princeton.edu/~schlegel/dust/dust.html}}. For stars, this value represents an upper limit to the \Nh}. We compared the distribution in \Nh\ versus hardness ratio of sources with (without) optical or infrared counterpart to expected values for stars (AGNs) (see Section~\ref{sec:xray-models}) and found similar trends (see Fig.~\ref{g:hrs_counterparts}). 

X-ray-to-optical flux ratios can also provide us with information on the nature of the sources. Whilst extragalactic sources have typical values $-1<\log($\Lx$/$L$_{\mathrm{R}})<1$, stars have lower X-ray-to-optical flux ratios, $\log($\Lx$/$L$_{\mathrm{R}})<-1$ \citep{maccacaroetal88-1}. We thus conclude that the majority of sources without an optical or infrared match have an extragalactic origin or are associated with high F$_{\mathrm{X}}$/F$_{\mathrm{opt}}$ galactic hard sources. 

There are a few soft sources without counterpart in either catalogue. At the limiting X-ray flux of our survey ($2\times10^{-15}$\,\ergcms), sources fainter than R$\,\sim\,20$ (R$\,\sim\,17$), corresponding to the USNO-B1.0 limiting magnitude (restricting to sources with \Pid\,$>\,0.9$), will have $\log($\Lx$/$L$_{\mathrm{R}})>-0.75$ ($\log($\Lx$/$L$_{\mathrm{R}})>-2$), i.~e. could be relatively distant late dMe stars too faint to be detected in the optical (see Fig.~\ref{g:satur_coronae}) \citep{morleyetal01-1}, but are also compatible with accreting binaries and with extragalactic objects in the high galactic latitude fields. A few sources with infrared or optical counterpart have higher HR4 value than expected for young active coronae (see Fig.~\ref{g:hrs_counterparts}). At the chosen cutoff for the individual identification probability, the number of spurious associations is lower than 2\%, so we expect a maximum of 5 spurious matches, a value not sufficient to explain the number of sources detected in the HR4 bands. Although in principle such sources could be AGNs with a bright optical or infrared catalogue counterpart, we have classified some of these hard sources as stars on the basis of their optical spectrum (see Section~\ref{sec:identifications}). 

\section{Source classification}
\label{sec:identifications}
We classified X-ray sources in three different ways: 
\begin{enumerate}
\item From our spectroscopic observations. In most cases the determination of the spectral type of active coronae was possible by fitting the observed spectra to template spectra (Section~\ref{sec:optical-data}). 
\item From our photometric observations. If the X-ray source position was coinciding with a source resolved in the optical (galaxy) we classified it as an extragalactic candidate. 
\item From cross-correlation with archival catalogues using XCat-DB developed in Strasbourg \citep{micheletal04-1}. The list of archival catalogues can be found under \url{http://xcatdb.u-strasbg.fr/2xmmidr3/catarch}. 
\end{enumerate}

Among the total 1319 detected sources we classified 316, with 275 having the best X-ray quality, i.e. sum\_flag\,=\,0 (see Table~\ref{t:iden}). We classified 280 X-ray sources on the basis of the optical spectra of their counterparts. In some cases no classification was possible due either to the optical faintness of the object or to the absence of absorption and/or emission lines in the spectra. We classified six X-ray sources as galaxies on the basis of their extension in the optical images. Finally, via cross-correlation with archival catalogues using XCat-DB we classified 30 sources. This SSC interface hosts all candidate identifications derived from the cross-correlation of EPIC source lists with more than 200 archival catalogues, as performed during the SSC pipeline processing \citep{watsonetal09-1}. All X-ray sources with their classification (when possible) will be available via the CDS and the XID result-DB\footnote{\fontsize{7}{7}{\url{http://saada.unistra.fr/xidresult/home}}}. From the 316 classified sources, 270 have a 2MASS counterpart and 296 an optical counterpart in either the SDSS-DR7, the USNO-B1.0 or the GSC\,2.3 catalogues. There are 20 classified sources by our photometric observations with no catalogue counterpart, these sources were too faint in optical to be detected by the USNO-B1.0 and GSC\,2.3 surveys and are outside the SDSS-DR7 footprint. The majority of these sources are classified as galaxies. 
\begin{table}
\begin{center}
\caption{Spectral types of the identified active coronae.}\label{t:spectral-types}
\begin{tabular}{lcc}
\hline
\hline
\noalign{\smallskip}
SpT   & Number & Fraction (\%)\\ 
\noalign{\smallskip}
\hline
\noalign{\smallskip}
A     & 13       &  5  \\
F     & 44       & 16  \\
G     & 64       & 24  \\
K     & 80       & 29  \\
M     & 71       & 26  \\
\noalign{\smallskip}
\hline
\end{tabular}
\end{center}
\end{table}

Optical classification statistics are presented in Table~\ref{t:iden}, separating the optically bright and faint samples defined in Section~\ref{sec:optical-data}. The majority of the classified sources are active coronae (AC), representing $\sim 18\%$ of the sources for the optically bright sample and up to $\sim 26\%$ for the optically faint sample. The total number of AC for each spectral type is given in Table~\ref{t:spectral-types}. The fraction of stars at each spectral type increases from 5\% for A stars to 29\% for K stars and then drops to about 26\% for M stars, reflecting either a deficit in M stars or an excess of G and K stars. We note that A stars are not expected to be strong X-ray emitters, and that if these sources are not spurious X-ray detections or wrong identifications, then their X-ray emission is likely to be related to a low-mass companion star \citep{delarosa11-1}.
A small fraction of the sources are CVs, $\gamma$-Cas-like objects, T~Tauri and Herbig-Ae stars. In the relatively high galactic latitude fields, due to the lower $\mathrm{N_{H}}$, some sources are identified as extragalactic objects. The fraction of identified AGN is larger towards higher Galactic latitudes, as expected. Around 80\% of the sources are unclassified for the optically bright sample, dropping to 60\% for the optically faint sample, where we could obtain optical spectra up to a limiting magnitude of about 21 in the R band. Spectral classifications, X-ray parameters and catalogue counterparts are given in Tables~8--33, available in the online version of the paper.
\begin{center}
\begin{figure*}
\includegraphics[width=0.5\linewidth,angle=0]{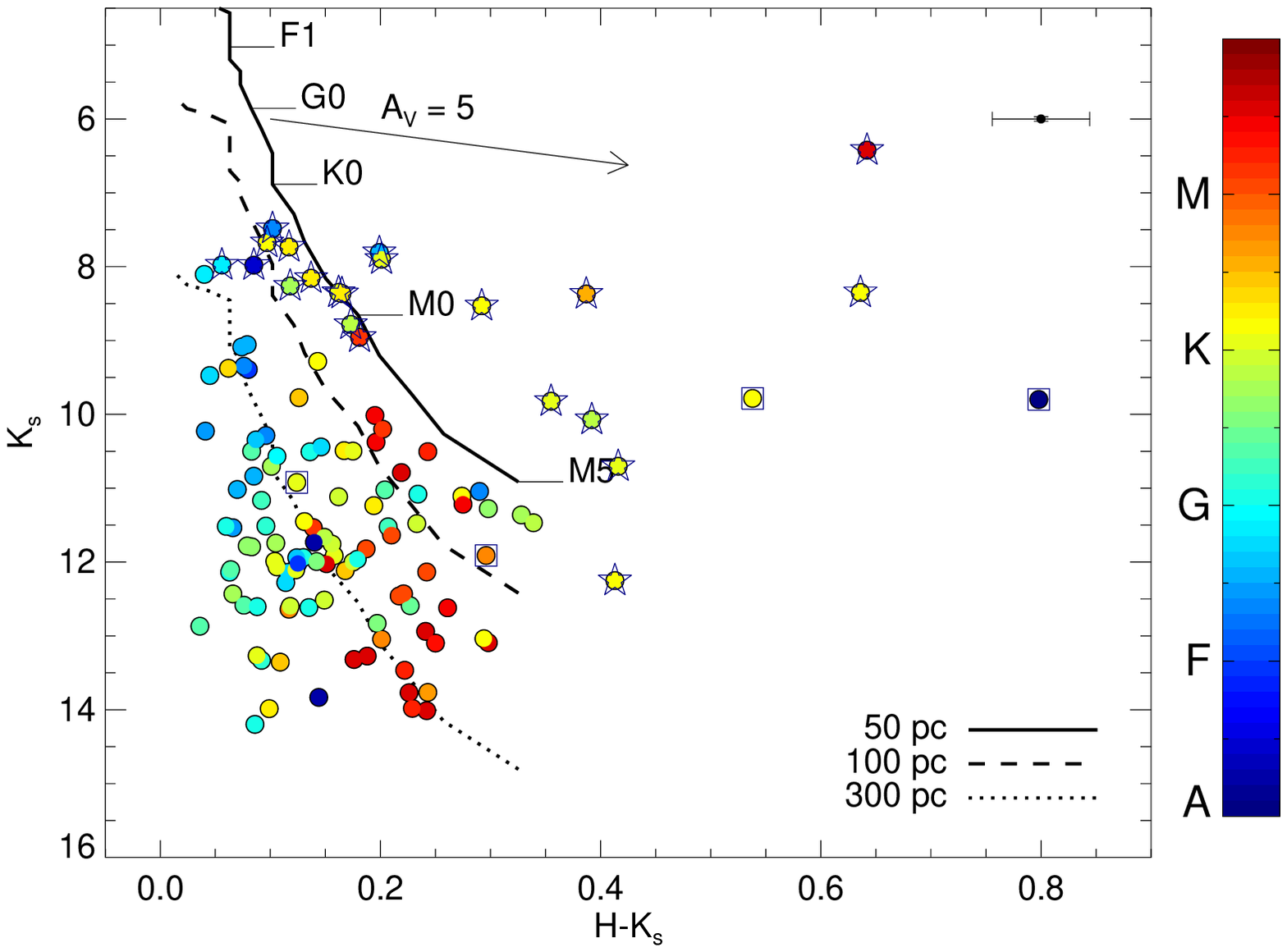}
\includegraphics[width=0.5\linewidth,angle=0]{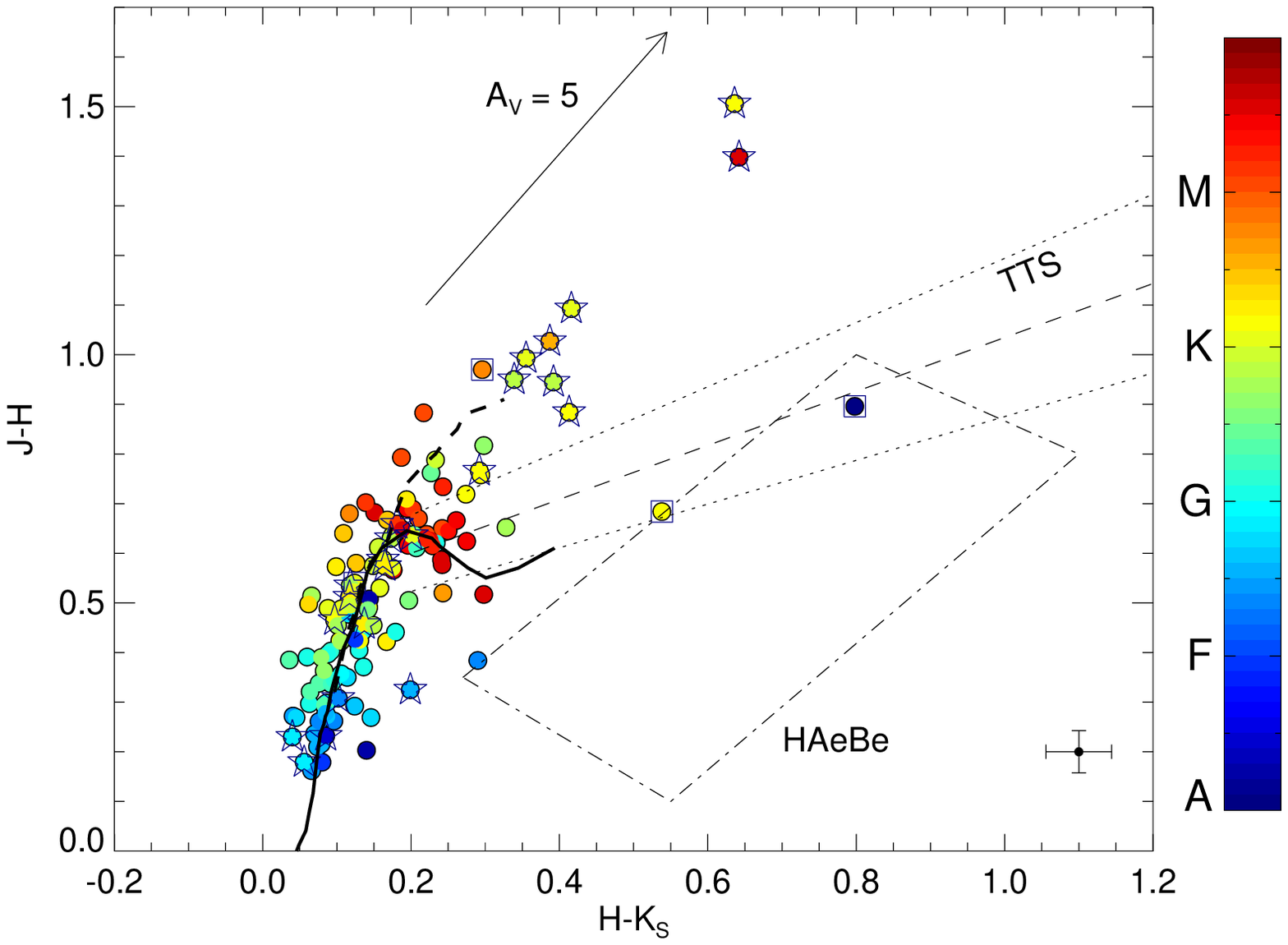}
\caption{Infrared colour-magnitude diagram (left) and colour-colour diagram (right) for the stellar content of our survey, with spectral type of the stars coded with colour. In the left panel the solid, dashed, and dotted lines show the ZAMS location from \cite{siessetal00-1} at a distance of 50, 100, and 300 pc respectively. Giant candidates, are over-plotted with a star-symbol. Pre-main sequence stars are shown with squares. We include the reddening vector for an extinction of A$_\mathrm{V}\,=\,5$ magnitudes. In the right panel, the thick black lines show the location of main sequence (solid) and giant (dashed) stars, the thin black dashed line and the thin black dotted lines show the location of classical T~Tauri stars (TTS) and associated errors respectively. The black parallelogram shows the location of Herbig AeBe stars (see text for details). }\label{g:CMD_giants}
\end{figure*}
\end{center}
\section{Stellar population}
\label{sec:stel_pop}
\subsection{Pre-main sequence, main sequence and giant stars}
\label{sec:giants}
While the majority of the Galactic X-ray-emitting stars are expected to be main sequence stars, a fraction of the stellar population will be evolved giants \citep{guilloutetal99-1}. 
To distinguish between dwarf and giant stars one can use the Balmer lines for stars hotter than about 10\,000 K, or Ca II H and K lines, the NaI doublet and Mg Ib lines for cooler stars, since their equivalent widths are highly dependent on surface gravity. Unfortunately, in most cases our spectral resolution was not high enough and/or our wavelength coverage was not sufficiently large so as to use such line diagnostics. We thus based our classification in dwarfs or giants on photometric analysis. We constructed infrared colour magnitude diagrams (CMD) in order to be able to distinguish giant stars from dwarf stars. In Fig.~\ref{g:CMD_giants} we show the locus of the stellar content of our survey (without correcting the magnitudes for extinction). Based on tabulated colours and absolute magnitudes for main sequence stars \citep{siessetal00-1} we calculated the expected locus of main sequence stars at limiting distances corresponding to five different intervals, one for each spectral type: 50 pc for M stars, 100 pc for K stars, 150 pc for G stars, 200 pc for F stars, and 300 pc for A stars. This main sequence represents a cut-off limit for dwarf stars. With a maximum apparent magnitude close to 8 in the K$_S$ band \citep{cutrietal03-1}, the nearest early M dwarf stars that are detected are going to be at a distance larger than $\sim50$ pc, while no A dwarf star can be at a closer distance than about 300 pc. We considered as giant candidate stars those stars that are found above or to the right of the main sequence in all possible combinations of infrared colour-magnitude diagrams. In the left panel of Fig.~\ref{g:CMD_giants} we show one of the CMDs with the corresponding main sequence cut-off. In the right panel, we show the colour-colour diagram. We include the location of the main sequence and the red giant branch from \cite{bessel+brett88-1}. We indicate the effect of reddening for $\mathrm{A}_\mathrm{V}=5$ transformed into infrared excess using the relations from \cite{mathis90-1}. We show the intrinsic colours of classical T~Tauri stars in Fig.~\ref{g:CMD_giants} \citep{meyeretal97-1}, and of Herbig-AeBe stars \citep{hernandezetal05-1}. We transformed all magnitudes and colours into the 2MASS photometric system using the colour transformation from \cite{carpenter01-1}.

Among the 125 stars with determined spectral type detected in the soft band with 2MASS photometry at high probability of identification, we find 98 main sequence stars (78.4\%), 23 giant candidates (18.4\%), and four pre-main sequence stars (3.2\%) (see Table~\ref{t:giants}). The spectrum of one of the two M giant candidates, 2XMM\,J184413.9+010026, shows a deficit of NaI at 5897\,\AA\ confirming a giant nature of the source. We considered the other M giant candidate, 2XMM\,J182845.5-111710, to be a dwarf since the X-ray luminosity is too high (see Section~\ref{sec:Xlum}) and its spectrum shows \Ha\ in emission, characteristic of dwarf M stars. 
There is an excess of yellow giant stars above the overall trend (see Table~\ref{t:giants}), corresponding to late type G and early K spectral types. This excess was first noticed by \cite{favataetal88-1} based on the Einstein Medium Sensitivity Survey, and confirmed by the larger number of sources of the Einstein Extended Medium Sensitivity Survey \citep{sciortinoetal95-1}. Later on, \cite{guilloutetal99-1} observed a similar excess based on cross-correlation of ROSAT with Tycho and Hipparcos stars, and associated it with the red clump. The yellow-excess was also noted in the Galactic Plane survey from \cite{morleyetal01-1}. As stars evolve off the main sequence, they slow their rotation, decreasing their X-ray luminosity. This age-rotation-luminosity relation implies that single evolved (off main-sequence) stars are not expected to be strong X-ray emitters. On the other hand, in close binaries the rotation of the stars are synchronised with the orbital period. This implies that stars in binary systems do not slow down their spin with age, but on the contrary they preserve their rotation with age, giving rise to X-ray emission through their magnetic activity \citep{frascaetal06-1}. It is thus very likely that the observed giant stars are evolved binary stars, i.~e. RS CVns. We discuss X-ray and infrared properties of these sources in Section~\ref{sec:infra_xray}.

\begin{figure}[t!]
\includegraphics[width=\linewidth,angle=0]{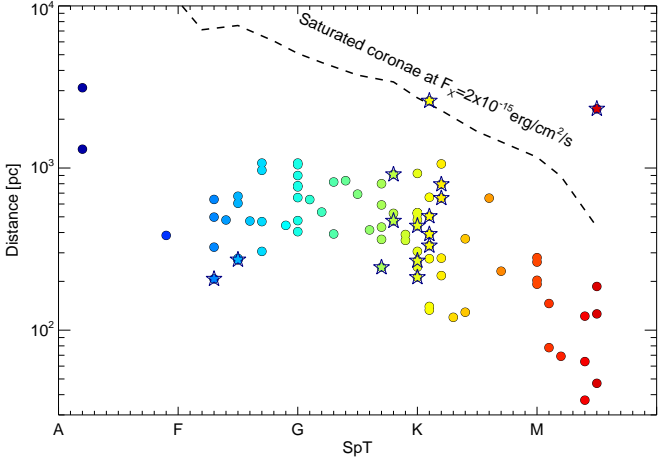}
\caption{Distance versus spectral type (color-coded symbols) for dwarf and giant (star symbol) stars. The limiting distance associated with the saturation level ($\log($\Lx$/$L$_\mathrm{bol})=-3$) of dwarf stars at the flux limit of our survey is indicated with a dashed line.}\label{g:satur_coronae}
\end{figure}
\subsection{Distance}
\label{sec:distances}

\begin{center}
\begin{figure*}
\includegraphics[width=\linewidth,angle=0]{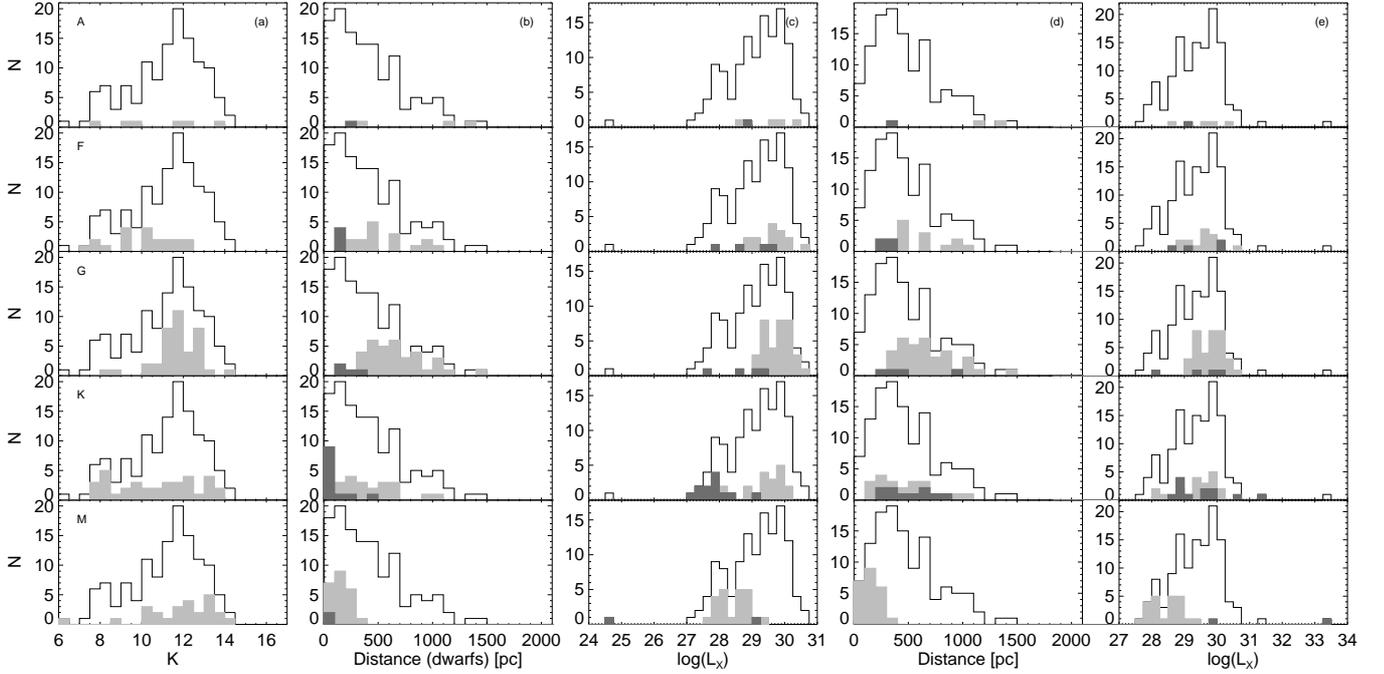}
\caption{a) Distribution of the K magnitude for all identified active coronae. b) Distribution of distances to the identified active coronae assuming they are all in the main-sequence and associated distribution of X-ray luminosities (c). d) Distribution of distances for main-sequence and giant star candidates and corresponding X-ray luminosities (e). We show distances only up to 2\,kpc, since only a few sources, with spectral types A and F, are found at larger distances. We overplot in light grey the distribution for each spectral type, A to M stars from top to bottom on top of the total distribution (white) for comparison. Giant star candidates are shown in dark grey.\label{g:distance2ACs}}
\end{figure*}
\end{center}

We computed the E(J\,--\,K$_\mathrm{s}$) excess assuming tabulated infrared colours for main-sequence and giant stars from \cite{siessetal00-1} and \cite{coveyetal07-1} respectively. We calculated the extinction in the K$_\mathrm{s}$ band using the relation $\mathrm{A_{\mathrm{K}}}\,=\,0.67\times$E(J\,--\,K$_\mathrm{s}$) from \cite{mathis90-1}. We corrected our magnitudes for extinction and using the appropriate tabulated absolute magnitudes M$_{\mathrm{J}}$ for each spectral type and luminosity class, we estimated the distance to the sources. For comparison, we also derived the distances to the sources assuming that all stars were on the main sequence (second panel in Fig.~\ref{g:distance2ACs}). 
Distances to the sources range from 50 pc to about 1\,kpc, with a few sources at larger distances. The distance up to which we can detect stars depends on the spectral type and luminosity class. 
Assuming saturated coronae, $\log($\Lx$/$L$_\mathrm{bol})=-3$, the X-ray flux limit of our survey gives us an upper distance limit for each spectral type (and luminosity class) up to which stars can be detected. The calculated distances for the stars in our survey are consistent with the maximum distance: while main-sequence F stars can be detected up to several kpc, M stars are detected only up to around 450\,pc (see Fig.~\ref{g:satur_coronae}). 

\begin{table} 
\begin{center}
\caption{PMS, MS and giant candidate stars.\label{t:giants}}
\begin{tabular}{lccc}
\hline\hline
\noalign{\smallskip}
SpT   & N$_{\mathrm{pre-MS}}$ & N$_{\mathrm{MS}}$  & N$_{\mathrm{Giant}}$ \\
\noalign{\smallskip}
\hline
\noalign{\smallskip}
A   & 1 &  4 &   1 \\
F   & 0 & 16 &   4 \\
G   & 0 & 35 &   4 \\
K   & 3 & 20 &  12 \\
M   & 0 & 23 &   2 \\
\noalign{\smallskip}
\hline
\noalign{\smallskip}
\end{tabular}
\end{center}
\end{table}

\subsection{X-ray luminosity}
\label{sec:Xlum}
We derived X-ray luminosities from the measured fluxes in the soft (0.5\,--\,2.0\,keV) band. We used the energy-to-flux conversion factor $1.75\times10^{−12}$\,\ergcms/pn counts s$^{-1}$, corresponding to a thermal plasma with kT\,=\,0.5\,keV and absorbed by \Nh$\sim 10^{21}$cm$^{-2}$ most frequent value for the stellar content of our survey (see Fig.~\ref{g:hrs_counterparts}). Details on the calculation are given in Section~\ref{sec:ecfs}. All luminosities from now on will refer to the energy band 0.5\,--\,2.0\,keV, unless specified otherwise. For giant candidate stars we obtained two values: a first one corresponding to the estimated distance if the source were a main sequence star, and a second one corresponding to the distance derived if it were a giant. The X-ray luminosity distributions are shown in the third and fifth panels of Fig.~\ref{g:distance2ACs}. Assuming all the stars in the sample are dwarf stars, a population of low X-ray luminosity ($<10^{28}$\,\ergs) K stars appears at very close distances (dark grey histograms). There is no observational evidence of an excess of low X-ray luminosity K stars within 70 parsecs to the Sun. It therefore seems more plausible that our sample contains two different evolutionary populations. This is consistent with the idea that not all identified stars are on the main sequence. Assuming that the giant star candidates classified in Section~\ref{sec:giants} are giants yields much more consistent results. X-ray luminosities are in the range $10^{28}$\,--\,$10^{31}$\,\ergs\ (see right panel in Fig.~\ref{g:distance2ACs}). 
The number of detected sources increases with the X-ray luminosity up to \Lx\,$=\,10^{30}$\,\ergs. Above that value the number of sources drops. This is a typical characteristic of an X-ray flux limited sample, where we preferentially detect sources with high X-ray luminosities. 

We find that, in general, early spectral type stars have higher X-ray luminosities than late spectral type stars, an effect usually attributed to their larger radii \citep{flemingetal89-1}. A similar trend has been observed by \cite{micelaetal88-1,barberaetal93-1,motchetal97-1,guilloutetal99-1} and \cite{zickgrafetal05-1}. Although such a trend has not been observed by \cite{wrightetal10-1}, other deep surveys such as that from \cite{coveyetal08-1} show the same behaviour, ruling out the possibility of being an effect associated to nearby stars as suggested by \cite{wrightetal10-1}. 

X-ray emission is known to decrease with the age of the stars \citep[][and references therein]{jacksonetal12-1}. Stars slow-down their rotation as they age due to magnetic braking \citep{kawaler1988,matt2012}. This implies that at high X-ray luminosities our survey should be dominated by fast rotators, either young stars or in binary systems, and at low X-ray luminosities the sample should contain an old-to-intermediate age population.

\begin{figure*}[t!]
\begin{center}
\includegraphics[width=\linewidth,angle=0]{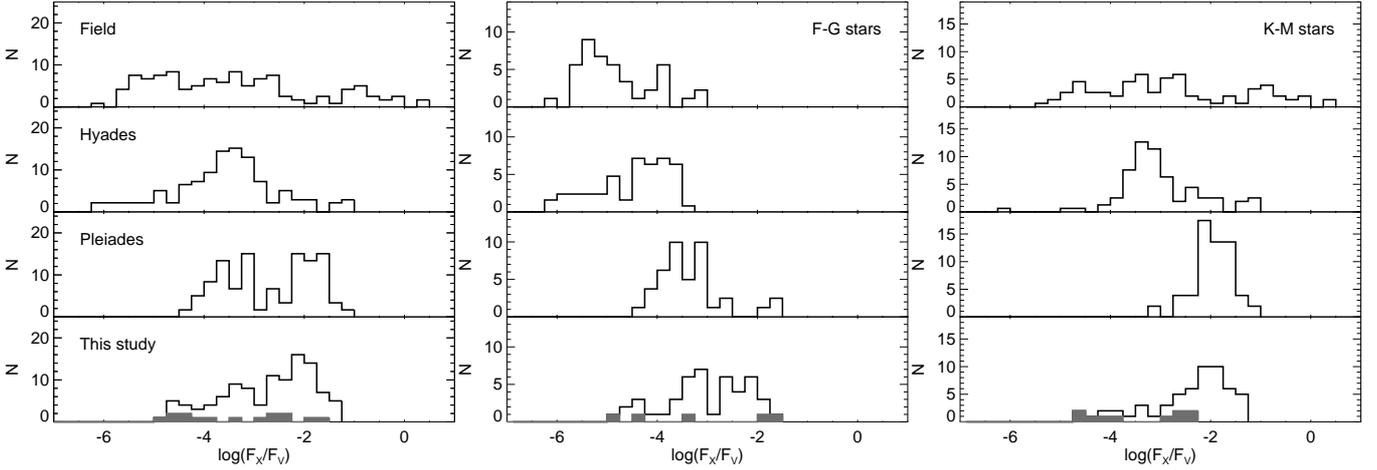}
\caption{X-ray-to-optical flux ratio ($\log$(F$_{\mathrm{X}}$/F$_\mathrm{V}$)) distribution of the stellar population of this survey compared to that of: field, Hyades and Pleiades stars representing old, intermediate and young populations respectively. We show the $\log$(F$_{\mathrm{X}}$/F$_\mathrm{V}$) distribution for spectral ranges A--M (left panel), F--G (middle panel), and K--M (right panel). In dark grey we show the location of giant star candidates. For comparison purposes, the total number of field, Hyades and Pleiades stars has been normalised to match the same number as our stellar content. }\label{g:logFxFv}
\end{center}
\end{figure*}

To assess better this statement we compared our X-ray-to-optical flux ratios with those of stars belonging to three different populations, corresponding to three different ages: i) field stars from \cite{schmitt+liefke04-1} representing an old population with ages older than about 1 Gyr; ii) the Hyades from \cite{sternetal95-1} as representative of an intermediate age of $\sim600$ Myr; and iii) the Pleiades from \cite{micelaetal96-1} representing a young population of about 100 Myr. We restrict the three samples to stars with luminosity class V and consider the same spectral range as covered in our sample, A--M, and divided in two spectral type ranges: F-G and K-M stars (see Fig.~\ref{g:logFxFv}). The X-ray-to-optical flux ratio $\log$(F$_{\mathrm{X}}$/F$_\mathrm{V}$) distribution of the identified stars in our survey is similar to that of the Pleiades with an extended tail at low values of $\log$(F$_{\mathrm{X}}$/F$_\mathrm{V}$), i.~e. is consistent with a young to intermediate age population, such as the one reported in \cite{koenigetal08-1}. There is a population of F--G stars younger than the Pleiades. In the disk X-ray surveys are dominated by young stars in the flux range covered by our survey \citep{guilloutetal96-1}, which explains the detection of this young population of stars.

\subsection{Notes on individual sources}
\label{sec:notes}
Among the identified sources we found CVs, T~Tauri stars, Herbig-Ae stars, and $\gamma$-Cas-like objects. We discuss here the most relevant features of these objects. 

\subsubsection{Cataclysmic variables}
Cataclysmic variables are semi-detached binary stars consisting of a white dwarf plus a low mass star where the low mass star is filling its Roche lobe and the transferred matter is accreted onto the white dwarf \citep[see][for a review]{warner1995}. If the white dwarf has a magnetic field, mass transfer can be channelled and accretion will take place on one or both poles of the white dwarf, labelling the CVs as polars or intermediate polars, for lower magnetic fields white dwarfs. They have orbital periods ranging from around 80 minutes and up to about one day \citep{ritter+kolb03-1,gaensickeetal09-1}.

\begin{figure}
\begin{center}
\includegraphics[width=\linewidth,angle=0]{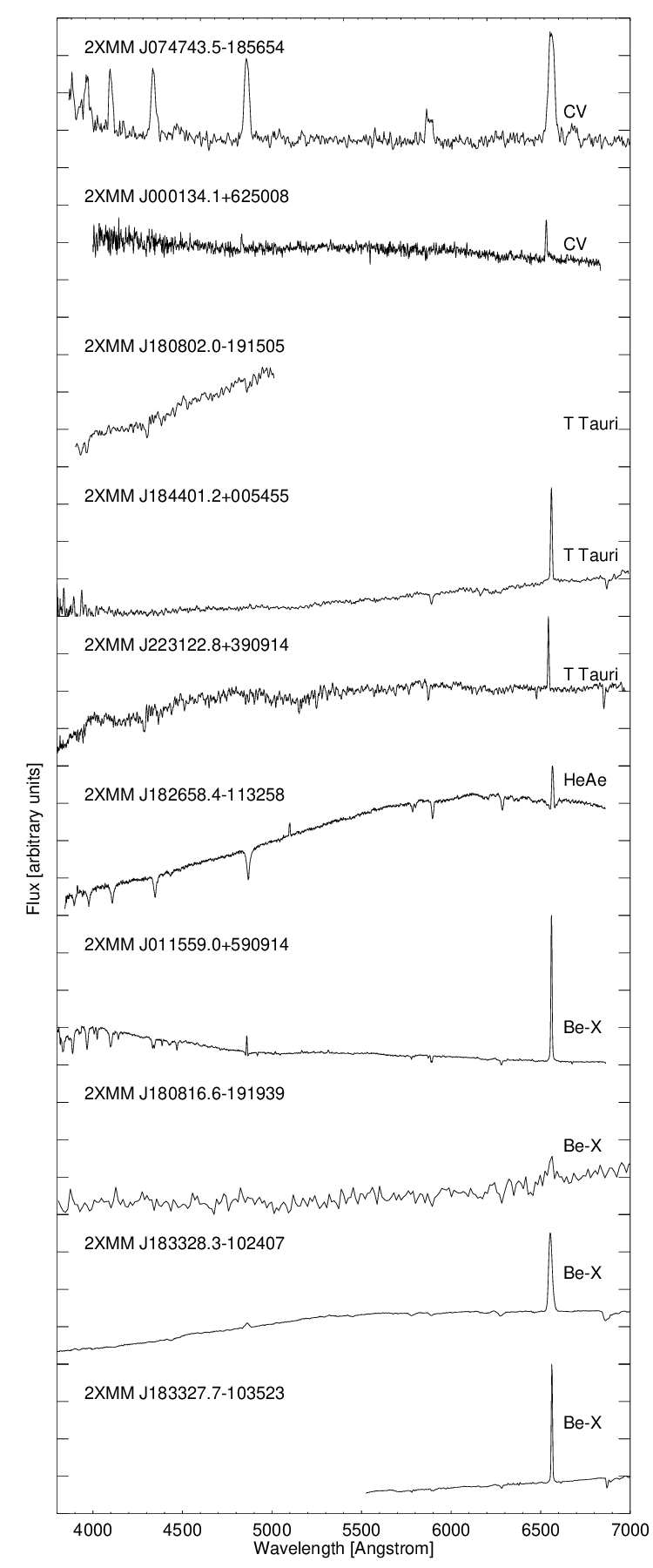}
\caption{Optical spectra of the two identified CVs, 2XMM\,J074743.5-185654 and 2XMM\,J000134.1+625008, the three identified T~Tauri stars, 2XMM\,J180802.0-191505, 2XMM\,J184401.2+005455, and 2XMM\,J223122.8+390914, the identified Herbig-Ae star, 2XMM\,J182658.4-113258, and the four massive X-ray binary candidates, 2XMM\,J183328.3-102407, 2XMM\,J011559.0+590914, 2XMM\,J183327.7-103523, and 2XMM\,J180816.6-191939, from top to bottom. \label{g:all_spec}}
\end{center}
\end{figure}

We identified two cataclysmic variables in this study: 2XMM\,J074743.5-185654, in field PKS~0743-19-off, and 2XMM\,J000134.1+625008, in field RXJ0002+6246. Both CVs display Balmer emission lines (see Fig.~\ref{g:all_spec}), and 2XMM\,J074743.5-185654 has some \HeI\ emission. They both lack the \HeII\ line at 4686\,\AA\ typical of polars and intermediate polars, indicating both CVs are non-magnetic. 

\begin{table}
\begin{center}
\caption{Main properties of the cataclysmic variables.}\label{t:ews}
\begin{tabular}{lccccc}
\hline
\hline
\noalign{\smallskip}
Name                   & EW(\Ha) [\AA] & EW(\Hb) [\AA] & V & D [kpc] \\
\hline
\noalign{\smallskip}
2XMM\,J074743.5-185654 & 140 & 65 & 20.4 & 0.6-2.0 \\
2XMM\,J000134.1+625008 &  6  &  2 & 17.8 & $<2$ \\
\noalign{\smallskip}
\hline
\end{tabular}
\end{center}
\end{table}

Equivalent widths of \Ha\ and \Hb\ emission lines are given in Table~\ref{t:ews}. Making use of the empirical EW(\Hb)--absolute magnitude relation from \cite{patterson84-1} we estimate the absolute magnitude of the accretion disk to be M$_\mathrm{V}\,\sim\,10.3$ for 2XMM\,J074743.5-185654. From the total Galactic absorption along the line of sight from \cite{schlegeletal98-1} and using the calibration from \cite{predehl+schmitt95-1} we derive an upper limit to A$_\mathrm{V}$ of 1.2. We estimate a magnitude of V=20.4 by folding the optical spectrum to the Johnson filter. The distance to the source is in the range 0.6--2.0\,kpc, depending on the assumed A$_{\mathrm{V}}$, implying \Lx$=4.8\times10^{30}$--$5.3\times10^{31}$\,\ergs\ (0.2\,--\,12\,keV). 

Source 2XMM\,J000134.1+625008 has a GSC-2.3 counterpart with magnitude V=17.82. Using the empirical relation from \cite{patterson84-1} would imply a very large distance of about 3.8\,kpc, incompatible with the small interstellar absorption visible in the slope of the optical spectrum. The spectral energy distribution of 2XMM\,J074743.5-185654 seems to consist of a blue continuum probably coming from the accretion disk and of a possible late K star continuum in the red. However, our spectrum is too noisy to detect metallic lines from the secondary star. The empirical EW(\Hb)--Mv relation from \cite{patterson84-1} is based on CVs in which the disk is dominating in the optical and therefore cannot probably be used in this particular case. Assuming a maximum X-ray luminosity of \Lx$<10^{32}$\,\ergs\ for non-magnetic CVs \citep{pretorius+knigge12-1}, the source must be closer than about 2\,kpc.

\subsubsection{T~Tauri stars}
T~Tauri stars are low mass (M$\le$2\Msun) pre-main sequence stars \citep{joy1945}. In \emph{Classical T~Tauri} stars (cTTS) the central object is accreting from a circumstellar disk, while in \emph{weak line T~Tauri} stars (wTTS) there is no longer evidence of accretion \citep[see][for a definition of the two sub-classes]{bertout1989,appenzeller1989}. They can be classified into weak or classical T~Tauri on the basis of the equivalent width of the \Ha\ emission line \citep{barradoetal03-1}. T~Tauri stars are young objects typically found in star forming regions. Their X-ray energy distribution can be described as thermal emission from an optically thin plasma with temperatures from few to several tens of MK and X-ray luminosities in the range $\times10^{28}-10^{31}$\,\ergs\ (0.5\,-\,10\,keV), mainly associated with the magnetically heated corona \citep[][and references therein]{guedel+naze09-1}.

\begin{figure*}[!t]
\begin{center}
\includegraphics[width=\linewidth,angle=0]{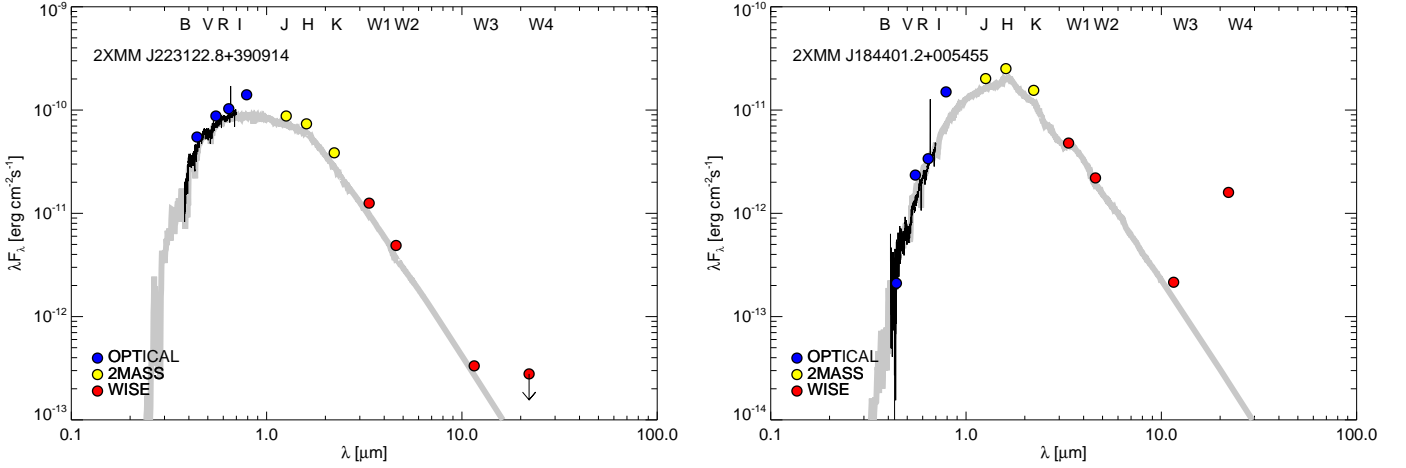}
\caption{Spectral energy distribution of two T~Tauri candidates: 2XMM\,J223122.8+390914 (left) and 2XMM\,J184401.2+005455 (right). We compare the SED of the T~Tauri stars with stellar atmosphere models (grey) from \cite{castelli+kurucz04-1} with effective temperatures $5200$ K and $4000$ K respectively, and $\log$(g)$\,=\,4.5$. The optical spectra obtained at the telescope are shown as a black line, and upper limits on the magnitude are indicated with an arrow. \label{g:SED}}
\end{center}
\end{figure*}

Source 2XMM\,J223122.8+390914, detected in field 3C449, has spectral type K0Ve. Its optical spectrum presents the \Ha\ line in emission, with an equivalent width of $\sim6$\,\AA, pointing to a wTTS nature \citep{barradoetal03-1}. It has an infrared counterpart in the WISE catalogue \citep{cutrietal12-1}, WISEP\,J223122.86+390913.7, which does not reveal any infrared excess with respect to a stellar atmosphere (see Fig.~\ref{g:SED}). Source 2XMM\,J223122.8+390914 is located at relatively high Galactic latitude ($l\,=\,95.3^\circ, b\,=\,-16.1^\circ$), far from any known star formation region, where T~Tauri stars are typically found. Nevertheless, the number of reported isolated T~Tauri stars is continuously increasing \citep{guilloutetal10-1}. The existence of T~Tauri stars far away from star forming regions containing molecular gas is a matter of lively debate. While kinematic studies suggest some of the stars classified as T~Tauri are in fact young main sequence objects \citep{bertout2006}, several mechanisms accounting for the apparent dispersal of T~Tauri stars have been proposed \citep{feigelsonetal96-1}. 

Source 2XMM\,J184401.2+005455, in field Ridge~2, with spectral type K7Ve, also shows the \Ha\ line in emission. The EW of the \Ha\ line is $\sim25$\,\AA, pointing to a cTTS nature for the source \citep{barradoetal03-1}. The optical images reveal that the source is close to a cloud. The source has a WISE counterpart, WISEP\,J184401.16+005456.3. The spectral energy distribution (SED), presented in Fig.~\ref{g:SED}, is consistent with a stellar atmosphere in the optical and NIR, and exhibits an excess in the 22 $\mu$m band. For comparison we show the SED based on a stellar model from \cite{castelli+kurucz04-1} with effective temperature $\mathrm{T_{\mathrm{eff}}}$\,=\,4000 K, and $\log$(g)$\,=\,4.5$. The most likely origin for the IR excess is cold dust emission from a transition disk, where the inner dust disk has dissipated while the disk at larger distance is still integral \citep[e.\,g.][]{fangetal09-1}. 
These two candidate T~Tauri stars are at a maximum distance of 300 pc, with maximum X-ray luminosities of $6.1\times10^{28}$ and $3.8\times10^{28}$\,\ergs\ (in the 0.5\,-\,2\,keV energy band), respectively. Their X-ray to bolometric luminosity ratio, $\log($\Lx$/$L$_\mathrm{bol})$, is below the saturation level. 

2MASS infrared photometry revealed an infrared excess ((H\,--\,K$_{\mathrm{S}}$)\,$>$\,0.5) for source 2XMM\,J180802.0-191505, in field WR110. The source has infrared colours within the T~Tauri region from \cite{meyeretal97-1} (see right panel in Fig.~\ref{g:CMD_giants}), consistent with a classical T~Tauri. From its optical spectrum we derived a K1V spectral type. The optical spectrum covers only up to 5000\,\AA, inspection of the \Ha\ line was therefore not possible, so we were not able to learn whether accretion is taking place, which would confirm a classical T~Tauri nature of the source. No emission from other Balmer lines was detected in the spectrum. After comparison of the WISE source catalogue with the 2MASS point source catalogue, we conclude that the angular resolution of WISE is too low to provide us a reliable photometry of this source in this crowded area.
The source was sufficiently bright in X-rays, which allowed us to analyse its X-ray spectrum. The X-ray spectrum is consistent with that of a young star, with a temperature of kT\,=\,$1.4_{-0.5}^{+0.6}$\,keV, and the derived \Nh\ is compatible with the reddening obtained from the optical spectrum, $\mathrm{A}_{\mathrm{V}}\,=\,2.17$. 

The lithium content can be used as a spectral type dependent proxy of the age of the stars. Unfortunately, at our spectral resolution we are not able to measure accurate Li abundances. High resolution spectra would be needed to make a clear statement about the evolutionary state of these stars.

\subsubsection{Herbig Ae stars}
The optical spectrum of source 2XMM\,J182658.4-113258  (see Fig.~\ref{g:all_spec}), located in field Ridge3, reveals an A0 spectral type star with a very strong and broad \Ha\ emission line. The X-ray source has an infrared counterpart, 2MASS\,18265832-1132585, with colour (H\,--\,K$_{\mathrm{S}}$)\,$=\,0.79 \pm 0.06$, showing a strong near infrared excess with respect to a main-sequence or giant A0 star (see Fig.~\ref{g:CMD_giants}). These two characteristics, \Ha\ emission plus infrared excess, are typical of Herbig-Ae (HeAe) stars \citep{herbig1960}, which are pre-main sequence stars of intermediate mass, i.e the high mass counterparts to T~Tauri stars. Broad \Ha\ emission and infrared excess is associated with dust emission of circumstellar material \citep{malfaitetal98-1,meeusetal98-1,meeusetal01-1}. Nevertheless we note that the K$_{\mathrm{S}}$ magnitude has a flag indicating bad quality (AAE, where E is standing for a bad PSF fitting), so one should be cautious with the conclusions. The origin of the X-ray emission in HeAe stars is still in debate. A stars are fully radiative stars, therefore not expected to maintain magnetic fields and coronal heating. Possible explanations, proposed earlier by \cite{zinneckeretal94-1} and \cite{stelzeretal06-1}, are stellar wind instabilities or presence of a T~Tauri star companion. 

\subsubsection{Massive X-ray binary candidates}
High mass X-ray binaries (HMXB) consist of a white dwarf (WD), a neutron star (NS) or a black hole and of an early type star dominating the optical emission. Most known HMXBs have a Be star as primary component. Although theoretical population studies predict Be+WD are more frequent than Be+NS \citep[e.\,g.][]{raguzova01-1}, there are only two candidate HMXBs with a WD known so far \citep{kahabkaetal06-1,sturmetal12-1}.
Unlike in CVs and LMXBs mass transfer in Be HMXBs occurs through an equatorially condensed decretion disk \citep{okazaki+negueruela01-1}. They usually have long orbital periods, of the order of days to hundreds of days. Be/X-ray binaries typically have X-ray luminosities in the range $10^{34}-10^{38}$\,\ergs\ \citep{grimmetal02-1}.

In four cases, we have detected X-ray emission from Be stars at levels significantly above that expected for single Be stars of the same spectral type \citep{cohenetal97-1,cohen00-1}. Their optical spectra are shown in Fig.~\ref{g:all_spec}. 

2XMM\,J183328.3-102407 (USNO\,0750-13549725 C), detected in field G21.5-09 (rev 60), is the brightest star in the cluster NGC\,6649. The spectral type of the source is B1-1.5IIIe, and it presents the \Ha\ line in emission with an EW of $\sim36$\,\AA. The source is identified with 2MASS\,18332830-1024087 with a probability above 99\%. We calculated the colour excess E(J-K$_\mathrm{s}$) with respect to a B1III star, and from its magnitude K$_\mathrm{s}=7.793 \pm 0.027$ we estimated a distance to the source of about 1.7\,kpc, consistent with that to the cluster \citep{walkeretal87-1}. The colour excess is E(B-V)$\sim1.2$ \citep{turner81-1}, and the X-ray luminosity is $\sim5\times10^{32}$\,\ergs\ (0.2\,--\,12\,keV). 

2XMM\,J183327.7-103523, aka SS\,397, detected in the field G21.5-09, has a spectral type B0Ve and also shows the \Ha\ line in emission, with an EW of $\sim34$\,\AA.  It has a 2MASS counterpart, 2MASS\,18332777-1035243, with an identification probability above 99\% and K magnitude of $8.267 \pm 0.026$, and an extinction $\mathrm{A}_\mathrm{V}\sim5.4$. The calculated distance of 1.5\,kpc implies an X-ray luminosity of $\sim4.4\times10^{32}$\,\ergs\ (0.2\,--\,12\,keV). The two sources above have already been reported in \cite{motchetal07-1}.

2XMM\,J011559.0+590914 (TYC\,3681-695-1), detected in the field of PSRJ0117+5914, has a counterpart in the 2MASS catalogue, 2MASS\,01155905+5909141, with an identification probability above 99\%. The estimated spectral type from the optical spectrum is B1-2III/Ve, presenting the \Ha\ line in emission with an EW $\sim31$\AA. The optical extinction is A$_{\mathrm{V}}\sim2.3$, and the 2MASS magnitude implies a distance to the system in the range of 1.9\,--\,3.8\,kpc. The X-ray luminosity is between $1.4\times10^{32}$\,\ergs\ and $5.8\times10^{32}$\,\ergs\ (0.2\,--\,12\,keV). 

For source 2XMM\,J180816.6-191939, located in the field of WR110, already reported as a Be/X-ray binary candidate by \cite{motchetal03-1}, we took several spectra, and combined them to improve the signal to noise ratio in order to facilitate the identification. The combined optical spectrum (see Fig.~\ref{g:all_spec}), shows the \Ha\ line in emission, with an EW of $\sim 50$\AA. The relatively noisy spectrum makes it hard to discern absorption lines that could help to characterise the source and estimate its spectral type, but the absence of the TiO molecular bands rules out a dMe star nature. Although it has a 2MASS counterpart with a low probability of it being the true association (\Pid$\sim$8\%), the presence of emission lines in the optical spectra confirm it as the right association. From the optical spectrum, we derived R\,=\,22 and R--I\,=\,2.6, indicating a high interstellar absorption. If the counterpart is an intrinsically blue object, A$_\mathrm{V}$ is of the order of 13.8, where E(B-V) was derived from the spectral fit with a B0V star \citep{castelli+kurucz04-1}. This value corresponds to \Nh\,$=\,2.47\times10^{22}$ cm$^{-2}$, half of the total expected Galactic absorption in that direction. Using the extinction maps from \cite{marshalletal06-1}, with an infrared extinction of A$_\mathrm{K}=1.23$ we estimated the distance to the source to be between 6 to 7\,kpc, ruling out a CV nature. At this distance, we estimate M$_\mathrm{V}$ to be in the range $-3.3$ to $-3.7$, absolute magnitude which is consistent with a B0V star. The X-ray luminosity is in the range \Lx$=2.4\times10^{32}$--$3.3\times10^{32}$\,\ergs\ (0.2\,--\,12\,keV).

Luminosities in this section were calculated in the 0.2\,--\,12\,keV energy band using the corresponding energy-to-flux conversion factor at the estimated galactic absorption and assuming a mekal with kT\,=\,8\,keV. All Be stars found in this study have X-ray luminosities $10^{32}\,<$\,\Lx\,$<\,10^{33}$\,\ergs, lower than those expected for neutron star accreting HMXBs \citep{grimmetal02-1}, but higher than those expected for single Be stars \citep{cohenetal97-1}. The observed X-ray luminosity of our Be stars is at least one order of magnitude below that detected from transient Be/X-ray binaries in quiescence or from members of the class of persistent Be/X-ray pulsars \citep{reigetal11-1}. The absence of recorded outburst from these Be stars together with their low steady Lx make it very likely that these sources belong to the new class of $\gamma$-Cas-like Be/X-ray systems \citep{motchetal07-1,oliveiraetal10-1,oliveiraetal11-1} in which X-ray emission arises either from the accretion onto a white dwarf or from magnetic interaction between the stellar photosphere and the inner part of the decretion disc. \cite{marco2007} have shown that the counterpart of 2XMM\,J183328.3-102407 is a blue straggler in the 50\,Myr old cluster NGC\,6649. A similar blue straggler nature has been established for other $\gamma$-Cas analogues \citep[see][and references therein]{lopes2007}. $\gamma$-Cas-like stars seem therefore preferentially created thanks to the high mass transfer occurring during the evolution of a massive close binary. The question remains open, however, whether the outstanding X-ray emission is related to the presence of a compact companion star remain or due to the high angular momentum transferred to the B star. 

\begin{figure*}
\includegraphics[width=0.5\linewidth,angle=0]{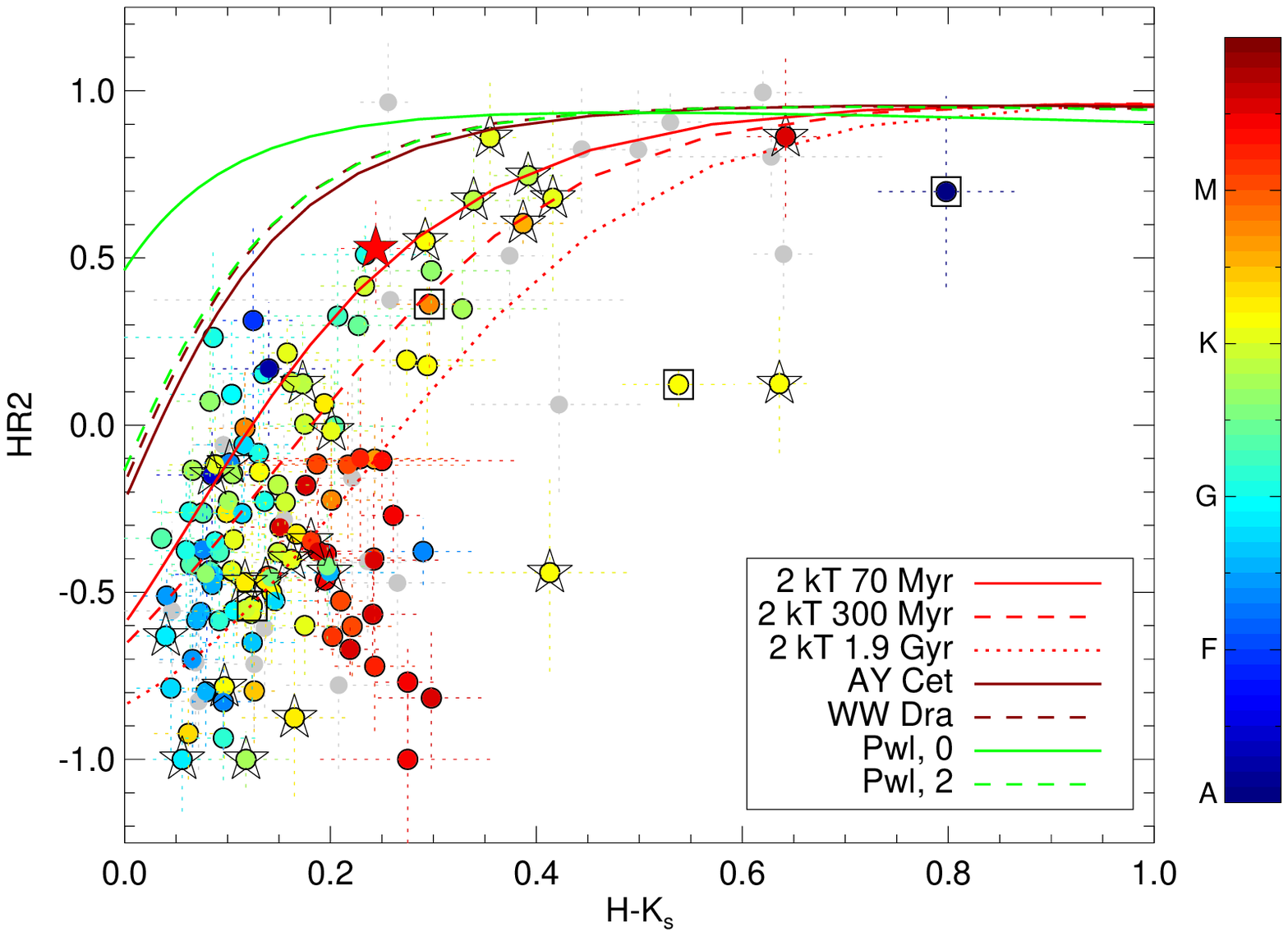}
\includegraphics[width=0.5\linewidth,angle=0]{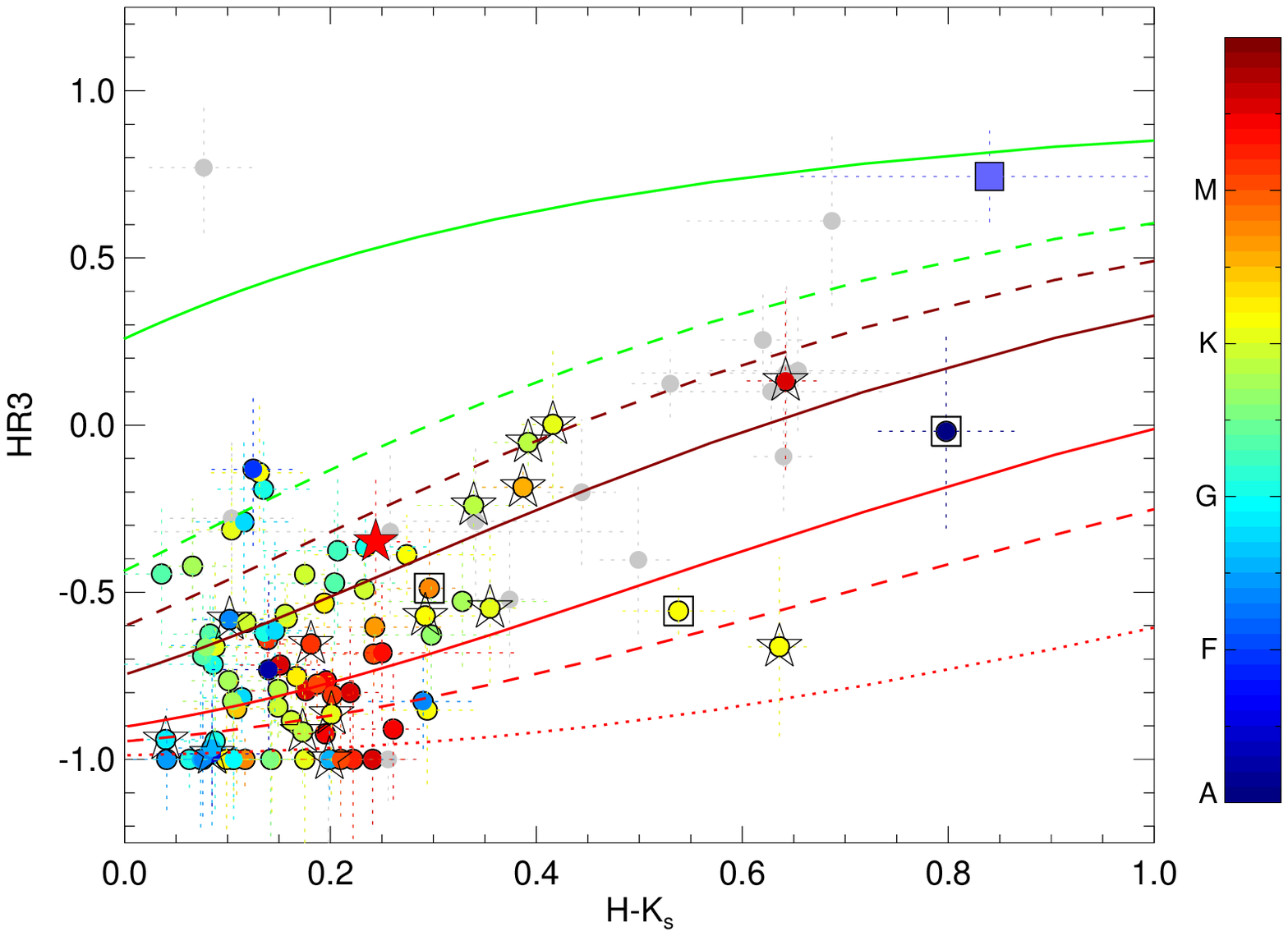}
\caption{X-ray-infrared colour-colour diagrams. Hardness ratio versus (H\,--\,K$_{\mathrm{S}}$) for all X-ray sources having a 2MASS counterpart with a probability higher than 90\%. Active coronae are represented by filled circles where colour indicates the spectral type, extragalactic objects with blue filled squares. Giant stars are overplotted as stars, pre-main sequence stars are highlighted by an open square and X-ray binaries (CVs, HMXBs) are plotted as a filled star in red. Lines are the expected colours for stars of different ages, main sequence and evolved binaries, and accreting objects (see legend and text). We only plot sources with errors lower than 0.3 in X-ray and infrared colours. \label{g:hr_hk}}
\end{figure*}

\section{Properties and characteristics of the sample}
\label{sec:properties}
We investigated the X-ray, infrared and optical properties of the sample with the aim of learning about the stellar content of our survey. 

\subsection{Infrared versus X-ray}
\label{sec:infra_xray}
The different hardness ratio versus (H\,--\,K$_{\mathrm{S}}$) colour distributions are shown in Fig.~\ref{g:hr_hk}. Since the colour (H\,--\,K$_{\mathrm{S}}$) remains within $0$ to $+0.1$ from A to K spectral types and all luminosity classes \citep[see][]{coveyetal07-1}, variations in colour mainly reflect variations in interstellar absorption \citep{motch+pakull12-1}, or intrinsic emission by circumstellar matter. Assuming intrinsic colour (H\,--\,K$_{\mathrm{S}}\,=\,0$), we calculated the expected hardness ratios as a function of the colour excess E(H\,--\,K$_{\mathrm{S}}$), through the expression \Nh\,$=\,3.5\times10^{22}\times$E(H\,--\,K$_{\mathrm{S}}$)\,cm$^{-2}$, for different kinds of objects: i) three different populations of active coronae: a young population of 70 Myr, an intermediate population of 300 Myr, and an old population of 1.9 Gyr, assuming 2--T thermal emission \citep{guedeletal97-1}; ii) for AY Cet, a typical BY\,Dra binary \citep{dempseyetal97-1}, and for the RS\,CVn star WW\,Dra \citep{dempseyetal93-1}; and iii) two power laws with photon indices $\Gamma$ of 0 and 2 (see Section~\ref{sec:xray-models}).

We limit to sources with X-ray HRs and infrared colours with errors smaller than 0.3. After this cut, we are left with a total of 159 sources, among which 124 are identified as stars, one as a Be/X, two as T~Tauri stars, one as HeAe star, two as AGNs, and the remaining 29 are not classified. Most of the stars have colours consistent with the expected values for active coronae younger than 2 Gyr and BY\,Dra or RS\,CVn binaries. 
The majority of the stars with HR2\,$>$\,0 have spectral types in the range G--K, with only a minor contribution from earlier and later spectral type stars. There is one M star (2XMM\,J184413.9+010026) with a very hard HR, classified in the previous section as a giant star candidate on the basis of its infrared colours. The lack of emission lines in its optical spectrum indicates that the star is not a symbiotic binary. Two among the three pre-main sequence stars found in our sample (2XMM\,J184401.2+005455 and 2XMM\,J182658.4-113258) and one Be/X-ray binary (2XMM\,J011559.0+590914) have HR2\,$>$\,0 (see left panel in Fig.~\ref{g:hr_hk}). 

Some stars have HRs significantly higher than expected for stars younger than 70 Myr but consistent with expected values for BY\,Dra or RS\,CVn binaries. The fraction of stars above the expected HR for 70 Myr old active coronae represents 3\% (0\%) of the identified stars in the HR2 bands, 37\% (8\%) in the HR3 bands, and up to 78\% (56\%) in the harder HR4 bands at a 1$\sigma$ (3$\sigma$) significance, and have infrared colours consistent with main sequence or evolved binary stars.

There are eight sources above the expected value for high accretion rate sources ($\Gamma\,=\,0$) (in either HR2\,--, HR3\,-- or HR4--(H\,--\,K$_{\mathrm{S}}$) diagram): the two identified AGNs, three stars, and three unidentified sources, likely to have extragalactic origin. One of the identified stars, 2XMM\,J174819.7-280727 with spectral type M6V \citep{rahartoetal84-1} is very close to the HMXB SAX\,J1748.2-2808 \citep{sidolietal06-1}. However, the two sources are well separated and furthermore 2XMM\,J174819.7-280727 is constantly detected during three observations. This implies that flaring from the M6V star is unlikely to account for the unusually hard X-ray spectrum. The other three identified stars displaying hard X-ray emission are 2XMM\,J092531.1-474851, and 2XMM\,J223021.2+392253. They have spectral types K7V, and G5V and 2MASS identification probabilities of \Pid\,$=\,0.99$, $0.96$, and $0.97$ respectively, and thus are not likely to be spurious associations. 

\begin{figure*}
\includegraphics[width=0.5\linewidth,angle=0]{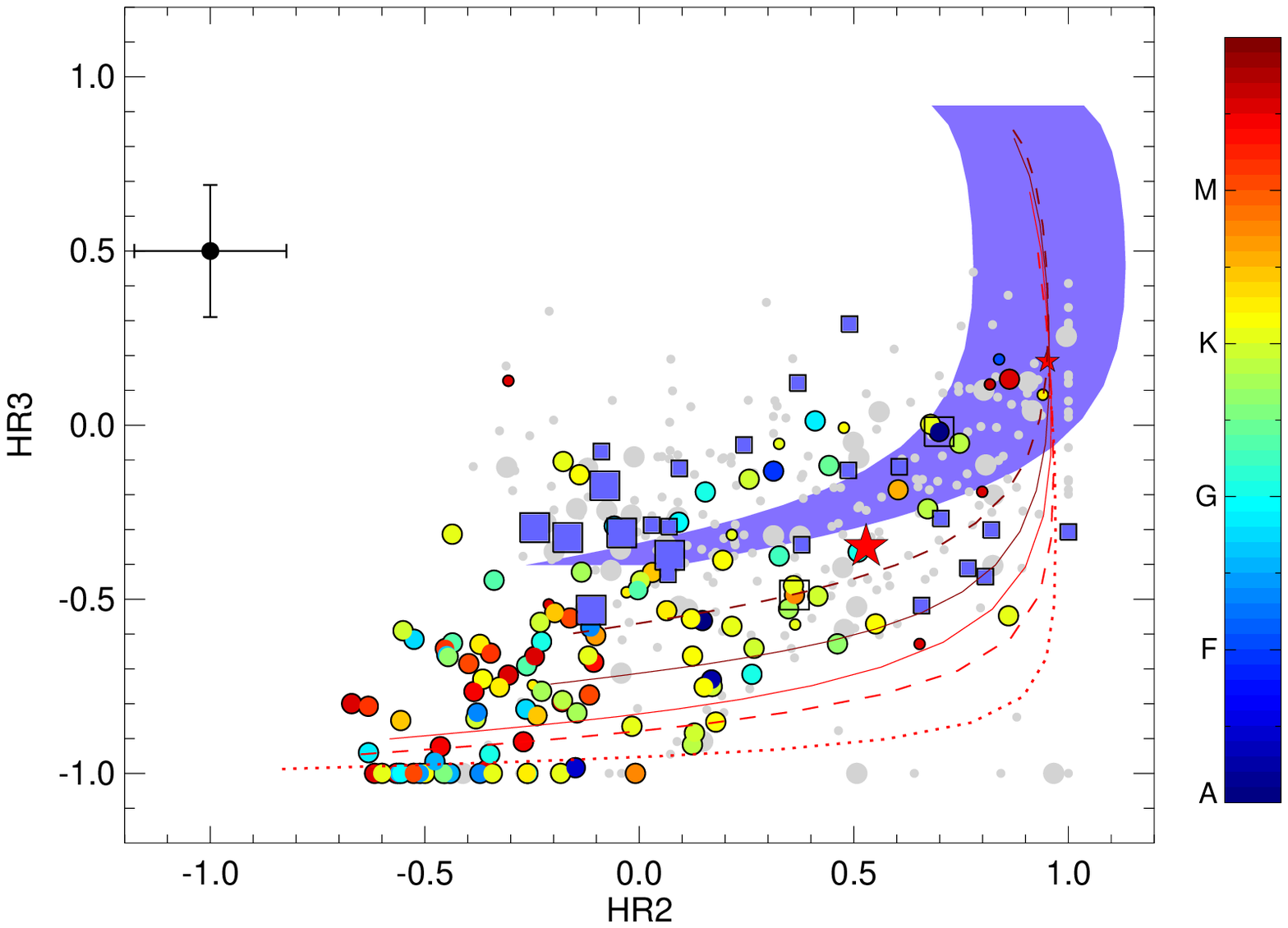}
\includegraphics[width=0.5\linewidth,angle=0]{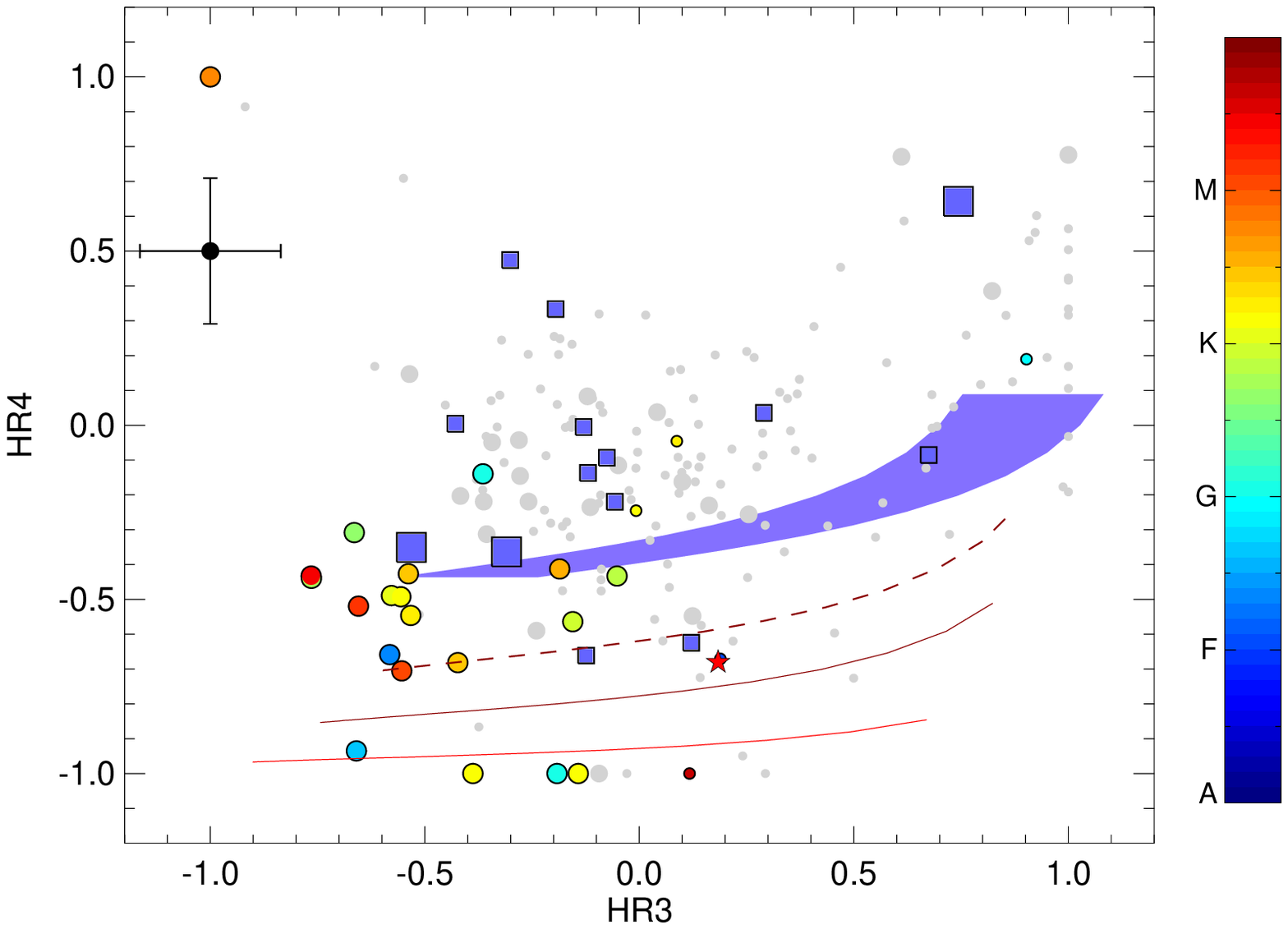}
\caption{Hardness ratio distribution for all detected X-ray sources (small symbols) and for those with infrared or optical counterpart with a probability higher than 90\% (big symbols). Grey are unidentified sources, coloured circles are active coronae with colour indicating their spectral type, blue filled squares are extragalactic objects, open squares highlight the pre-main sequence stars (T~Tauri and HeAe), and filled red star are X-ray binaries (CVs and HMXBs). We limit to sources with HR errors smaller than 0.3 and with $\mathrm{sum\_flag}\,=\,0$. The mean errors are plotted in the upper left corners. For comparison we show the expected hardness ratio values for a typical AGN (power law with photon index $\Gamma\,=\,1.9$) and for $10^{20}<$\,\Nh\,$<10^{23}$ cm$^{-2}$, as a shaded area with a width equal to the mean HR2 and HR3 errors in the left and right panels respectively. Lines indicate the position of stars of different ages and main-sequence and evolved binaries (see legend in Fig.~\ref{g:hr_hk}). \label{g:hr2_hr3_hr4}}
\end{figure*}

\subsection{X-ray colours}
The hardness ratio distributions (X-ray colour-colour diagrams) of X-ray sources are shown in Fig.~\ref{g:hr2_hr3_hr4}. We again restrict our analysis to sources with HR errors smaller than 0.3 and with the best quality flag. The left (right) panel of Fig.~\ref{g:hr2_hr3_hr4}, HR2--HR3 (HR3--HR4) diagram, is populated by 404 (193) sources, 37\% (23\%) have a bright optical or infrared counterpart (association probability $>90\%$). The majority of the sources with HR2 errors below 0.3 have HR3 below --0.25, consistent with a soft X-ray spectrum characteristic of active coronae. Among the sources with bright optical or infrared associations, 70\% have been classified as stars using optical spectroscopy, while the extragalactic sources represent 4\% of the bright optical/infrared associations. Thirteen spectroscopically identified sources have optical and infrared counterparts with an identification probability lower than 90\%. These relatively faint stars have late spectral types (one F, five K, and seven M stars). 

As stated in Section~\ref{sec:infra_xray}, a number of hard sources with bright optical or infrared counterparts have been classified as active coronae on the basis of their optical spectra (see right panel in Fig.~\ref{g:hr2_hr3_hr4}). We recall that, at the imposed 90\% threshold the number of expected spurious matches is lower than 2\%. There are 193 sources in the HR3 versus HR4 diagram (with $\sigma_{\mathrm{HR_{i}}}<0.3$), hence we expect a maximum of 4 hard sources associated with spurious matches, which is not enough to explain the identified coronae, with bright optical or infrared counterpart, detected in the hard bands. 

The majority of the unidentified sources do not have a bright optical or infrared counterpart, and have hardness ratios consistent with those expected for a typical AGN with a power law spectrum of photon index of $\Gamma\,=\,1.9$ absorbed by the full Galactic line of sight column density (blue region in Fig.~\ref{g:hr2_hr3_hr4}). A non negligible number of unidentified sources are found to be softer than expected if extragalactic. The nature of these sources is difficult to guess. A fraction among them could be distant dM stars that are too faint to be detected in infrared or optical surveys and fainter than our limit for spectroscopic identification (R$\sim$21). Assuming a saturated corona ($\log($\Lx$/$L$_\mathrm{bol})\,=\,-3$), at the limit of our X-ray survey (F$_{\mathrm{X}}\sim2\times10^{-15}$\,\ergcms), dM stars can only be detected up to about 450\,pc (see Fig.~\ref{g:satur_coronae}). 

\subsection{\lnls\ curves}
We computed the number of sources per square degree detected above a given flux (\lnls) in two energy bands: soft (0.5\,--\,2.0\,keV) and hard (2\,--\,12\,keV). We restricted our analysis to fields observed in full window mode and with the EPIC pn camera. Fields Ridge\,3 and Ridge\,4 cover nearly the same sky area. We only considered field Ridge\,3 in this analysis since it has the longest exposure time. This leaves us with a total of 18 fields.

\begin{figure}
\includegraphics[width=\linewidth,angle=0]{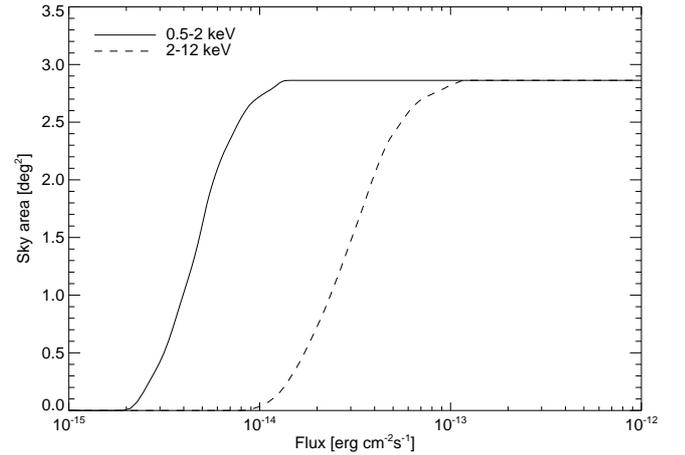}
\caption{Effective area of the survey in the soft (solid line) and the hard (dashed line) energy bands.\label{g:eff_area}}
\end{figure}

\subsubsection{Effective area of the survey}
\label{sec:eff_areas}
To estimate the effective area of our survey, we first built sensitivity maps following the method described by \cite{carreraetal07-1}. For each band and field, we created: i) exposure maps, that contain information on exposure times in each detector pixel taking into account the mirror vignetting, the detector efficiency, bad pixels and CCD gaps, and the field of view; and ii) background maps, which hold information on the counts in source free regions. We used the SAS tasks \verb eexpmap  and \verb esplinemap \, respectively. 
We calculated the count rate in the soft and the hard bands by adding the tabulated count rates in the bands 0.5\,--\,1.0 and 1.0\,--\,2.0\,keV, and 2.0\,--\,4.5 and 4.5\,--\,12.0\,keV respectively. We derived maximum likelihood (ML) values in the soft and hard energy bands making use of the listed individual ML, and using the formula given by \cite{watsonetal09-1} (see their Appendix). We limited our analysis to sources with ML $>$8, which corresponds to a $\sim 4\sigma$ detection. We chose a radius of 5.08 and 5.18 pixels \citep[see Table\,A.1 from][]{carreraetal07-1} for source extraction in the soft and the hard bands respectively. We then generated the sensitivity maps following the different steps listed by \cite{carreraetal07-1}. These sensitivity maps provide us with the minimum count rate that can be detected in each pixel at a given ML. The effective area over which we are sensitive to a given count rate is given by the sum of the number of pixels in the sensitivity map below a given count rate multiplied by the pixel area (1pix $=$ (4$\arcsec$)$^2$). The total area at a given count rate is then the sum of the areas of each field (see Fig.~\ref{g:eff_area}). 

With a ML $>8$, the probability of a spurious detection at this level is $\sim 0.0003$ per resolution element. The total area of the survey is $\sim 2.8$\,deg$^2$, and the individual detection cell used by the task \verb eexpmap  \, is $5\times5$ pixels. We thus expect a maximum of 27 spurious detections in each band, corresponding to around $6\%$ of the sources in the soft band and around $12\%$ of the sources in the hard band. 

\begin{figure*}[!t]
\includegraphics[width=0.5\linewidth,angle=0]{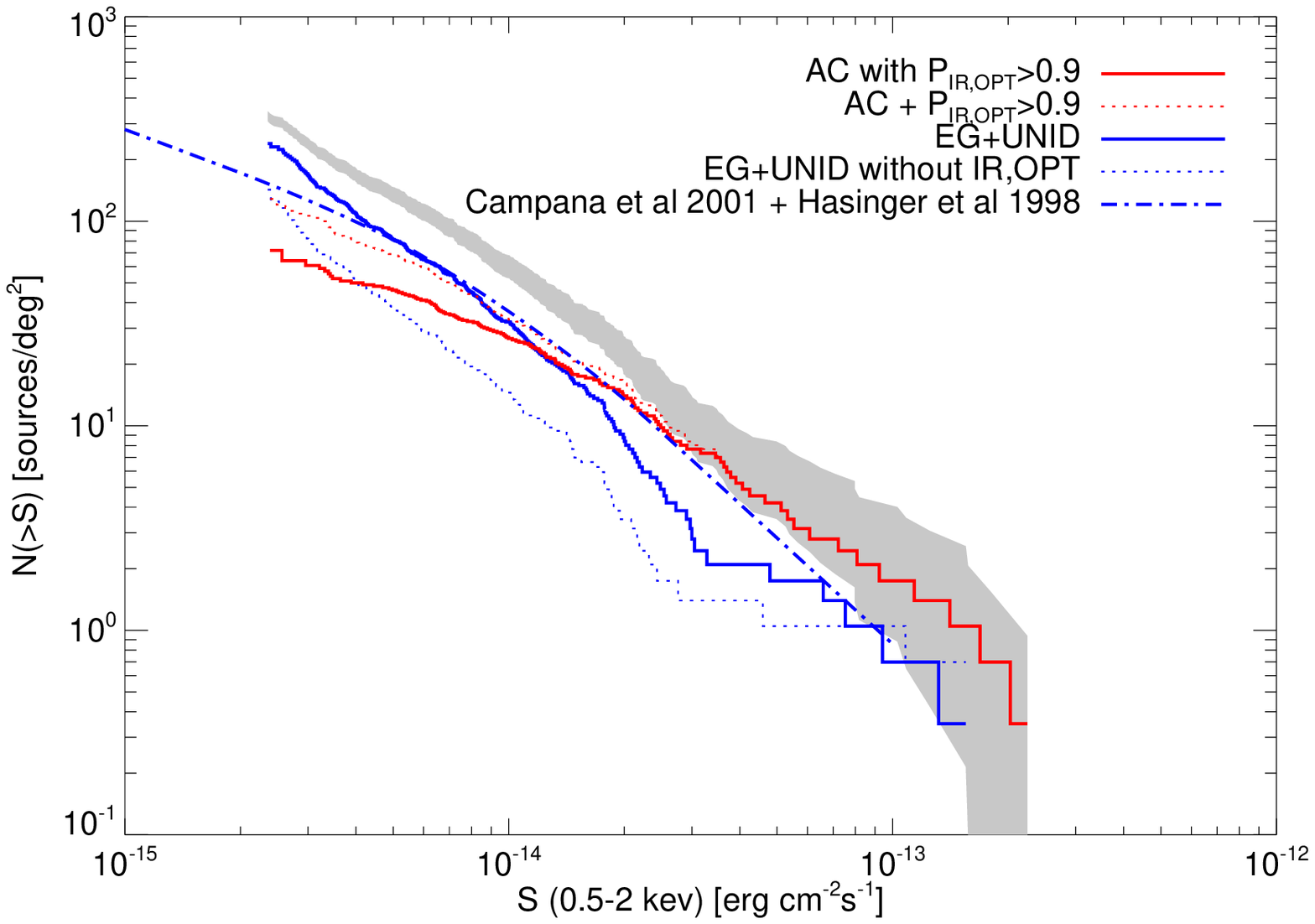}
\includegraphics[width=0.5\linewidth,angle=0]{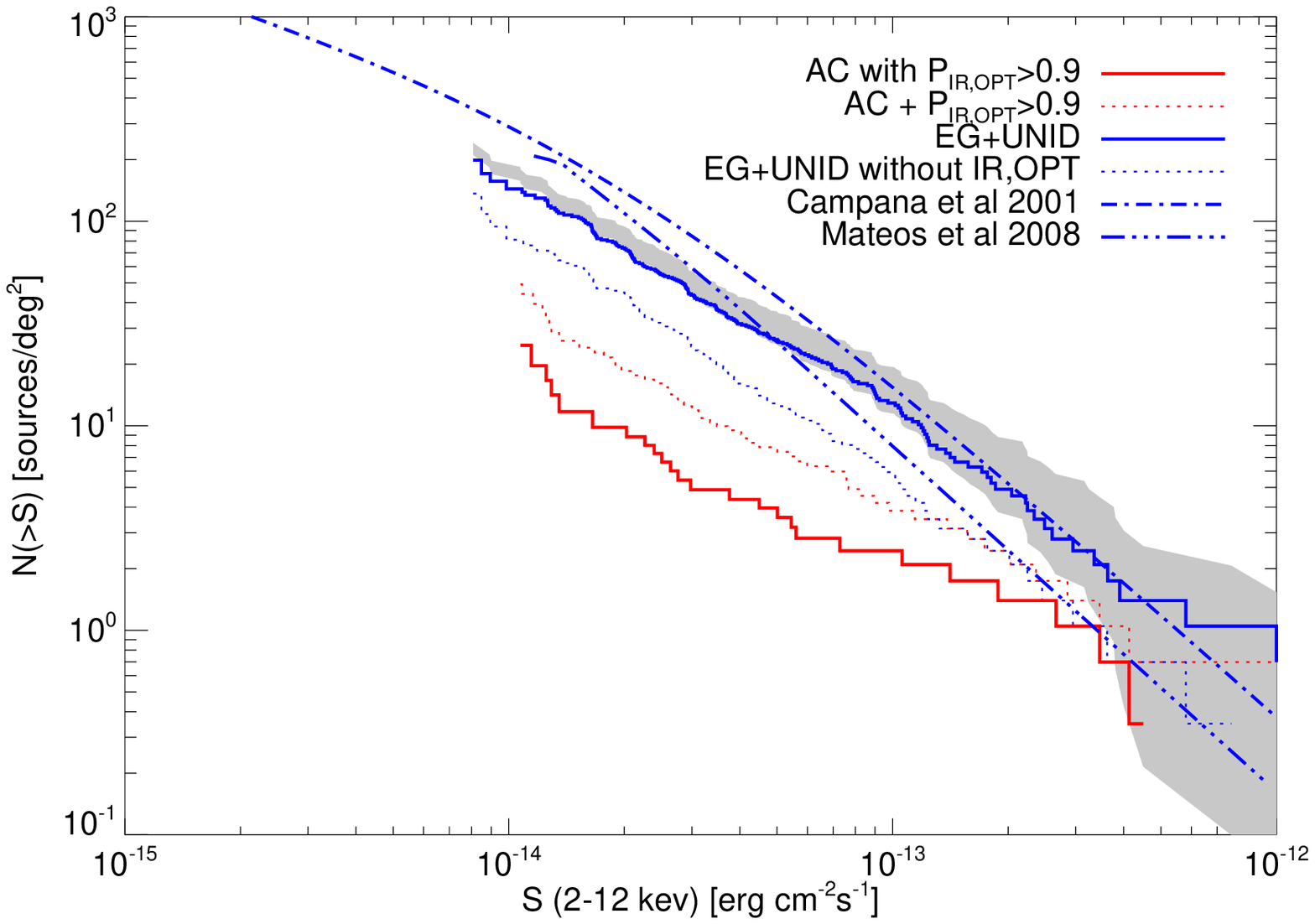}
\caption{\lnls\ curves in the soft band (0.5\,--\,2\,keV, left panel) and the hard band (2\,--\,12\,keV, right panel) for all detected X-ray sources (grey), identified active coronae with \Pid\,$>\,90\%$ (solid red lines) and unidentified plus extragalactic sources (solid blue blue). Sources either identified as active stars or with \Pid\ higher than 90\% are shown with red dotted lines, and sources either identified as extragalactic or unidentified sources with no infrared and no optical counterpart are shown with blue dotted lines. We show the expected contribution of extragalactic sources from \cite{campanaetal01-1} plus \cite{hasingeretal98-1} in the soft band, and from \cite{mateosetal08-1} and \cite{campanaetal01-1} in the hard band.\label{g:logN-logS}}
\end{figure*}

\subsubsection{Energy-to-flux conversion factors}
\label{sec:ecfs}
To convert count rates into fluxes we calculated energy-to-flux conversion factors using XSPEC. We used the EPIC pn response matrix version 6.7 in full frame mode, for spectra selected from single plus double-pixel events and for on-axis events (epn\_ff20\_sdY9\_v6.7.rmf). We restricted our calculation to the medium filter. For the soft band, since it is dominated by stars, we assumed a thin thermal spectrum, with kT\,=\,0.5\,keV absorbed by \Nh$\sim10^{21}$ cm$^{-2}$, which gives the energy-to-flux conversion factor $1.75\times10^{-12}$\,erg\,cm$^{-2}$\,s$^{-1}$\,/\,pn counts\,s$^{-1}$. We estimate that the difference in soft band energy-to-flux conversion factors introduced by selecting a different filter to the one used during the observations is lower than 2\% for the thin filter, while differences up to 23\% can be expected for the thick filter. Most of the fields were observed with medium filter, four fields with thin filter (GRB~001025, HT~Cas, PKS~0745-19, RX~J0925.7-4758), and only one field (Saturn) was observed with thick filter. For the hard band we assumed a power law with photon index equal to 1.7 typical of the AGN population dominating in the hard band. Since the energy-to-flux conversion factors depend on the Galactic absorption, we derived energy-to-flux conversion factors for each field, for the total Galactic absorption along the line of sight, thus assuming that most hard sources are indeed of extragalactic nature. The mean energy-to-flux conversion factor in the hard band is $\sim 10^{-11}$\,erg\,cm$^{-2}$\,s$^{-1}$\,/\,pn counts\,s$^{-1}$, corresponding to a mean \Nh\ of $4.5\times10^{22}$\,cm$^{-2}$. In the hard band, the choice of a different filter in the computation of the energy-to-flux conversion factors has a smaller impact that in the soft band, with a maximum relative error $\sim4\%$ for the thick filter.

\begin{table}
\begin{center}
\caption{Fraction of active coronae.\label{t:AC_count_fields}}
\begin{tabular}{lcccccc}
\hline\hline
\noalign{\smallskip}
Galactic & \multicolumn{2}{c}{Soft Band}     & \multicolumn{2}{c}{Hard Band}   & \multicolumn{2}{c}{Both Bands} \\
latitude & \multicolumn{2}{c}{[0.5\,--\,2\,keV]} & \multicolumn{2}{c}{[2\,--\,12\,keV]} & \multicolumn{2}{c}{}           \\  
\noalign{\smallskip}
\hline
\noalign{\smallskip}
                    & AC$^\dag$(\%) & opt/ir$^\ddag$(\%) &  AC$^\dag$(\%) & opt/ir$^\ddag$(\%) & AC$^\dag$(\%) & opt/ir$^\ddag$(\%) \\
\noalign{\smallskip}
\hline
\noalign{\smallskip}
$|b| \sim 15^\circ$ & 15 & 30     &   3 &  26    &  4 & 27 \\
$|b| \sim 3^\circ$  & 34 & 46     &  12 &  18    & 20 & 24 \\
$|b| \sim 1^\circ$  & 39 & 63     &  11 &  23    & 17 & 29 \\
$|b| \sim 0^\circ$  & 58 & 85     &  20 &  31    & 53 & 67 \\
\noalign{\smallskip}
\hline 
\noalign{\smallskip}
All$^\ast$ & $\frac{137}{460}$ & $\frac{216}{460}$ & $\frac{25}{228}$ & $\frac{54}{228}$ &  $\frac{23}{139}$ & $\frac{43}{139}$ \\
\noalign{\smallskip}
\hline
\noalign{\smallskip}
\end{tabular}
\end{center}
Notes: $^\dag$ AC are identified active coronae with infrared or optical counterpart with identification probability $P\ge90\%$. $^\ddag$ opt/ir are sources that have been either identified as active coronae on the basis of their optical spectra or have an infrared or optical counterpart with $P\ge90\%$, i.e. likely to be Galactic sources. \mbox{$^\ast$ number} of sources detected in each band.
\end{table}

\begin{figure*}
\begin{center}
\includegraphics[width=0.43\linewidth,angle=0]{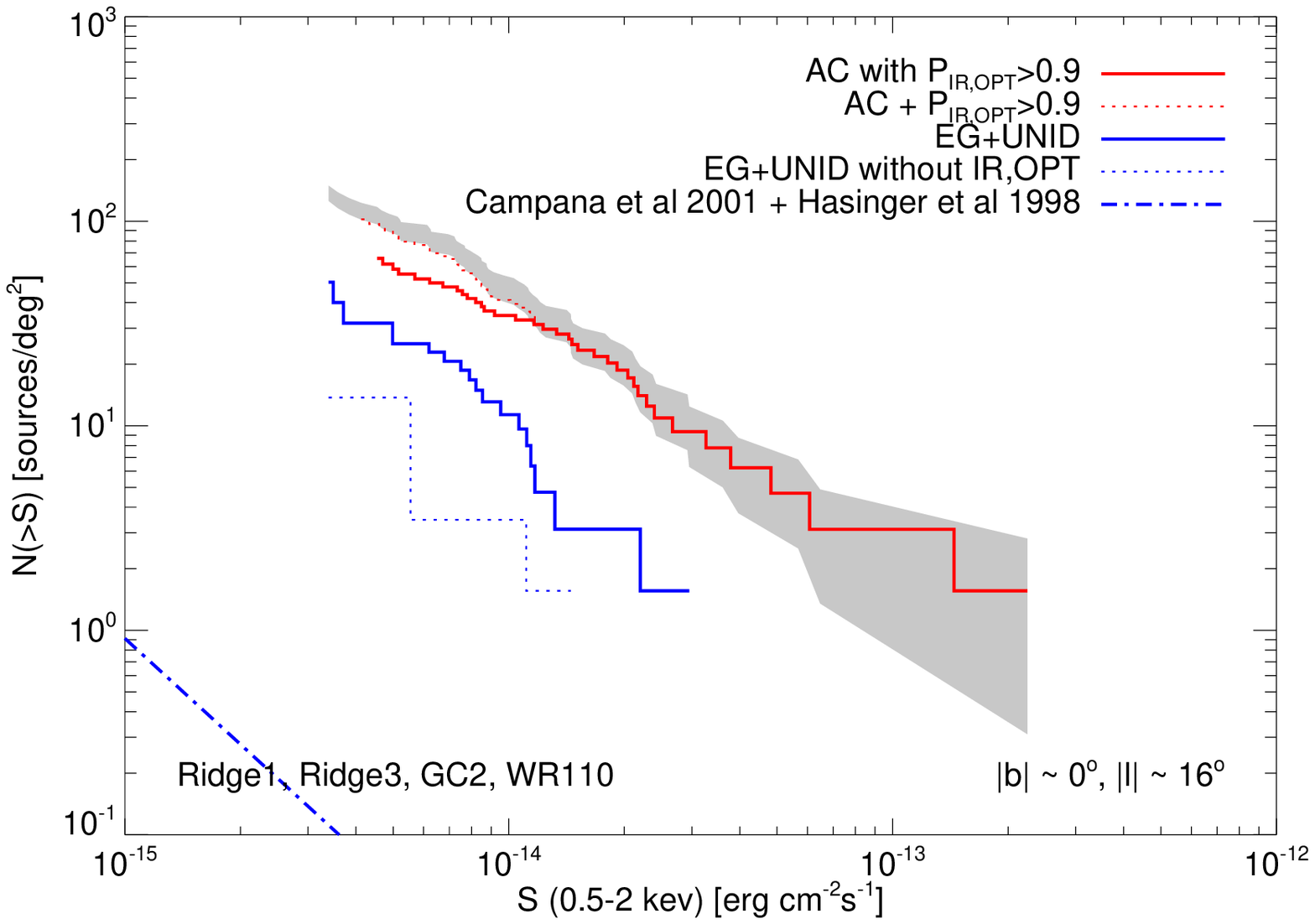}
\includegraphics[width=0.43\linewidth,angle=0]{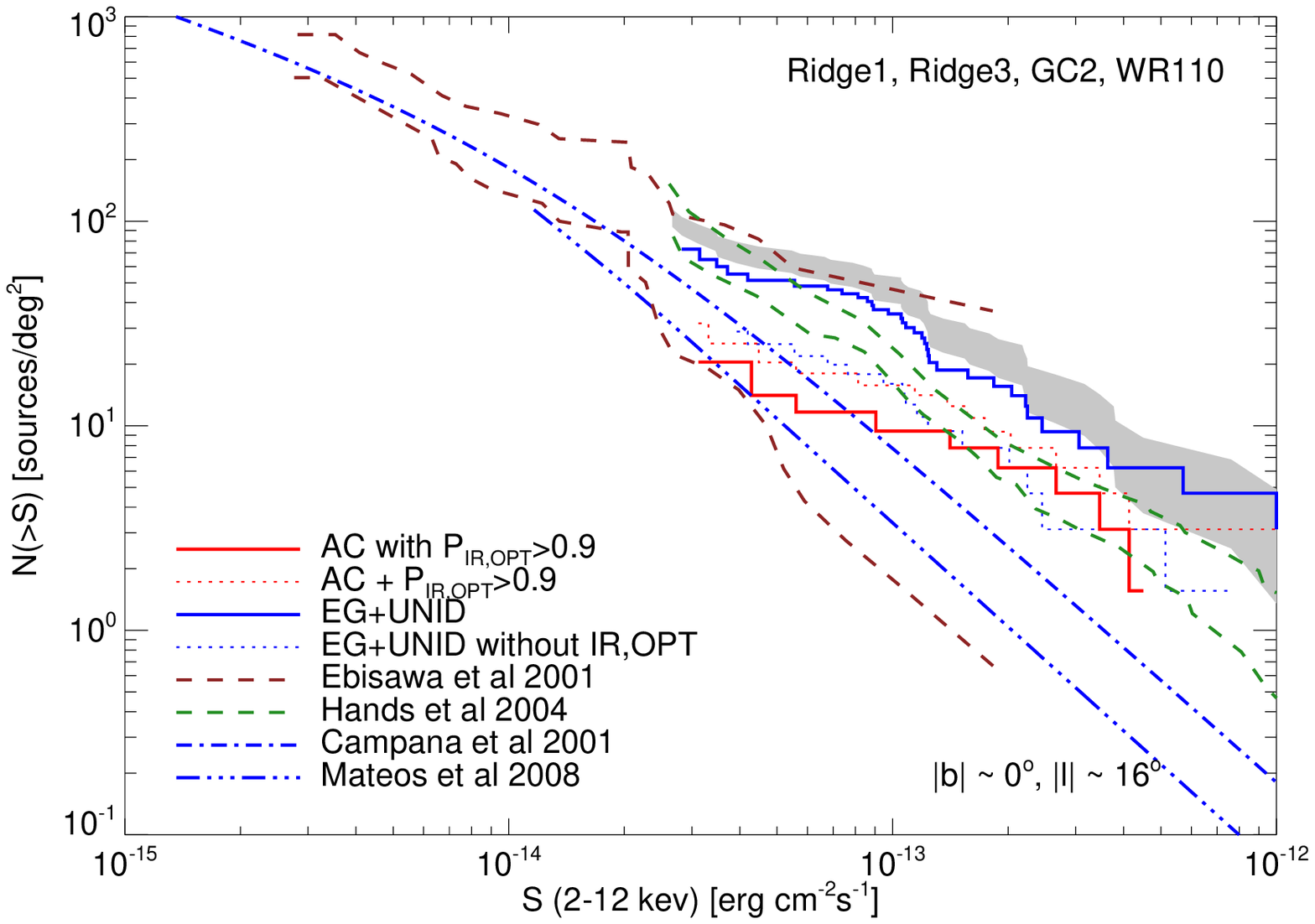}
\includegraphics[width=0.43\linewidth,angle=0]{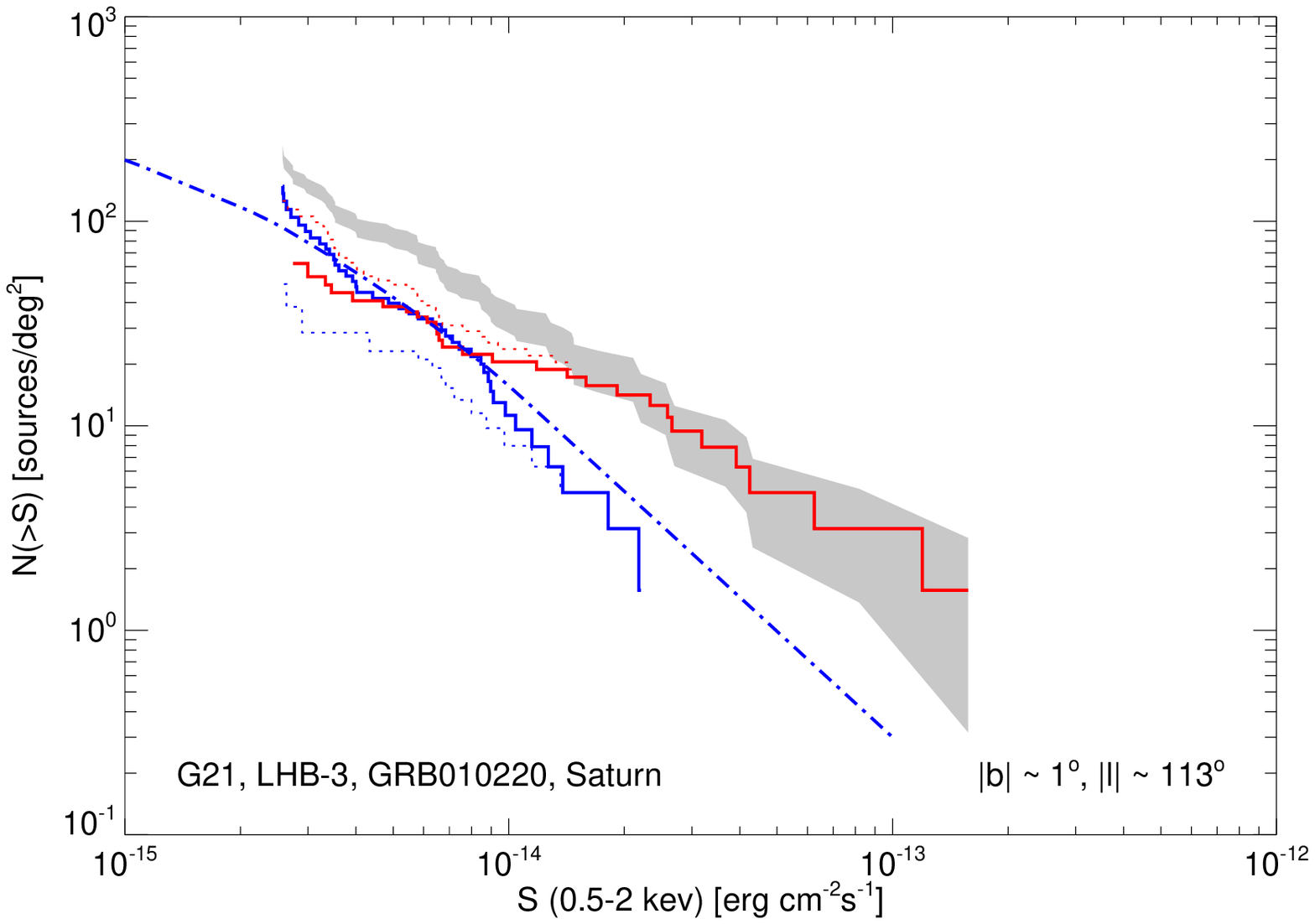}
\includegraphics[width=0.43\linewidth,angle=0]{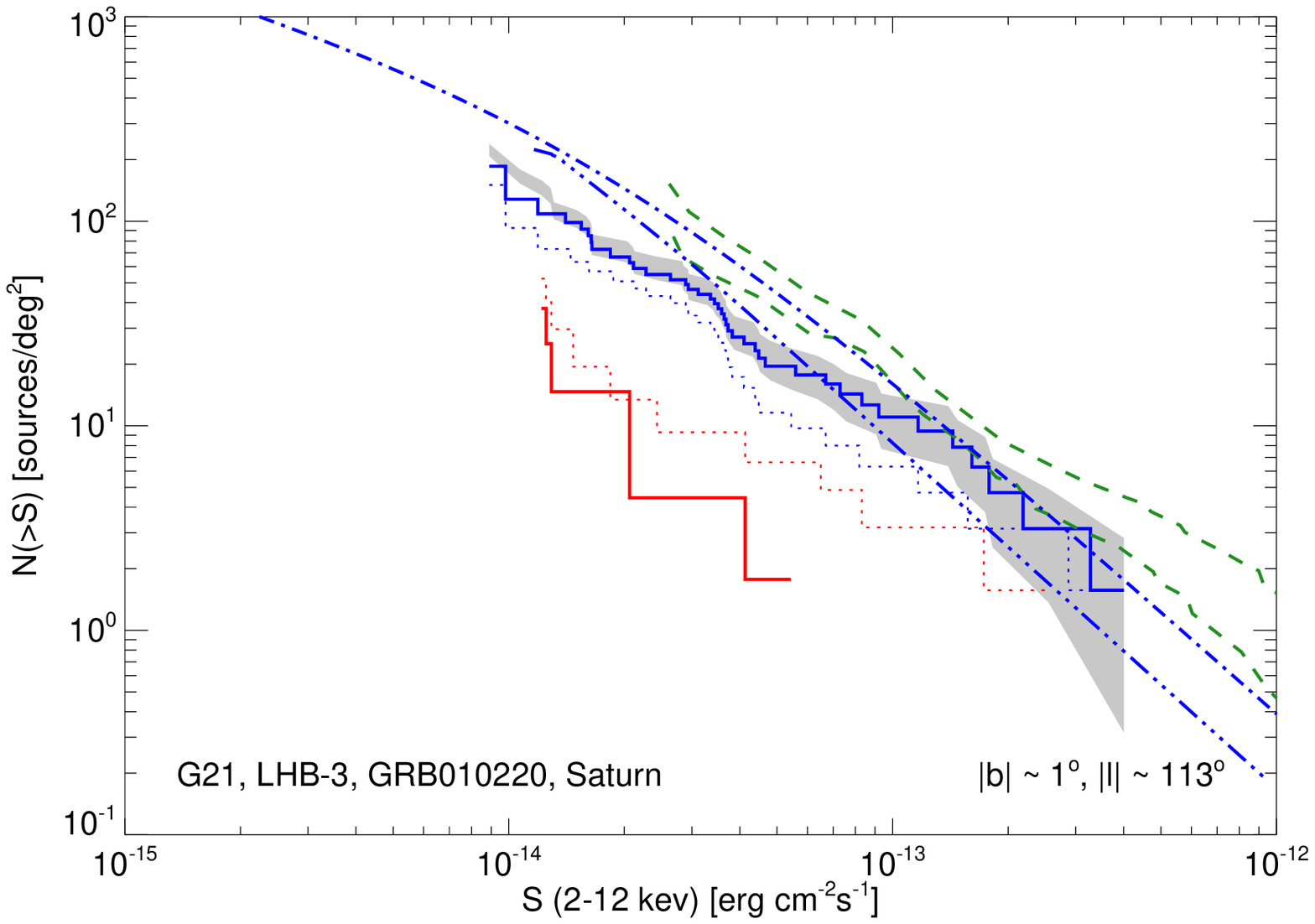}
\includegraphics[width=0.43\linewidth,angle=0]{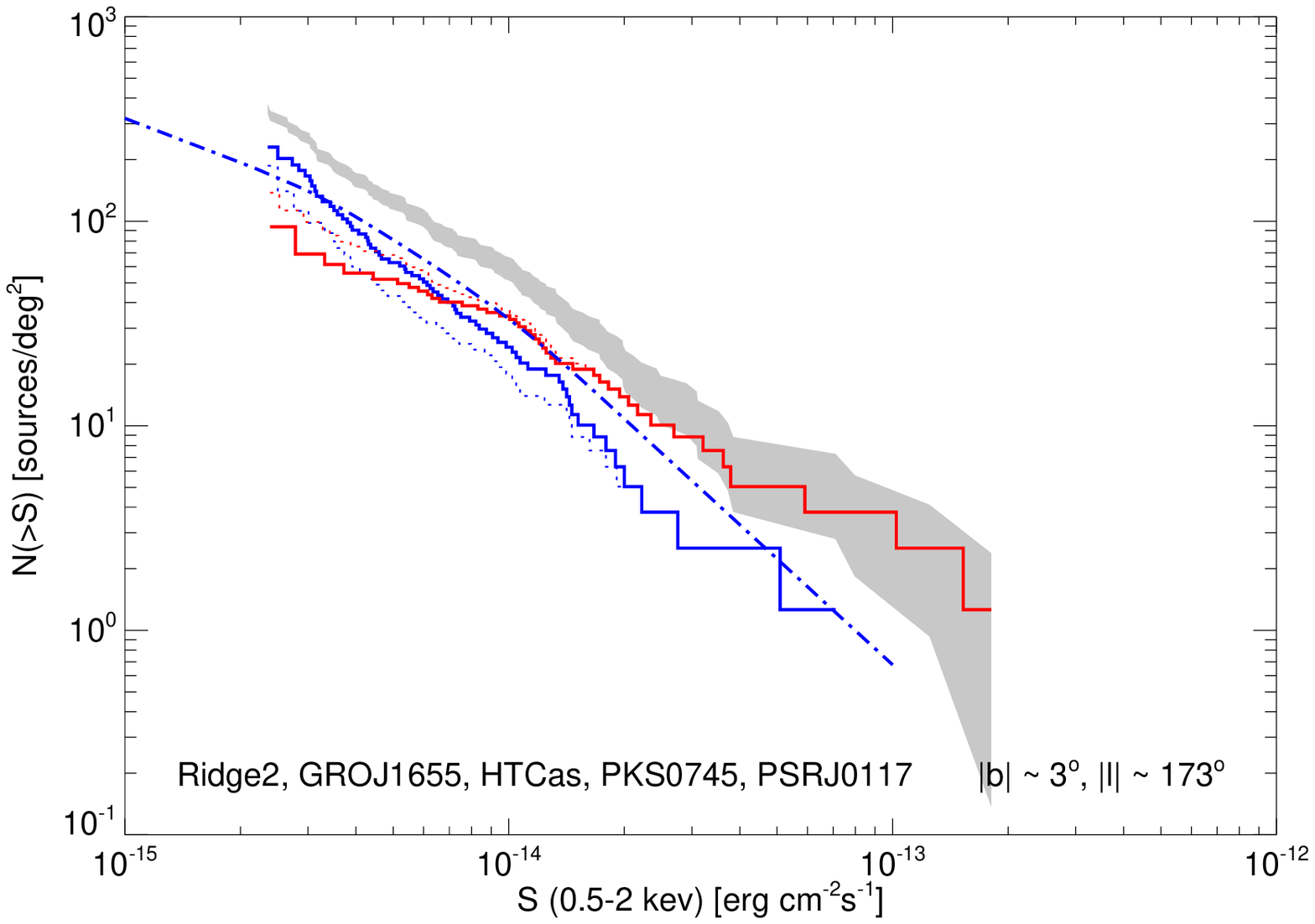}
\includegraphics[width=0.43\linewidth,angle=0]{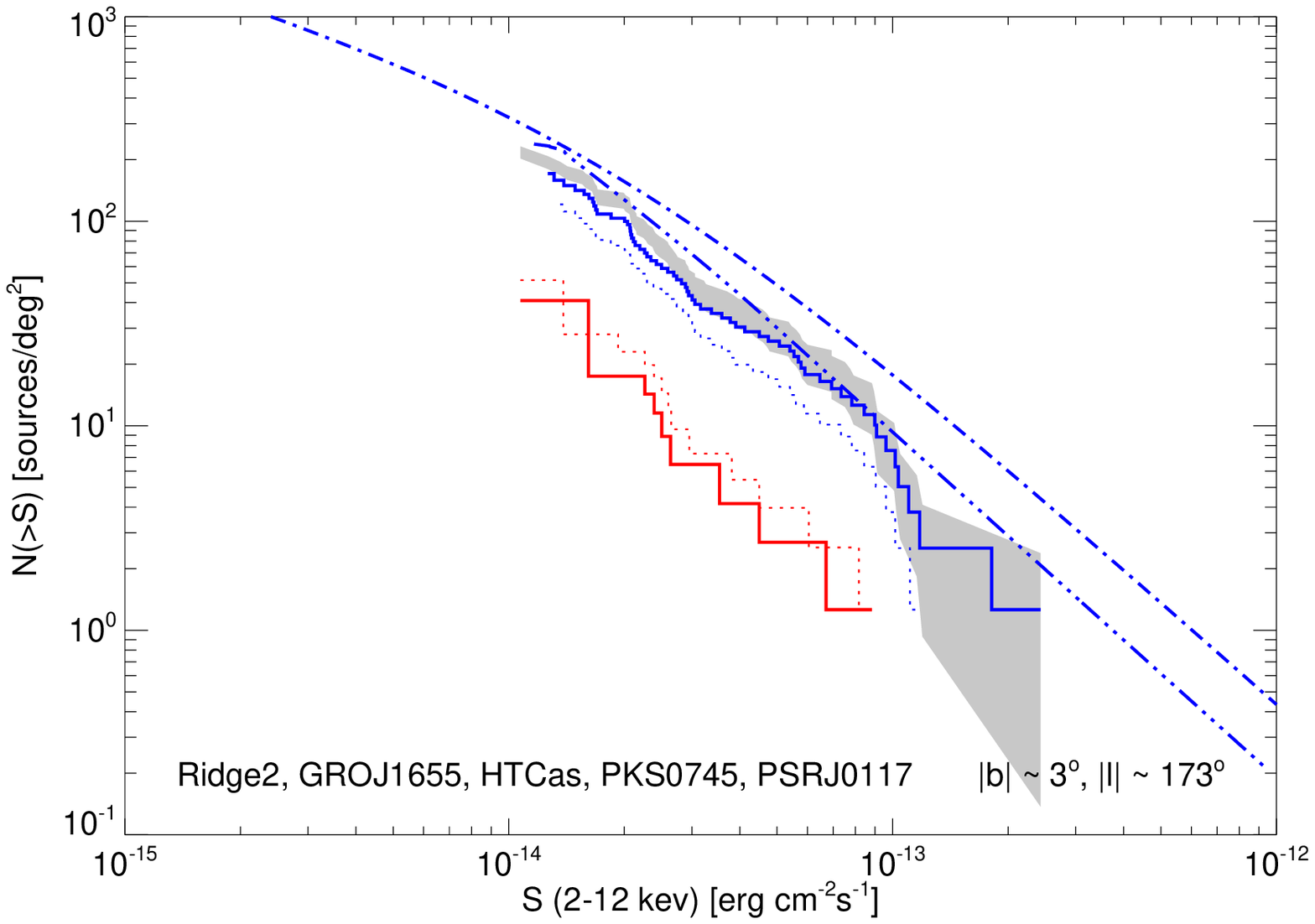}
\includegraphics[width=0.43\linewidth,angle=0]{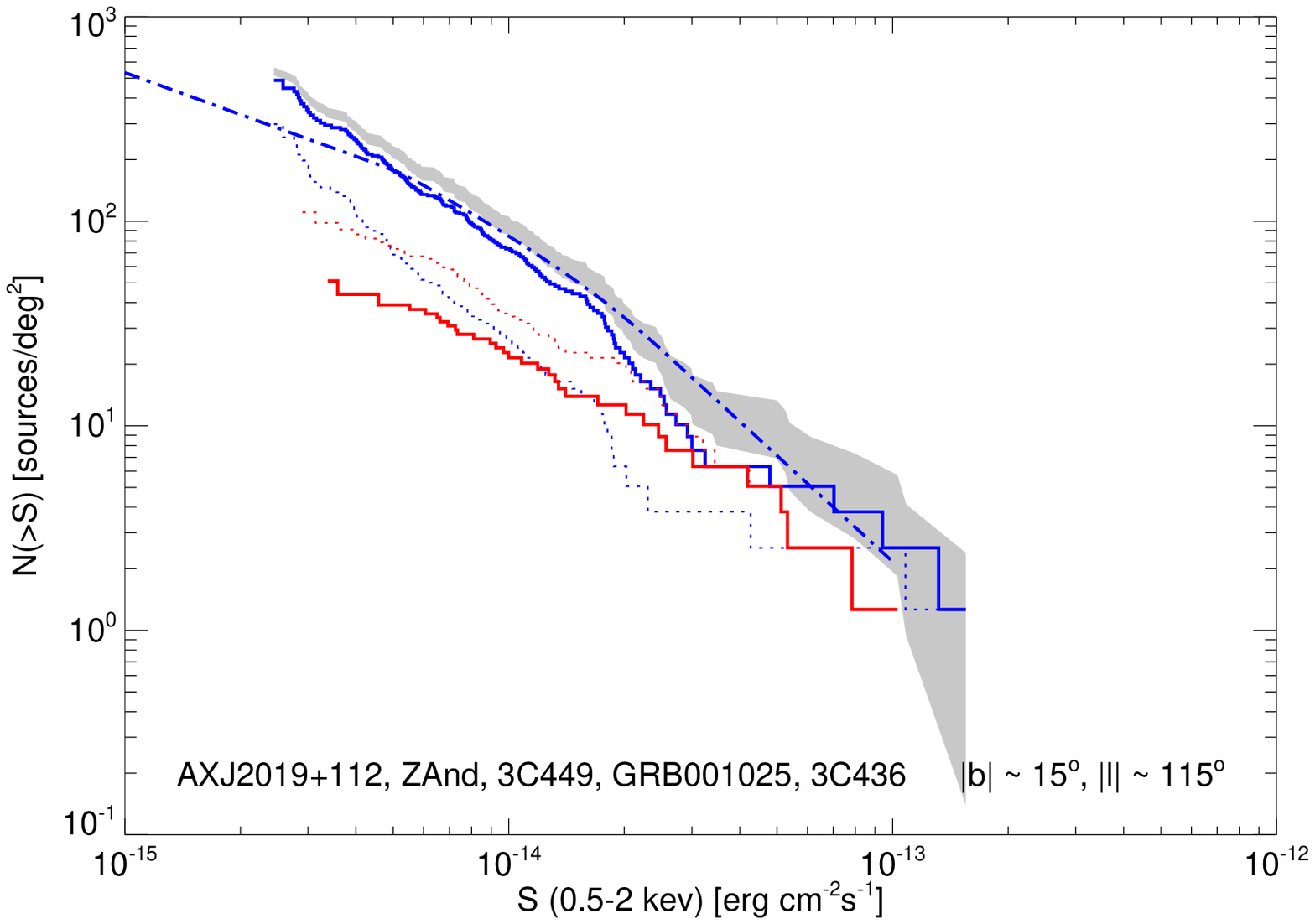}
\includegraphics[width=0.43\linewidth,angle=0]{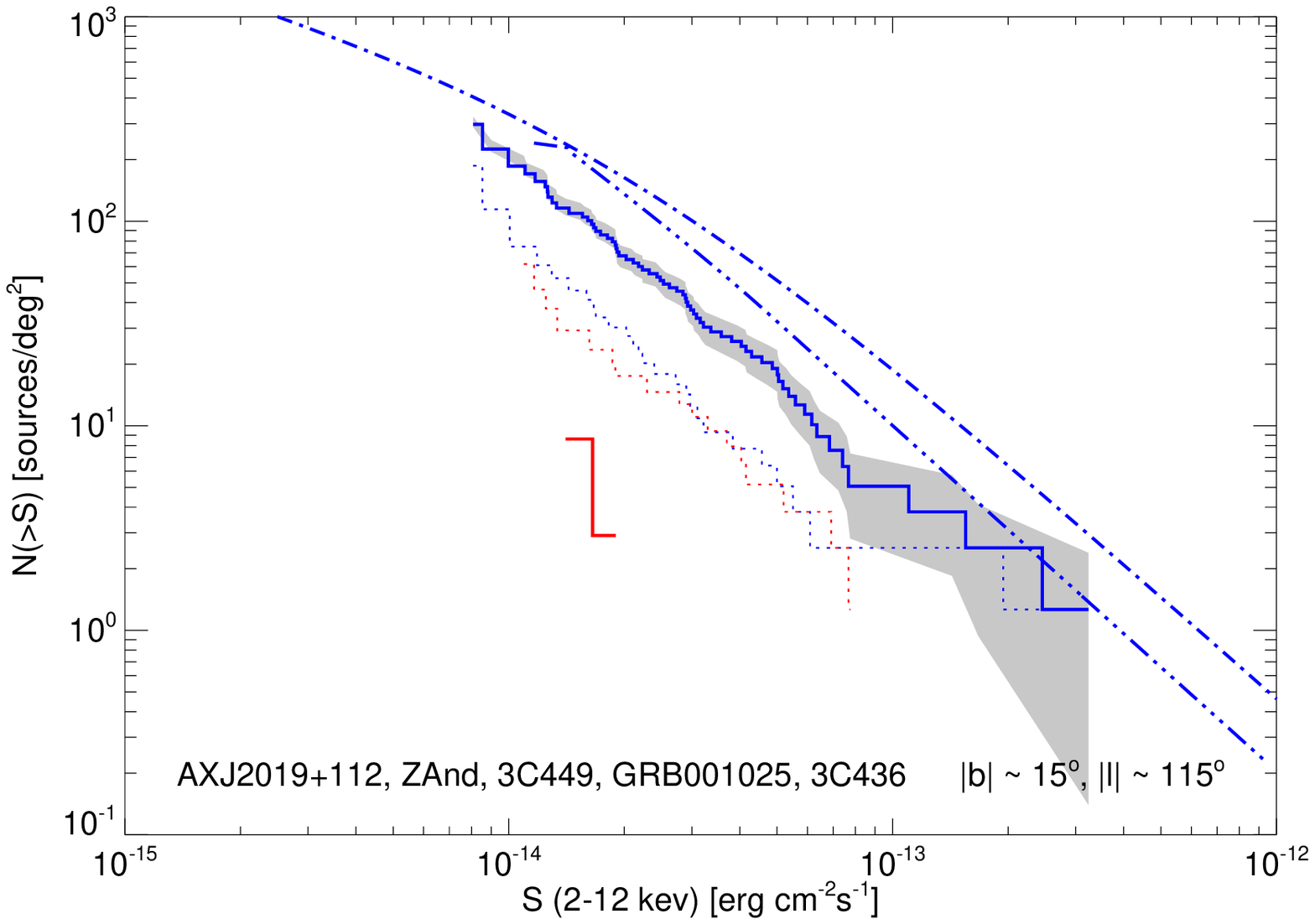}
\end{center}
\caption{\lnls\ curves in the soft band (0.5\,--\,2\,keV, left panel) and the hard band (2\,--\,12\,keV, right panel) for all detected X-ray sources (grey), identified active coronae with \Pid\,$>\,90\%$ (solid red lines) and unidentified plus extragalactic sources (solid blue blue). Sources either identified as active stars or with \Pid\,$>\,90\%$ are shown with red dotted lines, and sources either identified as extragalactic or unidentified sources with no infrared and no optical counterpart are shown with blue dotted lines. We show the expected contribution of extragalactic sources from \cite{campanaetal01-1} plus \cite{hasingeretal98-1} in the soft band and from \cite{mateosetal08-1} and \cite{campanaetal01-1} in the hard band. For the lower Galactic latitudes and in the hard band we compare our results with those of \cite{ebisawaetal01-1} (brown line) and \cite{handsetal04-1} (green line). \label{g:logN-logS_ALL}}
\end{figure*}
\subsubsection{\lnls}
\label{sec:logN_logS}
For each source flux we assigned an effective area. We calculated the number of sources above a given flux per unit sky area, N($>$S), as the sum of the inverse of the effective areas for sources with fluxes above that value. We excluded a few very faint sources to avoid problems occurring at extreme low effective areas. In Fig.~\ref{g:logN-logS} we show the \lnls\ curves in the soft and in the hard energy bands for all the sources in the 18 fields, for spectroscopically identified AC with high identification probability in either optical or infrared catalogues, and for extragalactic plus unidentified sources. In the soft band there is a total of 460 sources, among which 137 are spectroscopically classified as active coronae with an optical and infrared catalogue counterpart with high identification probability. In the hard band there are 228 sources, with 25 sources spectroscopically classified as AC. There are 139 sources in both energy bands, among which 23 are classified as AC. We also show the contribution of two other groups. A first group composed of sources either classified as AC on the basis of their optical spectra (some have \Pid\,$<\,90\%$) or with a bright optical and/or infrared catalogue counterpart, which are therefore likely to be stars. The second group is composed of sources either classified as extragalactic objects or without optical nor infrared catalogue counterpart, which are likely to be extragalactic. Stars are dominating the soft band, while extragalactic plus unidentified objects dominate the hard band. With a threshold of 90\% in the \Pid\ we expect the sample to be about 55\% complete, so we are missing a large number of the associations, we therefore expect the true fraction of spectroscopically classified sources with counterpart catalogue to be above the one shown in Fig.~\ref{g:logN-logS}.

The \lnls\ curve in the soft band, after subtraction of the stellar contribution, is steeper than the combined extragalactic contribution from \cite{campanaetal01-1} at low fluxes, based on Chandra Deep Field observations, plus \cite{hasingeretal98-1} at high fluxes, derived from ROSAT observations in the Lockman Hole\footnote{\fontsize{7}{7}{To compare with other publications, we applied a correction factor to convert fluxes from one band to another using PIMMS.}}. At the faintest fluxes we are subject to Eddington biases, associated with statistical flux variations, likely to introduce a fictitious further steepening of the \lnls.  

Extragalactic \lnls\ curves were corrected for the Galactic absorption in each field and we applied a correction factor to convert fluxes into the appropriate band. The extragalactic \lnls\ from \cite{campanaetal01-1} is based on ASCA and Chandra observations, while the one from \cite{mateosetal08-1} is based on XMM-Newton observations. Calibration uncertainties between different telescopes/instruments are expected to be lower for the latter, so we expect smaller deviations with our observations than with the former. 

In order to understand the contribution of the different populations to the \lnls\ curves we investigated the curves in four different Galactic latitude ranges corresponding to $|b|\,\sim\,0^\circ$, $1^\circ$, $3^\circ$, and $15^\circ$ (see Fig.~\ref{g:logN-logS_ALL}). 

At very low Galactic latitude, the soft band is dominated by stars. The number of stars increases towards lower Galactic latitudes. The fraction of identified stars (classified on the basis of their optical spectra and with \Pid\,$>\,90\%$) is 15\% at the highest Galactic latitude bin, and increases to 60\% at $|b|\,\sim\,0^\circ$ (see Table~\ref{t:AC_count_fields}). On the other hand the number of extragalactic plus unidentified sources increases as we move away from the Galactic Plane. At $|b|\,=\,15^\circ$ the number of extragalactic plus unidentified sources without optical nor infrared counterpart within the $3\sigma$ error circle begins to dominate very faint X-ray fluxes ($<2\times10^{-15}$\,\ergcms). This is consistent with the fact that \Nh\ is decreasing with increasing distance to the Galactic Plane, making it easier to detect AGN at soft energies. We are nevertheless aware that at high Galactic latitudes, where \Nh\ is low, some AGN may be bright enough to give a \Pid\,$>\,90\%$ crossmatch. 

The hard band is dominated by extragalactic plus unidentified sources. At high Galactic latitudes hard sources mainly have an extragalactic origin. In the Galactic Plane the number of hard sources is above the expected extragalactic contribution. This excess increases towards the Galactic Centre region.
A population of hard spectrum stars is present in all four \lnls\ curves. There are 25 stars detected in the hard band with spectral type known, corresponding to 10\% of the detected sources in the hard band. All of them but two are also detected in the soft band. Four hard sources were classified as giant candidates (see Section~\ref{sec:giants}): 2XMM\,174705.3-280859, 2XMM\,J180718.4-192454, 2XMM\,J180736.4-192658, and 2XMM\,J185139.1+001635. Among the 25 hard active coronae there are two early A stars, two F stars, nine G stars, eight K stars, and four M stars. 
With a threshold of ML=8 we expect around 27 spurious detections (see Section~\ref{sec:eff_areas}) in the hard band. We have computed sensitivity maps, effective areas and \lnls\  curves also for ML $>$ 15 ($5\sigma$ detection). Thirteen hard sources identified as AC remain at this ML threshold. At this ML cutoff level the expected number of spurious detection in our survey is about one and with an identification probability above 90\%, the number of false identifications is expected to be less than five (2\% of 228). This is not enough to explain the 13 hard active coronae. We conclude that there is indeed a significant population of hard X-ray emitting active coronae. 

\begin{center}
\begin{figure}
\includegraphics[width=\linewidth,angle=0]{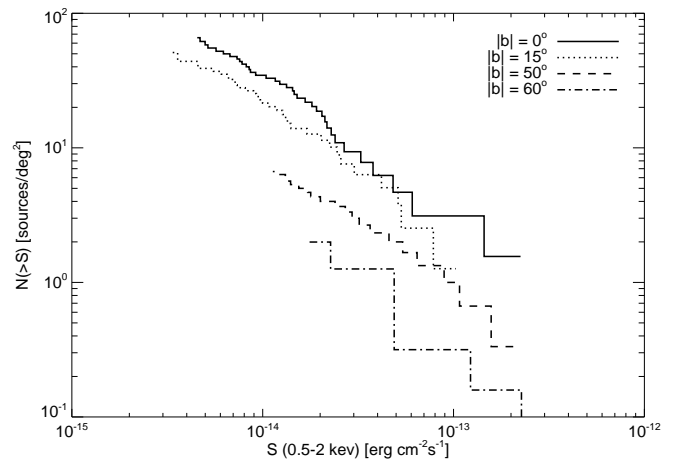}
\caption{\lnls\ curves in the soft band (0.5\,--\,2\,keV) for identified active coronae at four Galactic latitudes: $0^\circ$ and $15^\circ$ from this paper, $50^\circ$ and $60^\circ$ from \cite{barconsetal07-1} and \cite{lopez-santiagoetal07-1} respectively. \label{g:logN-logS_AC}}
\end{figure}
\end{center}
\subsubsection{Dependence of \lnls\ curves on Galactic latitude}
As mentioned above, the surface density (\lnls) of soft X-ray emitting stars varies with Galactic latitude. We compared our results with those found by \cite{barconsetal07-1} and by \cite{lopez-santiagoetal07-1} based on XMM-Newton observations of high Galactic latitude fields, with mean Galactic latitudes of $\sim50^\circ$ and $\sim 60^\circ$ respectively. We found that the number of sources per square degree varies steeply with Galactic latitude (see Fig. \ref{g:logN-logS_AC}), increasing by an order of magnitude from $b\,=\,60^\circ$ at $b\,=\,0^\circ$ for fluxes above $\sim2\times10^{-13}$\,\ergcms. 

We fitted these \lnls\ curves with a power-law function of the type $N(>S)=KS^{-\alpha}$ using a maximum likelihood technique \citep{crawfordetal70-1,murdochetal73-1}. We obtained a maximum likelihood for the following slopes of the curves:
\begin{equation}
\begin{array}{lr}
b\,\sim\,0^\circ, &\alpha\,=\,1.05\pm0.25  \\
b\,\sim\,15^\circ,& \alpha\,=\,0.76\pm0.21 \\
b\,\sim\,50^\circ,& \alpha\,=\,0.65\pm0.16 
\end{array}
\label{eq:mle}
\end{equation}
A Kolgomorov-Smirnov test (KS-test) allowed us to validate the power-law model. Due to the low number of sources in the highest Galactic latitude \lnls\ curve ($b\,=\,60^\circ$) we only performed a fit to the other three curves. We restricted to fluxes in the range $10^{-14}$\,--\,$2\times10^{-13}$\ergcms. The slope of the power-law varies with Galactic latitude, effect that is observed in X-ray models of the Galaxy \citep{guilloutetal96-1}, and which is due to the scale height of stars and the relative contribution of the different populations. At low Galactic latitudes, looking directly to the Galactic Plane, the number of stars steadily increases with decreasing limiting fluxes. On the contrary, at high Galactic latitudes, the number of sources is truncated due to the finite scale-height of stars. Young stars have smaller scale-height than old stars, therefore at low Galactic latitudes we are dominated by the young population over the whole range of observed fluxes, while at high Galactic latitudes we are dominated by young stars at high fluxes ($\ge 10^{-13}$\,\ergs\ in the 0.5\,--\,2\,keV band) and by old stars at fainter fluxes \citep{guilloutetal96-1}. Hence, the variation of the \lnls\ with Galactic latitude reflects an age-scale height dependence.

\begin{center}
\begin{figure}
\includegraphics[width=\linewidth,angle=0]{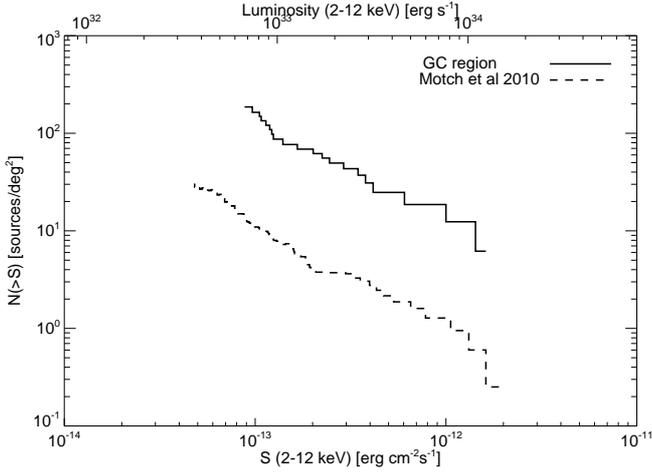}
\caption{\lnls\ curves in the Galactic Centre region for the hard band (2\,--\,12\,keV) corrected from the extragalactic contribution and compared with the results from \cite{motchetal10-1} based on the XGPS sample from \cite{handsetal04-1} at l$\sim20^\circ$.}\label{g:logN-logS_AC_hard}
\end{figure}
\end{center}
\subsubsection{The Galactic Centre region}
We clearly detect an excess of hard sources compared to the expected extragalactic contribution (see Section~\ref{sec:logN_logS}), the excess mainly coming from the Galactic Centre region (field GC2 in Table~\ref{t:iden}). We subtracted the extragalactic contribution from \cite{mateosetal08-1} from the total \lnls\ and compared our results with those from \cite{motchetal10-1} based on the XGPS sample from \cite{handsetal04-1} at ($l\,\sim\,20^\circ$, $b\,=\,0^\circ$). We found that the Galactic surface density of hard sources in the Galactic Centre region ($l\,=\,0.9^\circ$, $b\,=\,0^\circ$) is larger than at ($l\,=\,20^\circ$,$\,b=\,0^\circ$) by approximately a factor of ten, reaching a density of about 100 sources/deg$^2$ at a flux of $\sim10^{-13}$\,\ergcms\ (see Fig.~\ref{g:logN-logS_AC_hard}), but lower than at the Galactic Centre \citep{munoetal03-1}. This is consistent with the results from \cite{munoetal09-1} based on Chandra observations in the $3^\circ$ around Sgr~A$^*$, and with the results from \cite{hongetal09-1} also based on Chandra observations in the Galactic Bulge.

The increase of hard sources towards the Galactic Centre region might be associated with the Galactic Centre itself. If so, assuming a distance to the Galactic Centre of 8\,kpc, the obtained range of luminosities is $10^{33}$\,--\,$10^{34}$\,\ergs. It has been proposed that this concentration of sources is mainly due to an old population of CVs and a minor contribution of a young population of HMXBs at low mass transfer \citep{munoetal03-1,laycocketal05-1}. 

\section{Discussion and Conclusions}
\label{sec:concl}
We carried out optical spectroscopic follow-up observations of X-ray sources detected in 26 fields by XMM-Newton, at low and intermediate Galactic latitudes, distributed over a broad range in Galactic longitudes and covering a total area of 4\,deg$^2$. Our X-ray survey has a limiting X-ray flux of about $2\times10^{-15}$\,\ergcms\ and $1\times10^{-14}$\,\ergcms\ in the 0.5\,--\,2\,keV and in the 2\,--\,12\,keV energy bands respectively. 
A total of 1319 sources have an entry in the 2XMMi-DR3 catalogue. We cross-matched these X-ray sources with the SDSS-DR7, the GSC~2.3, the USNO-B1.0 and the 2MASS catalogues, finding matches for about 50\% of the sources. Twenty per cent of the matches are above a 90\% identification probability in at least one of the catalogues. At this threshold the number of spurious associations is expected to be less than 2\%, and the survey is about 50\% complete. Using our follow-up optical spectroscopic observations in combination with cross-correlation with a large range of archival catalogues we classified 316 X-ray sources. The XMM-Newton SSC survey of the Galactic Plane constitutes the largest sample of Galactic spectroscopically identified X-ray sources in this range of X-ray flux and Galactic latitudes.

In the soft band (0.5\,--\,2\,keV), the majority of the sources are positively identified as stars with spectral types in the range A--M. The number of detected X-ray emitting active corona increases with decreasing effective temperatures and presents an excess of G--K stars. Such a repartition in spectral types is similar to that observed in magnitude and X-ray flux limited samples extracted from the ROSAT all-sky survey at a factor 10 brigther X-ray flux \citep{motchetal97-1}. Making use of infrared colours we classify 23 stars as evolved giant candidates, being most of them K stars, and representing 18.4 \% of the sources with 2MASS counterpart and with individual probability higher than 90\%. Since giant stars are not expected to be strong X-ray emitters \citep{maggioetal90-1}, to maintain their X-ray luminosities with age the most likely scenario is that these stars are RS CVn, i.e. synchronised evolved binaries, where the rotation of stars has not spun-down with age. 

A handful of likely pre main sequence stars, T Tauri and Herbig Ae objects are also identified
spectroscopically. One of the T Tauri candidates appears far away from any known star forming region. 

The measured stellar F$_{\mathrm{X}}$/F$_\mathrm{V}$, X-ray and infrared colours are consistent with expected values for young (100 Myr) to intermediate age (600 Myr) active coronae with a small contribution of BY\,Dra and RS\,CVn like binaries. 
We find that the number of stars per square degree (\lnls) depends on the Galactic latitude, steadily increasing by one order of magnitude from $b\,\sim\,60$\degr\ to $b\,=\,0$\degr. X-ray stellar population modelling shows that this increase is due to a combination of age-X-ray luminosity and age-scale height dependencies \citep{guilloutetal96-1}. Since we detect and identify active coronae up to about distances of 1\,kpc, we can clearly witness the effect caused by the higher concentration of young and X-ray luminous stars at low Galactic latitudes. 

The overall distribution of X-ray detected stellar coronae in spectral types, X-ray spectra, flux and Galactic latitude contains important imprinted information on the local density, star formation rate of up to 2Gyr old stars and on the evolution with age of the X-ray luminosity, X-ray temperature and of the scale height. In a second paper, we will use X-ray stellar population modelling to constrain these stellar population parameters. 

In the hard band (2\,--\,12\,keV), most of the sources are consistent with an extragalactic nature. We identify, however, a genuine Galactic population of hard sources which accounts for 11\% of the detected hard sources. Among identified hard X-ray sources are some of the X-ray stars exhibiting the highest X-ray temperatures and therefore the highest X-ray luminosities. A small fraction of CVs also contributes to the galactic hard X-ray population. We report the discovery of four new $\gamma$-Cas analogs which assuming a typical X-ray temperature above $\sim$\,7\,keV, are conspicuous hard X-ray sources with (0.2-12\,keV) X-ray luminosities of a few 10$^{32}$\,\ergs. The equivalent widths of their \Ha\ emission line also spans a small range ($\sim30$\,\AA), typical of these objects. One of the candidates is a very faint R\,$\sim$\,22 highly reddened object. Although the number of $\gamma$-Cas objects still remains small, they appear to dominate by number the population of X-ray bright massive stars detected in the Galaxy. The mechanism giving rise to the X-ray emission of $\gamma$-Cas analogs remains unknown. Interestingly, one of the candidates found in our survey is the brightest star (star 9) in the NGC 6649 cluster and a blue straggler \citep{marco2007}. 

After removing the expected extragalactic contribution, we find that the population of Galactic hard sources, increases by a factor of ten from $l\,=\,20^\circ$ to $l\,=\,0.9^\circ$, reaching a surface density of about 100 sources deg$^{-2}$ at a flux of $\sim1.3\times10^{-13}$\,\ergcms, i.e. 10$^{33}$\,\ergs\ at a distance of 8\,kpc. We emphasis, however, that, at the exception of one of the $\gamma$-Cas analog candidate, 2XMM\,J180816.6-191939 which has an estimated distance of 6-7\,kpc, all other identified hard X-ray sources, stars and CVs are within a range of 1 or 2\,kpc. It is therefore unclear whether the local population we identify in this work is truely representative of the nature of the sources detected towards the Galactic Centre regions.   
 
\begin{acknowledgements}
%
The XMM-Newton SSC acknowledges sustained finantial support from CNES (France) and from Deutches Zemtrum f\"ur Luft und Raumfhart (Germany) DLR, under grant numbers FKZ 50 OX 0201 and 50 OX 0801.
%
%
This research has made use of the SIMBAD database, operated at CDS, Strasbourg, France 
This publication makes use of data products from the Two Micron All Sky Survey, which is a joint project of the University of Massachusetts and the Infrared Processing and Analysis Center/California Institute of Technology, funded by the National Aeronautics and Space Administration and the National Science Foundation
This research has made use of the USNO Image and Catalogue Archive operated by the United States Naval Observatory, Flagstaff Station (\url{http://www.nofs.navy.mil/data/fchpix/}).
The Guide Star Catalogue-II is a joint project of the Space Telescope Science Institute and the Osservatorio Astronomico di Torino. Space Telescope Science Institute is operated by the Association of Universities for Research in Astronomy, for the National Aeronautics and Space Administration under contract NAS5-26555. The participation of the Osservatorio Astronomico di Torino is supported by the Italian Council for Research in Astronomy. Additional support is provided by European Southern Observatory, Space Telescope European Coordinating Facility, the International GEMINI project and the European Space Agency Astrophysics Division.
Funding for the SDSS and SDSS-II has been provided by the Alfred P. Sloan Foundation, the Participating Institutions, the National Science Foundation, the U.S. Department of Energy, the National Aeronautics and Space Administration, the Japanese Monbukagakusho, the Max Planck Society, and the Higher Education Funding Council for England. The SDSS Web Site is \url{http://www.sdss.org/}.
%
This publication makes use of data products from the Wide-field Infrared Survey Explorer, which is a joint project of the University of California, Los Angeles, and the Jet Propulsion Laboratory/California Institute of Technology, funded by the National Aeronautics and Space Administration.
\end{acknowledgements}
\bibliographystyle{aa}
\bibliography{aamnem99,references}

\begin{thebibliography}{138}
\expandafter\ifx\csname natexlab\endcsname\relax\def\natexlab#1{#1}\fi

\bibitem[{{Abazajian} {et~al.}(2009){Abazajian}, {Adelman-McCarthy},
  {Ag{\"u}eros}, {Allam}, {Allende Prieto}, {An}, {Anderson}, {Anderson},
  {Annis}, {Bahcall}, \& et~al.}]{abazajianetal09-1}
{Abazajian}, K.~N., {Adelman-McCarthy}, J.~K., {Ag{\"u}eros}, M.~A., {et~al.}
  2009, \apjs, 182, 543

\bibitem[{{Appenzeller} \& {Mundt}(1989)}]{appenzeller1989}
{Appenzeller}, I. \& {Mundt}, R. 1989, \aapr, 1, 291

\bibitem[{{Barbera} {et~al.}(1993){Barbera}, {Micela}, {Sciortino}, {Harnden},
  \& {Rosner}}]{barberaetal93-1}
{Barbera}, M., {Micela}, G., {Sciortino}, S., {Harnden}, Jr., F.~R., \&
  {Rosner}, R. 1993, \apj, 414, 846

\bibitem[{{Barcons} {et~al.}(2007){Barcons}, {Carrera}, {Ceballos}, {Page},
  {Bussons-Gordo}, {Corral}, {Ebrero}, {Mateos}, {Tedds}, {Watson}, {Baskill},
  {Birkinshaw}, {Boller}, {Borisov}, {Bremer}, {Bromage}, {Brunner},
  {Caccianiga}, {Crawford}, {Cropper}, {Della Ceca}, {Derry}, {Fabian},
  {Guillout}, {Hashimoto}, {Hasinger}, {Hassall}, {Lamer}, {Loaring},
  {Maccacaro}, {Mason}, {McMahon}, {Mirioni}, {Mittaz}, {Motch}, {Negueruela},
  {Osborne}, {Panessa}, {P{\'e}rez-Fournon}, {Pye}, {Roberts}, {Rosen},
  {Schartel}, {Schurch}, {Schwope}, {Severgnini}, {Sharp}, {Stewart},
  {Szokoly}, {Ull{\'a}n}, {Ward}, {Warwick}, {Wheatley}, {Webb}, {Worrall},
  {Yuan}, \& {Ziaeepour}}]{barconsetal07-1}
{Barcons}, X., {Carrera}, F.~J., {Ceballos}, M.~T., {et~al.} 2007, \aap, 476,
  1191

\bibitem[{{Barcons} {et~al.}(2002){Barcons}, {Carrera}, {Watson}, {McMahon},
  {Aschenbach}, {Freyberg}, {Page}, {Page}, {Roberts}, {Turner}, {Barret},
  {Brunner}, {Ceballos}, {Della Ceca}, {Guillout}, {Hasinger}, {Maccacaro},
  {Mateos}, {Motch}, {Negueruela}, {Osborne}, {P{\'e}rez-Fournon}, {Schwope},
  {Severgnini}, {Szokoly}, {Webb}, {Wheatley}, \& {Worrall}}]{barconsetal02-1}
{Barcons}, X., {Carrera}, F.~J., {Watson}, M.~G., {et~al.} 2002, \aap, 382, 522

\bibitem[{{Barrado y Navascu{\'e}s} \& {Mart{\'{\i}}n}(2003)}]{barradoetal03-1}
{Barrado y Navascu{\'e}s}, D. \& {Mart{\'{\i}}n}, E.~L. 2003, \aj, 126, 2997

\bibitem[{{Bertout}(1989)}]{bertout1989}
{Bertout}, C. 1989, \araa, 27, 351

\bibitem[{{Bertout} \& {Genova}(2006)}]{bertout2006}
{Bertout}, C. \& {Genova}, F. 2006, \aap, 460, 499

\bibitem[{{Bessell} \& {Brett}(1988)}]{bessel+brett88-1}
{Bessell}, M.~S. \& {Brett}, J.~M. 1988, \pasp, 100, 1134

\bibitem[{{Campana} {et~al.}(2001){Campana}, {Moretti}, {Lazzati}, \&
  {Tagliaferri}}]{campanaetal01-1}
{Campana}, S., {Moretti}, A., {Lazzati}, D., \& {Tagliaferri}, G. 2001, \apjl,
  560, L19

\bibitem[{{Carpenter}(2001)}]{carpenter01-1}
{Carpenter}, J.~M. 2001, \aj, 121, 2851

\bibitem[{{Carrera} {et~al.}(2007){Carrera}, {Ebrero}, {Mateos}, {Ceballos},
  {Corral}, {Barcons}, {Page}, {Rosen}, {Watson}, {Tedds}, {Della Ceca},
  {Maccacaro}, {Brunner}, {Freyberg}, {Lamer}, {Bauer}, \&
  {Ueda}}]{carreraetal07-1}
{Carrera}, F.~J., {Ebrero}, J., {Mateos}, S., {et~al.} 2007, \aap, 469, 27

\bibitem[{{Castelli} \& {Kurucz}(2004)}]{castelli+kurucz04-1}
{Castelli}, F. \& {Kurucz}, R.~L. 2004, ArXiv Astrophysics e-prints

\bibitem[{{Cohen}(2000)}]{cohen00-1}
{Cohen}, D.~H. 2000, in Astronomical Society of the Pacific Conference Series,
  Vol. 214, IAU Colloq. 175: The Be Phenomenon in Early-Type Stars, ed. M.~A.
  {Smith}, H.~F. {Henrichs}, \& J.~{Fabregat}, 156

\bibitem[{{Cohen} {et~al.}(1997){Cohen}, {Cassinelli}, \&
  {Macfarlane}}]{cohenetal97-1}
{Cohen}, D.~H., {Cassinelli}, J.~P., \& {Macfarlane}, J.~J. 1997, \apj, 487,
  867

\bibitem[{{Covey} {et~al.}(2008){Covey}, {Ag{\"u}eros}, {Green}, {Haggard},
  {Barkhouse}, {Drake}, {Evans}, {Kashyap}, {Kim}, {Mossman}, {Pease}, \&
  {Silverman}}]{coveyetal08-1}
{Covey}, K.~R., {Ag{\"u}eros}, M.~A., {Green}, P.~J., {et~al.} 2008, \apjs,
  178, 339

\bibitem[{{Covey} {et~al.}(2007){Covey}, {Ivezi{\'c}}, {Schlegel},
  {Finkbeiner}, {Padmanabhan}, {Lupton}, {Ag{\"u}eros}, {Bochanski}, {Hawley},
  {West}, {Seth}, {Kimball}, {Gogarten}, {Claire}, {Haggard}, {Kaib},
  {Schneider}, \& {Sesar}}]{coveyetal07-1}
{Covey}, K.~R., {Ivezi{\'c}}, {\v Z}., {Schlegel}, D., {et~al.} 2007, \aj, 134,
  2398

\bibitem[{{Crawford} {et~al.}(1970){Crawford}, {Jauncey}, \&
  {Murdoch}}]{crawfordetal70-1}
{Crawford}, D.~F., {Jauncey}, D.~L., \& {Murdoch}, H.~S. 1970, \apj, 162, 405

\bibitem[{{Cutri} {et~al.}(2003){Cutri}, {Skrutskie}, {van Dyk}, {Beichman},
  {Carpenter}, {Chester}, {Cambresy}, {Evans}, {Fowler}, {Gizis}, {Howard},
  {Huchra}, {Jarrett}, {Kopan}, {Kirkpatrick}, {Light}, {Marsh}, {McCallon},
  {Schneider}, {Stiening}, {Sykes}, {Weinberg}, {Wheaton}, {Wheelock}, \&
  {Zacarias}}]{cutrietal03-1}
{Cutri}, R.~M., {Skrutskie}, M.~F., {van Dyk}, S., {et~al.} 2003, {2MASS All
  Sky Catalog of point sources.}, ed. {Cutri, R.~M., Skrutskie, M.~F., van Dyk,
  S., Beichman, C.~A., Carpenter, J.~M., Chester, T., Cambresy, L., Evans, T.,
  Fowler, J., Gizis, J., Howard, E., Huchra, J., Jarrett, T., Kopan, E.~L.,
  Kirkpatrick, J.~D., Light, R.~M., Marsh, K.~A., McCallon, H., Schneider, S.,
  Stiening, R., Sykes, M., Weinberg, M., Wheaton, W.~A., Wheelock, S., \&
  Zacarias, N.}

\bibitem[{{Cutri} {et~al.}(2012){Cutri}, {Wright}, {Conrow}, {Bauer},
  {Benford}, {Brandenburg}, {Dailey}, {Eisenhardt}, {Evans}, {Fajardo-Acosta},
  {Fowler}, {Gelino}, {Grillmair}, {Harbut}, {Hoffman}, {Jarrett},
  {Kirkpatrick}, {Leisawitz}, {Liu}, {Mainzer}, {Marsh}, {Masci}, {McCallon},
  {Padgett}, {Ressler}, {Royer}, {Skrutskie}, {Stanford}, {Wyatt}, {Tholen},
  {Tsai}, {Wachter}, {Wheelock}, {Yan}, {Alles}, {Beck}, {Grav}, {Masiero},
  {McCollum}, {McGehee}, {Papin}, \& {Wittman}}]{cutrietal12-1}
{Cutri}, R.~M., {Wright}, E.~L., {Conrow}, T., {et~al.} 2012, {Explanatory
  Supplement to the WISE All-Sky Data Release Products}, Tech. rep.

\bibitem[{{De Rosa} {et~al.}(2011){De Rosa}, {Bulger}, {Patience}, {Leland},
  {Macintosh}, {Schneider}, {Song}, {Marois}, {Graham}, {Bessell}, \&
  {Doyon}}]{delarosa11-1}
{De Rosa}, R.~J., {Bulger}, J., {Patience}, J., {et~al.} 2011, \mnras, 415, 854

\bibitem[{{Della Ceca} {et~al.}(2004){Della Ceca}, {Maccacaro}, {Caccianiga},
  {Severgnini}, {Braito}, {Barcons}, {Carrera}, {Watson}, {Tedds}, {Brunner},
  {Lehmann}, {Page}, {Lamer}, \& {Schwope}}]{dellacecaetal04-1}
{Della Ceca}, R., {Maccacaro}, T., {Caccianiga}, A., {et~al.} 2004, \aap, 428,
  383

\bibitem[{{Dempsey} {et~al.}(1997){Dempsey}, {Linsky}, {Fleming}, \&
  {Schmitt}}]{dempseyetal97-1}
{Dempsey}, R.~C., {Linsky}, J.~L., {Fleming}, T.~A., \& {Schmitt}, J.~H.~M.~M.
  1997, \apj, 478, 358

\bibitem[{{Dempsey} {et~al.}(1993){Dempsey}, {Linsky}, {Schmitt}, \&
  {Fleming}}]{dempseyetal93-1}
{Dempsey}, R.~C., {Linsky}, J.~L., {Schmitt}, J.~H.~M.~M., \& {Fleming}, T.~A.
  1993, \apj, 413, 333

\bibitem[{{den Herder} {et~al.}(2001){den Herder}, {Brinkman}, {Kahn},
  {Branduardi-Raymont}, {Thomsen}, {Aarts}, {Audard}, {Bixler}, {den Boggende},
  {Cottam}, {Decker}, {Dubbeldam}, {Erd}, {Goulooze}, {G{\"u}del}, {Guttridge},
  {Hailey}, {Janabi}, {Kaastra}, {de Korte}, {van Leeuwen}, {Mauche},
  {McCalden}, {Mewe}, {Naber}, {Paerels}, {Peterson}, {Rasmussen}, {Rees},
  {Sakelliou}, {Sako}, {Spodek}, {Stern}, {Tamura}, {Tandy}, {de Vries},
  {Welch}, \& {Zehnder}}]{denherder2001}
{den Herder}, J.~W., {Brinkman}, A.~C., {Kahn}, S.~M., {et~al.} 2001, \aap,
  365, L7

\bibitem[{{Ebisawa} {et~al.}(2001){Ebisawa}, {Maeda}, {Kaneda}, \&
  {Yamauchi}}]{ebisawaetal01-1}
{Ebisawa}, K., {Maeda}, Y., {Kaneda}, H., \& {Yamauchi}, S. 2001, Science, 293,
  1633

\bibitem[{{Ebisawa} {et~al.}(2005){Ebisawa}, {Tsujimoto}, {Paizis},
  {Hamaguchi}, {Bamba}, {Cutri}, {Kaneda}, {Maeda}, {Sato}, {Senda}, {Ueno},
  {Yamauchi}, {Beckmann}, {Courvoisier}, {Dubath}, \&
  {Nishihara}}]{ebisawaetal05-1}
{Ebisawa}, K., {Tsujimoto}, M., {Paizis}, A., {et~al.} 2005, \apj, 635, 214

\bibitem[{{Fang} {et~al.}(2009){Fang}, {van Boekel}, {Wang}, {Carmona},
  {Sicilia-Aguilar}, \& {Henning}}]{fangetal09-1}
{Fang}, M., {van Boekel}, R., {Wang}, W., {et~al.} 2009, \aap, 504, 461

\bibitem[{{Favata} {et~al.}(1988){Favata}, {Sciortino}, {Rosner}, \&
  {Vaiana}}]{favataetal88-1}
{Favata}, F., {Sciortino}, S., {Rosner}, R., \& {Vaiana}, G.~S. 1988, \apj,
  324, 1010

\bibitem[{{Feigelson}(1996)}]{feigelsonetal96-1}
{Feigelson}, E.~D. 1996, \apj, 468, 306

\bibitem[{{Fleming} {et~al.}(1989){Fleming}, {Gioia}, \&
  {Maccacaro}}]{flemingetal89-1}
{Fleming}, T.~A., {Gioia}, I.~M., \& {Maccacaro}, T. 1989, \apj, 340, 1011

\bibitem[{{Frasca} {et~al.}(2006){Frasca}, {Guillout}, {Marilli}, {Freire
  Ferrero}, {Biazzo}, \& {Klutsch}}]{frascaetal06-1}
{Frasca}, A., {Guillout}, P., {Marilli}, E., {et~al.} 2006, \aap, 454, 301

\bibitem[{{Gagn{\'e}} {et~al.}(2011){Gagn{\'e}}, {Fehon}, {Savoy}, {Cohen},
  {Townsley}, {Broos}, {Povich}, {Corcoran}, {Walborn}, {Remage Evans},
  {Moffat}, {Naz{\'e}}, \& {Oskinova}}]{gagneetal11-1}
{Gagn{\'e}}, M., {Fehon}, G., {Savoy}, M.~R., {et~al.} 2011, \apjs, 194, 5

\bibitem[{{G{\"a}nsicke} {et~al.}(2009){G{\"a}nsicke}, {Dillon}, {Southworth},
  {Thorstensen}, {Rodr{\'{\i}}guez-Gil}, {Aungwerojwit}, {Marsh}, {Szkody},
  {Barros}, {Casares}, {de Martino}, {Groot}, {Hakala}, {Kolb}, {Littlefair},
  {Mart{\'{\i}}nez-Pais}, {Nelemans}, \& {Schreiber}}]{gaensickeetal09-1}
{G{\"a}nsicke}, B.~T., {Dillon}, M., {Southworth}, J., {et~al.} 2009, \mnras,
  397, 2170

\bibitem[{{Giacconi} {et~al.}(1979){Giacconi}, {Branduardi}, {Briel},
  {Epstein}, {Fabricant}, {Feigelson}, {Forman}, {Gorenstein}, {Grindlay},
  {Gursky}, {Harnden}, {Henry}, {Jones}, {Kellogg}, {Koch}, {Murray},
  {Schreier}, {Seward}, {Tananbaum}, {Topka}, {Van Speybroeck}, {Holt},
  {Becker}, {Boldt}, {Serlemitsos}, {Clark}, {Canizares}, {Markert}, {Novick},
  {Helfand}, \& {Long}}]{giacconietal79-1}
{Giacconi}, R., {Branduardi}, G., {Briel}, U., {et~al.} 1979, \apj, 230, 540

\bibitem[{{Gilfanov}(2004)}]{gilfanov04-1}
{Gilfanov}, M. 2004, \mnras, 349, 146

\bibitem[{{Grimm} {et~al.}(2002){Grimm}, {Gilfanov}, \&
  {Sunyaev}}]{grimmetal02-1}
{Grimm}, H.-J., {Gilfanov}, M., \& {Sunyaev}, R. 2002, \aap, 391, 923

\bibitem[{{Grindlay} {et~al.}(2005){Grindlay}, {Hong}, {Zhao}, {Laycock}, {van
  den Berg}, {Koenig}, {Schlegel}, {Cohn}, {Lugger}, \&
  {Rogel}}]{grindlayetal05-1}
{Grindlay}, J.~E., {Hong}, J., {Zhao}, P., {et~al.} 2005, \apj, 635, 920

\bibitem[{{G{\"u}del} \& {Naz{\'e}}(2009)}]{guedel+naze09-1}
{G{\"u}del}, M. \& {Naz{\'e}}, Y. 2009, \aapr, 17, 309

\bibitem[{{Guedel} {et~al.}(1997){Guedel}, {Guinan}, \&
  {Skinner}}]{guedeletal97-1}
{Guedel}, M., {Guinan}, E.~F., \& {Skinner}, S.~L. 1997, \apj, 483, 947

\bibitem[{{Guillout} {et~al.}(2010){Guillout}, {Frasca}, {Klutsch}, {Marilli},
  \& {Montes}}]{guilloutetal10-1}
{Guillout}, P., {Frasca}, A., {Klutsch}, A., {Marilli}, E., \& {Montes}, D.
  2010, \aap, 520, A94

\bibitem[{{Guillout} {et~al.}(1996){Guillout}, {Haywood}, {Motch}, \&
  {Robin}}]{guilloutetal96-1}
{Guillout}, P., {Haywood}, M., {Motch}, C., \& {Robin}, A.~C. 1996, \aap, 316,
  89

\bibitem[{{Guillout} {et~al.}(1999){Guillout}, {Schmitt}, {Egret}, {Voges},
  {Motch}, \& {Sterzik}}]{guilloutetal99-1}
{Guillout}, P., {Schmitt}, J.~H.~M.~M., {Egret}, D., {et~al.} 1999, \aap, 351,
  1003

\bibitem[{{Guillout} {et~al.}(1998){Guillout}, {Sterzik}, {Schmitt}, {Motch},
  \& {Neuhaeuser}}]{guillout1998}
{Guillout}, P., {Sterzik}, M.~F., {Schmitt}, J.~H.~M.~M., {Motch}, C., \&
  {Neuhaeuser}, R. 1998, \aap, 337, 113

\bibitem[{{Hands} {et~al.}(2004){Hands}, {Warwick}, {Watson}, \&
  {Helfand}}]{handsetal04-1}
{Hands}, A.~D.~P., {Warwick}, R.~S., {Watson}, M.~G., \& {Helfand}, D.~J. 2004,
  \mnras, 351, 31

\bibitem[{{Hasinger} {et~al.}(1998){Hasinger}, {Burg}, {Giacconi}, {Schmidt},
  {Trumper}, \& {Zamorani}}]{hasingeretal98-1}
{Hasinger}, G., {Burg}, R., {Giacconi}, R., {et~al.} 1998, \aap, 329, 482

\bibitem[{{Herbig}(1960)}]{herbig1960}
{Herbig}, G.~H. 1960, \apjs, 4, 337

\bibitem[{{Hern{\'a}ndez} {et~al.}(2005){Hern{\'a}ndez}, {Calvet}, {Hartmann},
  {Brice{\~n}o}, {Sicilia-Aguilar}, \& {Berlind}}]{hernandezetal05-1}
{Hern{\'a}ndez}, J., {Calvet}, N., {Hartmann}, L., {et~al.} 2005, \aj, 129, 856

\bibitem[{{Hertz} \& {Grindlay}(1984)}]{hertz+grindlay84-1}
{Hertz}, P. \& {Grindlay}, J.~E. 1984, \apj, 278, 137

\bibitem[{{Hong}(2012)}]{hong12-1}
{Hong}, J. 2012, \mnras, 427, 1633

\bibitem[{{Hong} {et~al.}(2012){Hong}, {van den Berg}, {Grindlay}, {Servillat},
  \& {Zhao}}]{hongetal12-1}
{Hong}, J., {van den Berg}, M., {Grindlay}, J.~E., {Servillat}, M., \& {Zhao},
  P. 2012, \apj, 746, 165

\bibitem[{{Hong} {et~al.}(2009){Hong}, {van den Berg}, {Grindlay}, \&
  {Laycock}}]{hongetal09-1}
{Hong}, J.~S., {van den Berg}, M., {Grindlay}, J.~E., \& {Laycock}, S. 2009,
  \apj, 706, 223

\bibitem[{{Jackson} {et~al.}(2012){Jackson}, {Davis}, \&
  {Wheatley}}]{jacksonetal12-1}
{Jackson}, A.~P., {Davis}, T.~A., \& {Wheatley}, P.~J. 2012, \mnras, 422, 2024

\bibitem[{{Jacoby} {et~al.}(1984){Jacoby}, {Hunter}, \&
  {Christian}}]{jacobyetal84-1}
{Jacoby}, G.~H., {Hunter}, D.~A., \& {Christian}, C.~A. 1984, \apjs, 56, 257

\bibitem[{{Jansen} {et~al.}(2001){Jansen}, {Lumb}, {Altieri}, {Clavel}, {Ehle},
  {Erd}, {Gabriel}, {Guainazzi}, {Gondoin}, {Much}, {Munoz}, {Santos},
  {Schartel}, {Texier}, \& {Vacanti}}]{jansenetal01-1}
{Jansen}, F., {Lumb}, D., {Altieri}, B., {et~al.} 2001, \aap, 365, L1

\bibitem[{{Joy}(1945)}]{joy1945}
{Joy}, A.~H. 1945, \apj, 102, 168

\bibitem[{{Kahabka} {et~al.}(2006){Kahabka}, {Haberl}, {Payne}, \&
  {Filipovi{\'c}}}]{kahabkaetal06-1}
{Kahabka}, P., {Haberl}, F., {Payne}, J.~L., \& {Filipovi{\'c}}, M.~D. 2006,
  \aap, 458, 285

\bibitem[{{Kawaler}(1988)}]{kawaler1988}
{Kawaler}, S.~D. 1988, \apj, 333, 236

\bibitem[{{Koenig} {et~al.}(2008){Koenig}, {Grindlay}, {van den Berg},
  {Laycock}, {Zhao}, {Hong}, \& {Schlegel}}]{koenigetal08-1}
{Koenig}, X., {Grindlay}, J.~E., {van den Berg}, M., {et~al.} 2008, \apj, 685,
  463

\bibitem[{{Kogure}(2009)}]{kogure09-1}
{Kogure}, T. 2009, in Astronomical Society of the Pacific Conference Series,
  Vol. 404, The Eighth Pacific Rim Conference on Stellar Astrophysics: A
  Tribute to Kam-Ching Leung, ed. {S.~J.~Murphy \& M.~S.~Bessell}, 212

\bibitem[{{Lasker} {et~al.}(2008){Lasker}, {Lattanzi}, {McLean}, {Bucciarelli},
  {Drimmel}, {Garcia}, {Greene}, {Guglielmetti}, {Hanley}, {Hawkins},
  {Laidler}, {Loomis}, {Meakes}, {Mignani}, {Morbidelli}, {Morrison},
  {Pannunzio}, {Rosenberg}, {Sarasso}, {Smart}, {Spagna}, {Sturch},
  {Volpicelli}, {White}, {Wolfe}, \& {Zacchei}}]{laskeretal08-1}
{Lasker}, B.~M., {Lattanzi}, M.~G., {McLean}, B.~J., {et~al.} 2008, \aj, 136,
  735

\bibitem[{{Laycock} {et~al.}(2005){Laycock}, {Grindlay}, {van den Berg},
  {Zhao}, {Hong}, {Koenig}, {Schlegel}, \& {Persson}}]{laycocketal05-1}
{Laycock}, S., {Grindlay}, J., {van den Berg}, M., {et~al.} 2005, \apjl, 634,
  L53

\bibitem[{{Le Borgne} {et~al.}(2003){Le Borgne}, {Bruzual}, {Pell{\'o}},
  {Lan{\c c}on}, {Rocca-Volmerange}, {Sanahuja}, {Schaerer}, {Soubiran}, \&
  {V{\'{\i}}lchez-G{\'o}mez}}]{leborgneetal03-1}
{Le Borgne}, J.-F., {Bruzual}, G., {Pell{\'o}}, R., {et~al.} 2003, \aap, 402,
  433

\bibitem[{{Lopes de Oliveira} \& {Motch}(2011)}]{oliveiraetal11-1}
{Lopes de Oliveira}, R. \& {Motch}, C. 2011, \apjl, 731, L6

\bibitem[{{Lopes de Oliveira} {et~al.}(2007){Lopes de Oliveira}, {Motch},
  {Smith}, {Negueruela}, \& {Torrej{\'o}n}}]{lopes2007}
{Lopes de Oliveira}, R., {Motch}, C., {Smith}, M.~A., {Negueruela}, I., \&
  {Torrej{\'o}n}, J.~M. 2007, \aap, 474, 983

\bibitem[{{Lopes de Oliveira} {et~al.}(2010){Lopes de Oliveira}, {Smith}, \&
  {Motch}}]{oliveiraetal10-1}
{Lopes de Oliveira}, R., {Smith}, M.~A., \& {Motch}, C. 2010, \aap, 512, A22

\bibitem[{{L{\'o}pez-Santiago} {et~al.}(2007){L{\'o}pez-Santiago}, {Micela},
  {Sciortino}, {Favata}, {Caccianiga}, {Della Ceca}, {Severgnini}, \&
  {Braito}}]{lopez-santiagoetal07-1}
{L{\'o}pez-Santiago}, J., {Micela}, G., {Sciortino}, S., {et~al.} 2007, \aap,
  463, 165

\bibitem[{{Maccacaro} {et~al.}(1988){Maccacaro}, {Gioia}, {Wolter}, {Zamorani},
  \& {Stocke}}]{maccacaroetal88-1}
{Maccacaro}, T., {Gioia}, I.~M., {Wolter}, A., {Zamorani}, G., \& {Stocke},
  J.~T. 1988, \apj, 326, 680

\bibitem[{{Maggio} {et~al.}(1990){Maggio}, {Vaiana}, {Haisch}, {Stern},
  {Bookbinder}, {Harnden}, \& {Rosner}}]{maggioetal90-1}
{Maggio}, A., {Vaiana}, G.~S., {Haisch}, B.~M., {et~al.} 1990, \apj, 348, 253

\bibitem[{{Malfait} {et~al.}(1998){Malfait}, {Bogaert}, \&
  {Waelkens}}]{malfaitetal98-1}
{Malfait}, K., {Bogaert}, E., \& {Waelkens}, C. 1998, \aap, 331, 211

\bibitem[{{Marco} {et~al.}(2007){Marco}, {Negueruela}, \& {Motch}}]{marco2007}
{Marco}, A., {Negueruela}, I., \& {Motch}, C. 2007, in Astronomical Society of
  the Pacific Conference Series, Vol. 367, Massive Stars in Interactive
  Binaries, ed. N.~{St.-Louis} \& A.~F.~J. {Moffat}, 645

\bibitem[{{Marshall} {et~al.}(2006){Marshall}, {Robin}, {Reyl{\'e}},
  {Schultheis}, \& {Picaud}}]{marshalletal06-1}
{Marshall}, D.~J., {Robin}, A.~C., {Reyl{\'e}}, C., {Schultheis}, M., \&
  {Picaud}, S. 2006, \aap, 453, 635

\bibitem[{{Mateos} {et~al.}(2008){Mateos}, {Warwick}, {Carrera}, {Stewart},
  {Ebrero}, {Della Ceca}, {Caccianiga}, {Gilli}, {Page}, {Treister}, {Tedds},
  {Watson}, {Lamer}, {Saxton}, {Brunner}, \& {Page}}]{mateosetal08-1}
{Mateos}, S., {Warwick}, R.~S., {Carrera}, F.~J., {et~al.} 2008, \aap, 492, 51

\bibitem[{{Mathis}(1990)}]{mathis90-1}
{Mathis}, J.~S. 1990, \araa, 28, 37

\bibitem[{{Matt} {et~al.}(2012){Matt}, {MacGregor}, {Pinsonneault}, \&
  {Greene}}]{matt2012}
{Matt}, S.~P., {MacGregor}, K.~B., {Pinsonneault}, M.~H., \& {Greene}, T.~P.
  2012, \apjl, 754, L26

\bibitem[{{Meeus} {et~al.}(1998){Meeus}, {Waelkens}, \&
  {Malfait}}]{meeusetal98-1}
{Meeus}, G., {Waelkens}, C., \& {Malfait}, K. 1998, \aap, 329, 131

\bibitem[{{Meeus} {et~al.}(2001){Meeus}, {Waters}, {Bouwman}, {van den Ancker},
  {Waelkens}, \& {Malfait}}]{meeusetal01-1}
{Meeus}, G., {Waters}, L.~B.~F.~M., {Bouwman}, J., {et~al.} 2001, \aap, 365,
  476

\bibitem[{{Meyer} {et~al.}(1997){Meyer}, {Calvet}, \&
  {Hillenbrand}}]{meyeretal97-1}
{Meyer}, M.~R., {Calvet}, N., \& {Hillenbrand}, L.~A. 1997, \aj, 114, 288

\bibitem[{{Micela} {et~al.}(1996){Micela}, {Sciortino}, {Kashyap}, {Harnden},
  \& {Rosner}}]{micelaetal96-1}
{Micela}, G., {Sciortino}, S., {Kashyap}, V., {Harnden}, Jr., F.~R., \&
  {Rosner}, R. 1996, \apjs, 102, 75

\bibitem[{{Micela} {et~al.}(1988){Micela}, {Sciortino}, {Vaiana}, {Schmitt},
  {Stern}, {Harnden}, \& {Rosner}}]{micelaetal88-1}
{Micela}, G., {Sciortino}, S., {Vaiana}, G.~S., {et~al.} 1988, \apj, 325, 798

\bibitem[{{Michel} {et~al.}(2004){Michel}, {Herent}, {Motch}, {Pye}, \&
  {Watson}}]{micheletal04-1}
{Michel}, L., {Herent}, O., {Motch}, C., {Pye}, J., \& {Watson}, M.~G. 2004, in
  Astronomical Society of the Pacific Conference Series, Vol. 314, Astronomical
  Data Analysis Software and Systems (ADASS) XIII, ed. {F.~Ochsenbein,
  M.~G.~Allen, \& D.~Egret}, 570

\bibitem[{{Monet} {et~al.}(2003){Monet}, {Levine}, {Canzian}, {Ables}, {Bird},
  {Dahn}, {Guetter}, {Harris}, {Henden}, {Leggett}, {Levison}, {Luginbuhl},
  {Martini}, {Monet}, {Munn}, {Pier}, {Rhodes}, {Riepe}, {Sell}, {Stone},
  {Vrba}, {Walker}, {Westerhout}, {Brucato}, {Reid}, {Schoening}, {Hartley},
  {Read}, \& {Tritton}}]{monetetal03-1}
{Monet}, D.~G., {Levine}, S.~E., {Canzian}, B., {et~al.} 2003, \aj, 125, 984

\bibitem[{{Morley} {et~al.}(2001){Morley}, {Briggs}, {Pye}, {Favata}, {Micela},
  \& {Sciortino}}]{morleyetal01-1}
{Morley}, J.~E., {Briggs}, K.~R., {Pye}, J.~P., {et~al.} 2001, \mnras, 326,
  1161

\bibitem[{{Motch}(2006)}]{motch06-1}
{Motch}, C. 2006, in ESA Special Publication, Vol. 604, The X-ray Universe
  2005, ed. {A.~Wilson}, 383

\bibitem[{{Motch} {et~al.}(1997){Motch}, {Guillout}, {Haberl}, {Pakull},
  {Pietsch}, \& {Reinsch}}]{motchetal97-1}
{Motch}, C., {Guillout}, P., {Haberl}, F., {et~al.} 1997, \aap, 318, 111

\bibitem[{{Motch} {et~al.}(2003){Motch}, {Herent}, \&
  {Guillout}}]{motchetal03-1}
{Motch}, C., {Herent}, O., \& {Guillout}, P. 2003, Astronomische Nachrichten,
  324, 61

\bibitem[{{Motch} {et~al.}(2007){Motch}, {Lopes de Oliveira}, {Negueruela},
  {Haberl}, \& {Janot-Pacheco}}]{motchetal07-1}
{Motch}, C., {Lopes de Oliveira}, R., {Negueruela}, I., {Haberl}, F., \&
  {Janot-Pacheco}, E. 2007, in Astronomical Society of the Pacific Conference
  Series, Vol. 361, Active OB-Stars: Laboratories for Stellare and
  Circumstellar Physics, ed. {A.~T.~Okazaki, S.~P.~Owocki, \& S.~Stefl}, 117

\bibitem[{{Motch} \& {Pakull}(2012)}]{motch+pakull12-1}
{Motch}, C. \& {Pakull}, M.~W. 2012, \memsai, 83, 415

\bibitem[{{Motch} {et~al.}(2010){Motch}, {Warwick}, {Cropper}, {Carrera},
  {Guillout}, {Pineau}, {Pakull}, {Rosen}, {Schwope}, {Tedds}, {Webb},
  {Negueruela}, \& {Watson}}]{motchetal10-1}
{Motch}, C., {Warwick}, R., {Cropper}, M.~S., {et~al.} 2010, \aap, 523, A92

\bibitem[{{Muno} {et~al.}(2003){Muno}, {Baganoff}, {Bautz}, {Brandt}, {Broos},
  {Feigelson}, {Garmire}, {Morris}, {Ricker}, \& {Townsley}}]{munoetal03-1}
{Muno}, M.~P., {Baganoff}, F.~K., {Bautz}, M.~W., {et~al.} 2003, \apj, 589, 225

\bibitem[{{Muno} {et~al.}(2009){Muno}, {Bauer}, {Baganoff}, {Bandyopadhyay},
  {Bower}, {Brandt}, {Broos}, {Cotera}, {Eikenberry}, {Garmire}, {Hyman},
  {Kassim}, {Lang}, {Lazio}, {Law}, {Mauerhan}, {Morris}, {Nagata},
  {Nishiyama}, {Park}, {Ram{\`i}rez}, {Stolovy}, {Wijnands}, {Wang}, {Wang}, \&
  {Yusef-Zadeh}}]{munoetal09-1}
{Muno}, M.~P., {Bauer}, F.~E., {Baganoff}, F.~K., {et~al.} 2009, \apjs, 181,
  110

\bibitem[{{Murdoch} {et~al.}(1973){Murdoch}, {Crawford}, \&
  {Jauncey}}]{murdochetal73-1}
{Murdoch}, H.~S., {Crawford}, D.~F., \& {Jauncey}, D.~L. 1973, \apj, 183, 1

\bibitem[{{Okazaki} \& {Negueruela}(2001)}]{okazaki+negueruela01-1}
{Okazaki}, A.~T. \& {Negueruela}, I. 2001, in Astronomical Society of the
  Pacific Conference Series, Vol. 234, X-ray Astronomy 2000, ed. {R.~Giacconi,
  S.~Serio, \& L.~Stella}, 281

\bibitem[{{Pallavicini} {et~al.}(1981){Pallavicini}, {Golub}, {Rosner},
  {Vaiana}, {Ayres}, \& {Linsky}}]{pallavicini1981}
{Pallavicini}, R., {Golub}, L., {Rosner}, R., {et~al.} 1981, \apj, 248, 279

\bibitem[{{Patterson}(1984)}]{patterson84-1}
{Patterson}, J. 1984, \apjs, 54, 443

\bibitem[{{Pfahl} {et~al.}(2002){Pfahl}, {Rappaport}, \&
  {Podsiadlowski}}]{pfahl2002}
{Pfahl}, E., {Rappaport}, S., \& {Podsiadlowski}, P. 2002, \apjl, 571, L37

\bibitem[{{Pickles}(1998)}]{pickles98-1}
{Pickles}, A.~J. 1998, \pasp, 110, 863

\bibitem[{{Pineau} {et~al.}(2008){Pineau}, {Derriere}, {Michel}, \&
  {Motch}}]{pineauetal08-1}
{Pineau}, F.-X., {Derriere}, S., {Michel}, L., \& {Motch}, C. 2008, in
  Astronomical Society of the Pacific Conference Series, Vol. 394, Astronomical
  Data Analysis Software and Systems XVII, ed. {R.~W.~Argyle, P.~S.~Bunclark,
  \& J.~R.~Lewis}, 369

\bibitem[{{Pineau} {et~al.}(2010){Pineau}, {Derriere}, {Michel}, \&
  {Motch}}]{pineau2010}
{Pineau}, F.-X., {Derriere}, S., {Michel}, L., \& {Motch}, C. 2010, in
  Astronomical Society of the Pacific Conference Series, Vol. 434, Astronomical
  Data Analysis Software and Systems XIX, ed. Y.~{Mizumoto}, K.-I. {Morita}, \&
  M.~{Ohishi}, 369

\bibitem[{{Pineau} {et~al.}(2011){Pineau}, {Motch}, {Carrera}, {Della Ceca},
  {Derri{\`e}re}, {Michel}, {Schwope}, \& {Watson}}]{pineauetal11-1}
{Pineau}, F.-X., {Motch}, C., {Carrera}, F., {et~al.} 2011, \aap, 527, A126

\bibitem[{{Pires} {et~al.}(2009){Pires}, {Motch}, \&
  {Janot-Pacheco}}]{pires2009}
{Pires}, A.~M., {Motch}, C., \& {Janot-Pacheco}, E. 2009, \aap, 504, 185

\bibitem[{{Predehl} \& {Schmitt}(1995)}]{predehl+schmitt95-1}
{Predehl}, P. \& {Schmitt}, J.~H.~M.~M. 1995, \aap, 293, 889

\bibitem[{{Pretorius} \& {Knigge}(2012)}]{pretorius+knigge12-1}
{Pretorius}, M.~L. \& {Knigge}, C. 2012, \mnras, 419, 1442

\bibitem[{{Raguzova}(2001)}]{raguzova01-1}
{Raguzova}, N.~V. 2001, \aap, 367, 848

\bibitem[{{Raharto} {et~al.}(1984){Raharto}, {Hamajima}, {Ichikawa}, {Ishida},
  \& {Hidayat}}]{rahartoetal84-1}
{Raharto}, M., {Hamajima}, K., {Ichikawa}, T., {Ishida}, K., \& {Hidayat}, B.
  1984, Annals of the Tokyo Astronomical Observatory, 19, 469

\bibitem[{{Reig}(2011)}]{reigetal11-1}
{Reig}, P. 2011, \apss, 332, 1

\bibitem[{{Ritter} \& {Kolb}(2003)}]{ritter+kolb03-1}
{Ritter}, H. \& {Kolb}, U. 2003, \aap, 404, 301

\bibitem[{{Savage} \& {Mathis}(1979)}]{savage+mathis79-1}
{Savage}, B.~D. \& {Mathis}, J.~S. 1979, \araa, 17, 73

\bibitem[{{Sazonov} {et~al.}(2006){Sazonov}, {Revnivtsev}, {Gilfanov},
  {Churazov}, \& {Sunyaev}}]{sazonovetal06-1}
{Sazonov}, S., {Revnivtsev}, M., {Gilfanov}, M., {Churazov}, E., \& {Sunyaev},
  R. 2006, \aap, 450, 117

\bibitem[{{Schlegel} {et~al.}(1998){Schlegel}, {Finkbeiner}, \&
  {Davis}}]{schlegeletal98-1}
{Schlegel}, D.~J., {Finkbeiner}, D.~P., \& {Davis}, M. 1998, \apj, 500, 525

\bibitem[{{Schmitt} \& {Liefke}(2004)}]{schmitt+liefke04-1}
{Schmitt}, J.~H.~M.~M. \& {Liefke}, C. 2004, \aap, 417, 651

\bibitem[{{Schrijver}(1987)}]{schrijver1987}
{Schrijver}, C.~J. 1987, \aap, 172, 111

\bibitem[{{Sciortino} {et~al.}(1995){Sciortino}, {Favata}, \&
  {Micela}}]{sciortinoetal95-1}
{Sciortino}, S., {Favata}, F., \& {Micela}, G. 1995, \aap, 296, 370

\bibitem[{{Sidoli} {et~al.}(2006){Sidoli}, {Mereghetti}, {Favata},
  {Oosterbroek}, \& {Parmar}}]{sidolietal06-1}
{Sidoli}, L., {Mereghetti}, S., {Favata}, F., {Oosterbroek}, T., \& {Parmar},
  A.~N. 2006, \aap, 456, 287

\bibitem[{{Siess} {et~al.}(2000){Siess}, {Dufour}, \&
  {Forestini}}]{siessetal00-1}
{Siess}, L., {Dufour}, E., \& {Forestini}, M. 2000, \aap, 358, 593

\bibitem[{{Skinner} {et~al.}(2010){Skinner}, {Zhekov}, {G{\"u}del}, {Schmutz},
  \& {Sokal}}]{skinneretal10-1}
{Skinner}, S.~L., {Zhekov}, S.~A., {G{\"u}del}, M., {Schmutz}, W., \& {Sokal},
  K.~R. 2010, \aj, 139, 825

\bibitem[{{Stelzer} {et~al.}(2006){Stelzer}, {Micela}, {Hamaguchi}, \&
  {Schmitt}}]{stelzeretal06-1}
{Stelzer}, B., {Micela}, G., {Hamaguchi}, K., \& {Schmitt}, J.~H.~M.~M. 2006,
  \aap, 457, 223

\bibitem[{{Stern} {et~al.}(1995){Stern}, {Schmitt}, \&
  {Kahabka}}]{sternetal95-1}
{Stern}, R.~A., {Schmitt}, J.~H.~M.~M., \& {Kahabka}, P.~T. 1995, \apj, 448,
  683

\bibitem[{{Str{\"u}der} {et~al.}(2001){Str{\"u}der}, {Briel}, {Dennerl},
  {Hartmann}, {Kendziorra}, {Meidinger}, {Pfeffermann}, {Reppin}, {Aschenbach},
  {Bornemann}, {Br{\"a}uninger}, {Burkert}, {Elender}, {Freyberg}, {Haberl},
  {Hartner}, {Heuschmann}, {Hippmann}, {Kastelic}, {Kemmer}, {Kettenring},
  {Kink}, {Krause}, {M{\"u}ller}, {Oppitz}, {Pietsch}, {Popp}, {Predehl},
  {Read}, {Stephan}, {St{\"o}tter}, {Tr{\"u}mper}, {Holl}, {Kemmer}, {Soltau},
  {St{\"o}tter}, {Weber}, {Weichert}, {von Zanthier}, {Carathanassis}, {Lutz},
  {Richter}, {Solc}, {B{\"o}ttcher}, {Kuster}, {Staubert}, {Abbey}, {Holland},
  {Turner}, {Balasini}, {Bignami}, {La Palombara}, {Villa}, {Buttler},
  {Gianini}, {Lain{\'e}}, {Lumb}, \& {Dhez}}]{struederetal01-1}
{Str{\"u}der}, L., {Briel}, U., {Dennerl}, K., {et~al.} 2001, \aap, 365, L18

\bibitem[{{Sturm} {et~al.}(2012){Sturm}, {Haberl}, {Pietsch}, {Coe},
  {Mereghetti}, {La Palombara}, {Owen}, \& {Udalski}}]{sturmetal12-1}
{Sturm}, R., {Haberl}, F., {Pietsch}, W., {et~al.} 2012, \aap, 537, A76

\bibitem[{{Sugizaki} {et~al.}(2001){Sugizaki}, {Mitsuda}, {Kaneda},
  {Matsuzaki}, {Yamauchi}, \& {Koyama}}]{sugizakietal01-1}
{Sugizaki}, M., {Mitsuda}, K., {Kaneda}, H., {et~al.} 2001, \apjs, 134, 77

\bibitem[{{Tanaka} {et~al.}(1994){Tanaka}, {Inoue}, \& {Holt}}]{tanakaetal94-1}
{Tanaka}, Y., {Inoue}, H., \& {Holt}, S.~S. 1994, \pasj, 46, L37

\bibitem[{{Tauris} \& {van den Heuvel}(2006)}]{tauris2006}
{Tauris}, T.~M. \& {van den Heuvel}, E.~P.~J. 2006, {Formation and evolution of
  compact stellar X-ray sources}, ed. W.~H.~G. {Lewin} \& M.~{van der Klis},
  623--665

\bibitem[{{Tr\"umper}(1982)}]{truemper82-1}
{Tr\"umper}, J. 1982, Advances in Space Research, 2, 241

\bibitem[{{Turner}(1981)}]{turner81-1}
{Turner}, D.~G. 1981, \aj, 86, 231

\bibitem[{{Turner} {et~al.}(2001){Turner}, {Abbey}, {Arnaud}, {Balasini},
  {Barbera}, {Belsole}, {Bennie}, {Bernard}, {Bignami}, {Boer}, {Briel},
  {Butler}, {Cara}, {Chabaud}, {Cole}, {Collura}, {Conte}, {Cros}, {Denby},
  {Dhez}, {Di Coco}, {Dowson}, {Ferrando}, {Ghizzardi}, {Gianotti}, {Goodall},
  {Gretton}, {Griffiths}, {Hainaut}, {Hochedez}, {Holland}, {Jourdain},
  {Kendziorra}, {Lagostina}, {Laine}, {La Palombara}, {Lortholary}, {Lumb},
  {Marty}, {Molendi}, {Pigot}, {Poindron}, {Pounds}, {Reeves}, {Reppin},
  {Rothenflug}, {Salvetat}, {Sauvageot}, {Schmitt}, {Sembay}, {Short},
  {Spragg}, {Stephen}, {Str{\"u}der}, {Tiengo}, {Trifoglio}, {Tr{\"u}mper},
  {Vercellone}, {Vigroux}, {Villa}, {Ward}, {Whitehead}, \&
  {Zonca}}]{turner2001}
{Turner}, M.~J.~L., {Abbey}, A., {Arnaud}, M., {et~al.} 2001, \aap, 365, L27

\bibitem[{{van den Berg} {et~al.}(2012){van den Berg}, {Penner}, {Hong},
  {Grindlay}, {Zhao}, {Laycock}, \& {Servillat}}]{vandenberg2012}
{van den Berg}, M., {Penner}, K., {Hong}, J., {et~al.} 2012, \apj, 748, 31

\bibitem[{{Walker} \& {Laney}(1987)}]{walkeretal87-1}
{Walker}, A.~R. \& {Laney}, C.~D. 1987, \mnras, 224, 61

\bibitem[{{Walter} {et~al.}(2004){Walter}, {Courvoisier}, {Foschini}, {Lebrun},
  {Lund}, {Parmar}, {Rodriguez}, {Tomsick}, \& {Ubertini}}]{walteretal04-1}
{Walter}, R., {Courvoisier}, T.~J.-L., {Foschini}, L., {et~al.} 2004, in ESA
  Special Publication, Vol. 552, 5th INTEGRAL Workshop on the INTEGRAL
  Universe, ed. V.~{Schoenfelder}, G.~{Lichti}, \& C.~{Winkler}, 417--422

\bibitem[{{Warner}(1995)}]{warner1995}
{Warner}, B. 1995, {Cataclysmic Variable Stars}, ed. C.~Cambridge Univ.\~Press,
  1--100

\bibitem[{{Warwick} {et~al.}(2011){Warwick}, {P{\'e}rez-Ram{\'{\i}}rez}, \&
  {Byckling}}]{warwick2011}
{Warwick}, R.~S., {P{\'e}rez-Ram{\'{\i}}rez}, D., \& {Byckling}, K. 2011,
  \mnras, 413, 595

\bibitem[{{Watson} {et~al.}(2001){Watson}, {Augu{\`e}res}, {Ballet}, {Barcons},
  {Barret}, {Boer}, {Boller}, {Bromage}, {Brunner}, {Carrera}, {Cropper},
  {Denby}, {Ehle}, {Elvis}, {Fabian}, {Freyberg}, {Guillout}, {Hameury},
  {Hasinger}, {Hinshaw}, {Maccacaro}, {Mason}, {McMahon}, {Michel}, {Mirioni},
  {Mittaz}, {Motch}, {Olive}, {Osborne}, {Page}, {Pakull}, {Perry}, {Pierre},
  {Pietsch}, {Pye}, {Read}, {Roberts}, {Rosen}, {Sauvageot}, {Schwope},
  {Sekiguchi}, {Stewart}, {Stewart}, {Valtchanov}, {Ward}, {Warwick}, {West},
  {White}, \& {Worrall}}]{watsonetal01-1}
{Watson}, M.~G., {Augu{\`e}res}, J.-L., {Ballet}, J., {et~al.} 2001, \aap, 365,
  L51

\bibitem[{{Watson} {et~al.}(2009){Watson}, {Schr{\"o}der}, {Fyfe}, {Page},
  {Lamer}, {Mateos}, {Pye}, {Sakano}, {Rosen}, {Ballet}, {Barcons}, {Barret},
  {Boller}, {Brunner}, {Brusa}, {Caccianiga}, {Carrera}, {Ceballos}, {Della
  Ceca}, {Denby}, {Denkinson}, {Dupuy}, {Farrell}, {Fraschetti}, {Freyberg},
  {Guillout}, {Hambaryan}, {Maccacaro}, {Mathiesen}, {McMahon}, {Michel},
  {Motch}, {Osborne}, {Page}, {Pakull}, {Pietsch}, {Saxton}, {Schwope},
  {Severgnini}, {Simpson}, {Sironi}, {Stewart}, {Stewart}, {Stobbart}, {Tedds},
  {Warwick}, {Webb}, {West}, {Worrall}, \& {Yuan}}]{watsonetal09-1}
{Watson}, M.~G., {Schr{\"o}der}, A.~C., {Fyfe}, D., {et~al.} 2009, \aap, 493,
  339

\bibitem[{{Weisskopf} {et~al.}(2002){Weisskopf}, {Brinkman}, {Canizares},
  {Garmire}, {Murray}, \& {Van Speybroeck}}]{weisskopfetal02-1}
{Weisskopf}, M.~C., {Brinkman}, B., {Canizares}, C., {et~al.} 2002, \pasp, 114,
  1

\bibitem[{{Willems} \& {Kolb}(2003)}]{willems2003}
{Willems}, B. \& {Kolb}, U. 2003, \mnras, 343, 949

\bibitem[{{Wright} {et~al.}(2010){Wright}, {Drake}, \&
  {Civano}}]{wrightetal10-1}
{Wright}, N.~J., {Drake}, J.~J., \& {Civano}, F. 2010, \apj, 725, 480

\bibitem[{{Zickgraf} {et~al.}(2005){Zickgraf}, {Krautter}, {Reffert},
  {Alcal{\'a}}, {Mujica}, {Covino}, \& {Sterzik}}]{zickgrafetal05-1}
{Zickgraf}, F.-J., {Krautter}, J., {Reffert}, S., {et~al.} 2005, \aap, 433, 151

\bibitem[{{Zinnecker} \& {Preibisch}(1994)}]{zinneckeretal94-1}
{Zinnecker}, H. \& {Preibisch}, T. 1994, \aap, 292, 152

\end{thebibliography}

\Online

\addtolength{\tabcolsep}{0.75ex}
\begin{landscape}
\begin{table}
\begin{center}
\caption[Xray parameters for field 3C449]{
X-ray parameters for detected sources in field 3C449}
\label{t:xray-param}
\fontsize{7}{7}\selectfont
\begin{tabular}{lrrrcrrrcrrrcrrrcc} 
\hline\hline 
\noalign{\smallskip}
 2XMM  & r$_{90}$ & pn\_B1$^{*}$ & pn\_B2$^{**}$ & 2MASS & d$_{x-o}$ & P$_{id}$ & kmag & GSC  & d$_{x-o}$ & P$_{id}$ & V & USNO & d$_{x-o}$ & P$_{id}$ & R & Class$^\dag$ & SpT$^\ddag$ \\
 Name   & [\arcsec] & [cts ks$^{-1}$] & [cts ks$^{-1}$] & Name & [\arcsec]  &          &      & Name & [\arcsec]   &          &   & Name & [\arcsec]   &          &   &       &       \\
\noalign{\smallskip}
\hline\noalign{\smallskip}
\noalign{\smallskip}
 J223010.5+391803 &    2.37 &                     &                     &                      &        &         &         &      N2XK037841 &     5.4 &    0.07 &         &       1293-0507754 &     5.5 &    0.27 &   19.71 &            &         $^{       }$   \\
 J223016.2+392502 &    0.58 &    30.0$\pm$    2.3 &     3.0$\pm$    1.2 &    22301621+3925017  &    0.7 &    0.51 &   11.59 &      N2YH001089 &     0.3 &    1.00 &   14.34 &       1294-0502068 &     0.2 &    1.00 &   14.37 &       Star &   M2Ve  $^{    sp }$   \\
 J223021.2+392253 &    1.01 &     6.6$\pm$    1.1 &     1.2$\pm$    0.8 &    22302116+3922519  &    1.3 &    0.98 &   12.64 &      N2XM009117 &     1.3 &    0.98 &         &       1293-0507828 &     1.3 &    0.98 &   15.29 &       Star &   K7Ve  $^{    sp }$   \\
 J223032.9+392555 &    1.70 &     4.4$\pm$    1.2 &     0.9$\pm$    1.1 &    22303305+3925546  &    2.1 &    0.84 &   14.65 &      N2XM010425 &     2.1 &    0.91 &         &       1294-0502199 &     1.9 &    0.94 &   15.73 &            &         $^{       }$   \\
 J223036.1+392139 &    0.44 &    58.6$\pm$    2.5 &     0.5$\pm$    0.4 &    22303611+3921392  &    0.7 &    1.00 &    8.10 &      N2XM008981 &     0.6 &    0.50 &    9.27 &       1293-0507963 &     0.6 &    1.00 &    8.99 &       Star &    F8III  $^{     s }$   \\
 J223039.3+391940 &    1.83 &     2.0$\pm$    0.6 &     0.7$\pm$    0.5 &                      &        &         &         &      N2XN036915 &     3.9 &    0.30 &         &       1293-0507995 &     3.8 &    0.50 &   19.37 &            &         $^{       }$   \\
 J223041.5+391733 &    3.19 &     6.3$\pm$    1.3 &     1.8$\pm$    1.1 &                      &        &         &         &                 &         &         &         &                    &         &         &         &       EG   &    Gal  $^{     s }$   \\
 J223043.7+392257 &    1.27 &     2.4$\pm$    0.6 &                     &    22304371+3922579  &    0.2 &    1.00 &    7.68 &      N2XM001445 &     0.1 &    1.00 &   10.40 &       1293-0508029 &     0.1 &    1.00 &    9.71 &       Star &    K1III  $^{    sp }$   \\
 J223050.8+393323 &    2.08 &     2.6$\pm$    0.9 &     2.1$\pm$    1.5 &                      &        &         &         &      N2YH001423 &     4.3 &    0.26 &         &       1295-0508481 &     3.9 &    0.53 &   19.64 &            &         $^{       }$   \\
 J223057.6+392143 &    1.79 &     0.3$\pm$    0.4 &     0.1$\pm$    0.2 &                      &        &         &         &      N2XM030986 &     3.1 &    0.30 &         &       1293-0508125 &     3.3 &    0.44 &   19.70 &            &         $^{       }$   \\
 J223102.1+391838 &    0.53 &    19.3$\pm$    1.4 &     1.4$\pm$    0.5 &    22310213+3918391  &    0.5 &    1.00 &   12.14 &      N2XN025009 &     0.5 &    1.00 &         &       1293-0508160 &     0.5 &    1.00 &   15.09 &       Star &   M0Ve  $^{    sp }$   \\
 J223122.8+390914 &    1.93 &     3.4$\pm$    1.1 &     0.3$\pm$    0.8 &    22312286+3909137  &    0.7 &    0.99 &   10.92 &      N2XN000071 &     0.8 &    0.99 &   13.35 &       1291-0499913 &     0.6 &    0.99 &   12.99 &       TTS &   K0Ve  $^{    sp }$   \\
 J223123.5+393558 &    1.96 &    10.7$\pm$    1.8 &     3.9$\pm$    1.4 &                      &        &         &         &                 &         &         &         &                    &         &         &         &       EG   &    AGN  $^{    sp }$   \\
 J223125.0+391914 &    0.57 &    10.8$\pm$    1.1 &     4.9$\pm$    0.7 &                      &        &         &         &      N2XM007512 &     0.3 &    0.98 &         &       1293-0508332 &     0.2 &    0.98 &   18.90 &            &         $^{       }$   \\
 J223128.5+391051 &    1.42 &     6.9$\pm$    2.4 &     3.2$\pm$    1.9 &    22312847+3910520  &    1.3 &    0.93 &   14.76 &      N2XN024216 &     1.0 &    0.95 &         &       1291-0499944 &     1.4 &    0.94 &   16.54 &       Star &    K4V  $^{    sp }$   \\
 J223132.5+392458 &    1.99 &     3.1$\pm$    0.8 &     0.8$\pm$    0.5 &    22313239+3924583  &    1.8 &    0.93 &   14.20 &      N2XM009995 &     1.7 &    0.96 &         &       1294-0502686 &     1.7 &    0.96 &   14.99 &       Star &    G0V  $^{    sp }$   \\
 J223133.0+391419 &    1.35 &     3.2$\pm$    0.8 &                     &    22313317+3914201  &    1.7 &    1.00 &    9.28 &      N2XN000041 &     1.6 &    1.00 &   11.52 &       1292-0502438 &     1.6 &    1.00 &   10.46 &       Star &    K1V  $^{    sp }$   \\
 J223140.5+391539 &    1.24 &     3.0$\pm$    0.7 &     0.4$\pm$    0.4 &                      &        &         &         &      N2XN024784 &     2.6 &    0.62 &         &       1292-0502476 &     1.0 &    0.90 &   19.51 &            &         $^{       }$   \\
 J223150.1+392521 &    1.06 &     7.5$\pm$    1.2 &     1.8$\pm$    0.9 &    22315026+3925219  &    0.9 &    1.00 &   11.99 &      N2XM001392 &     1.0 &    1.00 &   13.59 &       1294-0502796 &     1.0 &    1.00 &   13.67 &       Star &    K0V  $^{    sp }$   \\
 J223200.3+393127 &    1.36 &    10.9$\pm$    2.9 &     1.0$\pm$    2.0 &                      &        &         &         &      N2XM012497 &     1.3 &    0.84 &         &       1295-0508917 &     1.7 &    0.81 &   19.73 &       Star &    G2V  $^{    sp }$   \\
 J223204.0+391435 &    1.85 &     1.4$\pm$    0.6 &     1.0$\pm$    0.6 &                      &        &         &         &                 &         &         &         &       1292-0502621 &     6.4 &    0.14 &   18.67 &            &         $^{       }$   \\
 J223204.0+392611 &    1.42 &     5.0$\pm$    1.0 &     0.4$\pm$    0.6 &    22320392+3926125  &    1.5 &    0.99 &   12.14 &      N2XM001365 &     1.6 &    0.99 &   13.28 &       1294-0502893 &     1.6 &    0.99 &   12.84 &       Star &    G0V  $^{    sp }$   \\
 J223204.3+390922 &    1.48 &                     &                     &                      &        &         &         &      N2XN024017 &     4.8 &    0.04 &         &       1291-0500156 &     5.1 &    0.04 &   19.11 &            &         $^{       }$   \\
 J223218.1+393006 &    1.63 &     5.6$\pm$    1.3 &     2.6$\pm$    1.5 &                      &        &         &         &      N2XM011980 &     2.0 &    0.77 &         &       1295-0509013 &     1.7 &    0.79 &   19.70 &            &         $^{       }$   \\
 J223221.1+393013 &    1.58 &     6.4$\pm$    1.4 &     3.4$\pm$    1.7 &                      &        &         &         &      N2XM012033 &     3.0 &    0.57 &         &       1295-0509027 &     3.1 &    0.59 &   19.42 &            &         $^{       }$   \\
 J223227.3+391355 &    0.75 &    45.4$\pm$    3.3 &    19.4$\pm$    3.0 &                      &        &         &         &      N2XM005797 &     2.4 &    0.36 &         &       1292-0502768 &     2.2 &    0.47 &   18.29 &       EG   &    AGN  $^{    sp }$   \\
 J223238.4+392409 &    2.61 &     3.3$\pm$    1.1 &     1.3$\pm$    1.0 &                      &        &         &         &      N2XM031494 &     2.5 &    0.35 &         &       1294-0503147 &     3.8 &    0.39 &   19.34 &            &         $^{       }$   \\
\noalign{\smallskip}\hline
\end{tabular}
\end{center}
\textbf{Notes.} 
$^{*}$ Count rate in the energy band 0.5--2.5 keV. $^{**}$ Count rate in the energy band 2--12 keV. $^\dag$ Class stands for the type of source. Classification can be \emph{Star} for active coronae, \emph{EG} for extragalactic sources, \emph{CV} for cataclysmic variables, \emph{TTS} for T~Tauri stars, or \emph{HMXB} for high mass X-ray binaries. The luminosity class corresponds to that derived in Section~\ref{sec:giants}.
$^\ddag$ Spectral type including information on how the source was identified, $^{sp}$ stands for spectroscopic identified sources, $^s$ for objects with SIMBAD identification, and $^{im}$ for sources found to be extended in the optical images. 
\end{table}
\end{landscape}
\normalsize

\begin{landscape}
\begin{table}
\begin{center}
\caption[Xray parameters for field 3C436]{
X-ray parameters for detected sources in field 3C436}
\label{t:xray-param}
\fontsize{7}{7}\selectfont
\begin{tabular}{lrrrcrrrcrrrcrrrcc} 
\hline\hline 
\noalign{\smallskip}
 2XMM  & r$_{90}$ & pn\_B1$^{*}$ & pn\_B2$^{**}$ & 2MASS & d$_{x-o}$ & P$_{id}$ & kmag & GSC  & d$_{x-o}$ & P$_{id}$ & V & USNO & d$_{x-o}$ & P$_{id}$ & R & Class$^\dag$ & SpT$^\ddag$ \\
 Name   & [\arcsec] & [cts ks$^{-1}$] & [cts ks$^{-1}$] & Name & [\arcsec]  &          &      & Name & [\arcsec]   &          &   & Name & [\arcsec]   &          &   &       &       \\
\noalign{\smallskip}
\hline\noalign{\smallskip}
\noalign{\smallskip}
 J214327.0+281603 &    1.59 &     1.5$\pm$    0.5 &     1.9$\pm$    0.9 &    21432699+2815580  &    5.2 &    0.12 &   16.08 &      N305023834 &     1.7 &    0.76 &         &       1182-0630006 &     2.0 &    0.45 &        &            &         $^{       }$   \\
 J214332.8+281213 &    0.75 &     7.8$\pm$    0.8 &     1.3$\pm$    0.6 &    21433285+2812134  &    0.6 &    0.99 &   13.77 &      N305022044 &     1.1 &    0.98 &   17.19 &       1182-0630057 &     0.8 &    0.98 &   15.85 &       Star &   K6Ve  $^{    sp }$   \\
 J214337.8+280537 &    1.21 &     2.1$\pm$    0.5 &     0.6$\pm$    0.5 &                      &        &         &         &      N305028465 &     1.8 &    0.71 &         &       1180-0668164 &     1.0 &    0.87 &   20.10 &            &         $^{       }$   \\
 J214350.1+275822 &    1.47 &     1.5$\pm$    0.6 &     2.4$\pm$    1.1 &                      &        &         &         &      N305016221 &     0.6 &    0.87 &         &       1179-0686320 &     2.6 &    0.46 &   19.37 &            &         $^{       }$   \\
 J214354.7+281022 &    0.86 &     3.7$\pm$    0.5 &     0.1$\pm$    0.2 &    21435472+2810236  &    1.4 &    0.95 &   13.98 &      N305021207 &     1.7 &    0.94 &   16.91 &       1181-0656366 &     1.6 &    0.93 &   15.90 &       Star &    K2V  $^{    sp }$   \\
 J214357.2+281957 &    1.27 &     2.1$\pm$    0.7 &     0.8$\pm$    0.7 &                      &        &         &         &      N305043888 &     0.6 &    0.82 &         &       1183-0599981 &     0.4 &    0.85 &   20.56 &            &         $^{       }$   \\
 J214400.7+281126 &    0.48 &    17.0$\pm$    0.9 &     3.4$\pm$    0.5 &                      &        &         &         &      N305021689 &     0.9 &    0.95 &   18.51 &       1181-0656437 &     0.5 &    0.98 &   18.59 &       EG   &    AGN  $^{    sp }$   \\
 J214400.8+280609 &    0.52 &    11.9$\pm$    0.8 &     3.7$\pm$    0.6 &                      &        &         &         &      N305019344 &     0.8 &    0.96 &         &       1181-0656435 &     1.1 &    0.90 &   19.93 &            &         $^{       }$   \\
 J214406.1+275958 &    0.68 &    10.1$\pm$    1.1 &     4.3$\pm$    0.9 &                      &        &         &         &      N305016825 &     0.8 &    0.94 &         &       1179-0686476 &     0.8 &    0.94 &   19.64 &            &         $^{       }$   \\
 J214412.0+281722 &    0.56 &    14.4$\pm$    1.1 &     4.7$\pm$    0.8 &                      &        &         &         &      N305024541 &     0.8 &    0.96 &         &       1182-0630418 &     0.8 &    0.95 &   19.37 &            &         $^{       }$   \\
 J214412.7+275954 &    1.62 &     2.4$\pm$    0.6 &     0.1$\pm$    0.3 &                      &        &         &         &      N305016816 &     1.7 &    0.77 &         &       1179-0686527 &     2.0 &    0.78 &   20.60 &            &         $^{       }$   \\
 J214414.6+280535 &    1.69 &     0.4$\pm$    0.2 &     1.5$\pm$    0.4 &    21441483+2805354  &    2.3 &    0.86 &   14.70 &      N305019137 &     2.4 &    0.96 &   16.69 &       1180-0668510 &     2.4 &    0.98 &   13.74 &            &         $^{       }$   \\
 J214421.1+280831 &    0.50 &    10.9$\pm$    0.8 &     2.2$\pm$    0.4 &                      &        &         &         &      N305020333 &     1.8 &    0.38 &         &       1181-0656642 &     1.6 &    0.54 &   18.44 &            &         $^{       }$   \\
 J214423.6+280320 &    0.99 &     3.8$\pm$    0.6 &                     &    21442364+2803205  &    0.4 &    1.00 &   12.94 &      N305018152 &     0.6 &    0.97 &   18.08 &       1180-0668591 &     0.5 &    0.97 &   17.09 &       Star &   M5Ve  $^{    sp }$   \\
 J214425.0+275821 &    0.98 &     4.4$\pm$    0.8 &     5.4$\pm$    1.3 &                      &        &         &         &      N305016199 &     1.2 &    0.94 &   18.70 &       1179-0686636 &     0.8 &    0.96 &   18.71 &            &         $^{       }$   \\
 J214427.1+282346 &    1.54 &     4.2$\pm$    0.9 &     1.0$\pm$    1.1 &    21442714+2823461  &    0.7 &    0.98 &   13.05 &      N305040365 &     4.9 &    0.98 &         &       1183-0600218 &     1.0 &    0.91 &   15.08 &       Star &   K7Ve  $^{    sp }$   \\
 J214427.6+282313 &    1.49 &     5.2$\pm$    1.0 &     3.4$\pm$    1.2 &                      &        &         &         &      N305036627 &     4.8 &    0.02 &   18.70 &       1183-0600226 &     0.7 &    0.79 &         &            &         $^{       }$   \\
 J214427.7+281153 &    1.44 &     1.7$\pm$    0.4 &                     &    21442773+2811529  &    0.9 &    0.97 &   13.87 &      N305021908 &     0.7 &    0.91 &   18.59 &       1181-0656708 &     0.7 &    0.91 &   18.41 &            &         $^{       }$   \\
 J214428.0+281846 &    1.39 &     2.7$\pm$    0.6 &     0.3$\pm$    0.5 &    21442813+2818457  &    0.9 &    0.97 &   13.93 &      N305025270 &     1.2 &    0.93 &   17.71 &       1183-0600230 &     1.3 &    0.96 &   16.20 &            &         $^{       }$   \\
 J214434.3+282306 &    1.44 &     7.6$\pm$    1.2 &     2.7$\pm$    1.4 &                      &        &         &         &      N305036605 &     2.7 &    0.49 &         &       1183-0600289 &     2.1 &    0.46 &   19.58 &            &         $^{       }$   \\
 J214436.2+280811 &    1.24 &     0.8$\pm$    0.4 &     1.4$\pm$    0.5 &                      &        &         &         &                 &         &         &         &       1181-0656777 &     4.3 &    0.32 &   19.45 &            &         $^{       }$   \\
 J214439.6+281405 &    0.49 &    19.8$\pm$    1.2 &    10.6$\pm$    1.1 &                      &        &         &         &      N305022954 &     0.2 &    0.97 &         &       1182-0630685 &     0.5 &    0.96 &   20.08 &            &         $^{       }$   \\
 J214440.8+275942 &    0.76 &                     &                     &                      &        &         &         &      N305016757 &     1.1 &    0.96 &         &       1179-0686754 &     0.7 &    0.98 &   19.81 &            &         $^{       }$   \\
 J214459.7+280717 &    0.80 &    14.2$\pm$    1.2 &     0.8$\pm$    0.6 &    21445977+2807167  &    0.3 &    1.00 &   12.28 &      N305000004 &     0.1 &    1.00 &   13.90 &       1181-0656966 &     0.3 &    1.00 &   13.79 &       Star &    F7V  $^{    sp }$   \\
 J214504.1+281639 &    1.49 &     4.5$\pm$    0.9 &     0.7$\pm$    0.8 &                      &        &         &         &      N305024225 &     0.3 &    0.89 &         &       1182-0630904 &     0.4 &    0.71 &   20.07 &            &         $^{       }$   \\
 J214506.7+280303 &    2.16 &                     &                     &                      &        &         &         &      N305018079 &     1.7 &    0.82 &         &       1180-0668938 &     1.2 &    0.85 &   18.85 &            &         $^{       }$   \\
 J214512.0+281614 &    1.15 &    11.3$\pm$    1.4 &     6.8$\pm$    1.8 &                      &        &         &         &      N305024004 &     1.6 &    0.81 &         &       1182-0630968 &     1.4 &    0.84 &   19.94 &            &         $^{       }$   \\
 J214517.9+281108 &    2.25 &     3.8$\pm$    0.9 &     1.2$\pm$    0.9 &    21451751+2811128  &    6.9 &    0.15 &   14.96 &      N305021627 &     7.3 &    0.02 &         &       1181-0657111 &     7.2 &    0.09 &   18.96 &            &         $^{       }$   \\
\noalign{\smallskip}\hline
\end{tabular}
\end{center}
\end{table}
\end{landscape}
\normalsize

\begin{landscape}
\begin{table}
\begin{center}
\caption[Xray parameters for field AXJ2019+112]{
X-ray parameters for detected sources in field AXJ2019+112}
\label{t:xray-param}
\fontsize{7}{7}\selectfont
\begin{tabular}{lrrrcrrrcrrrcrrrcc} 
\hline\hline 
\noalign{\smallskip}
 2XMM  & r$_{90}$ & pn\_B1$^{*}$ & pn\_B2$^{**}$ & 2MASS & d$_{x-o}$ & P$_{id}$ & kmag & GSC  & d$_{x-o}$ & P$_{id}$ & V & USNO & d$_{x-o}$ & P$_{id}$ & R & Class$^\dag$ & SpT$^\ddag$ \\
 Name   & [\arcsec] & [cts ks$^{-1}$] & [cts ks$^{-1}$] & Name & [\arcsec]  &          &      & Name & [\arcsec]   &          &   & Name & [\arcsec]   &          &   &       &       \\
\noalign{\smallskip}
\hline\noalign{\smallskip}
\noalign{\smallskip}
 J201848.3+112104 &    2.05 &     2.5$\pm$    0.8 &     2.4$\pm$    1.0 &                      &        &         &         &      N1YB031812 &     3.1 &    0.27 &         &       1013-0571884 &     3.5 &    0.26 &   20.10 &            &         $^{       }$   \\
 J201900.0+113335 &    1.86 &     3.1$\pm$    0.8 &     0.0$\pm$    0.5 &    20190010+1133350  &    1.1 &    0.73 &   15.48 &      N1YB038048 &     1.2 &    0.78 &   18.94 &       1015-0585360 &     0.8 &    0.77 &   18.68 &            &         $^{       }$   \\
 J201900.1+113507 &    1.12 &    30.7$\pm$    2.4 &     2.9$\pm$    1.1 &    20190019+1135077  &    0.6 &    1.00 &   10.50 &      N1YB000545 &     0.5 &    1.00 &   11.59 &       1015-0585363 &     0.5 &    1.00 &   10.81 &       Star &    G3V  $^{    sp }$   \\
 J201905.9+112833 &    1.58 &     2.0$\pm$    0.8 &     1.6$\pm$    0.7 &    20190591+1128371  &    3.3 &    0.34 &   13.84 &      N1YB035790 &     3.3 &    0.62 &   15.47 &       1014-0582124 &     3.3 &    0.56 &   15.03 &       Star &    G2V  $^{    sp }$   \\
 J201908.8+113542 &    1.58 &     5.4$\pm$    1.1 &                     &    20190893+1135426  &    0.8 &    1.00 &    9.09 &      N1YB000540 &     0.8 &    1.00 &   10.40 &       1015-0585506 &     0.8 &    1.00 &   10.01 &       Star &    F5V  $^{     s }$   \\
 J201909.4+113936 &    1.77 &                     &     1.4$\pm$    1.7 &    20190953+1139308  &    5.7 &    0.03 &   15.46 &      N1YB040486 &     5.9 &    0.02 &   18.89 &       1016-0593168 &     5.6 &    0.03 &   17.73 &            &         $^{       }$   \\
 J201909.7+113146 &    1.48 &     4.8$\pm$    0.9 &     0.8$\pm$    0.5 &    20190991+1131467  &    2.1 &    0.82 &   13.12 &      N1YB037233 &     2.3 &    0.80 &   15.79 &       1015-0585519 &     1.9 &    0.83 &   15.57 &       Star &    K2V  $^{    sp }$   \\
 J201918.0+112713 &    1.31 &     3.7$\pm$    0.7 &     3.9$\pm$    0.8 &                      &        &         &         &                 &         &         &         &                    &         &         &         &        EG  &    QSO  $^{     s }$   \\
 J201921.6+112354 &    1.14 &    11.8$\pm$    1.2 &     2.6$\pm$    0.7 &    20192160+1123543  &    0.7 &    0.99 &   11.91 &      N1YB033420 &     0.5 &    0.99 &   14.33 &       1013-0572442 &     0.3 &    1.00 &   14.13 &       Star &    K0V  $^{    sp }$   \\
 J201923.3+114051 &    1.94 &     6.0$\pm$    1.7 &     3.7$\pm$    1.7 &    20192355+1140507  &    3.8 &    0.12 &   15.84 &      N1YB040987 &     4.0 &    0.09 &   18.59 &       1016-0593424 &     3.6 &    0.29 &   17.60 &            &         $^{       }$   \\
 J201928.8+112852 &    2.30 &     2.9$\pm$    0.7 &                     &    20192896+1128529  &    1.4 &    0.70 &   15.68 &      N1YB049561 &     1.4 &         &         &                    &         &         &         &            &         $^{       }$   \\
 J201928.9+112930 &    1.09 &    15.0$\pm$    1.4 &     0.7$\pm$    0.6 &    20192899+1129299  &    0.9 &    0.97 &   13.32 &      N1YB036175 &     1.0 &    0.89 &   18.19 &       1014-0582534 &     0.7 &    0.94 &   17.08 &       Star &   M5Ve  $^{    sp }$   \\
 J201930.2+112449 &    1.15 &    14.5$\pm$    1.4 &     4.0$\pm$    0.9 &                      &        &         &         &                 &         &         &         &       1014-0582553 &     2.0 &    0.31 &   17.03 &            &         $^{       }$   \\
 J201931.2+112900 &    1.55 &     3.1$\pm$    0.7 &     1.0$\pm$    0.6 &    20193141+1128599  &    2.0 &    0.69 &   14.78 &      N1YB035965 &     2.2 &    0.70 &   18.51 &       1014-0582566 &     1.8 &    0.75 &   17.10 &            &         $^{       }$   \\
 J201942.5+111611 &    1.54 &     4.3$\pm$    1.2 &     2.8$\pm$    1.8 &    20194258+1116117  &    0.2 &    0.97 &   12.91 &      N1YB028900 &     0.1 &    0.92 &   16.54 &       1012-0569389 &     0.3 &    0.96 &   15.84 &            &         $^{       }$   \\
 J201942.8+114031 &    1.62 &    11.5$\pm$    2.2 &     5.0$\pm$    1.8 &    20194280+1140312  &    0.6 &    0.76 &   16.75 &      N1YB040859 &     0.1 &    0.83 &   19.15 &       1016-0593741 &     0.4 &    0.77 &   18.97 &            &         $^{       }$   \\
 J201944.5+113101 &    1.64 &     4.4$\pm$    1.0 &     1.5$\pm$    0.9 &                      &        &         &         &      N1YB036856 &     1.7 &    0.62 &         &       1015-0586123 &     1.3 &    0.70 &   19.40 &            &         $^{       }$   \\
 J201946.7+113454 &    4.52 &     4.6$\pm$    1.3 &     1.5$\pm$    1.0 &                      &        &         &         &      N1YB038700 &    12.9 &    0.03 &         &       1015-0586149 &    10.8 &    0.04 &   18.33 &            &         $^{       }$   \\
 J201953.9+113800 &    1.27 &    26.2$\pm$    3.0 &     4.4$\pm$    2.3 &    20195398+1138002  &    0.7 &    0.99 &   11.30 &      N1YB000529 &     0.8 &    0.99 &   13.18 &       1016-0593928 &     0.8 &    0.99 &   12.79 &       Star &    K0V  $^{    sp }$   \\
 J201959.1+113336 &    2.65 &     0.6$\pm$    2.0 &                     &    20195923+1133349  &    1.8 &    0.76 &   13.74 &      N1YB038025 &     1.8 &    0.62 &         &       1015-0586359 &     1.4 &    0.65 &   17.43 &            &         $^{       }$   \\
 J202000.8+112506 &    1.40 &     7.8$\pm$    1.5 &     8.2$\pm$    2.1 &    20200090+1125041  &    2.1 &    0.37 &   15.54 &      N1YB033993 &     1.9 &    0.56 &   18.80 &       1014-0582998 &     1.5 &    0.73 &   18.05 &            &         $^{       }$   \\
 J202005.9+113645 &    2.30 &     6.9$\pm$    1.9 &     2.0$\pm$    1.7 &                      &        &         &         &      N1YB039414 &     1.3 &    0.75 &   18.92 &       1016-0594132 &     1.5 &    0.54 &         &            &         $^{       }$   \\
 J202008.6+112118 &    2.20 &     3.5$\pm$    1.4 &     3.9$\pm$    2.2 &                      &        &         &         &      N1YB031980 &     3.4 &    0.25 &   18.73 &       1013-0573179 &     3.4 &    0.26 &   18.17 &            &         $^{       }$   \\
\noalign{\smallskip}\hline
\end{tabular}
\end{center}
\end{table}
\end{landscape}
\normalsize

\begin{landscape}
\begin{table}
\begin{center}
\caption[Xray parameters for field G21]{
X-ray parameters for detected sources in field G21.5-09 offset 2}
\label{t:xray-param}
\fontsize{7}{7}\selectfont
\begin{tabular}{lrrrcrrrcrrrcrrrcc} 
\hline\hline 
\noalign{\smallskip}
 2XMM  & r$_{90}$ & pn\_B1$^{*}$ & pn\_B2$^{**}$ & 2MASS & d$_{x-o}$ & P$_{id}$ & kmag & GSC  & d$_{x-o}$ & P$_{id}$ & V & USNO & d$_{x-o}$ & P$_{id}$ & R & Class$^\dag$ & SpT$^\ddag$ \\
 Name   & [\arcsec] & [cts ks$^{-1}$] & [cts ks$^{-1}$] & Name & [\arcsec]  &          &      & Name & [\arcsec]   &          &   & Name & [\arcsec]   &          &   &       &       \\
\noalign{\smallskip}
\hline\noalign{\smallskip}
\noalign{\smallskip}
 J183211.2-103442 &    1.09 &     4.1$\pm$    0.8 &     6.3$\pm$    1.4 &    18321132-1034443  &    1.5 &    0.63 &   14.40 &      S9MZ012436 &     1.2 &    0.89 &         &       0794-0408906 &     0.9 &    0.56 &         &            &         $^{       }$   \\
 J183218.1-102743 &    1.57 &     4.2$\pm$    0.9 &     1.3$\pm$    0.8 &    18321809-1027440  &    1.3 &    0.73 &   12.37 &      S9MZ016477 &     1.8 &    0.85 &         &       0795-0410775 &     1.4 &    0.87 &   16.92 &            &         $^{       }$   \\
 J183220.7-103509 &    0.54 &    10.0$\pm$    1.6 &   447.5$\pm$    8.4 &                      &        &         &         &                 &         &         &         &                    &         &         &         &       EG   &    HII  $^{     s }$   \\
 J183223.1-104349 &    0.90 &     8.4$\pm$    1.1 &     1.9$\pm$    1.1 &    18322304-1043495  &    1.5 &    0.40 &   13.60 &      S9MX019635 &     1.6 &    0.71 &         &       0792-0410638 &     1.3 &    0.77 &   19.87 &       Star &    M2V  $^{    sp }$   \\
 J183226.4-103848 &    1.59 &     1.8$\pm$    1.3 &     0.9$\pm$    1.8 &    18322648-1038460  &    2.2 &    0.92 &    9.84 &                 &         &         &         &                    &         &         &         &            &         $^{       }$   \\
 J183228.1-102709 &    0.80 &    10.6$\pm$    1.2 &     4.5$\pm$    1.1 &    18322816-1027103  &    0.7 &    0.89 &   12.22 &      S9MZ016804 &     0.7 &    0.91 &         &       0795-0410993 &     1.0 &    0.87 &   18.70 &       Star &    M4V  $^{    sp }$   \\
 J183231.4-103323 &    1.31 &     1.9$\pm$    0.5 &     0.5$\pm$    0.4 &    18323149-1033227  &    1.2 &    0.89 &   12.78 &      S9MZ013113 &     1.0 &    0.96 &   17.52 &       0794-0409293 &     1.3 &    0.92 &   16.91 &            &         $^{       }$   \\
 J183235.8-103206 &    1.90 &     1.0$\pm$    0.4 &     1.9$\pm$    0.6 &    18323597-1032058  &    1.6 &    0.87 &   11.91 &                 &         &         &         &                    &         &         &         &            &         $^{       }$   \\
 J183238.4-104908 &    1.15 &     4.8$\pm$    1.0 &    12.8$\pm$    2.2 &    18323848-1049084  &    0.4 &    0.77 &   13.43 &      S9MX057000 &     0.6 &    0.86 &         &       0791-0409361 &     0.3 &    0.78 &   18.96 &            &         $^{       }$   \\
 J183238.8-103458 &    1.02 &     1.6$\pm$    0.5 &     1.9$\pm$    0.6 &    18323890-1034577  &    1.1 &    0.98 &   10.70 &      S9MZ012261 &     0.8 &    0.98 &         &       0794-0409419 &     1.3 &    0.95 &   16.37 &       Star &    K0III  $^{    sp }$   \\
 J183239.1-103126 &    1.42 &     1.9$\pm$    0.4 &     1.3$\pm$    0.5 &    18323920-1031260  &    0.9 &    0.99 &    9.82 &      S9MZ014213 &     0.4 &    0.99 &   15.79 &       0794-0409425 &     1.2 &    0.99 &   14.66 &       Star &    K0III  $^{    sp }$   \\
 J183240.8-103945 &    1.87 &     0.4$\pm$    0.3 &     1.6$\pm$    0.6 &    18324068-1039496  &    4.0 &         &   14.70 &                 &         &         &         &                    &         &         &         &            &         $^{       }$   \\
 J183247.5-103038 &    1.08 &     2.3$\pm$    0.5 &     0.9$\pm$    0.4 &    18324750-1030383  &    0.3 &    0.99 &   11.30 &      S9MZ014672 &     0.5 &    0.99 &   17.27 &       0794-0409556 &     0.1 &    0.99 &   16.20 &            &         $^{       }$   \\
 J183248.3-103951 &    1.47 &     0.2$\pm$    0.2 &     1.3$\pm$    0.5 &    18324846-1039505  &    1.9 &    0.70 &   11.79 &                 &         &         &         &                    &         &         &         &            &         $^{       }$   \\
 J183258.0-104032 &    1.73 &     1.2$\pm$    0.4 &     1.0$\pm$    0.4 &                      &        &         &         &                 &         &         &         &       0793-0410458 &     5.6 &    0.05 &   19.75 &            &         $^{       }$   \\
 J183301.5-103527 &    1.46 &     0.6$\pm$    0.4 &     1.8$\pm$    0.7 &    18330130-1035282  &    4.5 &    0.36 &   13.38 &                 &         &         &         &                    &         &         &         &            &         $^{       }$   \\
 J183303.0-104214 &    1.21 &    10.3$\pm$    1.9 &     5.5$\pm$    2.6 &                      &        &         &         &      S9MX059820 &     1.3 &    1.00 &         &                    &         &         &         &       Star &    K4V  $^{    sp }$   \\
 J183305.9-103721 &    1.18 &     1.3$\pm$    0.4 &     0.8$\pm$    0.4 &    18330602-1037215  &    0.6 &    0.93 &   12.59 &      S9MX020860 &     0.3 &    0.98 &   17.48 &       0793-0410616 &     0.6 &    0.98 &   16.04 &       Star &    G4V  $^{    sp }$   \\
 J183308.9-103508 &    1.76 &     0.0$\pm$    0.2 &     0.2$\pm$    0.3 &    18330879-1035075  &    2.5 &    0.33 &   13.11 &                 &         &         &         &                    &         &         &         &            &         $^{       }$   \\
 J183315.7-102936 &    0.71 &     5.0$\pm$    0.8 &     7.9$\pm$    1.1 &    18331569-1029372  &    0.8 &    0.96 &   11.55 &      S9MX021773 &     1.2 &    0.77 &         &       0795-0411976 &     0.5 &    0.95 &   17.49 &            &         $^{       }$   \\
 J183327.7-103523 &    0.43 &    35.7$\pm$    1.9 &    57.8$\pm$    2.7 &    18332777-1035243  &    0.7 &    1.00 &    8.27 &      S9MX000062 &     0.8 &    1.00 &   11.96 &       0794-0410181 &     0.4 &    1.00 &   11.48 &       HMXB &   Be/X  $^{    sp }$   \\
 J183331.8-104028 &    1.12 &     3.3$\pm$    0.7 &     0.0$\pm$    0.4 &    18333177-1040288  &    0.8 &    0.99 &   10.51 &      S9MX020331 &     1.2 &    1.00 &   14.56 &       0793-0411091 &     0.7 &    1.00 &   14.04 &       Star &    M2V  $^{    sp }$   \\
\noalign{\smallskip}\hline
\end{tabular}
\end{center}
\end{table}
\end{landscape}
\normalsize

\begin{landscape}
\begin{table}
\begin{center}
\caption[Xray parameters for field GC2]{
X-ray parameters for detected sources in field GC2}
\label{t:xray-param}
\fontsize{7}{7}\selectfont
\begin{tabular}{lrrrcrrrcrrrcrrrcc} 
\hline\hline 
\noalign{\smallskip}
 2XMM  & r$_{90}$ & pn\_B1$^{*}$ & pn\_B2$^{**}$ & 2MASS & d$_{x-o}$ & P$_{id}$ & kmag & GSC  & d$_{x-o}$ & P$_{id}$ & V & USNO & d$_{x-o}$ & P$_{id}$ & R & Class$^\dag$ & SpT$^\ddag$ \\
 Name   & [\arcsec] & [cts ks$^{-1}$] & [cts ks$^{-1}$] & Name & [\arcsec]  &          &      & Name & [\arcsec]   &          &   & Name & [\arcsec]   &          &   &       &       \\
\noalign{\smallskip}
\hline\noalign{\smallskip}
\noalign{\smallskip}
 J174623.1-281156 &    1.94 &                     &                     &    17462327-2811532  &    3.3 &    0.86 &   10.90 &      S8DY022573 &     0.3 &    0.94 &   17.24 &       0618-0681779 &     0.7 &    0.94 &   16.20 &       Star &    K2V  $^{    sp }$   \\
 J174626.8-280413 &    1.34 &                     &                     &    17462682-2804152  &    1.3 &    0.98 &   10.48 &      S8DY029594 &     1.2 &    0.99 &   14.80 &       0619-0716672 &     1.5 &    0.97 &   14.29 &            &         $^{       }$   \\
 J174631.2-281028 &    0.85 &    13.5$\pm$    1.7 &     4.9$\pm$    1.6 &    17463125-2810287  &    0.6 &    1.00 &    8.53 &      S8DY024069 &     0.6 &    1.00 &   13.48 &       0618-0681890 &     0.5 &    1.00 &   12.70 &       Star &    K1III  $^{    sp }$   \\
 J174641.4-280811 &    1.89 &     0.3$\pm$    0.5 &     4.8$\pm$    1.4 &    17464164-2808110  &    2.5 &    0.19 &   11.37 &                 &         &         &         &                    &         &         &         &            &         $^{       }$   \\
 J174645.2-281547 &    0.58 &                     &    65.0$\pm$    4.7 &    17464524-2815476  &    0.2 &    1.00 &    7.18 &                 &         &         &         &                    &         &         &         &            &         $^{       }$   \\
 J174652.2-280909 &    1.79 &     0.4$\pm$    0.5 &     6.2$\pm$    1.6 &    17465241-2809149  &    5.9 &    0.02 &   11.78 &      S8DY025135 &     5.1 &    0.04 &   17.25 &       0618-0682213 &     4.6 &    0.31 &   16.32 &       Star &    F9V  $^{    sp }$   \\
 J174653.1-280203 &    1.53 &                     &     4.7$\pm$    1.4 &    17465284-2802027  &    3.8 &    0.26 &   12.39 &                 &         &         &         &                    &         &         &         &            &         $^{       }$   \\
 J174654.6-281658 &    0.62 &    20.6$\pm$    2.0 &    12.7$\pm$    2.1 &    17465462-2816580  &    0.6 &    1.00 &   11.08 &      S8DY016860 &     0.9 &    0.99 &   15.09 &       0617-0636761 &     0.6 &    1.00 &   14.30 &       Star &    G0V  $^{    sp }$   \\
 J174658.8-281423 &    2.38 &                     &     4.9$\pm$    1.3 &    17465885-2814264  &    2.9 &    0.06 &   13.35 &      S8DY019754 &     3.4 &    0.17 &         &       0617-0636840 &     7.5 &    0.30 &   14.90 &            &         $^{       }$   \\
 J174705.3-280859 &    0.38 &   127.8$\pm$    3.7 &    18.1$\pm$    1.6 &    17470538-2808594  &    0.5 &    1.00 &    7.48 &      S8DY000193 &     0.4 &    1.00 &    9.48 &       0618-0682463 &     0.4 &    1.00 &    9.05 &       Star &    F3III  $^{     s }$   \\
 J174707.8-280123 &    1.74 &     3.5$\pm$    0.9 &     0.6$\pm$    0.8 &    17470800-2801225  &    2.8 &    0.84 &   10.39 &      S8DY031922 &     2.6 &    0.96 &   13.55 &       0619-0717446 &     2.4 &    0.67 &   13.07 &       Star &    K3V  $^{    sp }$   \\
 J174714.8-280613 &    0.87 &     5.0$\pm$    0.8 &     2.4$\pm$    0.8 &    17471488-2806136  &    0.8 &    0.99 &   11.36 &      S8DY027712 &     0.4 &    1.00 &   15.38 &       0618-0682654 &     0.7 &    0.99 &   14.58 &       Star &    G7V  $^{    sp }$   \\
 J174715.1-281744 &    1.93 &     1.3$\pm$    0.6 &     2.9$\pm$    1.2 &    17471541-2817459  &    3.2 &    0.21 &   11.89 &                 &         &         &         &                    &         &         &         &            &         $^{       }$   \\
 J174717.7-275838 &    1.33 &     2.9$\pm$    0.9 &     0.6$\pm$    1.0 &    17471770-2758380  &    0.0 &    0.98 &   12.25 &      S8DY033929 &     0.3 &    0.99 &   15.76 &       0620-0729320 &     0.4 &    0.99 &   15.40 &       Star &    K1III  $^{    sp }$   \\
 J174717.8-281027 &    0.75 &     0.3$\pm$    0.3 &    15.3$\pm$    1.6 &    17471785-2810256  &    1.9 &    0.09 &   12.30 &                 &         &         &         &                    &         &         &         &            &         $^{       }$   \\
 J174718.0-281735 &    1.85 &     2.2$\pm$    0.8 &     2.0$\pm$    0.9 &    17471815-2817349  &    0.9 &    0.94 &   12.22 &                 &         &         &         &       0617-0637167 &     2.4 &    0.26 &         &            &         $^{       }$   \\
 J174720.5-281323 &    1.86 &     0.2$\pm$    0.3 &     3.5$\pm$    1.0 &    17472050-2813212  &    2.8 &    0.41 &   11.16 &                 &         &         &         &                    &         &         &         &            &         $^{       }$   \\
 J174726.5-281702 &    2.08 &     3.0$\pm$    0.9 &     0.4$\pm$    0.8 &    17472662-2817027  &    0.4 &    0.91 &   13.51 &      S8DY016600 &     0.3 &    0.95 &   17.35 &       0617-0637328 &     0.4 &    0.95 &   16.35 &       Star &    M1Ve $^{    sp }$   \\
 J174728.0-280421 &    1.58 &     0.8$\pm$    0.5 &     2.6$\pm$    0.9 &    17472808-2804227  &    1.0 &    0.92 &   12.06 &      S8DY047798 &     1.2 &    0.81 &         &       0619-0718002 &     1.0 &    0.85 &   17.60 &       Star &     F:  $^{    sp }$   \\
 J174729.8-281305 &    1.27 &     4.1$\pm$    0.9 &     0.8$\pm$    0.6 &    17472986-2813061  &    0.6 &    0.92 &   13.90 &      S8DY021179 &     0.4 &    0.98 &   16.46 &       0617-0637389 &     0.5 &    0.98 &   15.80 &       Star &    G9V  $^{    sp }$   \\
 J174730.8-281347 &    0.86 &     7.1$\pm$    1.0 &                     &    17473086-2813480  &    0.7 &    0.96 &   11.41 &      S8DY020377 &     0.6 &    0.97 &   17.42 &       0617-0637407 &     0.5 &    0.98 &   16.47 &       Star &    M5V  $^{    sp }$   \\
 J174732.7-282104 &    1.45 &     4.2$\pm$    1.3 &                     &    17473283-2821046  &    0.8 &    0.98 &   12.48 &      S8DY011343 &     0.6 &    1.00 &   14.82 &       0616-0602189 &     1.2 &    0.99 &   14.17 &            &         $^{       }$   \\
 J174733.5-281839 &    1.79 &     3.5$\pm$    1.1 &     1.5$\pm$    1.5 &    17473359-2818406  &    1.1 &    0.98 &   11.49 &      S8DY014563 &     1.1 &    1.00 &   13.47 &       0616-0602196 &     1.4 &    1.00 &   13.21 &            &         $^{       }$   \\
 J174735.5-282012 &    1.82 &     4.1$\pm$    1.1 &     0.8$\pm$    0.9 &    17473564-2820116  &    1.8 &    0.80 &   12.09 &      S8DY012417 &     1.9 &    0.82 &   16.94 &       0616-0602229 &     2.0 &    0.42 &         &       Star &      G  $^{     s }$   \\
 J174736.7-281445 &    1.95 &     2.8$\pm$    0.8 &     0.5$\pm$    0.6 &    17473672-2814458  &    0.8 &    0.98 &   11.22 &      S8DY019308 &     1.0 &    1.00 &   14.16 &       0617-0637522 &     0.9 &    1.00 &   13.79 &       Star &    F9V  $^{    sp }$   \\
 J174745.9-280732 &    1.63 &     2.6$\pm$    0.8 &     1.6$\pm$    0.9 &    17474629-2807356  &    5.4 &         &   12.45 &                 &         &         &         &                    &         &         &         &            &         $^{       }$   \\
 J174746.3-280655 &    2.11 &     0.5$\pm$    0.4 &     5.5$\pm$    1.4 &    17474628-2806569  &    1.4 &    0.76 &   11.22 &                 &         &         &         &                    &         &         &         &            &         $^{       }$   \\
 J174751.9-280248 &    1.39 &     0.6$\pm$    1.3 &     1.1$\pm$    2.5 &    17475200-2802476  &    1.1 &    0.71 &   12.62 &                 &         &         &         &                    &         &         &         &            &         $^{       }$   \\
 J174756.1-280508 &    1.10 &     6.8$\pm$    1.3 &     4.2$\pm$    1.2 &    17475617-2805084  &    0.7 &    0.94 &   12.55 &                 &         &         &         &       0619-0718676 &     1.0 &    0.75 &   19.07 &            &         $^{       }$   \\
 J174801.8-281710 &    1.05 &     4.5$\pm$    1.3 &     2.2$\pm$    2.1 &    17480185-2817106  &    0.6 &    1.00 &   11.52 &      S8DY016392 &     1.2 &    0.97 &   16.48 &       0617-0637905 &     1.7 &    0.86 &   15.10 &       Star &    M4Ve $^{    sp }$   \\
 J174810.7-281831 &    2.09 &     6.2$\pm$    1.6 &     0.6$\pm$    1.0 &    17481080-2818312  &    1.2 &    1.00 &    7.54 &      S8DY000265 &     1.3 &    1.00 &   10.73 &       0616-0602683 &     1.3 &    1.00 &    9.67 &            &         $^{       }$   \\
 J174819.7-280727 &    1.99 &                     &    15.1$\pm$    3.5 &    17481973-2807269  &    0.5 &    1.00 &    4.34 &      S8DY026529 &     0.9 &    0.99 &   16.09 &       0618-0683828 &     0.6 &    1.00 &   14.22 &       Star &    M6V  $^{     s }$   \\
\noalign{\smallskip}\hline
\end{tabular}
\end{center}
\end{table}
\end{landscape}
\normalsize

\begin{landscape}
\begin{table}
\begin{center}
\caption[Xray parameters for field GRB001025]{
X-ray parameters for detected sources in field GRB001025}
\label{t:xray-param}
\fontsize{7}{7}\selectfont
\begin{tabular}{lrrrcrrrcrrrcrrrcc} 
\hline\hline 
\noalign{\smallskip}
 2XMM  & r$_{90}$ & pn\_B1$^{*}$ & pn\_B2$^{**}$ & 2MASS & d$_{x-o}$ & P$_{id}$ & kmag & GSC  & d$_{x-o}$ & P$_{id}$ & V & USNO & d$_{x-o}$ & P$_{id}$ & R & Class$^\dag$ & SpT$^\ddag$ \\
 Name   & [\arcsec] & [cts ks$^{-1}$] & [cts ks$^{-1}$] & Name & [\arcsec]  &          &      & Name & [\arcsec]   &          &   & Name & [\arcsec]   &          &   &       &       \\
\noalign{\smallskip}
\hline\noalign{\smallskip}
\noalign{\smallskip}
 J083536.8-130647 &    2.95 &     4.5$\pm$    1.0 &     0.2$\pm$    0.9 &                      &        &         &         &      S5WJ005955 &     5.4 &    0.17 &         &       0768-0209716 &     7.1 &    0.29 &   20.80 &            &         $^{       }$   \\
 J083538.6-130356 &    1.60 &     5.4$\pm$    1.1 &     3.7$\pm$    1.6 &    08353868-1303546  &    1.9 &    0.92 &   14.47 &      S5WJ006888 &     1.5 &    0.95 &         &       0769-0217305 &     1.7 &    0.61 &   16.21 &            &         $^{       }$   \\
 J083544.1-125744 &    1.73 &     3.3$\pm$    0.9 &     0.2$\pm$    1.0 &    08354410-1257486  &    4.1 &    0.61 &   14.24 &      S5WI007738 &     4.1 &    0.53 &         &       0770-0229071 &     3.9 &    0.63 &   17.11 &            &         $^{       }$   \\
 J083546.3-130912 &    1.02 &    15.6$\pm$    2.9 &     9.6$\pm$    3.9 &                      &        &         &         &      S5WJ005116 &     0.1 &    0.97 &         &       0768-0209769 &     0.5 &    0.97 &   20.19 &            &         $^{       }$   \\
 J083555.1-125854 &    2.10 &     3.4$\pm$    1.2 &     0.2$\pm$    1.1 &    08355517-1258535  &    0.6 &    0.99 &   12.18 &      S5WI000795 &     0.4 &    0.99 &         &       0770-0229152 &     0.6 &    0.99 &   13.54 &            &         $^{       }$   \\
 J083603.5-131407 &    2.00 &     1.6$\pm$    0.6 &     3.5$\pm$    1.4 &                      &        &         &         &      S5WJ003325 &     5.5 &    0.04 &         &       0767-0198711 &     5.6 &    0.18 &   20.00 &            &         $^{       }$   \\
 J083605.7-125321 &    1.19 &     5.0$\pm$    1.0 &     2.2$\pm$    1.1 &                      &        &         &         &      S5WI009484 &     0.9 &    0.95 &         &       0771-0239435 &     0.8 &    0.96 &   19.54 &       EG   &     EG  $^{    sp }$   \\
 J083606.9-125348 &    0.82 &    10.7$\pm$    1.3 &     3.9$\pm$    1.4 &                      &        &         &         &      S5WI009304 &     1.4 &    0.82 &         &       0771-0239448 &     1.1 &    0.92 &         &       EG   &     EG  $^{    sp }$   \\
 J083609.1-130455 &    1.22 &                     &                     &                      &        &         &         &      S5WI005293 &     1.0 &    0.94 &         &       0769-0217581 &     0.6 &    0.98 &   20.16 &            &         $^{       }$   \\
 J083611.3-125840 &    0.41 &    61.8$\pm$    2.3 &    22.9$\pm$    1.7 &                      &        &         &         &      S5WI007472 &     0.4 &    0.99 &         &       0770-0229320 &     0.3 &    1.00 &   18.37 &       EG   &     EG  $^{    sp }$   \\
 J083614.8-125724 &    1.11 &     3.6$\pm$    0.7 &     1.7$\pm$    0.8 &                      &        &         &         &      S5WI007877 &     1.4 &    0.87 &         &       0770-0229371 &     1.2 &    0.92 &   20.21 &            &         $^{       }$   \\
 J083615.5-130341 &    1.09 &     3.0$\pm$    0.5 &     1.5$\pm$    0.6 &                      &        &         &         &      S5WI005740 &     1.5 &    0.91 &         &       0769-0217644 &     1.4 &    0.94 &   18.16 &            &         $^{       }$   \\
 J083616.3-130949 &    0.48 &    28.4$\pm$    1.5 &     1.9$\pm$    0.6 &    08361624-1309496  &    1.2 &    0.95 &   12.02 &      S5WI003854 &     1.2 &    0.98 &         &       0768-0210033 &     0.8 &    0.99 &   15.51 &       Star &     Me  $^{    sp }$   \\
 J083622.8-131502 &    1.68 &     2.5$\pm$    0.7 &     1.8$\pm$    1.0 &    08362279-1315034  &    1.1 &    1.00 &    9.67 &      S5WI000981 &     1.0 &    1.00 &   11.54 &       0767-0198871 &     1.0 &    1.00 &   10.95 &            &         $^{       }$   \\
 J083622.9-130903 &    0.74 &    39.2$\pm$    3.3 &     4.9$\pm$    2.1 &    08362291-1309049  &    1.4 &    1.00 &    6.89 &      S5WI000926 &     1.2 &    0.69 &    7.99 &       0768-0210115 &     1.2 &    1.00 &    7.68 &       Star &    F5V  $^{     s }$   \\
 J083624.2-125919 &    0.76 &     4.1$\pm$    0.6 &     1.8$\pm$    0.6 &                      &        &         &         &      S5WI007225 &     0.5 &    0.98 &         &       0770-0229459 &     0.2 &    0.99 &   20.23 &       EG   &     EG  $^{    sp }$   \\
 J083627.8-130200 &    0.89 &     2.4$\pm$    0.4 &     1.7$\pm$    0.5 &                      &        &         &         &      S5WI032790 &     1.3 &    0.82 &         &                    &         &         &         &       EG   &     EG  $^{    sp }$   \\
 J083631.6-131635 &    1.54 &     2.7$\pm$    0.7 &     1.4$\pm$    0.9 &                      &        &         &         &      S5WI002040 &     1.9 &    0.82 &         &       0767-0198941 &     1.6 &    0.87 &   18.57 &            &         $^{       }$   \\
 J083633.5-130033 &    1.38 &     1.0$\pm$    0.3 &     0.2$\pm$    0.3 &                      &        &         &         &      S5WI006798 &     4.9 &    0.01 &         &       0769-0217812 &     4.5 &    0.18 &   20.23 &            &         $^{       }$   \\
 J083639.4-125426 &    1.18 &     2.4$\pm$    0.6 &     1.0$\pm$    0.7 &                      &        &         &         &      S5WI009049 &     0.4 &    0.96 &         &       0770-0229590 &     1.2 &    0.90 &   20.23 &            &         $^{       }$   \\
 J083645.5-130401 &    1.24 &     0.2$\pm$    0.2 &     2.6$\pm$    0.5 &                      &        &         &         &      S5WI005638 &     0.9 &    0.92 &         &       0769-0217902 &     0.8 &    0.92 &   18.80 &            &         $^{       }$   \\
 J083645.6-125242 &    1.58 &     3.1$\pm$    0.7 &     1.0$\pm$    0.6 &                      &        &         &         &                 &         &         &         &       0771-0239798 &     3.9 &    0.28 &   20.23 &            &         $^{       }$   \\
 J083646.0-130716 &    1.87 &     0.4$\pm$    0.3 &     0.2$\pm$    0.3 &                      &        &         &         &      S5WI004612 &     6.1 &    0.02 &         &       0768-0210375 &     6.2 &    0.12 &   19.80 &            &         $^{       }$   \\
 J083646.3-125929 &    0.82 &     4.1$\pm$    0.6 &     1.6$\pm$    0.5 &                      &        &         &         &      S5WI007181 &     0.5 &    0.98 &         &       0770-0229655 &     0.6 &    0.98 &   19.47 &            &         $^{       }$   \\
 J083647.1-131652 &    0.84 &    10.2$\pm$    1.3 &     3.4$\pm$    1.3 &                      &        &         &         &                 &         &         &         &       0767-0199056 &     0.4 &    0.92 &         &       EG   &     EG  $^{    sp }$   \\
 J083648.5-130150 &    1.05 &     1.9$\pm$    0.4 &     0.4$\pm$    0.3 &    08364849-1301508  &    0.5 &    0.99 &   14.01 &      S5WI006369 &     0.7 &    0.97 &         &       0769-0217935 &     0.6 &    0.97 &   18.11 &       Star &     Me  $^{    sp }$   \\
 J083648.7-125729 &    1.35 &     2.2$\pm$    0.5 &     0.3$\pm$    0.4 &    08364883-1257297  &    0.6 &    0.99 &   13.09 &      S5WI007875 &     1.3 &    0.96 &         &       0770-0229672 &     0.7 &    0.92 &   17.20 &       Star &     Me  $^{    sp }$   \\
 J083649.7-130931 &    1.82 &     1.1$\pm$    0.4 &     0.7$\pm$    0.4 &                      &        &         &         &      S5WI003939 &     2.5 &    0.75 &         &       0768-0210423 &     3.7 &    0.59 &   18.60 &            &         $^{       }$   \\
 J083650.2-131249 &    1.93 &     1.3$\pm$    0.5 &     1.2$\pm$    0.7 &                      &        &         &         &                 &         &         &         &       0767-0199092 &     2.1 &    0.65 &   20.23 &            &         $^{       }$   \\
 J083700.8-131702 &    1.61 &     4.2$\pm$    0.9 &     2.2$\pm$    1.3 &    08370080-1317036  &    1.6 &    0.96 &   13.53 &      S5WI001935 &     1.5 &    0.97 &         &       0767-0199193 &     1.7 &    0.95 &   15.94 &            &    F/G  $^{    sp }$   \\
 J083702.0-130622 &    2.53 &     0.4$\pm$    0.4 &     0.3$\pm$    0.5 &                      &        &         &         &      S5WI004879 &     1.5 &    0.82 &         &       0768-0210540 &     1.2 &    0.89 &   20.19 &            &         $^{       }$   \\
 J083703.1-131439 &    1.75 &     3.1$\pm$    0.8 &     0.1$\pm$    0.3 &                      &        &         &         &      S5WI002500 &     0.6 &    0.90 &         &       0767-0199211 &     1.3 &    0.85 &   20.23 &            &         $^{       }$   \\
 J083708.8-130226 &    0.85 &     4.8$\pm$    0.8 &     0.7$\pm$    0.6 &                      &        &         &         &      S5WI006177 &     0.2 &    0.96 &         &       0769-0218101 &     0.3 &    0.97 &   19.94 &            &         $^{       }$   \\
 J083715.2-130923 &    1.00 &     3.9$\pm$    0.7 &     2.5$\pm$    0.8 &                      &        &         &         &      S5WI003980 &     0.4 &    0.96 &         &       0768-0210645 &     0.5 &    0.98 &   20.22 &            &         $^{       }$   \\
 J083715.5-131036 &    1.34 &     1.9$\pm$    0.6 &     1.6$\pm$    0.7 &                      &        &         &         &      S5WI003619 &     1.2 &    0.85 &         &       0768-0210648 &     1.2 &    0.83 &   20.19 &            &         $^{       }$   \\
 J083719.4-125623 &    1.58 &                     &                     &                      &        &         &         &      S5WI008287 &     1.4 &    0.87 &         &       0770-0229939 &     1.3 &    0.92 &   19.88 &            &         $^{       }$   \\
 J083721.5-125720 &    0.68 &     6.2$\pm$    2.9 &     2.5$\pm$    4.5 &                      &        &         &         &      S5WI007934 &     0.1 &    0.99 &         &       0770-0229958 &     0.5 &    0.98 &   19.73 &       EG   &     EG  $^{    sp }$   \\
 J083727.9-130012 &    1.51 &                     &                     &                      &        &         &         &      S5WI006959 &     2.4 &    0.63 &         &       0769-0218296 &     2.2 &    0.63 &   19.98 &            &         $^{       }$   \\
 J083732.5-130700 &    1.96 &                     &                     &    08373236-1307042  &    4.5 &    0.16 &   14.63 &      S5WI000899 &     4.2 &    0.44 &         &       0768-0210774 &     4.1 &    0.59 &   14.97 &            &         $^{       }$   \\
\noalign{\smallskip}\hline
\end{tabular}
\end{center}
\end{table}
\end{landscape}
\normalsize

\begin{landscape}
\begin{table}
\begin{center}
\caption[Xray parameters for field GROJ1655-40]{
X-ray parameters for detected sources in field GROJ1655-40}
\label{t:xray-param}
\fontsize{7}{7}\selectfont
\begin{tabular}{lrrrcrrrcrrrcrrrcc} 
\hline\hline 
\noalign{\smallskip}
 2XMM  & r$_{90}$ & pn\_B1$^{*}$ & pn\_B2$^{**}$ & 2MASS & d$_{x-o}$ & P$_{id}$ & kmag & GSC  & d$_{x-o}$ & P$_{id}$ & V & USNO & d$_{x-o}$ & P$_{id}$ & R & Class$^\dag$ & SpT$^\ddag$ \\
 Name   & [\arcsec] & [cts ks$^{-1}$] & [cts ks$^{-1}$] & Name & [\arcsec]  &          &      & Name & [\arcsec]   &          &   & Name & [\arcsec]   &          &   &       &       \\
\noalign{\smallskip}
\hline\noalign{\smallskip}
\noalign{\smallskip}
 J165256.2-394813 &    1.69 &                     &                     &    16525614-3948111  &    3.0 &    0.22 &   13.45 &      S8VD122070 &     4.3 &    0.19 &         &       0501-0492854 &     3.6 &    0.26 &   17.92 &            &         $^{       }$   \\
 J165301.1-395414 &    1.62 &     4.6$\pm$    1.0 &     2.1$\pm$    0.8 &    16530089-3954165  &    3.3 &    0.23 &   12.98 &      S8VD117711 &     4.7 &    0.01 &         &       0500-0495133 &     4.6 &    0.25 &   17.26 &            &         $^{       }$   \\
 J165304.2-395557 &    1.47 &     6.9$\pm$    1.2 &     1.2$\pm$    0.9 &                      &        &         &         &                 &         &         &         &       0500-0495183 &     1.1 &    0.98 &   12.12 &       Star &    M2V  $^{    sp }$   \\
 J165306.1-400008 &    1.67 &                     &                     &    16530610-4000033  &    4.9 &    0.04 &   13.07 &      S8VD112419 &     5.0 &         &         &       0499-0495184 &     5.0 &         &   18.66 &            &         $^{       }$   \\
 J165310.5-394449 &    1.13 &    12.6$\pm$    1.4 &     3.3$\pm$    0.9 &    16531037-3944492  &    0.9 &    0.91 &   11.11 &      S8VD040934 &     1.3 &    0.98 &   14.46 &       0502-0490447 &     0.8 &    0.98 &   13.44 &       Star &    K1V  $^{    sp }$   \\
 J165314.3-395805 &    2.15 &     1.0$\pm$    0.6 &     0.0$\pm$    0.6 &    16531441-3958024  &    2.0 &    0.34 &   13.44 &      S8VD114350 &     1.5 &    0.49 &         &                    &         &         &         &            &         $^{       }$   \\
 J165314.9-395309 &    1.38 &     5.9$\pm$    0.9 &     0.2$\pm$    0.3 &    16531480-3953088  &    0.3 &    0.88 &   12.92 &      S8VD039158 &     1.1 &    0.93 &   16.64 &       0501-0493194 &     0.5 &    0.92 &   15.88 &       Star &    K4V  $^{    sp }$   \\
 J165315.0-395509 &    1.84 &     4.2$\pm$    0.8 &     0.8$\pm$    0.7 &    16531489-3955070  &    0.9 &    0.79 &   13.42 &      S8VD038485 &     1.2 &    0.93 &   16.18 &       0500-0495363 &     1.0 &    0.90 &   15.74 &            &         $^{       }$   \\
 J165334.2-400005 &    1.49 &     4.8$\pm$    1.0 &     1.3$\pm$    0.7 &    16533391-4000054  &    1.6 &    0.96 &    9.06 &      S8VD000430 &     1.4 &    1.00 &   10.72 &       0499-0495689 &     1.4 &    1.00 &   10.47 &       Star &    F5V  $^{     s }$   \\
 J165342.7-395501 &    2.45 &     0.6$\pm$    0.4 &     0.6$\pm$    0.5 &    16534252-3954596  &    0.9 &    0.80 &   12.83 &      S8VD117190 &     0.9 &    0.89 &         &       0500-0495910 &     1.3 &    0.75 &   17.42 &            &         $^{       }$   \\
 J165348.8-395537 &    1.84 &     1.0$\pm$    0.4 &     0.5$\pm$    0.5 &    16534902-3955373  &    2.4 &    0.85 &   10.00 &      S8VD038316 &     2.2 &    0.92 &   14.42 &       0500-0496017 &     4.3 &    0.86 &         &            &         $^{       }$   \\
 J165351.4-393658 &    1.85 &     9.5$\pm$    3.1 &     2.5$\pm$    3.6 &    16535122-3936550  &    1.1 &    0.91 &   11.10 &      S8VH014628 &     1.0 &    0.95 &   15.80 &       0503-0487504 &     1.5 &    0.86 &   15.00 &            &         $^{       }$   \\
 J165354.2-395918 &    2.42 &     2.1$\pm$    0.6 &                     &    16535448-3959229  &    5.3 &    0.30 &   13.28 &      S8VD142309 &     4.4 &    0.14 &         &                    &         &         &         &            &         $^{       }$   \\
 J165358.3-394743 &    2.06 &     0.8$\pm$    0.4 &     0.2$\pm$    0.4 &    16535841-3947453  &    3.4 &    0.05 &   13.61 &      S8VD122478 &     3.0 &    0.05 &         &                    &         &         &         &       Star &      M  $^{     s }$   \\
 J165402.0-394630 &    1.34 &     0.6$\pm$    0.4 &     5.5$\pm$    0.9 &    16540206-3946305  &    0.5 &    0.91 &   12.65 &      S8VD123148 &     0.6 &    0.81 &         &                    &         &         &         &       EG   &    AGN  $^{    sp }$   \\
 J165402.6-393818 &    1.71 &     4.2$\pm$    1.1 &     2.6$\pm$    1.1 &                      &        &         &         &                 &         &         &         &                    &         &         &         &       Star &     K0  $^{     s }$   \\
 J165407.3-394601 &    1.43 &     3.6$\pm$    0.7 &     0.8$\pm$    0.5 &    16540733-3946006  &    1.0 &    0.97 &   10.29 &      S8VD000325 &     1.2 &    1.00 &   12.03 &       0502-0491615 &     0.6 &    1.00 &   12.31 &       Star &    F3V  $^{    sp }$   \\
 J165407.9-395738 &    1.65 &     2.6$\pm$    0.9 &     1.3$\pm$    0.9 &    16540782-3957388  &    0.7 &    0.88 &   12.70 &      S8VD037531 &     0.4 &    0.98 &   15.40 &       0500-0496309 &     1.5 &    0.87 &   14.72 &            &         $^{       }$   \\
 J165410.5-394204 &    1.34 &     7.6$\pm$    1.1 &     0.4$\pm$    0.5 &    16541072-3942036  &    0.8 &    0.89 &   12.35 &      S8VD041287 &     0.8 &    0.98 &   15.42 &       0502-0491662 &     1.3 &    0.89 &   15.05 &       Star &    K1V  $^{    sp }$   \\
 J165411.4-395236 &    2.00 &     1.1$\pm$    0.4 &     0.8$\pm$    0.4 &    16541121-3952363  &    2.9 &    0.53 &   11.46 &      S8VD039361 &     2.7 &    0.59 &   16.70 &       0501-0494271 &     2.6 &    0.52 &   16.04 &       Star &      G  $^{     s }$   \\
 J165412.4-395434 &    1.43 &     3.6$\pm$    0.7 &                     &    16541228-3954342  &    0.9 &    0.78 &   13.05 &      S8VD038715 &     1.2 &    0.91 &   16.32 &       0500-0496380 &     1.8 &    0.61 &   15.61 &       Star &    K0V  $^{    sp }$   \\
 J165418.3-394804 &    1.81 &     1.3$\pm$    0.5 &     3.0$\pm$    0.8 &    16541850-3948015  &    3.0 &    0.07 &   15.19 &      S8VD122289 &     3.0 &    0.07 &         &                    &         &         &         &            &         $^{       }$   \\
 J165420.0-394751 &    2.59 &     0.3$\pm$    0.4 &     0.7$\pm$    0.5 &    16542023-3947493  &    5.5 &    0.01 &   12.78 &      S8VD040423 &     5.2 &    0.28 &   14.81 &       0502-0491797 &     5.5 &    0.11 &   14.42 &            &         $^{       }$   \\
 J165421.3-394635 &    1.47 &     3.4$\pm$    0.7 &     1.1$\pm$    0.5 &    16542125-3946353  &    0.4 &    0.88 &   13.47 &      S8VD040651 &     0.4 &    0.97 &   16.68 &       0502-0491817 &     0.7 &    0.93 &   15.50 &       Star &    K0V  $^{    sp }$   \\
 J165421.8-395310 &    1.64 &     1.7$\pm$    0.5 &     1.2$\pm$    0.6 &    16542169-3953112  &    1.1 &    0.79 &   12.76 &      S8VD165306 &     1.5 &    0.59 &         &                    &         &         &         &            &         $^{       }$   \\
 J165423.5-394910 &    1.70 &     0.0$\pm$    0.2 &     3.7$\pm$    0.8 &    16542382-3949094  &    3.5 &    0.10 &   14.06 &                 &         &         &         &                    &         &         &         &            &         $^{       }$   \\
 J165424.0-394225 &    1.53 &     1.8$\pm$    0.7 &     3.6$\pm$    1.3 &    16542365-3942276  &    4.9 &         &   14.54 &                 &         &         &         &                    &         &         &         &            &         $^{       }$   \\
 J165426.3-394455 &    2.18 &     0.4$\pm$    0.4 &     1.3$\pm$    0.7 &    16542650-3944546  &    0.4 &    0.86 &   12.40 &      S8VD124019 &     0.8 &    0.65 &         &       0502-0491881 &     0.9 &    0.73 &         &            &         $^{       }$   \\
 J165427.0-395227 &    1.59 &     1.7$\pm$    0.5 &     1.3$\pm$    0.5 &    16542692-3952295  &    1.5 &    0.56 &   13.29 &      S8VD119313 &     1.4 &    0.67 &         &       0501-0494508 &     1.4 &    0.73 &         &            &         $^{       }$   \\
 J165428.3-400100 &    2.01 &     2.4$\pm$    1.3 &                     &    16542799-4001002  &    3.8 &    0.11 &   14.07 &      S8VD111646 &     6.4 &         &         &                    &         &         &         &            &         $^{       }$   \\
 J165438.2-400146 &    1.30 &    10.9$\pm$    1.6 &     1.9$\pm$    0.9 &    16543820-4001455  &    0.5 &    0.92 &   12.43 &      S8VD035512 &     0.8 &    0.98 &   15.13 &       0499-0496973 &     0.7 &    0.96 &   14.74 &       Star &    G7V  $^{    sp }$   \\
 J165440.5-394705 &    1.41 &     6.4$\pm$    1.0 &     1.7$\pm$    0.8 &    16544046-3947063  &    0.4 &    0.94 &   11.48 &      S8VD040569 &     0.3 &    0.97 &   16.23 &       0502-0492068 &     0.2 &    0.94 &   15.29 &       Star &    G9V  $^{    sp }$   \\
 J165440.5-394910 &    1.86 &     2.5$\pm$    0.7 &     0.7$\pm$    0.8 &    16544016-3949089  &    3.0 &    0.24 &   12.31 &      S8VD040199 &     2.8 &    0.82 &   14.05 &       0501-0494695 &     2.7 &    0.82 &   14.31 &       Star &    F5V  $^{    sp }$   \\
 J165445.1-400228 &    2.15 &     7.6$\pm$    1.5 &     1.3$\pm$    1.3 &    16544553-4002249  &    5.1 &    0.40 &   15.02 &      S8VD035135 &     3.3 &    0.74 &   15.54 &       0499-0497102 &     6.7 &    0.74 &   17.34 &            &         $^{       }$   \\
 J165445.9-394329 &    2.27 &     0.9$\pm$    0.7 &                     &    16544561-3943291  &    4.0 &    0.81 &   13.61 &      S8VD041126 &     3.9 &    0.58 &   17.15 &       0502-0492136 &     4.2 &    0.49 &   15.29 &            &         $^{       }$   \\
 J165500.1-394205 &    2.08 &     7.8$\pm$    1.7 &     1.2$\pm$    1.3 &    16550028-3942037  &    2.9 &    0.91 &    9.29 &      S8VD000306 &     2.9 &    1.00 &    9.99 &       0502-0492305 &     2.9 &    1.00 &    9.84 &            &         $^{       }$   \\
 J165508.2-395920 &    2.49 &     1.2$\pm$    0.9 &    14.8$\pm$    3.1 &    16550799-3959164  &    4.8 &    0.02 &   14.51 &      S8VD000423 &     8.1 &    0.11 &   12.69 &       0500-0497258 &     7.3 &    0.53 &         &            &         $^{       }$   \\
 J165515.6-394544 &    1.18 &    45.4$\pm$    3.1 &    11.3$\pm$    2.4 &    16551565-3945449  &    0.9 &    0.87 &   12.25 &      S8VD040806 &     1.6 &    0.93 &   15.75 &       0502-0492505 &     1.9 &    0.89 &   14.61 &       Star &    K4V  $^{    sp }$   \\
 J165519.0-395301 &    2.21 &     8.9$\pm$    1.9 &                     &    16551890-3953013  &    1.1 &    0.91 &   10.44 &      S8VD000365 &     1.2 &    0.98 &   12.47 &       0501-0495240 &     6.3 &    0.95 &   16.32 &            &         $^{       }$   \\
\noalign{\smallskip}\hline
\end{tabular}
\end{center}
\end{table}
\end{landscape}
\normalsize

\begin{landscape}
\begin{table}
\begin{center}
\caption[Xray parameters for field HTCas]{
X-ray parameters for detected sources in field HTCas}
\label{t:xray-param}
\fontsize{7}{7}\selectfont
\begin{tabular}{lrrrcrrrcrrrcrrrcc} 
\hline\hline 
\noalign{\smallskip}
 2XMM  & r$_{90}$ & pn\_B1$^{*}$ & pn\_B2$^{**}$ & 2MASS & d$_{x-o}$ & P$_{id}$ & kmag & GSC  & d$_{x-o}$ & P$_{id}$ & V & USNO & d$_{x-o}$ & P$_{id}$ & R & Class$^\dag$ & SpT$^\ddag$ \\
 Name   & [\arcsec] & [cts ks$^{-1}$] & [cts ks$^{-1}$] & Name & [\arcsec]  &          &      & Name & [\arcsec]   &          &   & Name & [\arcsec]   &          &   &       &       \\
\noalign{\smallskip}
\hline\noalign{\smallskip}
\noalign{\smallskip}
 J010844.0+600542 &    1.68 &     6.6$\pm$    1.2 &     1.8$\pm$    1.3 &                      &        &         &         &      NAMD035494 &     5.8 &         &   17.81 &                    &         &         &         &            &         $^{       }$   \\
 J010930.4+600639 &    2.09 &     3.1$\pm$    0.6 &                     &    01093093+6006419  &    4.0 &    0.72 &   11.73 &      NAMD035847 &     4.4 &    0.87 &   13.33 &       1501-0041748 &     4.7 &    0.65 &   13.06 &       Star &    F5V  $^{    sp }$   \\
 J010933.5+601619 &    1.52 &     6.1$\pm$    1.2 &     0.9$\pm$    1.0 &    01093377+6016205  &    2.0 &    0.97 &    9.77 &      NAMF000679 &     1.9 &    0.99 &   12.44 &       1502-0043542 &     1.9 &    0.99 &   11.92 &       Star &    K4V  $^{    sp }$   \\
 J010955.2+601852 &    2.82 &                     &                     &    01095635+6018496  &    8.9 &    0.02 &   14.58 &      NAMF016855 &     8.7 &         &   17.79 &       1503-0044677 &     8.5 &    0.01 &   17.38 &            &         $^{       }$   \\
 J010956.8+595336 &    1.57 &     4.8$\pm$    1.7 &     1.8$\pm$    2.4 &    01095677+5953369  &    0.7 &    0.81 &   14.83 &      NAMD046861 &     0.7 &    0.60 &         &       1498-0040815 &     1.1 &    0.56 &   19.16 &            &         $^{       }$   \\
 J010959.0+595359 &    1.99 &     3.5$\pm$    1.0 &     1.6$\pm$    1.5 &                      &        &         &         &      NAMD028376 &     4.2 &         &         &       1498-0040840 &     5.1 &         &   18.71 &       Star &    M2V  $^{    sp }$   \\
 J011001.9+601106 &    1.79 &     2.8$\pm$    0.6 &     0.1$\pm$    0.3 &    01100226+6011077  &    2.3 &    0.83 &   13.18 &      NAMD037374 &     2.3 &    0.66 &   18.13 &       1501-0042040 &     2.0 &    0.67 &   17.43 &       Star &    M4V  $^{    sp }$   \\
 J011008.9+595219 &    2.70 &     5.4$\pm$    1.3 &     3.0$\pm$    2.0 &    01100923+5952187  &    2.2 &    0.78 &   13.39 &      NAMD027075 &     1.8 &    0.82 &   17.15 &       1498-0040944 &     1.8 &    0.80 &   15.96 &       Star &    M3V  $^{    sp }$   \\
 J011009.1+600051 &    2.07 &     2.5$\pm$    0.6 &     1.3$\pm$    0.6 &    01100930+6000495  &    2.1 &    0.77 &   14.01 &      NAMD032753 &     2.0 &    0.75 &   17.54 &       1500-0040272 &     1.6 &    0.76 &   16.87 &            &         $^{       }$   \\
 J011017.6+600917 &    1.96 &     2.4$\pm$    0.5 &     1.0$\pm$    0.5 &                      &        &         &         &      NAMD036754 &     6.4 &         &         &       1501-0042168 &     6.1 &         &   19.03 &            &         $^{       }$   \\
 J011024.8+601119 &    1.49 &     4.1$\pm$    0.7 &     0.3$\pm$    0.3 &    01102501+6011193  &    0.9 &    0.80 &   14.66 &      NAMD049532 &     1.3 &    0.70 &         &                    &         &         &         &       Star &    M4V  $^{    sp }$   \\
 J011026.4+600352 &    1.71 &     2.5$\pm$    0.6 &     0.1$\pm$    0.3 &    01102672+6003513  &    2.5 &    0.71 &   13.48 &      NAMD034374 &     2.5 &    0.73 &   16.83 &       1500-0040432 &     2.2 &    0.75 &   16.21 &            &         $^{       }$   \\
 J011027.4+600039 &    1.16 &    20.0$\pm$    1.4 &     1.2$\pm$    0.5 &    01102768+6000390  &    1.6 &    0.92 &   11.53 &      NAMD032647 &     1.4 &    0.97 &   15.01 &       1500-0040445 &     1.2 &    0.97 &   14.47 &       Star &    M1V  $^{    sp }$   \\
 J011033.4+595853 &    1.97 &                     &                     &    01103408+5958544  &    4.8 &    0.14 &   13.19 &      NAMD031557 &     4.7 &    0.01 &   18.23 &       1499-0041427 &     4.7 &    0.01 &   17.75 &       Star &    M4V  $^{    sp }$   \\
 J011038.2+600146 &    1.27 &    14.1$\pm$    1.5 &     3.5$\pm$    1.1 &    01103854+6001456  &    2.7 &    0.84 &   10.00 &      NAMD000514 &     2.7 &    0.98 &   10.47 &       1500-0040559 &     2.7 &    0.99 &   10.34 &       Star &    A0V  $^{     s }$   \\
 J011056.5+595937 &    1.99 &     1.5$\pm$    0.7 &     2.8$\pm$    1.1 &    01105673+5959329  &    4.7 &    0.12 &   15.38 &      NAMD031903 &     4.7 &    0.02 &         &       1499-0041674 &     4.7 &    0.03 &   18.90 &            &         $^{       }$   \\
 J011056.6+600459 &    2.27 &     1.5$\pm$    0.5 &     1.5$\pm$    0.7 &    01105749+6004583  &    6.4 &    0.04 &   15.25 &      NAMD034868 &     6.5 &         &   18.23 &       1500-0040726 &     6.4 &    0.01 &   17.83 &            &         $^{       }$   \\
 J011105.0+601551 &    2.56 &                     &                     &    01110601+6015506  &    7.0 &    0.10 &   13.98 &      NAMF014008 &     6.9 &    0.04 &   17.38 &       1502-0044319 &     6.8 &    0.06 &   16.62 &            &         $^{       }$   \\
 J011123.3+600258 &    1.82 &     2.4$\pm$    0.7 &     1.4$\pm$    0.8 &    01112309+6002581  &    1.7 &    0.55 &   15.47 &                 &         &         &         &       1500-0040969 &     2.2 &    0.37 &         &            &         $^{       }$   \\
\noalign{\smallskip}\hline
\end{tabular}
\end{center}
\end{table}
\end{landscape}
\normalsize

\begin{landscape}
\begin{table}
\begin{center}
\caption[Xray parameters for field LHB-3]{
X-ray parameters for detected sources in field LHB-3}
\label{t:xray-param}
\fontsize{7}{7}\selectfont
\begin{tabular}{lrrrcrrrcrrrcrrrcc} 
\hline\hline 
\noalign{\smallskip}
 2XMM  & r$_{90}$ & pn\_B1$^{*}$ & pn\_B2$^{**}$ & 2MASS & d$_{x-o}$ & P$_{id}$ & kmag & GSC  & d$_{x-o}$ & P$_{id}$ & V & USNO & d$_{x-o}$ & P$_{id}$ & R & Class$^\dag$ & SpT$^\ddag$ \\
 Name   & [\arcsec] & [cts ks$^{-1}$] & [cts ks$^{-1}$] & Name & [\arcsec]  &          &      & Name & [\arcsec]   &          &   & Name & [\arcsec]   &          &   &       &       \\
\noalign{\smallskip}
\hline\noalign{\smallskip}
\noalign{\smallskip}
 J230733.5+613634 &    1.69 &                     &                     &    23073340+6136319  &    2.6 &    0.95 &   12.33 &      N190039257 &     2.5 &    0.97 &   15.37 &       1516-0352460 &     2.4 &    0.96 &   14.87 &       Star &   K1Ve  $^{    sp }$   \\
 J230804.6+613351 &    2.03 &     1.9$\pm$    0.6 &                     &    23080441+6133514  &    1.5 &    0.93 &   13.25 &      N190039027 &     1.5 &    0.95 &   17.30 &       1515-0347087 &     1.7 &    0.93 &   16.25 &            &         $^{       }$   \\
 J230817.5+614001 &    1.45 &     2.0$\pm$    0.5 &     0.1$\pm$    0.5 &    23081765+6140032  &    2.1 &    0.89 &   13.63 &      N193013396 &     2.4 &    0.93 &   16.93 &       1516-0352782 &     4.4 &    0.85 &   16.68 &            &         $^{       }$   \\
 J230825.6+613428 &    1.60 &     2.0$\pm$    0.5 &     0.7$\pm$    0.5 &    23082597+6134250  &    4.4 &    0.93 &   13.61 &      N190039115 &     1.7 &    0.96 &   17.63 &       1515-0347234 &     1.2 &    0.41 &         &            &         $^{       }$   \\
 J230837.3+613849 &    1.70 &     2.0$\pm$    0.5 &     0.3$\pm$    0.3 &    23083749+6138518  &    2.7 &    0.81 &   13.88 &      N190039407 &     2.7 &    0.89 &   17.10 &       1516-0352929 &     2.9 &    0.88 &   16.48 &            &         $^{       }$   \\
 J230855.0+613657 &    1.38 &                     &                     &    23085515+6136585  &    0.8 &    0.90 &   15.02 &      N19S028365 &     4.1 &    0.93 &   17.00 &       1516-0353040 &     1.6 &    0.63 &   16.66 &            &         $^{       }$   \\
 J230911.4+614327 &    0.40 &    89.6$\pm$    3.1 &     6.4$\pm$    1.0 &    23091139+6143277  &    0.3 &    1.00 &   10.70 &      N19U000026 &     0.4 &    1.00 &   12.72 &       1517-0359727 &     0.2 &    1.00 &   12.17 &       Star &    G7V  $^{    sp }$   \\
 J230915.4+612704 &    1.91 &     1.7$\pm$    0.6 &     1.8$\pm$    1.0 &    23091486+6127042  &    4.1 &    0.73 &   13.11 &      N19S027005 &     4.0 &    0.16 &         &       1514-0344726 &     4.3 &    0.22 &   19.13 &            &         $^{       }$   \\
 J230918.4+613623 &    1.86 &     2.3$\pm$    0.5 &                     &    23091838+6136209  &    2.3 &    0.89 &   13.29 &      N19S028341 &     2.3 &    0.90 &   17.18 &       1516-0353208 &     2.2 &    0.89 &   16.33 &            &         $^{       }$   \\
 J230920.1+613922 &    0.85 &     8.4$\pm$    1.1 &     0.2$\pm$    0.3 &    23092018+6139228  &    0.4 &    1.00 &    7.90 &      N19V000125 &     0.4 &    1.00 &    8.06 &       1516-0353240 &     0.4 &    1.00 &    8.06 &       Star &      A  $^{     s }$   \\
 J230923.2+612553 &    1.78 &     2.5$\pm$    0.8 &     0.6$\pm$    0.6 &    23092313+6125536  &    0.9 &    0.98 &   12.12 &      N19S026829 &     0.8 &    0.99 &   15.60 &       1514-0344777 &     0.9 &    0.98 &   15.10 &       Star &   K4Ve  $^{    sp }$   \\
 J230926.8+613550 &    1.51 &     1.3$\pm$    0.4 &                     &    23092660+6135509  &    1.7 &    0.98 &   12.10 &      N19S000005 &     1.4 &    0.99 &   14.07 &       1515-0347656 &     2.7 &    0.70 &   13.87 &       Star &    G3V  $^{    sp }$   \\
 J230933.2+614215 &    0.52 &    23.7$\pm$    1.7 &     3.3$\pm$    0.8 &    23093321+6142154  &    0.3 &    1.00 &   11.83 &      N19V004359 &     0.2 &    1.00 &   15.78 &       1517-0359907 &     0.5 &    1.00 &   14.48 &       Star &   M0Ve  $^{    sp }$   \\
 J230934.5+614614 &    2.00 &     0.8$\pm$    0.4 &     2.3$\pm$    0.8 &    23093459+6146178  &    3.6 &    0.58 &   14.16 &      N19U000756 &     3.8 &    0.10 &         &       1517-0359921 &     4.1 &    0.13 &   19.14 &            &         $^{       }$   \\
 J230939.8+613432 &    1.10 &     3.7$\pm$    0.6 &     0.7$\pm$    0.4 &    23093986+6134311  &    1.8 &    0.96 &   12.43 &      N19S028200 &     1.7 &    0.95 &   16.79 &       1515-0347770 &     1.5 &    0.95 &   16.08 &       Star &   M0Ve  $^{    sp }$   \\
 J230940.9+613510 &    1.04 &     1.7$\pm$    0.4 &     1.9$\pm$    0.5 &    23094084+6135119  &    1.6 &    0.93 &   13.23 &      N19S028257 &     2.2 &    0.83 &   17.62 &       1515-0347778 &     3.0 &    0.60 &   15.85 &            &         $^{       }$   \\
 J230943.4+614109 &    0.90 &     5.8$\pm$    1.3 &     0.6$\pm$    0.9 &    23094359+6141084  &    1.4 &    0.96 &   13.00 &      N19V004094 &     1.5 &    0.95 &   17.72 &       1516-0353465 &     1.2 &    0.96 &   16.25 &       Star &   K7Ve  $^{    sp }$   \\
 J230944.3+613435 &    1.17 &     3.3$\pm$    0.6 &     0.4$\pm$    0.3 &    23094418+6134353  &    1.0 &    1.00 &   10.51 &      N19S000009 &     0.7 &    1.00 &   13.01 &       1515-0347822 &     1.0 &    1.00 &   12.82 &       Star &    G0V  $^{    sp }$   \\
 J230955.7+613655 &    1.27 &     2.2$\pm$    0.5 &     0.1$\pm$    0.2 &    23095558+6136543  &    1.3 &    0.90 &   14.51 &      N19V003013 &     1.1 &    0.85 &         &       1516-0353561 &     1.4 &    0.80 &   18.81 &            &         $^{       }$   \\
 J230956.3+614006 &    1.06 &     3.9$\pm$    0.6 &     1.5$\pm$    0.4 &    23095635+6140082  &    1.3 &    0.98 &   12.62 &      N19V003918 &     1.5 &    0.98 &   15.83 &       1516-0353567 &     2.5 &    0.90 &   14.76 &       Star &    G0V  $^{    sp }$   \\
 J231001.0+613635 &    0.51 &    24.6$\pm$    1.5 &     1.4$\pm$    0.5 &    23100095+6136340  &    1.3 &    0.99 &   10.35 &      N19V000145 &     1.1 &    1.00 &   12.23 &       1516-0353591 &     1.1 &    1.00 &   11.99 &       Star &    F6V  $^{    sp }$   \\
 J231011.6+614904 &    1.56 &     5.6$\pm$    1.1 &                     &    23101154+6149049  &    0.8 &    0.99 &   11.02 &      N19U000010 &     0.8 &    1.00 &   12.66 &       1518-0362824 &     1.1 &    1.00 &   12.67 &       Star &    F5V  $^{    sp }$   \\
 J231012.5+613348 &    1.81 &     1.9$\pm$    0.5 &     0.5$\pm$    0.4 &    23101265+6133500  &    1.6 &    0.98 &   11.94 &      N19S028151 &     1.8 &    0.99 &   14.37 &       1515-0348118 &     1.8 &    0.99 &   14.38 &       Star &    G0V  $^{    sp }$   \\
 J231012.7+613239 &    1.63 &     1.3$\pm$    0.5 &     1.9$\pm$    0.8 &    23101298+6132414  &    2.6 &    0.89 &   13.07 &      N19S027998 &     2.9 &    0.49 &   18.81 &       1515-0348124 &     2.8 &    0.61 &   17.87 &            &         $^{       }$   \\
 J231021.4+614503 &    1.65 &     2.0$\pm$    0.6 &                     &    23102156+6145023  &    1.9 &    0.90 &   13.51 &      N19U000669 &     1.5 &    0.95 &   17.63 &       1517-0360246 &     2.1 &    0.82 &         &       Star &   K7Ve  $^{    sp }$   \\
 J231023.5+613515 &    1.42 &     2.4$\pm$    0.6 &     0.4$\pm$    0.5 &    23102343+6135147  &    1.3 &    1.00 &    9.53 &      N19V000158 &     1.1 &    1.00 &   11.06 &       1515-0348224 &     1.1 &    1.00 &   10.68 &       Star &    F3V  $^{    sp }$   \\
 J231028.2+614811 &    1.53 &     3.8$\pm$    0.9 &     0.2$\pm$    0.8 &    23102822+6148119  &    0.6 &    0.98 &   13.04 &      N19U001047 &     0.5 &    0.99 &   17.42 &       1518-0362905 &     0.9 &    0.98 &   16.68 &       Star &    K1V  $^{    sp }$   \\
 J231032.6+614328 &    1.25 &     2.7$\pm$    0.7 &     2.9$\pm$    1.1 &    23103252+6143300  &    2.0 &    0.96 &   11.98 &      N19V004808 &     1.9 &    0.78 &   18.75 &       1517-0360318 &     2.4 &    0.67 &   17.88 &            &         $^{       }$   \\
 J231036.8+613243 &    1.36 &     5.3$\pm$    0.9 &     1.3$\pm$    0.8 &    23103654+6132424  &    2.5 &    0.47 &   13.14 &      N19S028042 &     2.7 &    0.55 &   16.71 &       1515-0348345 &     1.7 &    0.28 &   14.97 &            &         $^{       }$   \\
 J231100.8+614028 &    0.91 &     8.0$\pm$    1.9 &     2.8$\pm$    2.2 &    23110073+6140272  &    1.5 &    0.99 &   11.61 &      N19V004104 &     1.3 &    0.99 &   16.41 &       1516-0353998 &     1.2 &    0.98 &   15.67 &       Star &   M2Ve  $^{    sp }$   \\
\noalign{\smallskip}\hline
\end{tabular}
\end{center}
\end{table}
\end{landscape}
\normalsize

\begin{landscape}
\begin{table}
\begin{center}
\caption[Xray parameters for field PKS0745-19 off]{
X-ray parameters for detected sources in field PKS0745-19-offset}
\label{t:xray-param}
\fontsize{7}{7}\selectfont
\begin{tabular}{lrrrcrrrcrrrcrrrcc} 
\hline\hline 
\noalign{\smallskip}
 2XMM  & r$_{90}$ & pn\_B1$^{*}$ & pn\_B2$^{**}$ & 2MASS & d$_{x-o}$ & P$_{id}$ & kmag & GSC  & d$_{x-o}$ & P$_{id}$ & V & USNO & d$_{x-o}$ & P$_{id}$ & R & Class$^\dag$ & SpT$^\ddag$ \\
 Name   & [\arcsec] & [cts ks$^{-1}$] & [cts ks$^{-1}$] & Name & [\arcsec]  &          &      & Name & [\arcsec]   &          &   & Name & [\arcsec]   &          &   &       &       \\
\noalign{\smallskip}
\hline\noalign{\smallskip}
\noalign{\smallskip}
 J074722.4-190702 &    1.29 &    16.5$\pm$    1.9 &     3.2$\pm$    1.1 &    07472249-1907028  &    0.3 &    1.00 &   12.00 &      S3NV029046 &     0.2 &    0.99 &         &       0708-0151420 &     2.7 &    0.92 &   13.68 &       Star &    G9V  $^{    sp }$   \\
 J074734.5-185656 &    0.79 &    17.6$\pm$    1.6 &    16.0$\pm$    2.0 &                      &        &         &         &                 &         &         &         &                    &         &         &         &       EG   &    AGN  $^{    sp }$   \\
 J074736.0-190554 &    1.68 &     2.6$\pm$    0.8 &     4.9$\pm$    1.6 &                      &        &         &         &                 &         &         &         &                    &         &         &         &       EG   &    Gal  $^{image }$   \\
 J074737.6-185944 &    1.12 &     4.2$\pm$    0.8 &     2.8$\pm$    1.0 &                      &        &         &         &      S3NV034178 &     2.7 &    0.11 &         &       0710-0150157 &     3.0 &    0.10 &   19.60 &            &         $^{       }$   \\
 J074739.6-190735 &    0.88 &     9.8$\pm$    1.3 &     0.7$\pm$    0.6 &    07473964-1907361  &    0.3 &    1.00 &   12.58 &      S3NV028660 &     0.3 &    1.00 &         &       0708-0151682 &     0.6 &    1.00 &   14.27 &       Star &    G3V  $^{    sp }$   \\
 J074740.3-190105 &    1.53 &     3.5$\pm$    0.7 &     1.9$\pm$    0.8 &    07474053-1901097  &    5.2 &    0.14 &   15.17 &      S3NV033288 &     5.4 &    0.01 &         &                    &         &         &         &            &         $^{       }$   \\
 J074742.2-190224 &    1.44 &     3.1$\pm$    0.8 &     0.8$\pm$    0.8 &                      &        &         &         &      S3NV032416 &     1.4 &    0.68 &         &       0709-0152008 &     1.3 &    0.71 &   19.90 &            &         $^{       }$   \\
 J074742.7-190749 &    2.04 &     2.2$\pm$    0.9 &     1.0$\pm$    1.1 &                      &        &         &         &      S3NV028339 &     6.0 &    0.01 &         &       0708-0151731 &     6.5 &    0.01 &   19.46 &            &         $^{       }$   \\
 J074743.5-185654 &    0.58 &    25.2$\pm$    1.7 &     7.3$\pm$    1.4 &                      &        &         &         &      S3NV036069 &     0.8 &    0.89 &         &       0710-0150259 &     1.5 &    0.64 &         &        CV  &     CV  $^{    sp }$   \\
 J074752.5-190428 &    1.50 &     3.3$\pm$    1.2 &     1.7$\pm$    1.3 &    07475247-1904248  &    3.3 &    0.77 &   13.04 &      S3NV031046 &     4.1 &    0.29 &         &       0709-0152175 &     3.8 &    0.47 &   15.27 &            &         $^{       }$   \\
 J074752.6-185713 &    1.42 &     2.0$\pm$    0.6 &     1.5$\pm$    0.8 &    07475277-1857099  &    4.2 &    0.16 &   14.91 &      S3NV035885 &     4.4 &    0.03 &         &       0710-0150416 &     4.0 &    0.23 &   17.12 &            &         $^{       }$   \\
 J074755.5-190647 &    0.90 &     9.7$\pm$    3.4 &     4.6$\pm$    3.5 &    07475552-1906480  &    0.5 &    0.99 &   13.11 &      S3NV029250 &     0.4 &    0.99 &         &       0708-0151945 &     0.6 &    0.98 &   14.89 &            &         $^{       }$   \\
 J074757.1-190243 &    1.25 &     1.2$\pm$    0.4 &     1.7$\pm$    0.5 &    07475724-1902411  &    3.2 &    0.26 &   15.52 &      S3NV032226 &     4.2 &    0.01 &         &       0709-0152254 &     4.2 &    0.04 &   18.22 &            &         $^{       }$   \\
 J074758.7-190602 &    0.86 &     2.9$\pm$    0.6 &     2.1$\pm$    0.7 &                      &        &         &         &                 &         &         &         &                    &         &         &         &       EG   &    AGN  $^{    sp }$   \\
 J074802.1-185952 &    1.57 &     0.3$\pm$    0.3 &     1.3$\pm$    0.5 &    07480243-1859498  &    5.1 &    0.15 &   14.86 &      S3NV034214 &     5.2 &    0.01 &         &       0710-0150592 &     5.4 &    0.17 &         &            &         $^{       }$   \\
 J074802.3-185820 &    1.56 &     3.1$\pm$    0.6 &                     &    07480248-1858198  &    1.8 &    0.98 &   11.94 &      S3NV000360 &     2.0 &    0.99 &         &       0710-0150598 &     2.1 &    0.98 &   13.55 &       Star &    F6V  $^{    sp }$   \\
 J074802.4-185745 &    1.87 &     0.1$\pm$    0.2 &     1.8$\pm$    0.7 &                      &        &         &         &      S3NV054772 &     2.2 &    0.33 &         &                    &         &         &         &       EG   &    Gal  $^{    sp }$   \\
 J074802.4-190251 &    1.06 &     2.0$\pm$    0.4 &     1.4$\pm$    0.5 &    07480246-1902517  &    0.9 &    0.95 &   15.01 &      S3NV032109 &     1.3 &    0.91 &         &       0709-0152357 &     1.3 &    0.89 &   16.68 &       Star &    G4V  $^{    sp }$   \\
 J074804.6-190814 &    0.99 &     1.3$\pm$    0.4 &     2.0$\pm$    0.6 &                      &        &         &         &                 &         &         &         &                    &         &         &         &       EG   &    AGN  $^{    sp }$   \\
 J074805.5-190506 &    0.68 &     3.6$\pm$    0.5 &     4.1$\pm$    0.6 &                      &        &         &         &                 &         &         &         &                    &         &         &         &       EG   &    AGN  $^{    sp }$   \\
 J074806.0-190856 &    0.59 &     8.0$\pm$    0.9 &     4.1$\pm$    0.7 &                      &        &         &         &                 &         &         &         &                    &         &         &         &       EG   &    AGN  $^{    sp }$   \\
 J074807.2-185640 &    2.22 &     1.2$\pm$    0.4 &     0.6$\pm$    0.6 &    07480737-1856385  &    2.5 &    0.88 &   13.60 &      S3NV036281 &     2.8 &    0.86 &         &       0710-0150697 &     3.0 &    0.84 &   15.37 &       Star &    G9V  $^{    sp }$   \\
 J074808.4-190528 &    1.88 &     0.5$\pm$    0.3 &     0.8$\pm$    0.4 &    07480860-1905315  &    3.6 &    0.27 &   14.83 &      S3NV030147 &     3.2 &    0.21 &         &       0709-0152479 &     3.5 &    0.39 &   18.59 &            &         $^{       }$   \\
 J074810.6-190417 &    1.60 &     0.6$\pm$    0.3 &     0.2$\pm$    0.2 &    07481067-1904122  &    5.3 &    0.15 &   14.53 &      S3NV031121 &     5.6 &    0.01 &         &       0709-0152527 &     5.0 &    0.09 &   16.42 &            &         $^{       }$   \\
 J074814.7-190536 &    1.18 &     1.3$\pm$    0.3 &     0.8$\pm$    0.3 &                      &        &         &         &      S3NV030067 &     0.8 &    0.74 &         &                    &         &         &         &       EG   &    AGN  $^{    sp }$   \\
 J074815.4-191415 &    1.11 &     3.0$\pm$    0.6 &     3.5$\pm$    0.8 &    07481542-1914175  &    1.8 &    0.73 &   15.52 &      S3NV022991 &     1.4 &    0.82 &         &       0707-0152048 &     1.7 &    0.67 &   18.38 &       Star &    K1V  $^{    sp }$   \\
 J074818.6-191730 &    0.55 &    21.2$\pm$    1.4 &     2.9$\pm$    0.9 &    07481863-1917317  &    1.1 &    0.98 &   11.76 &      S3NV020236 &     0.9 &    0.99 &         &       0707-0152103 &     1.2 &    0.96 &   14.36 &       Star &    G9V  $^{    sp }$   \\
 J074820.9-190721 &    1.52 &     1.0$\pm$    0.3 &     0.7$\pm$    0.3 &                      &        &         &         &      S3NV028800 &     3.7 &    0.33 &         &       0708-0152364 &     2.2 &    0.41 &         &       EG   &    Gal  $^{ image }$   \\
 J074821.3-191510 &    1.17 &     2.8$\pm$    0.5 &     0.7$\pm$    0.5 &    07482134-1915105  &    0.0 &    0.98 &   13.33 &      S3NV022264 &     0.1 &    0.98 &         &       0707-0152155 &     0.1 &    0.97 &   15.12 &       Star &    G1V  $^{    sp }$   \\
 J074821.9-190255 &    1.14 &     1.5$\pm$    0.3 &     1.0$\pm$    0.4 &    07482205-1902536  &    2.6 &    0.48 &   15.04 &      S3NV032061 &     2.7 &    0.35 &         &       0709-0152733 &     2.7 &    0.47 &   16.75 &       Star &    M5V  $^{    sp }$   \\
 J074822.1-190807 &    1.47 &     1.0$\pm$    0.3 &     0.3$\pm$    0.2 &    07482188-1908056  &    3.8 &    0.85 &   11.24 &      S3NV028312 &     3.6 &    0.93 &         &       0708-0152385 &     3.3 &    0.93 &   13.24 &       Star &    G9V  $^{    sp }$   \\
 J074823.1-190937 &    0.52 &    12.1$\pm$    0.8 &     0.3$\pm$    0.3 &    07482309-1909379  &    0.6 &    1.00 &   12.52 &      S3NV027003 &     0.3 &    1.00 &         &       0708-0152415 &     0.3 &    1.00 &   14.15 &       Star &    G9V  $^{    sp }$   \\
 J074823.3-191408 &    1.87 &     1.1$\pm$    0.4 &     1.7$\pm$    0.6 &                      &        &         &         &      S3NV023098 &     3.3 &    0.12 &         &                    &         &         &         &       EG   &    AGN  $^{    sp }$   \\
 J074824.5-190734 &    0.69 &     4.7$\pm$    0.6 &     0.8$\pm$    0.4 &    07482459-1907339  &    0.3 &    0.99 &   12.87 &      S3NV028656 &     0.3 &    1.00 &         &       0708-0152444 &     0.4 &    0.99 &   14.53 &       Star &    G3V  $^{    sp }$   \\
 J074825.0-185549 &    0.88 &     2.7$\pm$    0.6 &     7.5$\pm$    1.2 &                      &        &         &         &      S3NV036692 &     2.5 &    0.10 &         &       0710-0151034 &     0.2 &    0.89 &   20.28 &       EG   &    AGN  $^{    sp }$   \\
 J074826.0-191733 &    0.98 &     5.6$\pm$    0.8 &     0.0$\pm$    0.6 &    07482598-1917339  &    1.0 &    1.00 &    7.98 &      S3NV000651 &     1.0 &    0.55 &    9.33 &       0707-0152239 &     1.0 &    1.00 &    9.10 &       Star &  A4III  $^{     s }$   \\
 J074826.5-190216 &    1.25 &     1.3$\pm$    0.3 &     0.8$\pm$    0.3 &    07482637-1902149  &    3.0 &    0.43 &   14.74 &      S3NV032619 &     3.4 &    0.17 &         &       0709-0152805 &     2.6 &    0.75 &   15.45 &       Star &    F7V  $^{    sp }$   \\
 J074828.7-191245 &    1.16 &     1.8$\pm$    0.4 &     0.5$\pm$    0.3 &                      &        &         &         &                 &         &         &         &                    &         &         &         &       EG   &    Gal  $^{ image }$   \\
 J074829.0-191641 &    0.70 &     9.2$\pm$    0.9 &                     &    07482906-1916413  &    0.1 &    1.00 &   10.23 &      S3NV000630 &     0.0 &    1.00 &   11.82 &       0707-0152292 &     0.0 &    1.00 &   11.56 &       Star &    F4V  $^{    sp }$   \\
 J074829.3-191425 &    2.01 &     1.4$\pm$    0.4 &     0.0$\pm$    0.2 &    07482946-1914262  &    2.4 &    1.00 &    8.26 &      S3NV000593 &     2.4 &    1.00 &   10.80 &       0707-0152301 &     2.4 &    1.00 &   10.13 &       Star &    G7III  $^{    sp }$   \\
 J074830.1-190245 &    1.51 &     0.9$\pm$    0.3 &     0.8$\pm$    0.3 &    07483023-1902410  &    4.6 &    0.20 &   14.07 &      S3NV032232 &     4.6 &    0.10 &         &       0709-0152879 &     4.4 &    0.42 &   15.30 &       Star &    M5V  $^{    sp }$   \\
 J074832.1-185427 &    0.57 &    22.0$\pm$    2.1 &     1.4$\pm$    0.8 &    07483207-1854278  &    0.5 &    1.00 &   11.72 &      S3NU018280 &     0.4 &    1.00 &         &       0710-0151161 &     0.3 &    1.00 &   13.62 &       Star &    G2V  $^{    sp }$   \\
 J074836.0-185324 &    2.49 &                     &                     &    07483626-1853267  &    4.4 &    0.68 &   13.74 &      S3NU018812 &     4.1 &    0.77 &         &       0711-0152383 &     5.5 &    0.69 &         &            &         $^{       }$   \\
 J074836.9-191058 &    1.62 &     0.9$\pm$    0.3 &     1.2$\pm$    0.4 &    07483695-1910595  &    1.1 &    0.81 &   15.00 &      S3NV025814 &     5.1 &    0.02 &         &       0708-0152650 &     5.0 &    0.19 &   15.98 &            &         $^{       }$   \\
 J074837.8-185837 &    1.44 &     0.7$\pm$    0.3 &     0.5$\pm$    0.3 &                      &        &         &         &                 &         &         &         &       0710-0151250 &     3.9 &    0.16 &   17.44 &            &         $^{       }$   \\
 J074839.6-191301 &    1.38 &     0.2$\pm$    0.3 &     2.2$\pm$    0.5 &                      &        &         &         &      S3NV053803 &     1.6 &    0.57 &         &                    &         &         &         &       EG   &    Gal  $^{ image }$   \\
 J074841.3-190004 &    1.78 &     0.6$\pm$    0.3 &     0.5$\pm$    0.4 &    07484118-1900030  &    2.6 &    0.87 &   13.87 &      S3NU015531 &     2.9 &    0.85 &         &       0709-0153068 &     4.4 &    0.78 &         &            &         $^{       }$   \\
 J074841.5-190427 &    1.39 &     0.8$\pm$    0.3 &     0.1$\pm$    0.2 &    07484163-1904293  &    2.1 &    0.98 &   11.54 &      S3NU000818 &     2.0 &    1.00 &         &       0709-0153075 &     1.9 &    0.84 &   12.80 &       Star &    F4V  $^{    sp }$   \\
 J074842.7-190343 &    1.22 &     1.6$\pm$    0.4 &     2.8$\pm$    0.5 &    07484262-1903453  &    3.1 &    0.21 &   15.87 &      S3NU013927 &     3.0 &    0.12 &         &       0709-0153091 &     3.2 &    0.36 &   18.27 &            &         $^{       }$   \\
 J074843.8-190653 &    1.57 &     1.3$\pm$    0.3 &     0.1$\pm$    0.2 &                      &        &         &         &      S3NU050871 &     1.5 &    0.45 &         &       0708-0152762 &     1.6 &    0.56 &   20.39 &            &         $^{       }$   \\
 J074844.0-191029 &    0.39 &     0.3$\pm$    0.3 &     0.3$\pm$    0.4 &                      &        &         &         &                 &         &         &         &                    &         &         &         &       Star &  K1III  $^{     s }$   \\
 J074846.8-190338 &    1.20 &     1.7$\pm$    0.4 &     1.5$\pm$    0.4 &                      &        &         &         &      S3NU013964 &     1.1 &    0.69 &         &       0709-0153175 &     1.4 &    0.66 &   19.78 &       EG   &    Gal  $^{ image }$   \\
 J074847.1-190410 &    0.53 &     0.5$\pm$    0.3 &    14.0$\pm$    0.9 &    07484710-1904112  &    0.7 &    0.97 &   14.78 &      S3NU013757 &     0.8 &    0.93 &         &       0709-0153176 &     1.7 &    0.27 &   17.61 &       EG   &    AGN  $^{    sp }$   \\
 J074849.6-191704 &    0.79 &     5.8$\pm$    0.8 &     6.4$\pm$    1.2 &                      &        &         &         &                 &         &         &         &                    &         &         &         &       EG   &    AGN  $^{    sp }$   \\
 J074849.7-190900 &    1.54 &     0.8$\pm$    0.3 &     0.9$\pm$    0.4 &                      &        &         &         &                 &         &         &         &                    &         &         &         &       EG   &    AGN  $^{    sp }$   \\
 J074851.5-190746 &    1.50 &     1.2$\pm$    0.3 &     0.2$\pm$    0.3 &    07485130-1907493  &    4.6 &    0.97 &   14.23 &      S3NU012384 &     0.6 &    0.98 &         &       0708-0152846 &     0.5 &    0.97 &   14.52 &       Star &    K4V  $^{    sp }$   \\
 J074852.5-185856 &    0.86 &     4.1$\pm$    0.7 &     2.1$\pm$    0.6 &                      &        &         &         &                 &         &         &         &                    &         &         &         &       EG   &    AGN  $^{    sp }$   \\
 J074853.7-190127 &    2.15 &     1.0$\pm$    0.4 &     0.9$\pm$    0.5 &    07485392-1901245  &    3.4 &    0.44 &   15.18 &      S3NU014923 &     3.8 &    0.12 &         &       0709-0153296 &     3.5 &    0.35 &   19.58 &            &         $^{       }$   \\
 J074854.0-185629 &    1.79 &     2.1$\pm$    0.6 &     0.4$\pm$    0.4 &    07485400-1856300  &    0.6 &    0.94 &   13.79 &      S3NU017179 &     0.5 &    0.94 &         &       0710-0151522 &     0.5 &    0.93 &   15.36 &            &         $^{       }$   \\
 J074855.3-185823 &    0.81 &     6.5$\pm$    0.8 &     0.8$\pm$    0.6 &    07485529-1858224  &    1.1 &    0.98 &   13.35 &      S3NU016278 &     1.2 &    0.95 &         &       0710-0151548 &     1.0 &    0.96 &   16.02 &       Star &    K4V  $^{    sp }$   \\
 J074857.4-190632 &    0.37 &   103.0$\pm$    2.3 &     5.9$\pm$    0.7 &    07485744-1906330  &    0.5 &    1.00 &   10.38 &      S3NU012853 &     0.2 &    1.00 &         &       0708-0152958 &     0.7 &    1.00 &   14.32 &       Star &    M4V  $^{    sp }$   \\
 J074901.1-185827 &    0.76 &     6.8$\pm$    0.8 &     0.3$\pm$    0.5 &    07490118-1858280  &    0.5 &    1.00 &   11.76 &      S3NU016266 &     0.9 &    1.00 &         &       0710-0151662 &     1.2 &    0.99 &   13.93 &       Star &    G9V  $^{    sp }$   \\
 J074905.0-190033 &    1.25 &     3.5$\pm$    0.6 &     0.4$\pm$    0.3 &    07490523-1900349  &    2.8 &    0.78 &   13.19 &      S3NU054171 &     2.8 &    0.06 &         &       0709-0153481 &     0.9 &    0.85 &   18.65 &       Star &    M5V  $^{    sp }$   \\
 J074906.0-190013 &    1.73 &     0.3$\pm$    0.3 &     0.7$\pm$    0.5 &    07490611-1900085  &    4.6 &    0.14 &   16.30 &      S3NU015448 &     4.9 &    0.03 &         &       0709-0153501 &     4.7 &    0.18 &   18.08 &            &         $^{       }$   \\
 J074911.1-190521 &    1.68 &     1.3$\pm$    0.4 &     0.4$\pm$    0.4 &    07491150-1905234  &    5.4 &    0.16 &   14.46 &      S3NU013271 &     5.3 &    0.03 &         &       0709-0153602 &     5.4 &    0.13 &   16.32 &            &         $^{       }$   \\
\noalign{\smallskip}\hline
\end{tabular}
\end{center}
\end{table}
\end{landscape}
\normalsize

\begin{landscape}
\begin{table}
\begin{center}
\caption[Xray parameters for field PSRJ0117+5914]{
X-ray parameters for detected sources in field PSRJ0117+5914}
\label{t:xray-param}
\fontsize{7}{7}\selectfont
\begin{tabular}{lrrrcrrrcrrrcrrrcc} 
\hline\hline 
\noalign{\smallskip}
 2XMM  & r$_{90}$ & pn\_B1$^{*}$ & pn\_B2$^{**}$ & 2MASS & d$_{x-o}$ & P$_{id}$ & kmag & GSC  & d$_{x-o}$ & P$_{id}$ & V & USNO & d$_{x-o}$ & P$_{id}$ & R & Class$^\dag$ & SpT$^\ddag$ \\
 Name   & [\arcsec] & [cts ks$^{-1}$] & [cts ks$^{-1}$] & Name & [\arcsec]  &          &      & Name & [\arcsec]   &          &   & Name & [\arcsec]   &          &   &       &       \\
\noalign{\smallskip}
\hline\noalign{\smallskip}
\noalign{\smallskip}
 J011559.0+590914 &    1.44 &    17.7$\pm$    3.5 &     9.9$\pm$    4.6 &    01155905+5909141  &    0.2 &    0.99 &    9.49 &      NAMI000071 &     0.1 &    1.00 &   11.41 &       1491-0041464 &     0.1 &    1.00 &   11.11 &       HMXB  &   Be/X  $^{    sp }$   \\
 J011610.8+591341 &    1.79 &     5.2$\pm$    1.8 &     9.2$\pm$    3.0 &    01161131+5913387  &    4.2 &    0.55 &   14.96 &      NAMI032506 &     1.1 &    0.60 &         &       1492-0040572 &     1.2 &    0.42 &   17.37 &            &         $^{       }$   \\
 J011637.5+591434 &    3.08 &     2.9$\pm$    1.2 &     1.2$\pm$    1.5 &    01163885+5914375  &   10.3 &         &   15.03 &      NAMI033223 &    10.6 &    0.24 &         &       1492-0040833 &     2.6 &    0.25 &   19.23 &            &         $^{       }$   \\
 J011655.4+591807 &    1.64 &     7.4$\pm$    1.6 &     1.5$\pm$    0.9 &    01165553+5918054  &    2.5 &    0.86 &   12.29 &      NAMK004202 &     2.5 &    0.91 &   14.93 &       1493-0042033 &     4.6 &    0.94 &   15.91 &       Star &    K4V  $^{    sp }$   \\
 J011705.5+592228 &    2.22 &     1.5$\pm$    1.2 &     2.2$\pm$    1.7 &                      &        &         &         &      NAMK006268 &     0.6 &    0.53 &   18.54 &       1493-0042154 &     0.7 &    0.52 &   18.66 &            &         $^{       }$   \\
 J011734.6+591516 &    1.58 &     2.6$\pm$    0.9 &     4.3$\pm$    1.2 &    01173479+5915147  &    2.0 &    0.68 &   14.80 &      NAMI033702 &     2.2 &    0.70 &   17.49 &       1492-0041424 &     1.4 &    0.81 &   17.23 &       Star &    F0V  $^{    sp }$   \\
 J011747.5+591836 &    1.65 &     1.7$\pm$    0.8 &     3.8$\pm$    1.2 &    01174685+5918373  &    5.0 &    0.03 &   15.54 &      NAMI035913 &     5.0 &    0.01 &   17.32 &       1493-0042571 &     5.2 &    0.01 &   16.97 &            &         $^{       }$   \\
 J011749.3+591235 &    1.42 &     6.0$\pm$    1.2 &     1.6$\pm$    0.7 &    01174933+5912341  &    1.6 &    0.92 &   12.14 &      NAMI031660 &     1.6 &    0.97 &   14.38 &       1492-0041539 &     3.8 &    0.89 &   14.15 &       Star &    F7V  $^{    sp }$   \\
 J011750.6+592032 &    1.79 &    10.4$\pm$    4.2 &     0.2$\pm$    2.5 &                      &        &         &         &      NAMK005215 &     3.0 &    0.18 &         &       1493-0042609 &     2.7 &    0.31 &   19.06 &            &         $^{       }$   \\
 J011754.2+591001 &    1.47 &     9.9$\pm$    1.7 &     3.2$\pm$    1.2 &    01175433+5909598  &    1.5 &    0.95 &   11.45 &                 &         &         &         &                    &         &         &         &       Star &    K1V  $^{    sp }$   \\
 J011757.0+591149 &    1.21 &    21.9$\pm$    2.3 &     0.9$\pm$    0.7 &    01175713+5911480  &    1.9 &    0.93 &   11.74 &      NAMI000044 &     1.8 &    0.98 &   13.55 &       1491-0042638 &     1.5 &    0.98 &   13.35 &       Star &    G7V  $^{    sp }$   \\
 J011757.3+590835 &    1.78 &     3.8$\pm$    1.2 &     0.4$\pm$    0.6 &    01175724+5908338  &    1.3 &    0.61 &   15.59 &      NAMI049432 &     0.9 &    0.50 &         &                    &         &         &         &            &         $^{       }$   \\
 J011801.6+591420 &    1.73 &     4.1$\pm$    1.9 &     4.6$\pm$    2.3 &    01180198+5914176  &    3.5 &    0.70 &   13.06 &      NAMI032938 &     3.7 &    0.63 &   16.30 &       1492-0041699 &     3.3 &    0.70 &   15.92 &            &         $^{       }$   \\
 J011801.9+591757 &    2.18 &     1.8$\pm$    0.8 &     2.6$\pm$    1.1 &    01180162+5917587  &    2.9 &    0.34 &   16.55 &      NAMI035516 &     3.1 &    0.53 &   18.03 &       1492-0041691 &     3.5 &    0.46 &   17.74 &            &         $^{       }$   \\
 J011804.5+592442 &    2.13 &     3.0$\pm$    1.3 &     4.8$\pm$    2.1 &    01180421+5924383  &    5.3 &    0.76 &   15.02 &      NAMK007593 &     1.9 &    0.67 &         &       1494-0043767 &     1.2 &    0.50 &   17.37 &            &         $^{       }$   \\
 J011806.9+592641 &    1.33 &    15.9$\pm$    2.9 &    29.4$\pm$    4.8 &                      &        &         &         &                 &         &         &         &       1494-0043783 &     2.6 &    0.38 &   18.67 &       Star &    A3V  $^{    sp }$   \\
 J011820.1+592010 &    1.36 &    11.3$\pm$    2.1 &    11.9$\pm$    2.5 &    01182060+5920083  &    4.4 &         &   15.87 &      NAMI036823 &     4.2 &         &         &       1493-0042884 &     4.3 &         &   19.07 &            &         $^{       }$   \\
 J011836.8+591849 &    1.07 &    40.4$\pm$    3.6 &    32.0$\pm$    3.6 &    01183707+5918487  &    2.1 &    0.62 &   15.10 &      NAMI036019 &     1.9 &    0.61 &         &       1493-0043028 &     1.8 &    0.65 &   18.80 &            &         $^{       }$   \\
 J011842.9+591555 &    1.97 &     4.8$\pm$    1.5 &     0.8$\pm$    0.9 &    01184299+5915543  &    1.3 &    0.78 &   14.39 &      NAMI034108 &     1.4 &    0.75 &   17.88 &       1492-0042114 &     1.1 &    0.74 &   17.06 &            &         $^{       }$   \\
 J011849.0+591510 &    1.87 &     2.3$\pm$    1.0 &     3.5$\pm$    1.5 &                      &        &         &         &      NAMI033543 &     2.9 &    0.35 &   18.60 &       1492-0042172 &     2.7 &    0.36 &   18.59 &            &         $^{       }$   \\
 J011916.5+591037 &    2.02 &     9.3$\pm$    3.9 &     2.5$\pm$    3.4 &                      &        &         &         &      NAMI030086 &     1.7 &    0.31 &         &       1491-0043444 &     1.4 &    0.39 &   19.14 &            &         $^{       }$   \\
\noalign{\smallskip}\hline
\end{tabular}
\end{center}
\end{table}
\end{landscape}
\normalsize

\begin{landscape}
\begin{table}
\begin{center}
\caption[Xray parameters for field Ridge1]{
X-ray parameters for detected sources in field Ridge~1}
\label{t:xray-param}
\fontsize{7}{7}\selectfont
\begin{tabular}{lrrrcrrrcrrrcrrrcc} 
\hline\hline 
\noalign{\smallskip}
 2XMM  & r$_{90}$ & pn\_B1$^{*}$ & pn\_B2$^{**}$ & 2MASS & d$_{x-o}$ & P$_{id}$ & kmag & GSC  & d$_{x-o}$ & P$_{id}$ & V & USNO & d$_{x-o}$ & P$_{id}$ & R & Class$^\dag$ & SpT$^\ddag$ \\
 Name   & [\arcsec] & [cts ks$^{-1}$] & [cts ks$^{-1}$] & Name & [\arcsec]  &          &      & Name & [\arcsec]   &          &   & Name & [\arcsec]   &          &   &       &       \\
\noalign{\smallskip}
\hline\noalign{\smallskip}
\noalign{\smallskip}
 J185059.1+001650 &    1.62 &                     &     1.4$\pm$    2.1 &    18505896+0016476  &    4.6 &         &   13.91 &      N1NL009316 &     5.6 &     &         &       0902-0383927 &     4.4 &    0.02 &   18.21 &            &         $^{       }$   \\
 J185110.4+000815 &    1.37 &     5.0$\pm$    0.7 &     0.2$\pm$    0.3 &    18511045+0008152  &    0.3 &    0.98 &   11.03 &      N1NL000215 &     0.5 &    1.00 &   13.25 &       0901-0383704 &     0.5 &    1.00 &   12.99 &       Star &    K1V  $^{    sp }$   \\
 J185114.3-000004 &    1.07 &     1.3$\pm$    0.7 &    40.3$\pm$    2.7 &    18511447-0000036  &    1.5 &    0.64 &   11.80 &                 &         &         &         &                    &         &         &         &            &         $^{       }$   \\
 J185116.3+000515 &    1.77 &     0.1$\pm$    0.2 &     2.5$\pm$    0.7 &    18511640+0005138  &    2.0 &    0.58 &   12.40 &      N1NL024991 &     2.2 &    0.69 &         &       0900-0386154 &     2.2 &    0.70 &   17.28 &            &         $^{       }$   \\
 J185124.0-000349 &    2.51 &                     &                     &    18512423-0003435  &    6.3 &    0.37 &   12.19 &      S9LE138888 &     8.5 &    0.34 &   18.97 &       0899-0386249 &     1.3 &    0.26 &   18.15 &            &         $^{       }$   \\
 J185124.5+000430 &    1.63 &     0.5$\pm$    0.3 &     2.5$\pm$    0.6 &    18512445+0004337  &    3.8 &    0.36 &   10.30 &      N1NL023734 &     4.6 &         &         &       0900-0386440 &     4.3 &         &   18.71 &            &         $^{       }$   \\
 J185125.1+000742 &    1.11 &     9.7$\pm$    0.8 &     0.4$\pm$    0.3 &    18512525+0007422  &    2.3 &    0.69 &    8.73 &      N1NL000216 &     1.5 &    1.00 &   11.18 &       0901-0384185 &     1.5 &    1.00 &   10.80 &       Star &    F4V  $^{    sp }$   \\
 J185135.3+000928 &    1.17 &     9.7$\pm$    1.7 &     0.7$\pm$    1.0 &    18513544+0009288  &    1.1 &    1.00 &    8.93 &      N1NL000209 &     1.1 &    1.00 &   10.29 &       0901-0384552 &     1.1 &    1.00 &    9.83 &       Star &    G2V  $^{    sp }$   \\
 J185137.1+000936 &    1.99 &     1.4$\pm$    1.1 &     2.2$\pm$    1.3 &    18513704+0009407  &    4.6 &    0.08 &   11.96 &      N1NL032411 &     4.3 &    0.01 &         &       0901-0384607 &     5.2 &         &   18.28 &            &         $^{       }$   \\
 J185137.3-000330 &    1.46 &                     &                     &    18513733-0003314  &    0.8 &    0.91 &   11.05 &      S9LE139786 &     1.8 &    0.49 &         &       0899-0386994 &     1.5 &    0.92 &   15.04 &            &         $^{       }$   \\
 J185139.1+001635 &    1.08 &    16.6$\pm$    1.2 &    13.0$\pm$    1.3 &    18513925+0016350  &    2.2 &    0.92 &    8.37 &      N1NL009237 &     2.1 &    0.51 &   14.41 &       0902-0384825 &     1.8 &    0.98 &   13.51 &       Star &    K5III  $^{    sp }$   \\
 J185139.9+001308 &    1.12 &     9.6$\pm$    0.8 &     0.2$\pm$    0.3 &    18514005+0013084  &    2.2 &    0.95 &    9.22 &      N1NL000198 &     2.2 &    1.00 &   10.46 &       0902-0384844 &     2.1 &    1.00 &   10.10 &       Star &    F4V  $^{    sp }$   \\
 J185140.8+000557 &    1.62 &     1.1$\pm$    0.3 &     0.8$\pm$    0.4 &    18514092+0005574  &    0.5 &    0.82 &   12.12 &      N1NL026277 &     0.9 &    0.62 &         &       0900-0387044 &     0.5 &    0.70 &   17.73 &            &         $^{       }$   \\
 J185142.0+000021 &    1.54 &     2.6$\pm$    0.6 &     0.6$\pm$    0.4 &    18514181+0000226  &    3.6 &    0.34 &   12.30 &      N1NL015808 &     2.6 &    0.51 &   19.05 &       0900-0387085 &     2.5 &    0.54 &   18.42 &            &         $^{       }$   \\
 J185147.2+000924 &    1.46 &                     &     3.8$\pm$    0.6 &                      &        &         &         &                 &         &         &         &       0901-0384884 &     2.7 &    0.56 &   18.64 &            &         $^{       }$   \\
 J185147.6+000733 &    1.34 &     0.5$\pm$    0.3 &     3.3$\pm$    0.5 &    18514778+0007335  &    1.8 &    1.00 &    5.94 &      N1NL005952 &     1.8 &    0.62 &   17.40 &       0901-0384899 &     1.9 &    0.89 &   15.40 &       Star &   M5V+  $^{    sp }$   \\
 J185147.8+000130 &    1.51 &     2.2$\pm$    0.5 &     2.1$\pm$    0.7 &    18514792+0001304  &    1.5 &    0.76 &   12.01 &      N1NL017920 &     2.1 &    0.60 &         &       0900-0387281 &     1.7 &    0.69 &   16.77 &            &         $^{       }$   \\
 J185152.6+001927 &    1.72 &     2.0$\pm$    0.6 &     2.1$\pm$    1.0 &    18515276+0019273  &    2.4 &    0.79 &   10.79 &      N1NL009914 &     2.4 &    0.74 &   18.68 &       0903-0382702 &     2.5 &    0.72 &   16.97 &            &         $^{       }$   \\
 J185207.7+000609 &    1.95 &     2.1$\pm$    0.5 &     0.3$\pm$    0.4 &    18520785+0006100  &    1.5 &    0.97 &   10.42 &      N1NL000222 &     1.5 &    1.00 &   12.46 &       0901-0385400 &     1.3 &    1.00 &   12.43 &       Star &    F9V  $^{    sp }$   \\
 J185208.3+001507 &    1.69 &     0.4$\pm$    0.9 &     2.4$\pm$    2.3 &    18520846+0015048  &    3.2 &    0.40 &   11.67 &      N1NL000188 &     3.2 &    0.94 &   13.88 &       0902-0385423 &     3.2 &    0.94 &   13.50 &       Star &    K1V  $^{    sp }$   \\
 J185209.9+001207 &    1.50 &     0.5$\pm$    0.7 &     3.4$\pm$    1.3 &    18521006+0012073  &    2.4 &    0.91 &    7.85 &      N1NL035928 &     2.0 &    0.46 &         &       0902-0385469 &     1.9 &    0.58 &   17.53 &            &         $^{       }$   \\
 J185222.3+001127 &    2.64 &     0.2$\pm$    0.3 &     0.6$\pm$    0.5 &    18522230+0011266  &    1.2 &    0.29 &   11.88 &      N1NL035142 &     6.5 &    0.05 &         &       0901-0385782 &     6.5 &    0.04 &   16.17 &            &         $^{       }$   \\
 J185226.9+000353 &    2.20 &     2.6$\pm$    0.8 &     1.7$\pm$    1.0 &    18522711+0003543  &    2.7 &    0.39 &   12.16 &                 &         &         &         &       0900-0389003 &     0.7 &    0.07 &   21.00 &            &         $^{       }$   \\
 J185233.2+000638 &    1.19 &    16.8$\pm$    1.6 &    13.1$\pm$    1.9 &    18523311+0006362  &    3.0 &         &   13.94 &      N1NL027810 &     2.5 &    0.19 &   18.48 &       0901-0386136 &     2.8 &    0.41 &   16.42 &            &         $^{       }$   \\
 J185242.0+000715 &    2.95 &     3.6$\pm$    0.9 &                     &    18524201+0007205  &    5.3 &    0.88 &   10.26 &      N1NL000455 &     3.4 &    0.99 &   11.91 &       0901-0386495 &     3.4 &    0.99 &   11.28 &       Star &    K0V  $^{    sp }$   \\
\noalign{\smallskip}\hline
\end{tabular}
\end{center}
\end{table}
\end{landscape}
\normalsize

\begin{landscape}
\begin{table}
\begin{center}
\caption[Xray parameters for field Ridge2]{
X-ray parameters for detected sources in field Ridge~2}
\label{t:xray-param}
\fontsize{7}{7}\selectfont
\begin{tabular}{lrrrcrrrcrrrcrrrcc} 
\hline\hline 
\noalign{\smallskip}
 2XMM  & r$_{90}$ & pn\_B1$^{*}$ & pn\_B2$^{**}$ & 2MASS & d$_{x-o}$ & P$_{id}$ & kmag & GSC  & d$_{x-o}$ & P$_{id}$ & V & USNO & d$_{x-o}$ & P$_{id}$ & R & Class$^\dag$ & SpT$^\ddag$ \\
 Name   & [\arcsec] & [cts ks$^{-1}$] & [cts ks$^{-1}$] & Name & [\arcsec]  &          &      & Name & [\arcsec]   &          &   & Name & [\arcsec]   &          &   &       &       \\
\noalign{\smallskip}
\hline\noalign{\smallskip}
\noalign{\smallskip}
 J184343.2+005330 &    0.79 &    11.3$\pm$    1.4 &     2.3$\pm$    1.2 &    18434320+0053294  &    0.8 &    0.99 &   11.82 &      N1R4029991 &     0.9 &    1.00 &   16.35 &       0908-0372937 &     0.6 &    1.00 &   15.25 &       Star &   M0Ve  $^{    sp }$   \\
 J184351.8+004857 &    1.80 &     0.9$\pm$    0.6 &     1.2$\pm$    1.0 &    18435188+0048571  &    0.7 &    0.92 &   13.77 &      N1R4026955 &     0.4 &    0.99 &   18.54 &       0908-0373050 &     0.4 &    0.98 &   17.16 &       Star &   M5Ve  $^{    sp }$   \\
 J184353.9+005016 &    1.67 &     0.9$\pm$    0.4 &     1.8$\pm$    0.8 &    18435398+0050137  &    3.0 &    0.31 &   13.88 &      N1R4060211 &     2.6 &         &         &                    &         &         &         &            &         $^{       }$   \\
 J184401.2+005455 &    1.21 &     2.8$\pm$    0.6 &     0.8$\pm$    0.4 &    18440115+0054556  &    0.9 &    0.98 &   11.91 &      N1R4030948 &     0.8 &    1.00 &   17.36 &       0909-0376768 &     1.0 &    0.98 &   16.03 &       TTS &   K7Ve  $^{  sp }$   \\
 J184409.8+004736 &    1.35 &     3.9$\pm$    0.9 &     1.2$\pm$    0.9 &    18440983+0047354  &    0.6 &    0.97 &   12.46 &      N1R4026033 &     0.6 &    1.00 &   17.52 &       0907-0376900 &     0.3 &    1.00 &   15.72 &       Star &   M0Ve  $^{    sp }$   \\
 J184412.9+010106 &    1.59 &     1.8$\pm$    0.4 &     0.2$\pm$    0.3 &    18441291+0101068  &    0.1 &    1.00 &    9.39 &      N1R4000082 &     0.0 &    1.00 &   10.98 &       0910-0375578 &     0.0 &    1.00 &   10.49 &       Star &    A9V  $^{    sp }$   \\
 J184413.9+010026 &    1.38 &     0.8$\pm$    0.3 &     1.0$\pm$    0.4 &    18441389+0100268  &    0.5 &    1.00 &    6.42 &      N1R4034849 &     0.3 &    1.00 &   17.37 &       0910-0375599 &     0.7 &    0.99 &   15.51 &       Star &   M5III  $^{    sp }$   \\
 J184421.9+005333 &    1.09 &     2.3$\pm$    0.5 &     0.1$\pm$    0.2 &    18442195+0053342  &    0.7 &    1.00 &    9.35 &      N1R4000100 &     0.6 &    1.00 &   10.68 &       0908-0373556 &     0.6 &    1.00 &   10.28 &       Star &    F3V  $^{    sp }$   \\
 J184429.7+010546 &    1.28 &     3.2$\pm$    0.8 &     0.5$\pm$    0.7 &    18442956+0105465  &    2.8 &    0.89 &   13.03 &      N1R4038341 &     1.4 &    1.00 &   15.89 &       0910-0375912 &     1.6 &    0.99 &   14.89 &       Star &   M2Ve  $^{    sp }$   \\
 J184431.3+005542 &    2.80 &     0.2$\pm$    0.2 &     0.7$\pm$    0.4 &    18443159+0055453  &    4.7 &    0.29 &   13.89 &      N1R4064783 &     8.9 &    0.31 &         &       0909-0377245 &     4.9 &    0.72 &   17.71 &            &         $^{       }$   \\
 J184436.8+010242 &    1.33 &     1.8$\pm$    0.5 &     0.3$\pm$    0.4 &    18443698+0102420  &    1.6 &    0.74 &   12.75 &      N1R4036267 &     2.9 &    0.13 &         &       0910-0376036 &     1.5 &    0.79 &   18.25 &            &         $^{       }$   \\
 J184437.3+010959 &    1.27 &     7.1$\pm$    1.2 &     1.1$\pm$    0.9 &    18443741+0109595  &    0.5 &    0.99 &   11.16 &      N1R4000062 &     0.6 &    1.00 &   13.14 &       0911-0377504 &     0.7 &    1.00 &   12.65 &       Star &    G2V  $^{    sp }$   \\
 J184439.9+005709 &    1.03 &     1.9$\pm$    0.5 &     2.0$\pm$    0.6 &    18443986+0057088  &    1.2 &    0.86 &   12.67 &      N1R4032511 &     1.0 &    0.93 &         &       0909-0377370 &     1.0 &    0.92 &   18.32 &            &         $^{       }$   \\
 J184440.4+004904 &    0.92 &     7.5$\pm$    1.0 &     0.4$\pm$    0.4 &    18444046+0049048  &    0.9 &    1.00 &   10.44 &      N1R4000113 &     0.8 &    1.00 &   12.26 &       0908-0373785 &     0.8 &    1.00 &   11.71 &       Star &    F7V  $^{    sp }$   \\
 J184443.4+005908 &    2.08 &     1.8$\pm$    0.5 &     0.8$\pm$    0.5 &    18444347+0059054  &    3.2 &    0.20 &   12.81 &      N1R4033886 &     3.0 &    0.82 &   17.62 &       0909-0377420 &     2.8 &    0.79 &   16.29 &            &         $^{       }$   \\
 J184444.3+005123 &    0.97 &     7.1$\pm$    0.9 &     0.0$\pm$    0.5 &    18444442+0051232  &    1.3 &    0.89 &   12.56 &      N1R4028572 &     1.2 &    0.99 &   17.48 &       0908-0373846 &     1.0 &    1.00 &   15.96 &       Star &   M5Ve  $^{    sp }$   \\
 J184447.7+011131 &    0.68 &    71.1$\pm$    4.1 &     5.3$\pm$    1.9 &    18444773+0111320  &    0.5 &    1.00 &   10.49 &      N1R4000056 &     0.8 &    1.00 &   12.13 &       0911-0377699 &     1.7 &    1.00 &   11.33 &       Star &    K2V  $^{    sp }$   \\
 J184450.8+005242 &    1.70 &     1.8$\pm$    0.5 &                     &    18445096+0052394  &    3.2 &    0.59 &   11.33 &      N1R4029440 &     1.5 &    0.99 &   17.46 &       0908-0373969 &     1.8 &    0.52 &         &       Star &   M2Ve  $^{    sp }$   \\
 J184500.5+005330 &    2.39 &     3.5$\pm$    0.8 &     1.4$\pm$    0.9 &    18450047+0053340  &    3.8 &    0.60 &   13.21 &      N1R4030009 &     2.9 &    0.88 &   18.17 &       0908-0374143 &     1.9 &    0.70 &   16.35 &            &         $^{       }$   \\
 J184509.3+005012 &    1.57 &     3.7$\pm$    1.0 &     3.0$\pm$    1.3 &    18450929+0050109  &    1.7 &    0.99 &    8.79 &      N1NK046109 &     1.6 &    0.99 &   16.40 &       0908-0374277 &     1.8 &    0.99 &   14.88 &            &         $^{       }$   \\
 J184510.3+004826 &    1.72 &     5.2$\pm$    1.1 &                     &    18451046+0048275  &    2.2 &    0.99 &    9.47 &      N1NK000048 &     2.2 &    1.00 &   10.97 &       0908-0374297 &     2.2 &    1.00 &   10.67 &       Star &    F7V  $^{     s }$   \\
\noalign{\smallskip}\hline
\end{tabular}
\end{center}
\end{table}
\end{landscape}
\normalsize

\begin{landscape}
\begin{table}
\begin{center}
\caption[Xray parameters for field Ridge3]{
X-ray parameters for detected sources in field Ridge~3}
\label{t:xray-param}
\fontsize{7}{7}\selectfont
\begin{tabular}{lrrrcrrrcrrrcrrrcc} 
\hline\hline 
\noalign{\smallskip}
 2XMM  & r$_{90}$ & pn\_B1$^{*}$ & pn\_B2$^{**}$ & 2MASS & d$_{x-o}$ & P$_{id}$ & kmag & GSC  & d$_{x-o}$ & P$_{id}$ & V & USNO & d$_{x-o}$ & P$_{id}$ & R & Class$^\dag$ & SpT$^\ddag$ \\
 Name   & [\arcsec] & [cts ks$^{-1}$] & [cts ks$^{-1}$] & Name & [\arcsec]  &          &      & Name & [\arcsec]   &          &   & Name & [\arcsec]   &          &   &       &       \\
\noalign{\smallskip}
\hline\noalign{\smallskip}
\noalign{\smallskip}
 J182650.4-112632 &    1.96 &     2.5$\pm$    1.2 &     3.6$\pm$    1.7 &    18265065-1126375  &    5.9 &         &   12.83 &      S9JO083709 &     3.3 &    0.08 &         &                    &         &         &         &            &         $^{       }$   \\
 J182658.4-113258 &    1.59 &     2.9$\pm$    1.1 &     2.4$\pm$    1.2 &    18265832-1132585  &    1.5 &    0.96 &    9.80 &      S9JO000207 &     1.3 &    0.99 &         &       0784-0427781 &     5.0 &    0.99 &        &       HeAe &    A0V  $^{  sp }$   \\
 J182703.7-113713 &    0.92 &    13.8$\pm$    2.4 &     8.1$\pm$    2.2 &    18270375-1137135  &    0.1 &    0.95 &   11.68 &      S9JO041110 &     1.7 &    0.90 &         &       0783-0443340 &     1.7 &    0.89 &   15.69 &       Star &     G9  $^{     s }$   \\
 J182706.3-112436 &    0.94 &     0.2$\pm$    0.5 &    22.7$\pm$    3.1 &    18270617-1124385  &    2.8 &         &   13.27 &      S9MW001589 &     2.8 &    0.20 &         &       0785-0414570 &     2.8 &    0.25 &   16.79 &            &         $^{       }$   \\
 J182711.5-113256 &    1.91 &     0.9$\pm$    0.7 &     4.2$\pm$    1.5 &    18271134-1132578  &    3.5 &    0.04 &   14.49 &      S9JO044929 &     3.6 &    0.27 &         &       0784-0428047 &     3.4 &    0.42 &   18.16 &            &         $^{       }$   \\
 J182711.9-112805 &    1.58 &     0.6$\pm$    0.6 &     2.7$\pm$    1.1 &    18271176-1128058  &    3.4 &    0.11 &   12.76 &      S9JO050244 &     2.8 &    0.51 &         &       0785-0414707 &     2.9 &    0.48 &   17.90 &            &         $^{       }$   \\
 J182714.7-111814 &    1.49 &     4.8$\pm$    1.7 &     5.6$\pm$    2.0 &                      &        &         &         &      S9MW008373 &     2.5 &    0.33 &         &       0786-0413390 &     4.7 &    0.04 &   20.04 &       Star &    M6V  $^{    sp }$   \\
 J182718.0-112823 &    1.42 &     2.6$\pm$    0.8 &     2.5$\pm$    1.0 &    18271806-1128261  &    3.0 &    0.81 &   11.60 &      S9JO049977 &     1.3 &    0.92 &         &       0785-0414827 &     1.2 &    0.92 &   15.92 &       Star &    F8V  $^{    sp }$   \\
 J182726.8-112040 &    1.23 &     8.3$\pm$    1.8 &     3.4$\pm$    1.6 &    18272674-1120398  &    1.1 &    0.92 &   11.52 &      S9MW005936 &     1.3 &    0.99 &         &       0786-0413664 &     1.1 &    0.99 &   14.39 &       Star &    G2V  $^{    sp }$   \\
 J182728.5-113741 &    1.04 &    11.3$\pm$    2.0 &     2.0$\pm$    1.1 &    18272856-1137400  &    1.6 &    0.91 &   11.02 &      S9JO000332 &     3.0 &    0.90 &         &       0783-0443878 &     2.7 &    0.86 &   13.51 &       Star &    G3V  $^{    sp }$   \\
 J182730.5-113512 &    1.92 &     4.8$\pm$    1.2 &     1.6$\pm$    1.0 &    18273027-1135116  &    4.2 &    0.79 &   10.56 &      S9JO000268 &     4.4 &    0.98 &         &       0784-0428420 &     6.2 &    0.97 &         &       Star &    A2V  $^{    sp }$   \\
 J182732.2-113357 &    0.82 &     4.6$\pm$    1.1 &    12.3$\pm$    1.9 &    18273223-1133571  &    0.5 &    0.99 &    9.88 &      S9JO076856 &     0.4 &    0.93 &         &       0784-0428444 &     0.9 &    0.93 &   17.72 &            &         $^{       }$   \\
 J182734.2-112303 &    1.39 &     5.0$\pm$    1.3 &     1.0$\pm$    1.3 &    18273432-1123045  &    1.5 &    0.57 &   13.35 &      S9MW042864 &     1.4 &    0.55 &         &       0786-0413863 &     5.2 &    0.05 &   18.86 &       Star &    M5V  $^{    sp }$   \\
 J182736.5-113751 &    1.65 &     2.8$\pm$    1.1 &     0.5$\pm$    0.9 &    18273637-1137497  &    3.0 &    0.51 &   13.34 &      S9JO040369 &     5.4 &    0.88 &         &       0783-0444064 &     5.2 &    0.89 &   17.41 &            &         $^{       }$   \\
 J182740.3-113953 &    0.73 &    36.9$\pm$    3.7 &     2.8$\pm$    1.8 &    18274040-1139532  &    0.7 &    0.95 &   10.03 &      S9JO038707 &     1.2 &    0.84 &         &                    &         &         &         &       Star &    M5V  $^{    sp }$   \\
 J182741.2-112716 &    1.13 &     6.9$\pm$    1.3 &     1.8$\pm$    0.9 &    18274138-1127132  &    3.2 &    0.95 &   11.97 &      S9JO051242 &     1.5 &    0.97 &         &       0785-0415307 &     1.4 &    0.97 &   15.24 &       Star &    K0V  $^{    sp }$   \\
 J182744.6-113957 &    1.37 &    11.9$\pm$    2.2 &     0.3$\pm$    0.7 &    18274465-1139576  &    0.3 &    1.00 &    8.78 &      S9JO000395 &     0.4 &    1.00 &         &       0783-0444254 &     0.6 &    1.00 &   11.31 &       Star &    G8III  $^{    sp }$   \\
 J182749.2-112137 &    1.64 &     6.0$\pm$    1.7 &     1.1$\pm$    1.2 &    18274911-1121402  &    3.4 &    0.77 &   11.24 &      S9MW004873 &     1.9 &    0.57 &         &       0786-0414235 &     1.8 &    0.62 &   17.71 &       Star &    M4V  $^{    sp }$   \\
 J182749.5-113725 &    1.24 &     5.0$\pm$    1.4 &     8.1$\pm$    2.3 &    18274944-1137264  &    1.6 &    0.80 &   10.97 &      S9JO040657 &     1.4 &    0.71 &         &       0783-0444363 &     1.2 &    0.77 &   18.35 &            &         $^{       }$   \\
 J182819.0-113021 &    2.03 &     5.9$\pm$    1.9 &     1.0$\pm$    1.1 &    18281911-1130202  &    2.0 &    0.76 &   10.32 &      S9JO079893 &     5.0 &    0.01 &         &                    &         &         &         &            &         $^{       }$   \\
 J182824.8-112719 &    1.98 &     2.4$\pm$    2.0 &    14.2$\pm$    4.2 &    18282509-1127191  &    3.7 &    0.20 &   11.35 &      S9JO050970 &     5.6 &    0.14 &         &       0785-0416368 &     4.3 &    0.24 &   16.93 &            &         $^{       }$   \\
\noalign{\smallskip}\hline
\end{tabular}
\end{center}
\end{table}
\end{landscape}
\normalsize

\begin{landscape}
\begin{table}
\begin{center}
\caption[Xray parameters for field Ridge4]{
X-ray parameters for detected sources in field Ridge~4}
\label{t:xray-param}
\fontsize{7}{7}\selectfont
\begin{tabular}{lrrrcrrrcrrrcrrrcc} 
\hline\hline 
\noalign{\smallskip}
 2XMM  & r$_{90}$ & pn\_B1$^{*}$ & pn\_B2$^{**}$ & 2MASS & d$_{x-o}$ & P$_{id}$ & kmag & GSC  & d$_{x-o}$ & P$_{id}$ & V & USNO & d$_{x-o}$ & P$_{id}$ & R & Class$^\dag$ & SpT$^\ddag$ \\
 Name   & [\arcsec] & [cts ks$^{-1}$] & [cts ks$^{-1}$] & Name & [\arcsec]  &          &      & Name & [\arcsec]   &          &   & Name & [\arcsec]   &          &   &       &       \\
\noalign{\smallskip}
\hline\noalign{\smallskip}
\noalign{\smallskip}
 J182811.6-110436 &    1.47 &     3.5$\pm$    1.2 &     7.7$\pm$    1.9 &                      &        &         &         &                 &         &         &         &                    &         &         &         &       Star &    M2V  $^{     s }$   \\
 J182813.6-110112 &    1.65 &     1.6$\pm$    1.1 &    11.9$\pm$    2.8 &    18281374-1101113  &    1.8 &    0.53 &   11.47 &                 &         &         &         &                    &         &         &         &            &         $^{       }$   \\
 J182816.1-111127 &    1.85 &     2.9$\pm$    1.0 &                     &    18281645-1111277  &    3.9 &    0.29 &   12.77 &      S9MW014554 &     1.5 &    0.62 &         &       0788-0410068 &     1.6 &    0.63 &   18.48 &            &         $^{       }$   \\
 J182819.8-111756 &    2.02 &     4.5$\pm$    2.0 &     7.7$\pm$    2.9 &    18281958-1117529  &    5.1 &    0.22 &   10.52 &      S9MW047577 &     3.7 &    0.04 &         &       0787-0411021 &     3.6 &    0.09 &   18.43 &            &         $^{       }$   \\
 J182822.9-105955 &    2.24 &     0.2$\pm$    0.7 &    12.0$\pm$    3.3 &    18282307-1059563  &    1.5 &    0.82 &   11.07 &      S9MW022732 &     1.6 &    0.94 &         &       0790-0407829 &     1.7 &    0.92 &   15.36 &            &         $^{       }$   \\
 J182827.5-111749 &    1.25 &     4.5$\pm$    1.9 &    49.2$\pm$    6.7 &    18282760-1117485  &    2.0 &    0.09 &   11.95 &                 &         &         &         &                    &         &         &         &            &         $^{       }$   \\
 J182832.1-112009 &    2.31 &                     &                     &    18283173-1120066  &    6.4 &    0.27 &   10.59 &      S9MW006275 &     5.6 &    0.21 &         &       0786-0415215 &     3.1 &    0.48 &   18.23 &            &         $^{       }$   \\
 J182845.5-111710 &    1.03 &   287.8$\pm$   12.9 &    25.8$\pm$    5.3 &    18284546-1117112  &    1.2 &    0.98 &    8.96 &      S9MW000461 &     0.6 &    1.00 &         &       0787-0411608 &     1.2 &    1.00 &   11.77 &       Star &    M1e  $^{     s }$   \\
\noalign{\smallskip}\hline
\end{tabular}
\end{center}
\end{table}
\end{landscape}
\normalsize

\begin{landscape}
\begin{table}
\begin{center}
\caption[Xray parameters for field Saturn]{
X-ray parameters for detected sources in field Saturn}
\label{t:xray-param}
\fontsize{7}{7}\selectfont
\begin{tabular}{lrrrcrrrcrrrcrrrcc} 
\hline\hline 
\noalign{\smallskip}
 2XMM  & r$_{90}$ & pn\_B1$^{*}$ & pn\_B2$^{**}$ & 2MASS & d$_{x-o}$ & P$_{id}$ & kmag & GSC  & d$_{x-o}$ & P$_{id}$ & V & USNO & d$_{x-o}$ & P$_{id}$ & R & Class$^\dag$ & SpT$^\ddag$ \\
 Name   & [\arcsec] & [cts ks$^{-1}$] & [cts ks$^{-1}$] & Name & [\arcsec]  &          &      & Name & [\arcsec]   &          &   & Name & [\arcsec]   &          &   &       &       \\
\noalign{\smallskip}
\hline\noalign{\smallskip}
\noalign{\smallskip}
 J055449.8+221341 &    1.56 &     2.9$\pm$    0.7 &     0.4$\pm$    0.5 &    05544979+2213421  &    1.3 &    1.00 &   11.52 &      NA68000286 &     1.4 &    1.00 &   13.42 &       1122-0115462 &     1.6 &    1.00 &   12.69 &       Star &    G0V  $^{    sp }$   \\
 J055452.2+220557 &    0.73 &     5.1$\pm$    0.9 &     9.7$\pm$    1.5 &    05545226+2205570  &    1.0 &    0.98 &   14.23 &      NA68004885 &     1.1 &    0.92 &         &       1120-0103285 &     1.3 &    0.63 &   18.90 &            &         $^{       }$   \\
 J055458.2+221758 &    1.88 &     1.9$\pm$    0.7 &     2.4$\pm$    1.0 &    05545839+2217598  &    2.0 &    0.44 &   15.44 &      NA68011496 &     1.7 &    0.89 &   18.51 &       1123-0112485 &     2.0 &    0.88 &   17.29 &            &         $^{       }$   \\
 J055504.8+220617 &    1.22 &     2.8$\pm$    0.6 &     0.1$\pm$    0.4 &    05550465+2206156  &    2.8 &    0.42 &   14.13 &      NA68005051 &     2.6 &    0.17 &         &       1121-0112930 &     2.7 &    0.25 &   18.11 &            &         $^{       }$   \\
 J055512.1+220523 &    1.82 &     1.1$\pm$    0.4 &     0.4$\pm$    0.4 &    05551224+2205240  &    1.1 &    1.00 &   11.26 &      NA68004633 &     1.2 &    0.98 &   16.04 &       1120-0103505 &     1.2 &    0.96 &   14.62 &            &         $^{       }$   \\
 J055513.6+215601 &    1.58 &     3.7$\pm$    0.9 &     0.8$\pm$    1.3 &    05551351+2156008  &    1.9 &    1.00 &    8.16 &      NA68000511 &     1.8 &    1.00 &   10.75 &       1119-0101543 &     1.8 &    1.00 &   10.00 &       Star &    K2III  $^{    sp }$   \\
 J055514.2+221849 &    1.65 &     2.3$\pm$    0.7 &     0.3$\pm$    0.6 &    05551422+2218487  &    0.5 &    0.97 &   13.90 &      NA68012008 &     0.5 &    0.92 &   18.90 &       1123-0112707 &     0.1 &    0.91 &   17.92 &            &         $^{       }$   \\
 J055518.4+215539 &    0.54 &    46.6$\pm$    3.3 &     0.5$\pm$    1.2 &    05551838+2155386  &    0.8 &    1.00 &   10.02 &      NA68032153 &     0.6 &    1.00 &   14.76 &       1119-0101592 &     1.3 &    0.75 &   14.08 &       Star &   M4Ve  $^{    sp }$   \\
 J055526.1+221248 &    1.93 &     1.5$\pm$    0.4 &                     &    05552636+2212497  &    2.9 &    0.87 &   13.47 &      NA68008500 &     3.0 &    0.89 &   16.53 &       1122-0115847 &     3.0 &    0.88 &   15.36 &            &         $^{       }$   \\
 J055530.7+221724 &    1.95 &     0.3$\pm$    0.4 &     2.6$\pm$    0.8 &    05553085+2217235  &    2.1 &    0.92 &   13.56 &      NA68011155 &     3.1 &    0.91 &   15.18 &       1122-0115910 &     2.6 &    0.59 &   14.09 &            &         $^{       }$   \\
 J055533.5+221507 &    1.40 &     2.1$\pm$    0.5 &     0.7$\pm$    0.4 &    05553345+2215064  &    2.2 &    0.86 &   13.64 &      NA68009771 &     2.0 &    0.93 &   16.93 &       1122-0115958 &     2.2 &    0.87 &   15.55 &            &         $^{       }$   \\
 J055536.2+220504 &    1.50 &     1.0$\pm$    0.3 &     0.6$\pm$    0.3 &    05553616+2205052  &    1.0 &    0.96 &   14.04 &      NA68004474 &     0.8 &    0.97 &   17.49 &       1120-0103782 &     1.3 &    0.96 &   16.28 &            &         $^{       }$   \\
 J055544.8+221028 &    1.13 &     1.5$\pm$    0.4 &     0.6$\pm$    0.3 &    05554484+2210289  &    0.7 &    0.92 &   15.60 &      NA68007224 &     0.7 &    0.85 &         &       1121-0113406 &     0.6 &    0.83 &   19.10 &            &         $^{       }$   \\
 J055551.2+220714 &    1.25 &     1.1$\pm$    0.3 &     1.7$\pm$    0.5 &    05555106+2207119  &    3.3 &    0.98 &   13.83 &      NA68000376 &     0.2 &    1.00 &   14.16 &       1121-0113474 &     0.3 &    1.00 &   13.42 &       Star &    A2V  $^{    sp }$   \\
 J055553.9+220033 &    0.86 &     8.3$\pm$    1.0 &     0.4$\pm$    0.4 &    05555404+2200330  &    0.8 &    1.00 &   11.12 &      NA68000457 &     0.8 &    1.00 &   14.17 &       1120-0103973 &     0.9 &    1.00 &   13.11 &       Star &    G9V  $^{    sp }$   \\
 J055605.7+221727 &    2.19 &     2.0$\pm$    0.7 &     0.2$\pm$    0.5 &    05560562+2217303  &    3.6 &    0.65 &   14.41 &      NA68011225 &     3.7 &    0.51 &   17.87 &       1122-0116389 &     3.9 &    0.51 &   16.88 &            &         $^{       }$   \\
 J055617.5+221352 &    2.12 &     1.3$\pm$    0.6 &                     &    05561773+2213516  &    3.2 &    0.74 &   14.47 &      NA68009082 &     3.4 &    0.70 &   17.78 &       1122-0116544 &     3.0 &    0.75 &   16.82 &            &         $^{       }$   \\
 J055619.7+221115 &    0.70 &    15.4$\pm$    1.5 &     0.3$\pm$    0.5 &    05561977+2211157  &    0.4 &    1.00 &   10.79 &      NA68007675 &     1.0 &    1.00 &   15.70 &       1121-0113833 &     0.7 &    0.98 &   14.23 &       Star &   M5Ve  $^{    sp }$   \\
\noalign{\smallskip}\hline
\end{tabular}
\end{center}
\end{table}
\end{landscape}
\normalsize

\begin{landscape}
\begin{table}
\begin{center}
\caption[Xray parameters for field WR110]{
X-ray parameters for detected sources in field WR110}
\label{t:xray-param}
\fontsize{7}{7}\selectfont
\begin{tabular}{lrrrcrrrcrrrcrrrcc} 
\hline\hline 
\noalign{\smallskip}
 2XMM  & r$_{90}$ & pn\_B1$^{*}$ & pn\_B2$^{**}$ & 2MASS & d$_{x-o}$ & P$_{id}$ & kmag & GSC  & d$_{x-o}$ & P$_{id}$ & V & USNO & d$_{x-o}$ & P$_{id}$ & R & Class$^\dag$ & SpT$^\ddag$ \\
 Name   & [\arcsec] & [cts ks$^{-1}$] & [cts ks$^{-1}$] & Name & [\arcsec]  &          &      & Name & [\arcsec]   &          &   & Name & [\arcsec]   &          &   &       &       \\
\noalign{\smallskip}
\hline\noalign{\smallskip}
\noalign{\smallskip}
 J180656.4-192251 &    2.38 &     8.9$\pm$    1.4 &                     &    18065671-1922528  &    4.5 &    0.97 &    8.71 &      S9JJ000159 &     1.0 &    1.00 &    9.51 &       0706-0533589 &     1.0 &    1.00 &    9.12 &       Star &    F6V  $^{     s }$   \\
 J180656.5-192711 &    2.76 &     4.7$\pm$    1.2 &                     &    18065650-1927118  &    0.5 &    0.96 &   10.11 &      S9JJ000169 &     0.7 &    1.00 &   12.49 &       0705-0521645 &     0.7 &    0.99 &   12.29 &            &         $^{       }$   \\
 J180658.2-192724 &    2.23 &     3.5$\pm$    1.1 &     4.7$\pm$    1.8 &    18065837-1927222  &    3.2 &    0.14 &   13.01 &      S9JJ001828 &     4.1 &    0.27 &   17.49 &       0705-0521680 &     3.7 &    0.42 &   16.92 &            &         $^{       }$   \\
 J180701.9-192042 &    1.52 &     5.2$\pm$    1.1 &     2.6$\pm$    1.5 &    18070222-1920392  &    4.5 &    0.70 &   11.85 &      S9JJ003336 &     4.3 &    0.68 &   13.93 &       0706-0533710 &     3.4 &    0.88 &   13.21 &       Star &    K0V  $^{    sp }$   \\
 J180706.8-192559 &    0.59 &    32.3$\pm$    2.2 &     0.6$\pm$    0.6 &    18070680-1926006  &    1.2 &    0.97 &   10.20 &      S9JJ002115 &     1.3 &    0.99 &   13.98 &       0705-0521914 &     1.9 &    0.71 &   13.20 &       Star &    M1V  $^{    sp }$   \\
 J180712.5-191337 &    0.96 &    35.1$\pm$    3.9 &     0.5$\pm$    1.6 &    18071239-1913385  &    1.8 &    0.98 &    9.24 &      S9JJ000130 &     2.0 &    1.00 &   11.46 &       0707-0527031 &     2.1 &    1.00 &   10.80 &       Star &    G6V  $^{    sp }$   \\
 J180713.4-191729 &    1.16 &    10.1$\pm$    1.4 &                     &    18071344-1917271  &    2.1 &    1.00 &    7.81 &      S9JJ000139 &     2.1 &    1.00 &    9.81 &       0707-0527053 &     2.1 &    1.00 &    9.37 &       Star &    F5III  $^{     s }$   \\
 J180718.4-192454 &    1.03 &     6.4$\pm$    1.3 &     4.6$\pm$    1.5 &    18071844-1924550  &    1.1 &    0.93 &   11.47 &      S9JJ002310 &     1.2 &    0.96 &   16.73 &       0705-0522249 &     1.6 &    0.89 &   15.23 &       Star &    G8III  $^{    sp }$   \\
 J180718.8-191428 &    1.72 &     2.7$\pm$    0.9 &     0.8$\pm$    0.8 &    18071921-1914285  &    5.8 &    0.77 &   10.65 &      S9JJ004927 &     0.9 &    0.84 &         &       0707-0527165 &     0.6 &    0.91 &   15.52 &            &         $^{       }$   \\
 J180721.0-191855 &    1.60 &     3.6$\pm$    0.8 &     0.4$\pm$    0.6 &    18072079-1918575  &    3.5 &    0.91 &   11.57 &      S9JJ003814 &     0.6 &    1.00 &   13.29 &       0706-0534184 &     0.5 &    0.99 &   12.67 &       Star &    K0V  $^{    sp }$   \\
 J180722.1-191446 &    1.50 &     4.0$\pm$    1.0 &     0.4$\pm$    0.8 &    18072226-1914436  &    3.1 &    0.40 &   12.50 &      S9JJ004839 &     3.2 &    0.73 &   15.77 &       0707-0527234 &     3.0 &    0.77 &   14.96 &       Star &    K2V  $^{    sp }$   \\
 J180722.6-192038 &    0.93 &     8.4$\pm$    1.0 &     0.7$\pm$    0.5 &    18072264-1920395  &    1.2 &    0.82 &   12.82 &      S9JJ003407 &     1.1 &    0.96 &   15.45 &       0706-0534225 &     1.2 &    0.95 &   14.86 &       Star &    G9V  $^{    sp }$   \\
 J180724.7-191921 &    1.86 &     2.2$\pm$    0.6 &                     &    18072451-1919267  &    5.5 &         &   12.91 &      S9JJ003663 &     4.2 &    0.02 &         &       0706-0534286 &     4.3 &    0.15 &   18.55 &            &         $^{       }$   \\
 J180727.9-193226 &    1.22 &     6.6$\pm$    1.3 &     2.4$\pm$    1.3 &    18072780-1932283  &    1.9 &    0.74 &   11.82 &      S9JI012181 &     1.9 &    0.95 &   14.97 &       0704-0545117 &     2.2 &    0.94 &   13.76 &            &         $^{       }$   \\
 J180732.4-191950 &    1.63 &     3.0$\pm$    0.7 &     1.5$\pm$    0.7 &    18073237-1919485  &    2.2 &    0.40 &   12.28 &      S9JJ003572 &     2.9 &    0.41 &   17.79 &       0706-0534471 &     2.6 &    0.52 &   16.96 &       Star &    K0V  $^{    sp }$   \\
 J180733.3-193045 &    0.99 &     0.5$\pm$    0.4 &     7.5$\pm$    1.3 &    18073311-1930446  &    3.4 &         &   14.27 &                 &         &         &         &       0704-0545257 &     3.3 &    0.05 &   19.21 &            &         $^{       }$   \\
 J180733.4-192645 &    2.10 &     2.3$\pm$    0.6 &     0.5$\pm$    0.5 &    18073337-1926432  &    2.2 &    0.29 &   13.52 &      S9JI016736 &     2.4 &    0.70 &   17.18 &       0705-0522635 &     1.7 &    0.77 &   16.48 &            &         $^{       }$   \\
 J180736.4-192658 &    0.49 &    12.2$\pm$    1.1 &    13.4$\pm$    1.3 &    18073644-1926581  &    0.3 &    1.00 &   10.07 &      S9JI016580 &     0.1 &    1.00 &   16.12 &       0705-0522725 &     0.4 &    0.99 &   15.05 &       Star &    G8V  $^{    sp }$   \\
 J180736.6-192815 &    2.16 &     1.4$\pm$    0.5 &     0.6$\pm$    0.5 &    18073674-1928178  &    2.5 &    0.20 &   12.87 &      S9JI015476 &     2.1 &    0.66 &   17.84 &       0705-0522722 &     6.4 &    0.64 &   18.65 &            &         $^{       }$   \\
 J180736.7-192943 &    1.01 &     4.8$\pm$    0.8 &     0.8$\pm$    0.5 &    18073664-1929442  &    1.5 &    0.92 &   11.28 &      S9JI014351 &     1.5 &    0.93 &   16.13 &       0705-0522730 &     1.2 &    0.96 &   14.98 &       Star &    G6V  $^{    sp }$   \\
 J180739.2-192347 &    1.27 &     2.4$\pm$    0.5 &     0.2$\pm$    0.2 &    18073928-1923474  &    0.1 &    0.49 &   12.60 &      S9JJ002526 &     3.9 &     &         &       0706-0534633 &     3.4 &    0.18 &     &       Star &    M5V  $^{    sp }$   \\
 J180739.6-192040 &    1.27 &     2.8$\pm$    0.6 &     0.4$\pm$    0.4 &    18073959-1920418  &    1.7 &    0.97 &    8.35 &      S9JJ003390 &     0.7 &    0.96 &   16.06 &       0706-0534647 &     0.2 &    0.94 &   14.95 &       Star &    K1III  $^{    sp }$   \\
 J180742.6-193716 &    2.99 &                     &     7.6$\pm$    2.1 &    18074287-1937199  &    4.4 &    0.02 &   12.97 &      S9JI008118 &     5.5 &         &   17.51 &       0703-0530878 &     5.0 &    0.14 &   16.10 &            &         $^{       }$   \\
 J180744.5-193305 &    2.00 &     1.6$\pm$    0.7 &                     &    18074469-1933061  &    2.5 &    0.88 &   11.63 &      S9JI011709 &     2.3 &    0.97 &   13.77 &       0704-0545527 &     2.6 &    0.97 &   13.46 &            &         $^{       }$   \\
 J180748.6-191834 &    0.93 &     1.1$\pm$    0.5 &     6.6$\pm$    1.0 &                      &        &         &         &      S9JJ003922 &     1.8 &    0.56 &         &       0706-0534870 &     1.9 &    0.56 &   17.91 &            &         $^{       }$   \\
 J180748.8-192036 &    0.97 &     4.1$\pm$    1.1 &     2.6$\pm$    1.0 &    18074865-1920355  &    2.7 &    0.11 &   12.00 &      S9JJ003372 &     2.9 &    0.10 &   17.06 &       0706-0534868 &     2.9 &    0.21 &   16.36 &            &         $^{       }$   \\
 J180749.4-191724 &    1.80 &     1.4$\pm$    1.3 &     3.6$\pm$    1.9 &    18074949-1917240  &    0.7 &    0.72 &   12.10 &      S9JJ004215 &     0.7 &    0.84 &   17.76 &       0707-0527721 &     0.3 &    0.83 &   17.05 &            &         $^{       }$   \\
 J180749.5-192606 &    0.68 &    10.5$\pm$    1.3 &     1.0$\pm$    0.5 &    18074953-1926066  &    0.4 &    0.99 &   11.66 &      S9JI017225 &     0.4 &    1.00 &   14.18 &       0705-0522959 &     0.3 &    1.00 &   13.94 &       Star &    G8V  $^{    sp }$   \\
 J180750.1-191749 &    1.38 &     0.8$\pm$    0.7 &     1.6$\pm$    1.3 &    18075014-1917517  &    2.3 &    0.50 &   12.38 &      S9JJ004086 &     2.3 &    0.75 &   17.20 &       0707-0527728 &     2.4 &    0.66 &   16.29 &            &         $^{       }$   \\
 J180752.8-192624 &    1.06 &     2.6$\pm$    0.5 &     0.3$\pm$    0.3 &    18075285-1926241  &    0.5 &    0.98 &   11.73 &      S9JI000004 &     1.0 &    1.00 &   13.03 &       0705-0523029 &     1.3 &    1.00 &   12.76 &       Star &    A2V  $^{    sp }$   \\
 J180754.1-192035 &    0.97 &     1.3$\pm$    0.4 &     3.0$\pm$    0.6 &    18075414-1920345  &    0.6 &    0.91 &   10.82 &      S9JJ003380 &     3.1 &    0.01 &         &       0706-0535004 &     3.2 &    0.12 &   17.21 &            &         $^{       }$   \\
 J180758.4-192523 &    1.53 &     0.3$\pm$    0.3 &     1.3$\pm$    0.5 &    18075855-1925283  &    5.0 &    0.02 &   13.33 &      S9JI017708 &     4.7 &    0.02 &   17.66 &       0705-0523129 &     4.9 &    0.11 &   17.04 &            &         $^{       }$   \\
 J180758.7-193650 &    2.51 &     5.5$\pm$    1.1 &     1.0$\pm$    1.0 &    18075858-1936518  &    2.8 &    0.99 &    7.90 &      S9JI000019 &     2.4 &    1.00 &   11.47 &       0703-0531159 &     2.4 &    1.00 &   10.73 &       Star &    K0III  $^{    sp }$   \\
 J180802.0-191505 &    0.53 &    22.6$\pm$    1.5 &     4.8$\pm$    0.9 &    18080208-1915048  &    0.6 &    1.00 &    9.78 &      S9JJ004782 &     0.6 &    1.00 &   13.62 &       0707-0527902 &     0.3 &    1.00 &   13.20 &       TTS &    K1V  $^{    sp }$   \\
 J180803.1-192526 &    1.78 &     1.5$\pm$    0.4 &     0.1$\pm$    0.2 &    18080321-1925269  &    1.5 &    0.63 &   13.31 &      S9JI017776 &     5.2 &    0.66 &         &       0705-0523206 &     1.6 &    0.62 &   18.24 &            &         $^{       }$   \\
 J180804.3-191704 &    0.75 &     8.1$\pm$    0.9 &     0.3$\pm$    0.3 &    18080434-1917043  &    0.6 &    0.99 &   11.04 &      S9JJ004316 &     0.5 &    1.00 &   13.21 &       0707-0527931 &     0.7 &    1.00 &   12.67 &       Star &    F3V  $^{    sp }$   \\
 J180804.4-192453 &    0.61 &     8.4$\pm$    0.9 &     6.3$\pm$    0.8 &    18080444-1924533  &    0.4 &    0.89 &   13.42 &                 &         &         &         &       0705-0523232 &     1.5 &    0.85 &        &       Star &  K3III  $^{    sp }$   \\
 J180805.5-192305 &    1.04 &     3.1$\pm$    0.5 &     1.2$\pm$    0.4 &    18080562-1923050  &    1.6 &    0.88 &   11.87 &      S9JI019784 &     1.3 &    0.96 &   15.60 &       0706-0535244 &     1.1 &    0.97 &   14.56 &       Star &    A5V  $^{    sp }$   \\
 J180808.9-193553 &    1.89 &     3.6$\pm$    0.9 &     0.4$\pm$    0.9 &    18080902-1935573  &    4.3 &    0.69 &   13.13 &      S9JI009470 &     2.2 &    0.92 &   15.63 &       0704-0546102 &     2.2 &    0.91 &   14.66 &            &         $^{       }$   \\
 J180809.4-191848 &    1.77 &     0.5$\pm$    0.5 &     1.3$\pm$    0.7 &    18080923-1918443  &    5.2 &    0.27 &   12.85 &                 &         &         &         &                    &         &         &         &            &         $^{       }$   \\
 J180814.4-192659 &    0.92 &     4.4$\pm$    0.6 &     0.4$\pm$    0.4 &    18081452-1926589  &    0.5 &    0.93 &   13.10 &      S9JI016605 &     0.7 &    0.91 &         &       0705-0523450 &     0.4 &    0.92 &   17.22 &       Star &    M3V  $^{    sp }$   \\
 J180816.6-191939 &    1.12 &     1.3$\pm$    0.4 &     2.2$\pm$    0.6 &    18081689-1919395  &    2.8 &    0.08 &   13.15 &                 &         &         &         &                    &         &         &         &       HMXB  &   Be/X  $^{    sp }$   \\
 J180819.4-192252 &    1.32 &     0.5$\pm$    0.3 &     1.5$\pm$    0.5 &    18081960-1922531  &    1.7 &    0.80 &   10.84 &                 &         &         &         &       0706-0535464 &     4.5 &    0.18 &   16.85 &            &         $^{       }$   \\
 J180819.8-191407 &    2.18 &     2.7$\pm$    0.7 &     1.2$\pm$    0.8 &    18081956-1914065  &    3.5 &    0.78 &   10.83 &      S9JJ005058 &     3.8 &    0.76 &   15.81 &       0707-0528144 &     8.1 &    0.76 &   18.52 &       Star &    K0V  $^{    sp }$   \\
 J180822.4-191813 &    0.63 &    12.5$\pm$    1.1 &     2.6$\pm$    0.7 &    18082246-1918128  &    0.4 &    0.99 &   11.24 &      S9JI023953 &     0.3 &    1.00 &   14.90 &       0706-0535521 &     0.2 &    1.00 &   14.20 &       Star &    K2V  $^{    sp }$   \\
 J180822.5-193501 &    3.14 &     0.6$\pm$    0.5 &     2.6$\pm$    1.3 &    18082265-1934551  &    6.3 &    0.45 &   12.20 &      S9JI010280 &     6.4 &    0.62 &   14.86 &       0704-0546399 &     9.3 &    0.64 &   17.47 &            &         $^{       }$   \\
 J180822.8-193121 &    1.87 &     1.3$\pm$    0.5 &                     &    18082305-1931186  &    3.7 &    0.51 &   13.35 &      S9JI013195 &     0.8 &    0.88 &   16.46 &       0704-0546421 &     1.1 &    0.88 &   15.52 &            &         $^{       }$   \\
 J180825.7-192026 &    1.76 &     0.2$\pm$    0.3 &     2.4$\pm$    0.7 &    18082554-1920235  &    4.0 &    0.11 &   11.88 &                 &         &         &         &                    &         &         &         &            &         $^{       }$   \\
 J180827.2-192406 &    2.14 &     1.2$\pm$    1.9 &     0.1$\pm$    0.8 &    18082739-1923593  &    7.4 &         &   12.88 &                 &         &         &         &                    &         &         &         &            &         $^{       }$   \\
 J180828.1-193140 &    1.78 &     0.4$\pm$    0.4 &     3.5$\pm$    1.1 &    18082799-1931443  &    4.4 &    0.61 &   13.58 &                 &         &         &         &                    &         &         &         &            &         $^{       }$   \\
 J180829.9-192716 &    1.83 &     0.8$\pm$    0.5 &     1.6$\pm$    0.7 &    18083028-1927143  &    5.5 &    0.69 &   13.30 &      S9JI016420 &     3.6 &    0.18 &         &       0705-0523807 &     3.2 &    0.41 &   18.47 &            &         $^{       }$   \\
 J180833.6-192318 &    1.11 &     4.3$\pm$    0.7 &     0.3$\pm$    0.4 &    18083376-1923169  &    1.8 &    0.99 &    8.35 &      S9JI000446 &     2.0 &    1.00 &   11.49 &       0706-0535794 &     2.0 &    1.00 &   10.46 &       Star &    K0III  $^{    sp }$   \\
 J180834.0-193217 &    2.10 &                     &     2.2$\pm$    0.8 &    18083388-1932206  &    4.2 &    0.13 &   12.77 &                 &         &         &         &                    &         &         &         &            &         $^{       }$   \\
\noalign{\smallskip}\hline
\end{tabular}
\end{center}
\end{table}
\end{landscape}
\normalsize

\begin{landscape}
\begin{table}
\begin{center}
\caption[Xray parameters for field ZAnd]{
X-ray parameters for detected sources in field Z~And}
\label{t:xray-param}
\fontsize{7}{7}\selectfont
\begin{tabular}{lrrrcrrrcrrrcrrrcc} 
\hline\hline 
\noalign{\smallskip}
 2XMM  & r$_{90}$ & pn\_B1$^{*}$ & pn\_B2$^{**}$ & 2MASS & d$_{x-o}$ & P$_{id}$ & kmag & GSC  & d$_{x-o}$ & P$_{id}$ & V & USNO & d$_{x-o}$ & P$_{id}$ & R & Class$^\dag$ & SpT$^\ddag$ \\
 Name   & [\arcsec] & [cts ks$^{-1}$] & [cts ks$^{-1}$] & Name & [\arcsec]  &          &      & Name & [\arcsec]   &          &   & Name & [\arcsec]   &          &   &       &       \\
\noalign{\smallskip}
\hline\noalign{\smallskip}
\noalign{\smallskip}
 J233223.2+485251 &    2.75 &     3.4$\pm$    1.1 &     0.1$\pm$    1.3 &                      &        &         &         &      N16Z023098 &     4.1 &    0.28 &         &       1388-0496401 &     3.9 &    0.47 &   19.15 &            &         $^{       }$   \\
 J233229.3+484757 &    1.95 &     3.3$\pm$    0.9 &     1.9$\pm$    0.9 &                      &        &         &         &      N16Z020914 &     4.1 &    0.28 &         &       1387-0498129 &     3.8 &    0.52 &   18.02 &            &         $^{       }$   \\
 J233245.9+484354 &    1.06 &     7.0$\pm$    1.2 &                     &    23324587+4843533  &    1.3 &    1.00 &   10.50 &      N16Z000014 &     1.3 &    1.00 &   13.32 &       1387-0498227 &     1.3 &    1.00 &   12.37 &       Star &    K0V  $^{    sp }$   \\
 J233247.1+484326 &    1.83 &     2.7$\pm$    0.8 &     0.5$\pm$    0.7 &    23324713+4843247  &    1.8 &    1.00 &   10.63 &      N16Z000019 &     1.7 &    1.00 &   12.16 &       1387-0498236 &     1.3 &    1.00 &   11.44 &            &         $^{       }$   \\
 J233254.5+484831 &    0.61 &    24.9$\pm$    1.7 &    10.6$\pm$    1.4 &    23325455+4848309  &    0.3 &    0.99 &   14.70 &      N16Z021191 &     0.5 &    0.98 &   18.64 &       1388-0496609 &     0.3 &    0.99 &   17.35 &       EG   &     EG  $^{    sp }$   \\
 J233309.4+484933 &    1.82 &     0.7$\pm$    0.4 &     1.5$\pm$    0.6 &                      &        &         &         &      N16Z021729 &     6.2 &    0.01 &   18.90 &                    &         &         &         &            &         $^{       }$   \\
 J233315.1+485635 &    1.45 &     2.0$\pm$    0.7 &     0.2$\pm$    0.6 &    23331494+4856350  &    1.8 &    1.00 &    7.98 &      N16Z000541 &     2.0 &    0.60 &    9.13 &       1389-0500243 &     2.0 &    1.00 &    8.83 &       Star &    F8III  $^{     s }$   \\
 J233315.9+484651 &    0.81 &     5.2$\pm$    0.8 &     0.1$\pm$    0.3 &    23331597+4846505  &    0.8 &    1.00 &   12.62 &      N16Z020379 &     0.8 &    0.98 &   17.78 &       1387-0498432 &     0.6 &    0.98 &   16.96 &       Star &    M4V  $^{    sp }$   \\
 J233317.0+484138 &    1.78 &     1.9$\pm$    0.6 &     1.3$\pm$    0.6 &                      &        &         &         &      N16Z017706 &     1.2 &    0.80 &         &       1386-0498636 &     1.5 &    0.79 &   19.06 &            &         $^{       }$   \\
 J233324.8+485804 &    1.91 &     2.9$\pm$    1.0 &     0.2$\pm$    0.7 &    23332495+4858048  &    1.1 &    0.92 &   13.47 &      N16Z025328 &     1.1 &    0.89 &   17.88 &       1389-0500345 &     1.3 &    0.85 &   16.84 &       Star &    M2V  $^{    sp }$   \\
 J233329.1+485728 &    1.79 &     2.2$\pm$    0.8 &     0.8$\pm$    1.0 &    23332918+4857261  &    2.6 &    0.65 &   15.01 &      N16Z025047 &     2.5 &    0.60 &         &       1389-0500379 &     2.3 &    0.56 &   16.72 &            &         $^{       }$   \\
 J233335.1+485424 &    0.85 &     6.3$\pm$    1.0 &     2.0$\pm$    0.8 &                      &        &         &         &      N16Z023816 &     1.6 &    0.67 &         &       1389-0500419 &     1.0 &    0.87 &   19.35 &            &         $^{       }$   \\
 J233340.9+485715 &    0.78 &    18.3$\pm$    1.9 &     2.0$\pm$    1.1 &    23334111+4857164  &    1.5 &    0.99 &   11.51 &      N16Z025017 &     1.6 &    0.98 &   14.67 &       1389-0500476 &     1.6 &    0.97 &   14.03 &       Star &    K0V  $^{    sp }$   \\
 J233349.7+484142 &    2.25 &     1.2$\pm$    0.5 &     0.5$\pm$    0.5 &                      &        &         &         &                 &         &         &         &       1386-0498858 &     7.0 &    0.23 &   18.21 &            &         $^{       }$   \\
 J233349.9+483651 &    1.55 &     4.1$\pm$    1.0 &     1.9$\pm$    1.4 &    23334989+4836502  &    1.4 &    1.00 &    9.38 &      N16Z000079 &     1.7 &    1.00 &   11.77 &       1386-0498849 &     1.7 &    1.00 &   11.19 &       Star &    K3V  $^{    sp }$   \\
 J233359.8+485253 &    1.12 &     2.0$\pm$    0.7 &     2.2$\pm$    0.8 &                      &        &         &         &      N16Z023126 &     1.5 &    0.64 &         &       1388-0497068 &     1.6 &    0.73 &   19.36 &            &         $^{       }$   \\
 J233401.3+484811 &    0.84 &     7.5$\pm$    1.0 &     0.8$\pm$    0.6 &    23340142+4848111  &    1.2 &    1.00 &   11.99 &      N16Z021069 &     1.2 &    0.99 &   14.87 &       1388-0497082 &     1.0 &    0.99 &   14.18 &       Star &    G5V  $^{    sp }$   \\
 J233402.8+485110 &    0.67 &   326.4$\pm$    7.4 &   153.9$\pm$    6.0 &                      &        &         &         &                 &         &         &         &                    &         &         &         &       EG   &  ClGal  $^{     s }$   \\
 J233404.9+485720 &    1.09 &     8.2$\pm$    1.4 &     1.5$\pm$    1.0 &    23340500+4857208  &    0.3 &    1.00 &   12.11 &      N16Z025026 &     0.4 &    1.00 &   14.59 &       1389-0500672 &     0.4 &    0.99 &   14.25 &       Star &    K0V  $^{    sp }$   \\
 J233412.0+484338 &    1.98 &     1.6$\pm$    0.6 &     0.8$\pm$    0.7 &                      &        &         &         &      N16Z018779 &     6.7 &    0.01 &   19.27 &       1387-0498827 &     7.3 &    0.10 &   19.28 &            &         $^{       }$   \\
 J233412.4+483833 &    0.91 &    11.2$\pm$    1.7 &     0.1$\pm$    0.6 &    23341251+4838333  &    1.0 &    0.99 &   12.60 &      N16Z015905 &     1.0 &    0.99 &   14.55 &       1386-0499016 &     1.0 &    0.99 &   14.19 &       Star &    G0V  $^{    sp }$   \\
 J233419.3+485114 &    1.09 &     5.7$\pm$    0.9 &     0.5$\pm$    0.7 &    23341929+4851141  &    0.6 &    0.99 &   13.27 &      N16Z022458 &     0.4 &    0.99 &   15.70 &       1388-0497231 &     0.5 &    0.99 &   15.30 &       Star &    K0V  $^{    sp }$   \\
 J233420.5+484418 &    1.28 &     2.3$\pm$    0.7 &     1.8$\pm$    0.9 &                      &        &         &         &      N16Z019141 &     2.0 &    0.63 &         &       1387-0498896 &     2.5 &    0.55 &   19.18 &            &         $^{       }$   \\
 J233422.4+484520 &    1.94 &     5.3$\pm$    2.2 &     0.3$\pm$    1.4 &    23342256+4845208  &    1.1 &    1.00 &   10.54 &      N16Z000642 &     1.1 &    1.00 &   11.87 &       1387-0498910 &     1.1 &    1.00 &   11.32 &       Star &    F3V  $^{    sp }$   \\
 J233428.9+483905 &    1.37 &                     &                     &    23342881+4839051  &    1.6 &    0.91 &   14.47 &      N16Z016219 &     1.6 &    0.95 &   16.37 &       1386-0499132 &     1.6 &    0.96 &   15.79 &       Star &    F6V  $^{    sp }$   \\
 J233431.4+485519 &    0.74 &     4.8$\pm$    1.1 &    10.3$\pm$    2.1 &                      &        &         &         &                 &         &         &         &       1389-0500877 &     1.7 &    0.90 &   18.93 &            &         $^{       }$   \\
 J233443.8+484109 &    1.62 &                     &                     &    23344381+4841070  &    2.2 &    0.95 &   12.10 &      N16Z017450 &     2.3 &    0.79 &   18.21 &       1386-0499238 &     2.1 &    0.85 &   16.84 &            &         $^{       }$   \\
 J233444.2+485625 &    1.05 &     9.4$\pm$    1.6 &     3.7$\pm$    1.8 &                      &        &         &         &      N16Z024658 &     2.5 &    0.37 &         &       1389-0500968 &     1.3 &    0.74 &   19.34 &            &         $^{       }$   \\
 J233447.0+485434 &    1.41 &     5.6$\pm$    1.2 &     2.6$\pm$    1.7 &    23344688+4854368  &    2.8 &    0.96 &   11.51 &      N16Z000567 &     2.8 &    0.98 &   13.41 &       1389-0500983 &     3.1 &    0.95 &   13.02 &       Star &    G1V  $^{    sp }$   \\
 J233500.4+484601 &    1.11 &    13.8$\pm$    2.0 &     0.8$\pm$    1.0 &    23350049+4846006  &    1.1 &    1.00 &   11.22 &      N16Z020023 &     1.1 &    0.98 &   15.95 &       1387-0499194 &     0.9 &    0.99 &   15.09 &       Star &    M4V  $^{    sp }$   \\
 J233509.9+485114 &    2.24 &     3.7$\pm$    1.1 &     3.5$\pm$    2.3 &    23350993+4851114  &    3.2 &    1.00 &    8.36 &      N16Z022625 &     4.2 &    0.70 &         &       1388-0497584 &     3.1 &    1.00 &   10.69 &       Star &    K2III  $^{    sp }$   \\
\noalign{\smallskip}\hline
\end{tabular}
\end{center}
\end{table}
\end{landscape}
\normalsize

\begin{landscape}
\begin{table}
\begin{center}
\caption[Xray parameters for field grb010220]{
X-ray parameters for detected sources in field GRB010220}
\label{t:xray-param}
\fontsize{7}{7}\selectfont
\begin{tabular}{lrrrcrrrcrrrcrrrcc} 
\hline\hline 
\noalign{\smallskip}
 2XMM  & r$_{90}$ & pn\_B1$^{*}$ & pn\_B2$^{**}$ & 2MASS & d$_{x-o}$ & P$_{id}$ & kmag & GSC  & d$_{x-o}$ & P$_{id}$ & V & USNO & d$_{x-o}$ & P$_{id}$ & R & Class$^\dag$ & SpT$^\ddag$ \\
 Name   & [\arcsec] & [cts ks$^{-1}$] & [cts ks$^{-1}$] & Name & [\arcsec]  &          &      & Name & [\arcsec]   &          &   & Name & [\arcsec]   &          &   &       &       \\
\noalign{\smallskip}
\hline\noalign{\smallskip}
\noalign{\smallskip}
 J023523.0+614950 &    1.73 &    14.6$\pm$    2.4 &                     &    02352318+6149508  &    1.3 &    0.96 &   11.63 &      NAWC018003 &     1.5 &    0.96 &   15.49 &       1518-0076033 &     1.5 &    0.95 &   14.72 &       Star &    M0V  $^{    sp }$   \\
 J023538.4+614542 &    2.03 &    12.0$\pm$    1.9 &     0.1$\pm$    0.5 &    02353855+6145438  &    1.5 &    0.97 &   10.84 &      NAWC000260 &     1.4 &    1.00 &   11.91 &       1517-0076673 &     1.4 &    1.00 &   11.22 &       Star &    F5V  $^{    sp }$   \\
 J023542.2+615241 &    1.88 &     7.2$\pm$    1.7 &     1.0$\pm$    1.7 &    02354293+6152437  &    5.4 &    0.38 &   11.59 &      NAWC020661 &     5.5 &    0.43 &   14.59 &       1518-0076103 &     5.5 &    0.37 &   14.38 &       Star &    K0V  $^{    sp }$   \\
 J023550.9+614424 &    1.51 &    14.9$\pm$    1.9 &     1.7$\pm$    1.1 &    02355110+6144270  &    2.9 &    0.86 &   11.86 &      NAWC012302 &     3.1 &    0.94 &   14.82 &       1517-0076766 &     3.4 &    0.81 &   14.38 &       Star &    K1V  $^{    sp }$   \\
 J023608.0+613451 &    1.53 &    20.9$\pm$    2.9 &                     &    02360804+6134515  &    0.4 &    0.99 &   12.07 &      NAWC002906 &     0.4 &    1.00 &   14.50 &       1515-0079451 &     0.7 &    0.99 &   14.41 &       Star &    K0V  $^{    sp }$   \\
 J023611.4+614047 &    2.46 &     3.3$\pm$    1.1 &     1.1$\pm$    1.1 &    02361162+6140496  &    2.6 &    0.68 &   13.52 &      NAWC008548 &     2.6 &    0.70 &   17.29 &       1516-0078393 &     3.0 &    0.59 &   16.31 &       Star &    M1V  $^{    sp }$   \\
 J023639.4+613945 &    2.23 &     3.3$\pm$    1.0 &     0.1$\pm$    0.5 &    02363879+6139491  &    5.8 &    0.16 &   12.78 &      NAWC007370 &     5.7 &    0.09 &   17.13 &       1516-0078612 &     5.9 &    0.07 &   15.80 &       Star &    M0V  $^{    sp }$   \\
 J023645.0+613638 &    1.50 &     0.9$\pm$    0.7 &    33.1$\pm$    3.9 &    02364508+6136376  &    0.8 &    0.93 &   14.47 &      NAWC004176 &     1.0 &    0.87 &         &       1516-0078665 &     1.2 &    0.80 &   17.74 &            &         $^{       }$   \\
 J023647.0+613922 &    1.60 &     9.8$\pm$    1.6 &                     &    02364707+6139227  &    0.3 &    1.00 &   10.57 &      NAWC000369 &     0.5 &    1.00 &   12.84 &       1516-0078681 &     0.5 &    1.00 &   12.74 &       Star &    F9V  $^{    sp }$   \\
 J023708.5+615709 &    2.50 &     7.9$\pm$    2.1 &     0.6$\pm$    1.6 &    02370828+6157038  &    6.2 &    0.55 &   15.52 &      NAWC000927 &     6.6 &    0.93 &   12.90 &       1519-0073694 &     6.3 &    0.66 &   13.28 &       Star &    F9V  $^{    sp }$   \\
 J023714.7+614540 &    2.15 &     3.7$\pm$    0.9 &     0.4$\pm$    0.5 &    02371514+6145368  &    4.9 &    0.99 &    7.73 &      NAWC000270 &     4.7 &    0.99 &   10.48 &       1517-0077442 &     4.7 &    0.99 &    9.71 &       Star &    K2III  $^{    sp }$   \\
 J023719.8+614100 &    2.45 &     2.6$\pm$    0.9 &     0.8$\pm$    0.8 &    02372020+6141040  &    4.1 &    0.33 &   13.79 &      NAWC008819 &     4.2 &    0.10 &   18.35 &       1516-0078995 &     4.2 &    0.06 &   17.24 &       Star &    M2V  $^{    sp }$   \\
 J023726.4+614538 &    2.37 &     0.0$\pm$    0.2 &     6.2$\pm$    1.4 &    02372712+6145332  &    7.1 &    0.06 &   14.27 &      NAWC013396 &     7.3 &    0.02 &   17.36 &       1517-0077545 &     6.9 &    0.08 &   16.33 &            &         $^{       }$   \\
 J023748.6+613715 &    2.23 &     4.8$\pm$    1.4 &     1.7$\pm$    1.3 &    02374860+6137161  &    0.6 &    0.99 &   11.96 &      NAWC004740 &     0.6 &    1.00 &   14.71 &       1516-0079220 &     1.1 &    0.99 &   14.48 &       Star &    G0V  $^{    sp }$   \\
 J023756.8+615857 &    3.43 &     9.0$\pm$    2.9 &     7.0$\pm$    4.7 &    02375691+6159005  &    2.9 &    0.72 &   12.64 &      NAWC025258 &     8.6 &    0.76 &   19.00 &       1519-0074134 &     8.5 &    0.70 &   18.29 &            &         $^{       }$   \\
 J023757.1+614907 &    1.81 &     5.9$\pm$    1.4 &     4.7$\pm$    1.7 &                      &        &         &         &                 &         &         &         &       1518-0077146 &     3.9 &    0.22 &   19.61 &            &         $^{       }$   \\
\noalign{\smallskip}\hline
\end{tabular}
\end{center}
\end{table}
\end{landscape}
\normalsize

\begin{landscape}
\begin{table}
\begin{center}
\caption[Xray parameters for field rxj0925.7–4758]{
X-ray parameters for detected sources in field RXJ0925.7-–4758}
\label{t:xray-param}
\fontsize{7}{7}\selectfont
\begin{tabular}{lrrrcrrrcrrrcrrrcc} 
\hline\hline 
\noalign{\smallskip}
 2XMM  & r$_{90}$ & pn\_B1$^{*}$ & pn\_B2$^{**}$ & 2MASS & d$_{x-o}$ & P$_{id}$ & kmag & GSC  & d$_{x-o}$ & P$_{id}$ & V & USNO & d$_{x-o}$ & P$_{id}$ & R & Class$^\dag$ & SpT$^\ddag$ \\
 Name   & [\arcsec] & [cts ks$^{-1}$] & [cts ks$^{-1}$] & Name & [\arcsec]  &          &      & Name & [\arcsec]   &          &   & Name & [\arcsec]   &          &   &       &       \\
\noalign{\smallskip}
\hline\noalign{\smallskip}
\noalign{\smallskip}
 J092432.8-475917 &    1.51 &                     &                     &    09243283-4759163  &    0.9 &    0.95 &   13.49 &      S5MW007845 &     0.6 &    0.90 &         &       0420-0186803 &     0.5 &    0.93 &   18.26 &            &         $^{       }$   \\
 J092439.8-475218 &    1.49 &     1.9$\pm$    0.5 &     0.1$\pm$    0.4 &                      &        &         &         &      S5MW009708 &     3.7 &    0.03 &         &       0421-0198115 &     3.4 &    0.20 &   16.76 &            &         $^{       }$   \\
 J092448.1-474928 &    1.18 &     2.7$\pm$    0.5 &     0.9$\pm$    0.7 &    09244827-4749265  &    2.7 &    0.51 &   13.75 &      S5MW010259 &     2.6 &    0.77 &         &       0421-0198198 &     2.7 &    0.72 &   16.81 &       Star &    M0V  $^{    sp }$   \\
 J092451.8-475905 &    0.69 &     8.5$\pm$    0.8 &                     &    09245177-4759045  &    1.4 &    1.00 &    8.14 &      S5MW000060 &     1.5 &    1.00 &   10.82 &       0420-0187014 &     1.5 &    1.00 &   10.13 &       Star &    K2V  $^{    sp }$   \\
 J092454.9-475427 &    1.72 &     1.3$\pm$    0.4 &     0.9$\pm$    0.4 &    09245501-4754274  &    0.6 &    0.98 &   12.02 &      S5MW009194 &     0.7 &    0.99 &         &       0420-0187059 &     5.2 &    0.80 &   13.09 &       Star &    A9V  $^{    sp }$   \\
 J092458.0-480412 &    3.64 &                     &                     &                      &        &         &         &      S5MW006353 &     7.3 &    0.05 &         &       0419-0178555 &     7.7 &    0.27 &   18.39 &            &         $^{       }$   \\
 J092500.5-480443 &    1.24 &                     &                     &    09250046-4804432  &    0.8 &    0.97 &   13.30 &      S5MW006271 &     0.8 &    0.96 &         &       0419-0178577 &     0.9 &    0.96 &   17.18 &            &         $^{       }$   \\
 J092500.6-474613 &    1.12 &     1.9$\pm$    0.5 &     3.9$\pm$    1.0 &    09250038-4746125  &    2.6 &    0.53 &   15.16 &      S5MW010884 &     3.0 &    0.05 &         &       0422-0215654 &     2.4 &    0.43 &   19.54 &            &         $^{       }$   \\
 J092506.0-475205 &    1.41 &     0.5$\pm$    0.2 &     1.5$\pm$    0.5 &                      &        &         &         &                 &         &         &         &       0421-0198378 &     2.0 &    0.87 &   19.77 &            &         $^{       }$   \\
 J092507.7-481018 &    1.23 &                     &                     &    09250779-4810187  &    0.1 &    0.99 &   12.37 &      S5MW004269 &     0.0 &    1.00 &         &       0418-0170896 &     0.4 &    0.99 &   15.36 &           &         $^{       }$   \\
 J092512.5-480832 &    1.22 &                     &                     &    09251244-4808321  &    1.3 &    0.73 &   15.15 &      S5MW004905 &     1.7 &    0.89 &         &       0418-0170950 &     1.2 &    0.96 &   17.44 &           &         $^{       }$   \\
 J092513.6-475152 &    1.29 &     1.0$\pm$    0.3 &     0.4$\pm$    0.3 &    09251379-4751521  &    1.1 &    0.96 &   13.83 &      S5MW009760 &     2.3 &    0.51 &         &       0421-0198455 &     1.4 &    0.61 &   18.90 &            &         $^{       }$   \\
 J092516.4-475422 &    1.83 &     1.0$\pm$    0.3 &     0.1$\pm$    0.2 &    09251653-4754209  &    1.8 &    1.00 &    9.47 &      S5MW009209 &     1.6 &    1.00 &         &       0420-0187264 &     2.2 &    0.92 &   17.21 &            &         $^{       }$   \\
 J092518.2-474727 &    1.64 &     1.4$\pm$    0.4 &     0.3$\pm$    0.4 &    09251826-4747282  &    0.8 &    0.95 &   13.98 &      S5MW010620 &     0.9 &    0.97 &         &       0422-0215893 &     1.1 &    0.97 &   16.99 &       Star &    M2V  $^{    sp }$   \\
 J092525.4-480654 &    1.78 &                     &                     &    09252566-4806532  &    2.4 &    0.98 &   11.15 &      S5MW000109 &     2.3 &    1.00 &         &       0418-0171071 &     6.4 &    0.65 &    0.00 &            &         $^{       }$   \\
 J092527.3-474755 &    0.41 &    41.0$\pm$    1.5 &     6.9$\pm$    0.8 &                      &        &         &         &      S5MW010573 &     1.5 &    0.97 &         &                    &         &         &         &       Star &    K4V  $^{    sp }$   \\
 J092527.3-480202 &    1.19 &     1.6$\pm$    0.3 &     0.1$\pm$    0.2 &    09252730-4802027  &    0.8 &    1.00 &   10.07 &      S5MW000076 &     0.7 &    1.00 &   11.96 &       0419-0178866 &     0.7 &    1.00 &   11.44 &            &         $^{       }$   \\
 J092531.1-474851 &    1.60 &     1.5$\pm$    0.4 &     0.9$\pm$    0.4 &    09253133-4748512  &    1.8 &    0.97 &   12.83 &      S5MW010350 &     1.7 &    0.99 &         &       0421-0198636 &     1.8 &    0.79 &   15.82 &       Star &    G5V  $^{    sp }$   \\
 J092541.3-474759 &    1.76 &                     &                     &                      &        &         &         &      S5MW059247 &     1.4 &    0.83 &         &       0422-0216190 &     1.2 &    0.83 &   19.34 &            &         $^{       }$   \\
 J092541.7-475310 &    1.72 &     0.9$\pm$    0.3 &     0.0$\pm$    0.1 &    09254173-4753086  &    2.3 &    0.98 &   11.94 &      S5MW000037 &     2.1 &    1.00 &         &       0421-0198735 &     5.1 &    1.00 &    0.00 &            &         $^{       }$   \\
 J092550.0-480150 &    0.54 &    10.1$\pm$    0.6 &     0.7$\pm$    0.2 &    09254994-4801502  &    0.8 &    0.98 &   13.02 &      S5MW007085 &     1.0 &    0.99 &         &       0419-0179060 &     0.6 &    0.99 &   16.93 &            &         $^{       }$   \\
 J092554.4-475017 &    1.07 &                     &                     &    09255437-4750172  &    0.8 &    1.00 &   10.31 &      S5MW000031 &     0.8 &    1.00 &   11.31 &       0421-0198878 &     0.8 &    1.00 &   11.23 &            &         $^{       }$   \\
 J092555.6-474618 &    1.48 &                     &                     &    09255552-4746167  &    2.2 &    0.99 &   10.57 &      S5MW000021 &     2.1 &    1.00 &   11.65 &       0422-0216378 &     2.1 &    1.00 &   11.33 &            &         $^{       }$   \\
 J092556.2-480932 &    1.28 &     1.5$\pm$    0.4 &     0.4$\pm$    0.3 &    09255625-4809330  &    0.5 &    0.98 &   13.27 &      S5MW004521 &     0.5 &    0.97 &         &       0418-0171364 &     0.5 &    0.97 &   17.23 &       Star &    M5V  $^{    sp }$   \\
 J092556.3-480421 &    0.88 &     2.8$\pm$    0.4 &     2.1$\pm$    0.3 &    09255633-4804224  &    0.6 &    1.00 &   10.01 &      S5MW006316 &     0.0 &    1.00 &         &       0419-0179120 &     0.3 &    1.00 &   15.11 &            &         $^{       }$   \\
 J092556.5-475927 &    1.60 &     0.2$\pm$    0.3 &                     &    09255661-4759287  &    1.5 &    0.99 &   11.36 &      S5MW007778 &     1.8 &    0.99 &         &       0420-0187660 &     2.3 &    0.98 &   14.17 &            &         $^{       }$   \\
 J092559.0-480036 &    2.26 &     1.0$\pm$    0.3 &                     &    09255917-4800369  &    1.7 &    0.74 &   15.02 &      S5MW058072 &     1.5 &    0.67 &         &       0419-0179140 &     2.2 &    0.76 &   20.32 &            &         $^{       }$   \\
 J092603.1-480610 &    1.52 &     1.5$\pm$    0.3 &     0.7$\pm$    0.3 &    09260337-4806143  &    4.6 &    0.43 &   14.55 &      S5MW005660 &     4.5 &    0.01 &         &       0418-0171417 &     5.0 &    0.01 &   19.63 &            &         $^{       }$   \\
 J092616.4-475149 &    1.54 &                     &                     &    09261654-4751510  &    2.1 &    0.68 &   14.72 &                 &         &         &         &                    &         &         &         &            &         $^{       }$   \\
 J092616.9-475235 &    1.84 &                     &                     &    09261713-4752377  &    3.1 &    0.28 &   15.42 &      S5MW009530 &     3.2 &    0.05 &         &       0421-0199110 &     3.7 &    0.01 &   19.57 &            &         $^{       }$   \\
 J092620.3-475822 &    0.52 &     9.6$\pm$    0.6 &     0.4$\pm$    0.2 &    09262038-4758226  &    0.2 &    1.00 &   12.47 &      S5MW008063 &     0.4 &    1.00 &         &       0420-0187910 &     0.7 &    1.00 &   14.86 &       Star &    G9V  $^{    sp }$   \\
 J092620.6-480330 &    0.55 &     9.8$\pm$    0.6 &     0.6$\pm$    0.2 &    09262069-4803306  &    0.5 &    1.00 &   12.60 &      S5MW006574 &     0.5 &    1.00 &         &       0419-0179349 &     0.7 &    1.00 &   14.94 &       Star &    G9V  $^{    sp }$   \\
 J092621.7-475504 &    1.17 &                     &                     &    09262177-4755043  &    0.0 &    0.99 &   13.34 &      S5MW008973 &     0.2 &    0.99 &         &       0420-0187921 &     2.5 &    0.85 &   17.37 &            &         $^{       }$   \\
 J092622.6-475551 &    0.47 &    17.1$\pm$    0.8 &     1.5$\pm$    0.3 &    09262272-4755508  &    0.7 &    0.99 &   11.78 &      S5MW000043 &     0.8 &    1.00 &         &       0420-0187929 &     0.9 &    1.00 &   13.93 &       Star &    G6V  $^{    sp }$   \\
 J092626.2-475633 &    1.35 &     1.5$\pm$    0.4 &     0.1$\pm$    0.3 &    09262624-4756318  &    1.2 &    0.99 &   11.79 &      S5MW000046 &     1.1 &    1.00 &         &       0420-0187971 &     1.0 &    1.00 &   13.70 &       Star &    G6V  $^{    sp }$   \\
 J092640.0-475110 &    0.97 &                     &                     &    09264001-4751095  &    0.8 &    1.00 &   11.40 &      S5MW000034 &     0.9 &    1.00 &         &       0421-0199313 &     0.5 &    1.00 &   13.33 &            &         $^{       }$   \\
 J092642.4-480845 &    1.53 &                     &                     &    09264247-4808445  &    1.2 &    1.00 &   11.27 &      S5MW000123 &     1.1 &    1.00 &         &       0418-0171773 &     2.5 &    0.82 &   12.43 &            &         $^{       }$   \\
 J092649.7-475855 &    1.90 &                     &                     &    09264982-4758531  &    2.1 &    0.84 &   14.04 &      S5MW007879 &     2.9 &    0.51 &         &       0420-0188178 &     2.9 &    0.52 &   17.60 &            &         $^{       }$   \\
\noalign{\smallskip}\hline
\end{tabular}
\end{center}
\end{table}
\end{landscape}
\normalsize

\begin{landscape}
\begin{table}
\begin{center}
\caption[Xray parameters for field ARLac]{
X-ray parameters for detected sources in field ARLac}
\label{t:xray-param}
\fontsize{7}{7}\selectfont
\begin{tabular}{lrrrcrrrcrrrcrrrcc} 
\hline\hline 
\noalign{\smallskip}
 2XMM  & r$_{90}$ & pn\_B1$^{*}$ & pn\_B2$^{**}$ & 2MASS & d$_{x-o}$ & P$_{id}$ & kmag & GSC  & d$_{x-o}$ & P$_{id}$ & V & USNO & d$_{x-o}$ & P$_{id}$ & R & Class$^\dag$ & SpT$^\ddag$ \\
 Name   & [\arcsec] & [cts ks$^{-1}$] & [cts ks$^{-1}$] & Name & [\arcsec]  &          &      & Name & [\arcsec]   &          &   & Name & [\arcsec]   &          &   &       &       \\
\noalign{\smallskip}
\hline\noalign{\smallskip}
\noalign{\smallskip}
 J220742.2+455258 &    2.22 &                     &                     &    22074198+4552576  &    2.8 &    0.82 &   12.21 &      N2TW034514 &     2.6 &    0.60 &   17.27 &       1358-0468831 &     2.9 &    0.54 &   16.61 &       Star &   M4Ve  $^{    sp }$   \\
 J220752.3+454742 &    1.14 &                     &                     &                      &        &         &         &      N2TW030293 &     0.5 &    0.76 &         &       1357-0480103 &     0.4 &    0.80 &   18.97 &            &         $^{       }$   \\
 J220755.6+454838 &    1.51 &                     &                     &                      &        &         &         &      N2TW072280 &     3.4 &    0.05 &   19.54 &       1358-0469023 &     4.1 &    0.02 &   18.71 &            &         $^{       }$   \\
 J220802.2+455600 &    2.21 &                     &                     &    22080238+4556026  &    2.1 &    0.96 &   10.81 &      N2TW001240 &     2.2 &    0.99 &   12.63 &       1359-0465243 &     2.2 &    0.96 &   12.11 &       Star &    G9V  $^{    sp }$   \\
 J220804.8+454140 &    1.37 &                     &                     &                      &        &         &         &      N2X4071640 &     1.1 &    0.56 &         &       1356-0483745 &     1.9 &    0.53 &   19.10 &            &         $^{       }$   \\
 J220805.6+454805 &    1.32 &                     &                     &    22080555+4548054  &    1.1 &    0.98 &   11.71 &      N2X4056378 &     0.7 &    0.98 &   14.33 &       1358-0469168 &     1.1 &    0.97 &   14.06 &       Star &   K0Ve  $^{    sp }$   \\
 J220808.7+453929 &    1.63 &                     &                     &                      &        &         &         &      N2X4055750 &     3.2 &    0.27 &   18.58 &       1356-0483804 &     3.2 &    0.26 &   18.14 &            &         $^{       }$   \\
 J220814.1+454034 &    1.36 &                     &                     &    22081450+4540347  &    3.5 &    0.11 &   15.74 &      N2X4055852 &     4.0 &    0.01 &   18.81 &       1356-0483885 &     3.4 &    0.09 &   18.34 &            &         $^{       }$   \\
 J220822.6+454535 &    1.85 &                     &                     &    22082281+4545330  &    3.1 &    0.53 &   14.94 &      N2X4056238 &     2.9 &    0.68 &   17.01 &       1357-0480507 &     2.5 &    0.83 &   15.43 &       Star &    K0V  $^{    sp }$   \\
 J220833.1+454204 &    2.59 &                     &                     &    22083387+4542007  &    8.3 &    0.88 &   14.53 &      N2X4029403 &     8.8 &    0.98 &   18.16 &       1357-0480660 &     6.1 &    0.94 &   18.43 &            &         $^{       }$   \\
 J220836.0+454632 &    2.37 &                     &                     &                      &        &         &         &      N2X4071910 &     7.1 &    0.04 &         &       1357-0480711 &     5.7 &    0.02 &   18.74 &            &         $^{       }$   \\
 J220837.8+453129 &    1.23 &                     &                     &    22083774+4531279  &    2.3 &    0.98 &   10.06 &      N2X4027141 &     2.4 &    0.92 &   15.11 &       1355-0486243 &     2.1 &    0.92 &   14.41 &       Star &   M4Ve  $^{    sp }$   \\
 J220842.0+453716 &    2.96 &                     &                     &                      &        &         &         &      N2X4028358 &     5.3 &    0.08 &   18.81 &       1356-0484255 &     5.4 &    0.11 &   19.09 &            &         $^{       }$   \\
 J220855.0+453448 &    2.40 &                     &                     &                      &        &         &         &      N2X4072926 &     5.4 &         &         &       1355-0486459 &     2.2 &    0.37 &   19.25 &            &         $^{       }$   \\
 J220856.1+455510 &    1.76 &                     &                     &                      &        &         &         &      N2X4031802 &     1.5 &    0.50 &         &       1359-0465995 &     0.9 &    0.45 &   19.17 &            &         $^{       }$   \\
 J220912.6+454021 &    1.47 &                     &                     &                      &        &         &         &      N2X4028954 &     4.9 &         &   18.54 &                    &         &         &         &            &         $^{       }$   \\
 J220916.5+453255 &    2.37 &                     &                     &    22091597+4532508  &    7.9 &         &   14.63 &      N2X4027369 &     7.3 &    0.01 &   17.16 &       1355-0486726 &     7.7 &         &   16.37 &            &         $^{       }$   \\
 J220924.8+454449 &    1.88 &                     &                     &    22092428+4544499  &    5.9 &    0.39 &   13.66 &      N2X4029911 &     5.1 &    0.20 &   15.74 &       1357-0481411 &     3.8 &    0.11 &   18.74 &            &         $^{       }$   \\
 J220929.1+455206 &    2.26 &                     &                     &                      &        &         &         &      N2X4031245 &     0.8 &    0.49 &         &       1358-0470349 &     1.4 &    0.39 &   19.21 &            &         $^{       }$   \\
 J220932.5+454217 &    2.81 &                     &                     &    22093291+4542123  &    6.5 &    0.03 &   15.63 &      N2X4029287 &     6.8 &    0.02 &   18.33 &       1357-0481525 &     6.0 &    0.06 &   17.88 &            &         $^{       }$   \\
 J220932.8+454730 &    2.04 &                     &                     &    22093282+4547334  &    2.7 &    0.45 &   15.30 &      N2X4030439 &     2.6 &    0.46 &   18.36 &       1357-0481524 &     2.9 &    0.41 &   17.83 &            &         $^{       }$   \\
 J220942.7+454639 &    1.80 &                     &                     &    22094274+4546393  &    0.7 &    1.00 &    9.16 &      N2X4000715 &     0.7 &    1.00 &    9.73 &       1357-0481668 &     0.7 &    1.00 &    9.58 &       Star &    A5V  $^{    sp }$   \\
\noalign{\smallskip}\hline
\end{tabular}
\end{center}
\end{table}
\end{landscape}
\normalsize

\begin{landscape}
\begin{table}
\begin{center}
\caption[Xray parameters for field Geminga]{
X-ray parameters for detected sources in field Geminga}
\label{t:xray-param}
\fontsize{7}{7}\selectfont
\begin{tabular}{lrrrcrrrcrrrcrrrcc} 
\hline\hline 
\noalign{\smallskip}
 2XMM  & r$_{90}$ & pn\_B1$^{*}$ & pn\_B2$^{**}$ & 2MASS & d$_{x-o}$ & P$_{id}$ & kmag & GSC  & d$_{x-o}$ & P$_{id}$ & V & USNO & d$_{x-o}$ & P$_{id}$ & R & Class$^\dag$ & SpT$^\ddag$ \\
 Name   & [\arcsec] & [cts ks$^{-1}$] & [cts ks$^{-1}$] & Name & [\arcsec]  &          &      & Name & [\arcsec]   &          &   & Name & [\arcsec]   &          &   &       &       \\
\noalign{\smallskip}
\hline\noalign{\smallskip}
\noalign{\smallskip}
 J063259.6+174637 &    2.25 &                     &                     &    06325942+1746402  &    4.9 &    0.14 &   15.86 &      N8L5012571 &     5.0 &    0.08 &   18.10 &       1077-0136581 &     5.2 &    0.08 &   18.61 &            &         $^{       }$   \\
 J063307.4+175236 &    1.96 &                     &                     &                      &        &         &         &      N8L5015706 &     1.3 &    0.62 &   18.61 &       1078-0139187 &     2.0 &    0.46 &   18.94 &            &         $^{       }$   \\
 J063313.4+175432 &    1.49 &                     &                     &    06331327+1754321  &    1.4 &    1.00 &    8.65 &      N8L5000363 &     1.3 &    1.00 &   11.53 &       1079-0140062 &     1.3 &    1.00 &   10.95 &       Star &    K2V  $^{    sp }$   \\
 J063314.2+173539 &    1.21 &                     &                     &    06331408+1735389  &    2.2 &    0.35 &   15.07 &      N8L5006823 &     2.3 &    0.20 &   17.86 &                    &         &         &         &       Star &      K  $^{    sp }$   \\
 J063317.3+175212 &    2.40 &                     &                     &    06331730+1752098  &    2.4 &    0.83 &   13.10 &      N8L5015536 &     2.3 &    0.92 &   13.92 &       1078-0139361 &     2.4 &    0.93 &   13.86 &            &         $^{       }$   \\
 J063318.9+175521 &    1.27 &                     &                     &    06331896+1755215  &    1.2 &    0.95 &   11.58 &      N8L5017134 &     1.1 &    0.98 &   14.31 &       1079-0140185 &     2.7 &    0.96 &         &       Star &    G0V  $^{    sp }$   \\
 J063319.4+175035 &    1.84 &                     &                     &    06331927+1750390  &    4.2 &    0.26 &   14.93 &      N8L5014677 &     4.3 &    0.07 &   18.57 &       1078-0139386 &     4.4 &    0.06 &   18.88 &            &         $^{       }$   \\
 J063322.6+173430 &    1.68 &                     &                     &    06332285+1734348  &    4.9 &    0.09 &   14.96 &      N8L5006467 &     5.7 &         &   17.21 &       1075-0135475 &     5.3 &    0.02 &   18.43 &            &         $^{       }$   \\
 J063324.0+175307 &    1.63 &                     &                     &                      &        &         &         &      N8L5015966 &     3.9 &     &         &       1078-0139446 &     3.4 &    0.04 &   19.05 &            &         $^{       }$   \\
 J063324.3+174806 &    1.85 &                     &                     &                      &        &         &         &      N8L5013314 &     3.5 &    0.07 &         &       1078-0139440 &     5.0 &    0.43 &   19.10 &            &         $^{       }$   \\
 J063330.3+175300 &    2.31 &                     &                     &    06333026+1753004  &    1.0 &    0.75 &   14.82 &      N8L5015828 &     6.9 &    0.71 &         &       1078-0139517 &     7.3 &    0.61 &   19.23 &            &         $^{       }$   \\
 J063334.0+175512 &    1.86 &                     &                     &    06333392+1755104  &    3.4 &    0.53 &   14.64 &      N8L5017008 &     3.3 &    0.41 &   17.99 &       1079-0140436 &     3.6 &    0.27 &   17.71 &            &         $^{       }$   \\
 J063335.4+173408 &    1.73 &                     &                     &    06333529+1734091  &    2.3 &    0.83 &   13.71 &      N8L5006301 &     1.8 &    0.88 &   17.11 &       1075-0135634 &     2.2 &    0.43 &   18.08 &            &         $^{       }$   \\
 J063335.5+173927 &    1.23 &                     &                     &    06333542+1739278  &    1.5 &    0.98 &   10.02 &      N8L5008231 &     1.5 &    0.99 &   13.65 &       1076-0135209 &     1.8 &    0.96 &   13.09 &       Star &    G6V  $^{    sp }$   \\
 J063339.0+175933 &    1.30 &                     &                     &                      &        &         &         &      N8L5019086 &     1.3 &    0.80 &   18.55 &       1079-0140504 &     1.9 &    0.31 &   18.87 &            &         $^{       }$   \\
 J063341.8+175408 &    1.98 &                     &                     &                      &        &         &         &      N8L5016459 &     5.0 &         &         &       1079-0140559 &     5.1 &    0.06 &   18.85 &            &         $^{       }$   \\
 J063342.7+175017 &    1.73 &                     &                     &                      &        &         &         &      N8L5014405 &     4.5 &    0.01 &         &       1078-0139723 &     4.4 &    0.12 &   19.27 &            &         $^{       }$   \\
 J063344.6+175644 &    1.28 &                     &                     &    06334452+1756438  &    1.6 &    0.73 &   15.40 &      N8L5044039 &     1.5 &    0.67 &         &       1079-0140604 &     1.6 &    0.71 &   19.24 &            &         $^{       }$   \\
 J063344.8+174552 &    2.45 &                     &                     &                      &        &         &         &      N8L5012036 &     8.3 &         &         &       1077-0137200 &     7.2 &         &   19.20 &            &         $^{       }$   \\
 J063347.8+173520 &    1.97 &                     &                     &                      &        &         &         &      N8L5006655 &     5.3 &    0.01 &   18.35 &       1075-0135812 &     5.6 &    0.01 &   19.14 &            &         $^{       }$   \\
 J063349.2+174732 &    1.46 &                     &                     &    06334914+1747311  &    0.9 &    1.00 &    8.94 &      N8L5000416 &     0.8 &    1.00 &   11.56 &       1077-0137269 &     0.8 &    1.00 &   11.11 &       Star &    K2V  $^{    sp }$   \\
 J063350.2+173444 &    1.78 &                     &                     &    06335008+1734448  &    1.4 &    0.90 &   13.34 &      N8L5006493 &     1.2 &    0.94 &   15.71 &       1075-0135849 &     1.7 &    0.90 &   15.48 &            &         $^{       }$   \\
 J063351.2+173826 &    1.39 &                     &                     &    06335106+1738254  &    3.1 &    0.20 &   13.37 &      N8L5007772 &     2.8 &    0.24 &   17.39 &                    &         &         &         &       Star &   M2Ve  $^{    sp }$   \\
 J063357.3+174050 &    1.32 &                     &                     &                      &        &         &         &      N8L5008976 &     3.1 &    0.02 &   17.74 &                    &         &         &         &       Star &   M4Ve  $^{    sp }$   \\
 J063400.3+173624 &    1.38 &                     &                     &                      &        &         &         &      N8L5007043 &     1.6 &    0.62 &         &       1076-0135568 &     2.4 &    0.16 &   18.89 &            &         $^{       }$   \\
 J063402.6+175308 &    1.35 &                     &                     &    06340266+1753066  &    1.2 &    0.94 &   13.43 &      N8L5015919 &     1.0 &    0.98 &   15.90 &       1078-0140019 &     1.1 &    0.97 &   15.69 &            &         $^{       }$   \\
 J063406.7+175945 &    1.02 &                     &                     &    06340671+1759451  &    0.1 &    1.00 &   10.01 &      N8L5019184 &     0.3 &    1.00 &   14.60 &       1079-0140931 &     0.1 &    0.99 &   14.08 &       Star &   M4Ve  $^{    sp }$   \\
 J063410.3+174949 &    1.74 &                     &                     &    06341057+1749476  &    4.0 &    0.33 &   15.34 &      N8L5014246 &     3.6 &    0.67 &   16.62 &       1078-0140135 &     3.3 &    0.43 &   15.83 &            &         $^{       }$   \\
 J063413.0+175652 &    2.21 &                     &                     &    06341318+1756519  &    2.8 &    0.45 &   15.89 &      N8L5017821 &     2.5 &    0.59 &   17.96 &       1079-0141016 &     2.1 &    0.57 &   18.37 &            &         $^{       }$   \\
 J063415.6+174000 &    2.09 &                     &                     &    06341534+1739546  &    7.0 &         &   15.67 &      N8L5008390 &     6.6 &    0.46 &         &       1076-0135782 &     7.0 &    0.89 &   18.26 &            &         $^{       }$   \\
 J063419.4+174039 &    1.64 &                     &                     &    06341951+1740371  &    2.8 &    0.92 &   11.29 &      N8L5008865 &     2.4 &    0.96 &   14.40 &       1076-0135839 &     2.2 &    0.94 &   13.89 &            &         $^{       }$   \\
 J063419.8+173415 &    1.84 &                     &                     &                      &        &         &         &      N8L5006261 &     2.5 &    0.34 &         &       1075-0136304 &     3.4 &    0.25 &   18.85 &            &         $^{       }$   \\
 J063425.1+174212 &    1.09 &                     &                     &    06342536+1742110  &    3.1 &    0.48 &   10.46 &      N8L5000446 &     3.2 &    0.97 &   11.11 &       1077-0137788 &     3.2 &    0.99 &   11.02 &       Star &    A7V  $^{    sp }$   \\
 J063431.2+174645 &    2.42 &                     &                     &    06343131+1746406  &    4.4 &    0.64 &   13.24 &      N8L5012515 &     4.6 &    0.74 &   14.56 &       1077-0137855 &     4.6 &    0.76 &   14.11 &            &         $^{       }$   \\
 J063434.2+174728 &    1.83 &                     &                     &    06343417+1747276  &    1.3 &    0.82 &   14.20 &      N8L5012877 &     1.1 &    0.81 &         &       1077-0137888 &     1.3 &    0.67 &   18.83 &            &         $^{       }$   \\
 J063437.3+173937 &    2.69 &                     &                     &    06343704+1739355  &    4.7 &    0.71 &   12.99 &      N8L5008207 &     4.7 &    0.86 &   14.08 &       1076-0136067 &     5.1 &    0.79 &   14.03 &            &         $^{       }$   \\
 J063446.4+174355 &    2.50 &                     &                     &                      &        &         &         &      N8L5010882 &     5.0 &    0.09 &   18.15 &       1077-0138063 &     4.8 &    0.12 &   18.65 &            &         $^{       }$   \\
 J063449.3+174355 &    1.96 &                     &                     &                      &        &         &         &      N8L5010861 &     5.0 &    0.01 &         &       1077-0138095 &     5.5 &    0.02 &   19.07 &            &         $^{       }$   \\
\noalign{\smallskip}\hline
\end{tabular}
\end{center}
\end{table}
\end{landscape}
\normalsize

\begin{landscape}
\begin{table}
\begin{center}
\caption[Xray parameters for field PSR0656+14]{
X-ray parameters for detected sources in field PSR0656+14}
\label{t:xray-param}
\fontsize{7}{7}\selectfont
\begin{tabular}{lrrrcrrrcrrrcrrrcc} 
\hline\hline 
\noalign{\smallskip}
 2XMM  & r$_{90}$ & pn\_B1$^{*}$ & pn\_B2$^{**}$ & 2MASS & d$_{x-o}$ & P$_{id}$ & kmag & GSC  & d$_{x-o}$ & P$_{id}$ & V & USNO & d$_{x-o}$ & P$_{id}$ & R & Class$^\dag$ & SpT$^\ddag$ \\
 Name   & [\arcsec] & [cts ks$^{-1}$] & [cts ks$^{-1}$] & Name & [\arcsec]  &          &      & Name & [\arcsec]   &          &   & Name & [\arcsec]   &          &   &       &       \\
\noalign{\smallskip}
\hline\noalign{\smallskip}
\noalign{\smallskip}
 J065909.7+141815 &    2.79 &                     &                     &                      &        &         &         &      N8NG016348 &     6.5 &    0.06 &   18.03 &       1043-0124959 &     6.1 &    0.07 &   17.46 &            &         $^{       }$   \\
 J065911.5+141114 &    1.42 &                     &                     &                      &        &         &         &      N8NG013375 &     5.0 &    0.02 &   15.16 &       1041-0123358 &     4.5 &    0.46 &   14.34 &       Star &   K4Ve  $^{    sp }$   \\
 J065920.2+140910 &    2.10 &                     &                     &    06592030+1409103  &    1.4 &    0.93 &   12.99 &      N8NG012537 &     1.5 &    0.91 &   14.93 &       1041-0123456 &     1.2 &    0.88 &   15.12 &            &         $^{       }$   \\
 J065948.6+141917 &    1.98 &                     &                     &                      &        &         &         &      N8NG016828 &     5.7 &    0.01 &   17.98 &       1043-0125399 &     5.8 &    0.02 &   16.86 &            &         $^{       }$   \\
 J065950.7+142201 &    1.92 &                     &                     &                      &        &         &         &      N8NG018137 &     0.7 &    0.64 &         &       1043-0125421 &     1.1 &    0.67 &   20.25 &            &         $^{       }$   \\
 J065956.9+141219 &    1.56 &                     &                     &    06595708+1412228  &    3.8 &    0.93 &   10.41 &      N8NG000520 &     3.5 &    0.99 &   12.04 &       1042-0123543 &     3.7 &    0.92 &   12.10 &       Star &    G9V  $^{    sp }$   \\
 J065959.2+140319 &    1.56 &                     &                     &                      &        &         &         &      N8NG031476 &     1.2 &    0.56 &         &       1040-0123513 &     0.6 &    0.68 &   20.06 &       Star &    G0V  $^{    sp }$   \\
 J070001.0+141210 &    2.50 &                     &                     &    07000164+1412093  &    7.9 &    0.01 &   14.91 &      N8NG013701 &     8.3 &    0.38 &   18.77 &       1042-0123602 &     8.1 &         &   18.34 &            &         $^{       }$   \\
 J070003.9+141047 &    2.23 &                     &                     &    07000357+1410476  &    5.0 &    0.21 &   15.35 &      N8NG013163 &     4.3 &    0.40 &   17.24 &       1041-0123938 &     4.6 &    0.32 &   16.83 &            &         $^{       }$   \\
 J070006.2+141315 &    2.95 &                     &                     &    07000650+1413133  &    4.8 &    0.41 &   14.87 &      N8NG032672 &     8.8 &    0.49 &         &       1042-0123658 &     8.2 &    0.33 &   14.53 &            &         $^{       }$   \\
 J070008.7+140205 &    3.39 &                     &                     &                      &        &         &         &      N8NG031372 &     9.1 &    0.35 &         &       1040-0123627 &     9.6 &         &   18.59 &            &         $^{       }$   \\
 J070009.6+142213 &    1.22 &                     &                     &    07000973+1422159  &    3.2 &    0.18 &   16.27 &      N8NG018269 &     2.8 &    0.46 &   17.72 &       1043-0125642 &     2.9 &    0.38 &   17.14 &       Star &    K5V  $^{    sp }$   \\
 J070015.3+142109 &    1.37 &                     &                     &                      &        &         &         &      N8NG033935 &     1.0 &    0.69 &         &       1043-0125715 &     0.4 &    0.72 &   19.23 &            &         $^{       }$   \\
 J070019.7+141211 &    1.32 &                     &                     &    07001983+1412119  &    1.4 &    0.95 &   13.52 &      N8NG013730 &     1.5 &    0.87 &   17.54 &       1042-0123794 &     1.3 &    0.85 &   16.76 &       Star &   M0Ve  $^{    sp }$   \\
 J070020.1+140700 &    2.60 &                     &                     &    07002034+1406582  &    3.3 &    0.72 &   13.94 &      N8NG011674 &     3.5 &    0.55 &   17.35 &       1041-0124133 &     3.3 &    0.52 &   16.56 &            &         $^{       }$   \\
 J070026.9+142044 &    2.03 &                     &                     &    07002693+1420443  &    0.6 &    0.97 &   11.63 &      N8NG017571 &     0.4 &    0.98 &   13.95 &       1043-0125854 &     0.6 &    0.97 &   13.71 &            &         $^{       }$   \\
 J070029.5+140851 &    1.54 &                     &                     &    07002926+1408501  &    4.6 &    0.01 &   15.41 &      N8NG012365 &     4.4 &    0.07 &   18.15 &       1041-0124255 &     4.2 &    0.14 &   17.77 &            &         $^{       }$   \\
 J070032.8+140714 &    1.48 &                     &                     &    07003287+1407122  &    2.0 &    0.96 &   11.58 &      N8NG000600 &     1.9 &    0.98 &   13.16 &       1041-0124287 &     1.7 &    0.97 &   13.21 &       Star &    G0V  $^{    sp }$   \\
\noalign{\smallskip}\hline
\end{tabular}
\end{center}
\end{table}
\end{landscape}
\normalsize

\begin{landscape}
\begin{table}
\begin{center}
\caption[Xray parameters for field RXJ0002+6246]{
X-ray parameters for detected sources in field RX~J0002+6246}
\label{t:xray-param}
\fontsize{7}{7}\selectfont
\begin{tabular}{lrrrcrrrcrrrcrrrcc} 
\hline\hline 
\noalign{\smallskip}
 2XMM  & r$_{90}$ & pn\_B1$^{*}$ & pn\_B2$^{**}$ & 2MASS & d$_{x-o}$ & P$_{id}$ & kmag & GSC  & d$_{x-o}$ & P$_{id}$ & V & USNO & d$_{x-o}$ & P$_{id}$ & R & Class$^\dag$ & SpT$^\ddag$ \\
 Name   & [\arcsec] & [cts ks$^{-1}$] & [cts ks$^{-1}$] & Name & [\arcsec]  &          &      & Name & [\arcsec]   &          &   & Name & [\arcsec]   &          &   &       &       \\
\noalign{\smallskip}
\hline\noalign{\smallskip}
\noalign{\smallskip}
 J000100.9+624803 &    2.85 &                     &                     &    00010072+6248002  &    3.2 &    0.41 &   14.43 &      NAKK099897 &     3.4 &    0.13 &         &       1527-0000767 &     5.2 &    0.15 &         &            &         $^{       }$   \\
 J000105.3+624933 &    3.30 &                     &                     &    00010531+6249344  &    0.7 &    0.77 &   12.85 &      NAKK101247 &     9.4 &    0.66 &         &       1528-0000834 &     0.4 &    0.59 &   16.03 &            &         $^{       }$   \\
 J000114.0+625231 &    1.28 &                     &                     &    00011391+6252302  &    1.0 &    1.00 &    7.94 &      NAKK000440 &     0.8 &    1.00 &   10.47 &       1528-0000936 &     0.8 &    1.00 &    9.76 &            &         $^{       }$   \\
 J000130.0+625236 &    1.68 &                     &                     &    00013004+6252367  &    0.2 &    0.96 &   12.33 &      NAKK038949 &     0.3 &    0.97 &   15.65 &       1528-0001138 &     5.4 &    0.94 &   17.88 &       Star &    K2V  $^{    sp }$   \\
 J000132.4+624326 &    2.60 &                     &                     &    00013333+6243272  &    6.2 &    0.50 &   11.71 &      NAKK000650 &     6.3 &    0.67 &   13.47 &       1527-0001201 &     6.8 &    0.50 &   17.77 &       Star &    F3V  $^{    sp }$   \\
 J000134.1+625008 &    1.22 &                     &                     &                      &        &         &         &      NAKK035163 &     1.1 &    0.85 &   17.82 &       1528-0001181 &     0.8 &    0.88 &   17.48 &       CV  &     CV  $^{    sp }$   \\
 J000135.8+625723 &    1.89 &                     &                     &    00013575+6257262  &    3.2 &    0.77 &   12.97 &      NAKK045931 &     3.2 &    0.72 &   16.62 &       1529-0001203 &     3.4 &    0.63 &   15.82 &            &         $^{       }$   \\
 J000150.5+625749 &    2.98 &                     &                     &    00015040+6257460  &    3.8 &    0.20 &   15.32 &      NAKK046355 &     4.2 &    0.09 &   18.81 &       1529-0001381 &     4.0 &    0.11 &   18.09 &            &         $^{       }$   \\
 J000203.2+624436 &    1.74 &                     &                     &    00020300+6244377  &    1.7 &    0.88 &   13.02 &      NAKK027914 &     1.6 &    0.92 &   14.72 &       1527-0001596 &     0.9 &    0.54 &         &            &         $^{       }$   \\
 J000203.9+625601 &    2.68 &                     &                     &                      &        &         &         &      NAKK043581 &     5.2 &    0.05 &   18.49 &       1529-0001571 &     7.8 &    0.04 &   19.32 &            &         $^{       }$   \\
 J000208.3+625347 &    2.22 &                     &                     &    00020815+6253486  &    1.3 &    0.57 &   15.23 &      NAKK126605 &     1.3 &         &         &       1528-0001589 &     4.3 &    0.14 &         &            &         $^{       }$   \\
 J000222.1+624510 &    1.79 &                     &                     &    00022277+6245112  &    4.7 &    0.07 &   15.22 &      NAKK028607 &     4.9 &    0.05 &   17.70 &       1527-0001817 &     3.7 &    0.06 &   18.64 &            &         $^{       }$   \\
 J000238.7+623357 &    2.67 &                     &                     &    00023971+6233599  &    7.2 &    0.11 &   14.61 &      NAKK087638 &     7.5 &    0.01 &         &       1525-0002122 &     7.3 &    0.02 &   16.78 &            &         $^{       }$   \\
 J000245.8+625846 &    2.33 &                     &                     &    00024589+6258472  &    0.8 &    0.80 &   13.93 &      NAKK111713 &     0.9 &    0.72 &         &       1529-0002088 &     7.4 &    0.47 &   15.40 &            &         $^{       }$   \\
 J000248.0+625149 &    1.81 &                     &                     &    00024843+6251514  &    3.4 &    0.95 &   10.45 &      NAKK124515 &     4.1 &    0.75 &         &       1528-0002133 &     3.3 &    0.52 &   15.83 &       Star &    F6V  $^{    sp }$   \\
 J000248.4+624014 &    2.72 &                     &                     &    00024908+6240218  &    8.7 &    0.06 &   14.95 &      NAKK000736 &     8.6 &    0.21 &   13.14 &       1526-0002314 &     8.5 &    0.10 &   18.71 &            &         $^{       }$   \\
 J000257.2+624230 &    1.90 &                     &                     &    00025683+6242318  &    2.9 &    0.42 &   14.73 &      NAKK094747 &     2.9 &    0.51 &   17.15 &       1527-0002332 &     6.4 &    0.34 &   17.80 &            &         $^{       }$   \\
 J000257.9+623547 &    2.47 &                     &                     &                      &        &         &         &      NAKK088977 &     5.0 &    0.01 &         &       1525-0002369 &     4.3 &    0.11 &   19.27 &            &         $^{       }$   \\
 J000309.2+625404 &    2.20 &                     &                     &    00030922+6254083  &    3.6 &    0.42 &   14.95 &      NAKK106321 &     3.5 &    0.32 &   17.51 &       1529-0002370 &     3.5 &    0.28 &   16.77 &            &         $^{       }$   \\
 J000317.9+623824 &    2.39 &                     &                     &    00031845+6238252  &    3.2 &    0.58 &   13.73 &      NAKK091105 &     3.2 &    0.15 &         &       1526-0002787 &     3.8 &    0.22 &   16.85 &       Star &    M5V  $^{    sp }$   \\
 J000323.6+625418 &    2.03 &                     &                     &    00032372+6254192  &    0.7 &    0.95 &   12.50 &      NAKK126633 &     6.6 &    0.96 &         &       1529-0002562 &     1.1 &    0.62 &   15.19 &       Star &    K2V  $^{    sp }$   \\
 J000326.5+625629 &    1.77 &                     &                     &    00032679+6256308  &    1.8 &    0.76 &   13.91 &      NAKK131894 &     5.7 &    0.80 &         &       1529-0002601 &     2.2 &    0.73 &   16.20 &            &         $^{       }$   \\
 J000335.7+624736 &    1.48 &                     &                     &    00033537+6247384  &    3.8 &    0.12 &   15.34 &      NAKK099381 &     3.6 &    0.08 &   18.52 &       1527-0002884 &     3.7 &    0.08 &   17.94 &            &         $^{       }$   \\
 J000336.0+625156 &    3.07 &                     &                     &    00033643+6252057  &    9.2 &    0.14 &   15.43 &      NAKK103938 &     3.4 &    0.19 &   18.50 &       1528-0002787 &     9.5 &    0.20 &   19.24 &            &         $^{       }$   \\
 J000340.4+624745 &    2.42 &                     &                     &    00034054+6247469  &    1.7 &    0.74 &   13.80 &      NAKK099689 &     0.8 &    0.86 &   16.68 &       1527-0002955 &     1.2 &    0.60 &   15.61 &            &         $^{       }$   \\
 J000351.7+625449 &    2.29 &                     &                     &                      &        &         &         &      NAKK107162 &     6.2 &         &         &       1529-0002895 &     7.5 &         &   19.42 &            &         $^{       }$   \\
 J000400.2+625330 &    1.26 &                     &                     &    00040061+6253275  &    4.1 &         &   15.45 &      NAKK105549 &     4.2 &         &         &       1528-0003134 &     3.7 &         &   18.87 &            &         $^{       }$   \\
 J000404.5+624326 &    1.18 &                     &                     &    00040446+6243254  &    0.8 &    0.99 &   11.32 &      NAKK095530 &     0.9 &    0.98 &   14.83 &       1527-0003230 &     0.5 &    0.98 &   14.07 &       Star &    M0V  $^{    sp }$   \\
 J000405.4+624428 &    1.91 &                     &                     &    00040535+6244299  &    1.2 &    0.87 &   13.69 &      NAKK096482 &     1.1 &    0.86 &   16.84 &       1527-0003241 &     1.5 &    0.84 &   16.17 &       Star &    G0V  $^{    sp }$   \\
 J000419.3+625342 &    2.46 &                     &                     &    00042000+6253375  &    6.2 &    0.48 &   14.13 &      NAKK105786 &     6.2 &    0.37 &         &       1528-0003413 &     7.1 &    0.56 &         &            &         $^{       }$   \\
\noalign{\smallskip}\hline
\end{tabular}
\end{center}
\end{table}
\end{landscape}
\normalsize

\begin{landscape}
\begin{table}
\begin{center}
\caption[Xray parameters for field SSCyg]{
X-ray parameters for detected sources in field SS~Cyg}
\label{t:xray-param}
\fontsize{7}{7}\selectfont
\begin{tabular}{lrrrcrrrcrrrcrrrcc} 
\hline\hline 
\noalign{\smallskip}
 2XMM  & r$_{90}$ & pn\_B1$^{*}$ & pn\_B2$^{**}$ & 2MASS & d$_{x-o}$ & P$_{id}$ & kmag & GSC  & d$_{x-o}$ & P$_{id}$ & V & USNO & d$_{x-o}$ & P$_{id}$ & R & Class$^\dag$ & SpT$^\ddag$ \\
 Name   & [\arcsec] & [cts ks$^{-1}$] & [cts ks$^{-1}$] & Name & [\arcsec]  &          &      & Name & [\arcsec]   &          &   & Name & [\arcsec]   &          &   &       &       \\
\noalign{\smallskip}
\hline\noalign{\smallskip}
\noalign{\smallskip}
 J214158.1+432518 &    2.25 &                     &                     &    21415811+4325110  &    7.1 &    0.68 &   15.50 &      N2TU118332 &     1.7 &    0.91 &   16.10 &       1334-0441110 &     7.5 &    0.87 &   18.03 &       Star &   M4Ve  $^{    sp }$   \\
 J214222.8+432332 &    1.92 &                     &                     &    21422285+4323330  &    0.5 &    0.97 &   11.88 &      N2TU116949 &     0.6 &    0.92 &   16.16 &       1333-0464128 &     0.8 &    0.88 &   15.90 &       Star &   M2Ve  $^{    sp }$   \\
 J214229.5+432533 &    2.07 &                     &                     &    21422942+4325320  &    2.1 &    0.96 &   11.29 &      N2TU118768 &     2.0 &    0.95 &   14.77 &       1334-0441611 &     1.7 &    0.56 &   14.23 &       Star &   M0Ve  $^{    sp }$   \\
 J214229.8+434615 &    2.36 &                     &                     &                      &        &         &         &      N2U5023566 &     7.0 &         &   18.52 &       1337-0436148 &     7.2 &         &   18.10 &            &         $^{       }$   \\
 J214232.5+432722 &    1.40 &                     &                     &    21423257+4327228  &    0.2 &    0.99 &   11.48 &      N2U5002936 &     0.5 &    0.99 &   13.97 &       1334-0441661 &     0.6 &    0.98 &   13.72 &       Star &   K3Ve  $^{    sp }$   \\
 J214309.4+432335 &    1.83 &                     &                     &    21430939+4323345  &    1.3 &    0.98 &   10.75 &      N2TU000683 &     1.1 &    0.99 &   12.98 &       1333-0464948 &     6.1 &    0.98 &   17.70 &            &         $^{       }$   \\
 J214314.4+432352 &    2.14 &                     &                     &                      &        &         &         &      N2TU117668 &     6.2 &    0.44 &   18.14 &       1333-0465038 &     6.0 &    0.44 &   18.48 &            &         $^{       }$   \\
 J214316.6+433801 &    2.61 &                     &                     &    21431647+4337568  &    5.2 &    0.10 &   17.02 &      N2U5014287 &     5.2 &    0.06 &   18.45 &       1336-0438409 &     5.3 &    0.07 &   18.50 &            &         $^{       }$   \\
 J214319.9+433635 &    1.77 &                     &                     &    21432012+4336388  &    4.2 &    0.93 &    9.92 &      N2U5000783 &     4.2 &    0.96 &   11.70 &       1336-0438459 &     4.2 &    0.97 &   11.57 &       Star &    F7V  $^{    sp }$   \\
 J214320.0+433435 &    1.22 &                     &                     &    21432016+4334346  &    1.2 &    0.99 &   10.80 &                 &         &         &         &       1335-0436830 &     2.3 &    0.86 &   16.38 &       EG   &    Gal  $^{     s }$   \\
 J214326.9+433320 &    1.61 &                     &                     &    21432700+4333182  &    2.5 &    0.63 &   11.64 &      N2U5000808 &     1.4 &    1.00 &   11.70 &       1335-0436934 &     3.6 &    0.81 &    9.87 &       EG   &    Gal  $^{ image }$   \\
 J214344.0+433506 &    1.76 &                     &                     &    21434411+4335064  &    0.4 &    0.97 &   11.92 &      N2U5011105 &     0.4 &    0.96 &   15.00 &       1335-0437233 &     0.5 &    0.94 &   14.72 &       Star &   K4Ve  $^{    sp }$   \\
 J214345.3+432744 &    2.26 &                     &                     &    21434545+4327485  &    3.9 &    0.22 &   15.80 &      N2U5003467 &     5.7 &    0.15 &   18.66 &       1334-0442898 &     5.8 &    0.12 &   18.83 &            &         $^{       }$   \\
\noalign{\smallskip}\hline
\end{tabular}
\end{center}
\end{table}
\end{landscape}
\normalsize

\begin{landscape}
\begin{table}
\begin{center}
\caption[Xray parameters for field psrj2043]{
X-ray parameters for detected sources in field PSRJ2043+2740}
\label{t:xray-param}
\fontsize{7}{7}\selectfont
\begin{tabular}{lrrrcrrrcrrrcrrrcc} 
\hline\hline 
\noalign{\smallskip}
 2XMM  & r$_{90}$ & pn\_B1$^{*}$ & pn\_B2$^{**}$ & 2MASS & d$_{x-o}$ & P$_{id}$ & kmag & GSC  & d$_{x-o}$ & P$_{id}$ & V & USNO & d$_{x-o}$ & P$_{id}$ & R & Class$^\dag$ & SpT$^\ddag$ \\
 Name   & [\arcsec] & [cts ks$^{-1}$] & [cts ks$^{-1}$] & Name & [\arcsec]  &          &      & Name & [\arcsec]   &          &   & Name & [\arcsec]   &          &   &       &       \\
\noalign{\smallskip}
\hline\noalign{\smallskip}
\noalign{\smallskip}
 J204244.6+274528 &    1.09 &                     &                     &    20424479+2745290  &    2.0 &    1.00 &    6.97 &      N333000391 &     2.1 &    0.92 &   10.48 &       1177-0649471 &     2.1 &    1.00 &    9.65 &       Star &    K5V  $^{     s }$   \\
 J204245.9+273738 &    2.58 &                     &                     &                      &        &         &         &      N333056243 &     3.9 &    0.20 &         &       1176-0636927 &     3.4 &    0.18 &   19.41 &            &         $^{       }$   \\
 J204255.9+273628 &    2.52 &                     &                     &                      &        &         &         &      N333054632 &     7.6 &    0.00 &         &       1176-0637156 &     7.6 &    0.00 &   18.49 &            &         $^{       }$   \\
 J204258.2+274351 &    1.50 &                     &                     &    20425847+2743514  &    2.9 &    0.76 &   13.21 &      N333064028 &     3.0 &    0.84 &         &       1177-0649784 &     2.8 &    0.89 &   13.69 &         EG &    Gal  $^{     s }$   \\
 J204258.9+273622 &    1.80 &                     &                     &    20425922+2736228  &    3.0 &    0.84 &   12.39 &      N333054712 &     3.3 &    0.59 &         &       1176-0637225 &     3.2 &    0.63 &   16.35 &       Star &   M5Ve  $^{    sp }$   \\
 J204259.5+274153 &    2.53 &                     &                     &    20430004+2741479  &    8.5 &    0.60 &   15.57 &      N333061372 &     5.2 &    0.48 &         &       1176-0637235 &     4.0 &    0.43 &   16.44 &            &         $^{       }$   \\
 J204305.5+274004 &    2.51 &                     &                     &                      &        &         &         &      N333059288 &     0.8 &    0.32 &         &       1176-0637376 &     0.8 &    0.34 &   19.34 &            &         $^{       }$   \\
 J204307.5+274405 &    2.72 &                     &                     &                      &        &         &         &      N333108742 &     3.9 &    0.04 &         &       1177-0650007 &     7.2 &    0.10 &   19.30 &            &         $^{       }$   \\
 J204308.7+273225 &    1.82 &                     &                     &    20430910+2732287  &    6.2 &    0.09 &   15.33 &      N333049680 &     6.3 &    0.10 &         &       1175-0619288 &     4.7 &    0.05 &   16.75 &            &         $^{       }$   \\
 J204322.5+274506 &    1.84 &                     &                     &                      &        &         &         &      N333108998 &     6.2 &    0.10 &         &       1177-0650344 &     3.0 &    0.13 &   19.32 &            &         $^{       }$   \\
 J204325.8+273331 &    2.19 &                     &                     &    20432529+2733316  &    7.2 &    0.00 &   15.79 &                 &         &         &         &       1175-0619714 &     7.0 &    0.00 &   19.74 &            &         $^{       }$   \\
 J204325.8+273721 &    2.17 &                     &                     &                      &        &         &         &      N333056066 &     1.7 &    0.31 &         &       1176-0637852 &     1.6 &    0.30 &   19.45 &            &         $^{       }$   \\
 J204326.4+272853 &    1.85 &                     &                     &                      &        &         &         &      N333044801 &     5.8 &    0.00 &         &       1174-0630822 &     5.7 &    0.00 &   19.09 &            &         $^{       }$   \\
 J204326.5+274407 &    2.05 &                     &                     &    20432685+2744088  &    3.9 &    0.95 &   10.19 &      N333141226 &     4.0 &    0.60 &         &       1177-0650455 &     4.0 &    0.98 &   11.06 &       Star &    G0V  $^{     s }$   \\
 J204326.7+274444 &    1.20 &                     &                     &    20432691+2744459  &    2.3 &    0.89 &   12.48 &      N333065021 &     3.0 &    0.91 &         &       1177-0650459 &     3.3 &    0.69 &   13.83 &       Star &    F6V  $^{    sp }$   \\
 J204327.0+274307 &    1.81 &                     &                     &                      &        &         &         &      N333062974 &     2.8 &    0.19 &         &       1177-0650468 &     2.8 &    0.20 &   19.35 &            &         $^{       }$   \\
 J204328.2+274141 &    1.95 &                     &                     &    20432821+2741381  &    3.5 &    0.31 &   15.71 &      N333061271 &     4.6 &    0.17 &         &       1176-0637893 &     4.3 &    0.23 &   16.91 &            &         $^{       }$   \\
 J204330.4+273215 &    2.18 &                     &                     &    20432998+2732182  &    7.3 &    0.03 &   13.06 &      N333049631 &     7.3 &    0.02 &         &       1175-0619821 &     7.5 &    0.01 &   14.78 &            &         $^{       }$   \\
 J204331.0+274200 &    1.92 &                     &                     &                      &        &         &         &      N333061782 &     5.2 &    0.00 &         &       1177-0650537 &     5.3 &    0.00 &   19.68 &            &         $^{       }$   \\
 J204336.4+273054 &    1.69 &                     &                     &    20433644+2730541  &    0.1 &    0.68 &   15.45 &      N333047716 &     0.5 &    0.69 &         &       1175-0619971 &     3.9 &    0.63 &   16.83 &            &         $^{       }$   \\
 J204337.9+273700 &    1.65 &                     &                     &    20433812+2737006  &    2.1 &    0.92 &   12.82 &      N333055746 &     2.2 &    0.94 &         &       1176-0638121 &     6.4 &    0.87 &   19.04 &       Star &    G3V  $^{    sp }$   \\
 J204338.1+273423 &    1.28 &                     &                     &    20433820+2734233  &    0.7 &    0.77 &   16.13 &      N333105950 &     0.6 &    0.61 &         &       1175-0620009 &     0.6 &    0.56 &   19.53 &            &         $^{       }$   \\
 J204340.5+273513 &    1.93 &                     &                     &                      &        &         &         &      N333106236 &     5.8 &    0.15 &         &       1175-0620059 &     3.7 &    0.24 &   17.68 &            &         $^{       }$   \\
 J204341.7+273133 &    1.97 &                     &                     &                      &        &         &         &      N333139456 &     2.5 &    0.27 &         &       1175-0620085 &     3.9 &    0.37 &   18.74 &            &         $^{       }$   \\
 J204342.7+275205 &    1.04 &                     &                     &    20434291+2752052  &    2.4 &    0.98 &    9.17 &      N332000188 &     2.6 &    0.97 &   12.26 &       1178-0653933 &     2.6 &    0.94 &   11.51 &       Star &   K4Ve  $^{    sp }$   \\
 J204346.4+273334 &    2.16 &                     &                     &                      &        &         &         &      N333051568 &     2.9 &    0.31 &         &       1175-0620190 &     3.0 &    0.40 &   17.58 &            &         $^{       }$   \\
 J204346.9+273424 &    2.33 &                     &                     &    20434708+2734175  &    6.7 &    0.01 &   15.47 &      N333052707 &     6.6 &    0.32 &         &       1175-0620198 &     6.4 &    0.00 &   18.23 &            &         $^{       }$   \\
 J204347.2+273515 &    1.04 &                     &                     &    20434731+2735154  &    1.2 &    0.80 &   15.06 &      N333053493 &     1.4 &    0.53 &         &       1175-0620205 &     1.8 &    0.48 &   19.08 &            &         $^{       }$   \\
 J204349.1+274403 &    1.80 &                     &                     &    20434938+2744024  &    3.0 &    0.76 &   13.13 &      N332013378 &     3.2 &    0.85 &         &       1177-0650972 &     3.7 &    0.79 &   18.41 &       Star &    F6V  $^{    sp }$   \\
 J204351.0+272940 &    2.51 &                     &                     &    20435100+2729348  &    6.0 &    0.03 &   15.49 &      N333046015 &     5.9 &    0.04 &         &       1174-0631348 &     5.7 &    0.06 &   16.96 &            &         $^{       }$   \\
 J204353.7+273229 &    2.26 &                     &                     &                      &        &         &         &      N333049933 &     4.5 &    0.09 &         &       1175-0620369 &     4.1 &    0.12 &   17.70 &            &         $^{       }$   \\
 J204359.5+274227 &    1.86 &                     &                     &    20435975+2742248  &    3.3 &    0.58 &   14.18 &      N332012428 &     3.2 &    0.21 &         &       1177-0651195 &     3.4 &    0.23 &   18.39 &            &         $^{       }$   \\
 J204400.5+273243 &    1.88 &                     &                     &    20440093+2732440  &    5.3 &    0.01 &   15.13 &      N333050325 &     5.6 &    0.01 &         &       1175-0620525 &     5.5 &    0.01 &   17.31 &            &         $^{       }$   \\
 J204402.5+272911 &    1.75 &                     &                     &                      &        &         &         &      N333113398 &     2.3 &    0.41 &         &       1174-0631595 &     2.7 &    0.17 &   19.52 &            &         $^{       }$   \\
 J204403.7+273939 &    2.54 &                     &                     &    20440366+2739444  &    5.5 &    0.14 &   15.04 &      N332010986 &     5.5 &    0.01 &         &       1176-0638711 &     6.9 &    0.01 &    0.00 &            &         $^{       }$   \\
 J204408.1+274629 &    3.19 &                     &                     &                      &        &         &         &      N332014982 &     9.0 &    0.00 &         &       1177-0651349 &     9.2 &    0.00 &   17.87 &            &         $^{       }$   \\
 J204410.4+274850 &    2.08 &                     &                     &                      &        &         &         &      N332016421 &     6.1 &    0.00 &         &       1178-0654565 &     5.9 &    0.01 &   18.26 &            &         $^{       }$   \\
 J204416.5+274735 &    1.56 &                     &                     &    20441669+2747361  &    1.9 &    1.00 &    8.96 &      N332015927 &     2.0 &    0.55 &         &       1177-0651549 &     1.9 &    1.00 &    9.78 &       Star &    F2V  $^{     s }$   \\
 J204417.1+273613 &    1.35 &                     &                     &    20441726+2736119  &    2.6 &    0.93 &   11.65 &      N332000374 &     4.2 &    0.90 &         &       1176-0639040 &     3.9 &    0.67 &    0.00 &       Star &    G1V  $^{    sp }$   \\
 J204425.0+274236 &    2.32 &                     &                     &                      &        &         &         &      N332012562 &     8.0 &    0.00 &         &       1177-0651762 &     7.9 &    0.00 &   18.16 &            &         $^{       }$   \\
 J204428.2+273334 &    1.91 &                     &                     &                      &        &         &         &      N332008019 &     2.7 &    0.34 &         &       1175-0621138 &     5.9 &    0.34 &   19.18 &            &         $^{       }$   \\
 J204440.2+274231 &    3.07 &                     &                     &    20444049+2742309  &    3.8 &    0.86 &   12.28 &      N332084691 &     6.7 &    0.53 &         &       1177-0652143 &     3.8 &    0.55 &   16.19 &       Star &   M2Ve  $^{    sp }$   \\
\noalign{\smallskip}\hline
\end{tabular}
\end{center}
\end{table}
\end{landscape}
\normalsize

\end{document}